\documentstyle[preprint,aps,eqsecnum,epsf,rotate]{revtex}
\def\faka{\left[{k\cos\theta\over q}-{1\over 2}\right]^2}
\def\fakb{\left[{k\cos\theta\over q}-{1\over2}+{g\over 2q^2}\right]^2}
\def\inq#1{\int \limits_{#1}}

\newcommand{\vS}[0]{{\bf S}}
\newcommand{\bfnabla}[0]{\mbox{\boldmath $\nabla$}}

\def\gvect#1{\mbox{\boldmath $#1$}}
\begin{document}

\setcounter{page}{1}

%\begin{titlepage}
%%%%%%%%%%%%%%%%%%%%%%%%%%%%%%%%%%%%%%%%%%%%%%%%%%%%%%%%%%%%%%%%%%%%%%%%%%%%%%%%

\title{\huge Critical Dynamics of Magnets}

\author{
 By {\sc E. Frey} and {\sc F. Schwabl} \\
Technische Universit\"at M\"unchen, \\
Institut f\"ur Theoretische Physik, \\
D-85747 Garching, Germany}
\vfill
\date{\it \today}
\maketitle

\begin{abstract}
We review our current understanding of the critical dynamics of magnets above 
and below the transition temperature with focus on the effects due to the 
dipole--dipole interaction present in all real magnets. Significant progress in 
our understanding of real ferromagnets in the vicinity of the critical point 
has been made in the last decade through improved experimental techniques and 
theoretical advances in taking into account realistic spin-spin interactions. 
We start our review with a discussion of the theoretical results for the 
critical dynamics based on recent renormalization group, mode coupling and
spin wave theories. A detailed comparison is made of the theory with 
experimental results obtained by different measuring techniques, such as 
neutron scattering, hyperfine interaction, muon--spin--resonance, 
electron--spin--resonance, and magnetic relaxation, in various materials.
Furthermore we discuss the effects of dipolar interaction on the critical 
dynamics of three--dimensional isotropic antiferromagnets and uniaxial 
ferromagnets. Special attention is also paid to a discussion of the 
consequences of dipolar anisotropies on the existence of magnetic order and 
the spin--wave spectrum in two--dimensional ferromagnets and antiferromagnets.
We close our review with a formulation of critical dynamics in terms of 
nonlinear Langevin equations.
\end{abstract}
%\end{titlepage}

\tableofcontents
\markboth{Contents}{}

%%%%%%%%%%%%%%%%%%%%%%%%%%%%%%%%%%%%%%%%%%%%%%%%%%%%%%%%%%%%%%%%%%%%%%%%%%%%%%%%
\section {INTRODUCTION}

Remarkable progress has been achieved in our qualitative and quantitative 
understanding of the critical dynamics of magnetic materials. This is partly
due to the advances and new developments in experimental techniques such as
neutron scattering, electron spin resonance and hyperfine interaction probes.
Simultaneously, on the theoretical side, important developments were provided 
by dynamical scaling theory, mode coupling theory and the renormalization 
group theory. The theoretical progress has been mainly promoted by
including effects of magnetic interactions, such as dipole--dipole and 
spin--orbit interaction, on top of the exchange interaction. 
These interactions, which are present in all real magnetic materials, were 
found to change the critical dynamics quite drastically. Around 1986 the 
experimental situation was quite puzzling showing both quite excellent
agreement but also large discrepancies with the theories existing at that time. 
Motivated by this seemingly contradictory situation anewed theoretical interest 
on the subject of critical dynamics was promoted.

In this review we shall mostly be interested in the effects of the
dipole--dipole interaction on the critical dynamics of isotropic 
ferromagnetic materials in three dimensions. The static properties
are reviewed only in as much as they are needed for the dynamics or if
they emerge naturally together with the dynamical results.
We will also discuss other 
dipolar systems, such as dipolar antiferromagnets and uniaxial ferromagnets
in three dimensions, as well as the consequences of dipolar interactions in
two--dimensional magnets. We do not consider pure dipolar systems such as the
nuclear magnets $Cu$, $Ag$, and $Au$.
We aim at giving an up-to-date account of the
situation in the field, where we try to give attention to the theoretical as
well as the experimental progress. In order to make the discussion as
self-contained as possible we give brief discussions of the various
theoretical concepts where needed. 

Ferromagnets have always played a special role in the field of critical 
phenomena. Several simplified models, such as the Ising~\cite{Ising25}, 
Heisenberg~\cite{Heisenberg28}, and Hubbard~\cite{Hubbard79} model have been
developed in order to understand the fundamental questions of the
statistical behavior of magnetically ordered systems. 
The attraction of these models resides in the simplicity of their form and
complexity of behavior they are capable to describe. Over the last
two decades various approaches to study the above models have been
intensively developed, which lead to a profound understanding of the 
statistical mechanics of magnetic systems on the basis of the above
simplified models. However, one has to keep in mind that for a more
realistic description of magnetic materials it is not sufficient to consider
solely the electrostatic interaction between the electrons in the partly
filled shells leading in conjunction with the Pauli principle to the exchange
interaction. There are magnetic interactions, such as the mutual
interaction of the electron spins (dipole--dipole interaction) and the
interaction of the spin and orbital moments of electrons (spin-orbit
interaction) which may lead to magneto-crystalline anisotropies.

The exchange interaction is responsible for 
the phenomenon of magnetic ordering itself (at least above two dimensions). 
The characteristic feature of the exchange forces is their isotropic 
and short--range nature. They do not impose any definite orientation
of the magnetic moments with respect to the crystallographic axis. 
Magneto-crystalline anisotropies are due to relativistic or magnetic forces 
(spin--spin dipole, quadrupole, etc.; spin--orbital, orbital--orbital).
Microscopic models of various magnetic interactions have shown that, in the
majority of cases, the spin--orbital interaction is the basic one responsible 
for the magneto-crystalline anisotropy (see e.g. Ref.~\cite{Turov65,Morrish65}). 
It relates the directions of the spin
magnetic moments of atoms through their orbital states with the
crystallographic axes. Irrespective of the microscopic nature of the
magnetic anisotropy forces, their macroscopic manifestation in a crystal
seems to be determined mainly by the type of symmetry of the lattice. 
A special role among the magnetic forces is played by the dipole--dipole
interaction between the magnetic moments of the electrons, because it is of 
long range and in cubic crystals leads to an anisotropy with respect to the 
wave vector but to leading order not to the crystallographic axes.

A Hamiltonian which takes into account exchange interaction as well as
dipole--dipole interaction between the magnetic moments of the electrons
has first been given and studied by Holstein and Primakoff~\cite{Holstein40}.
The dipolar interaction is usually two to three orders of magnitude weaker 
than the exchange interaction, and it can therefore (in many cases) be 
considered to be a small perturbation. However, because of its long--range 
nature, the effects of the dipole--dipole interaction become significant at 
least very close to the critical temperature $T_c$ and for small wave vectors
${\bf q}$, as will become clear in this review. Note, that in particular it 
leads to the 
appearance of the demagnetization factors. Using renormalization
group theory it has been shown by Fisher and Aharony~\cite{f73,af73,a73} that the 
dipolar interaction is a relevant perturbation with respect to the Heisenberg model.
They find that the short--range Heisenberg 
fixed point of the renormalization group is unstable against perturbations 
resulting from the dipolar interaction, and the asymptotic critical behavior
is characterized by a new dipolar critical fixed point. Furthermore, the 
long--range nature of the dipolar interaction reflects itself in the 
Fourier transform containing a contribution of the singular direction dependent
form $q^\alpha q^\beta / q^2$. An immediate consequence is the fact that 
longitudinal fluctuations are reduced in comparison to the two transverse ones
and do not have a divergent susceptibility any more at $T_c$. Here, by 
longitudinal and transverse we refer to the direction of the wave vector ${\bf q}$.
A consequence of this reduction of the number of effective order parameter components
is a change in the static critical indices. Hence in a system with dominating 
exchange interaction there is a crossover from Heisenberg critical behavior to 
dipolar critical behavior. For instance the effective temperature dependent critical exponent
of the static susceptibility $\gamma$ approaches the Heisenberg value then goes
through a minimum until finally it ends up at the dipolar limiting value.

In general the long--range dipolar interaction is of importance whenever 
fluctuations get large. This is the case in the vicinity of critical points 
and in systems of reduced dimensionality. In the vicinity of critical points 
longitudinal fluctuations are suppressed and rotational invariance is 
destroyed. This leads to modified static critical behavior and to drastic 
changes in the dynamics. In systems of reduced dimensionality which on the 
basis of short--range interactions would not have a phase transition at a 
finite temperature due to the large fluctuations destroying the order 
parameter, the dipolar interaction suppresses these fluctuations and thereby 
allows a finite order parameter. With the detection of high--$T_c$ 
superconductors and their fascinating magnetic properties, the study of 
mechanisms which lead to phase transitions in such quasi--two--dimensional 
systems (inter--plane interaction, anisotropy, dipolar interaction, etc.) is of 
prime importance.

In this paper we review the critical dynamics of magnetic systems.
As a reference and a simpler situation to start with we treat
in chapter 2 first isotropic ferromagnets without dipolar 
interaction. This allows us to introduce the main theoretical 
concepts such as dynamic scaling, mode coupling and dynamic 
renormalization group theory in a quite elementary and hopefully 
pedagogical way.

In chapter 3 we describe the theoretical results on the dynamics of
dipolar ferromagnets with an emphasis on the mode coupling theory
for the paramagnetic phase. A detailed analysis of the
consequences of the dipolar interaction on the functional form of 
the dynamic scaling laws, critical exponents and the line shape and 
line width crossover will be given. For the ferromagnetic phase we 
give some results based on spin--wave theory and we also
comment on some recent theoretical approaches, which go beyond the 
linearized spin--wave theory.

In chapter 4 the theoretical results are then compared with the 
findings from a variety of experimental techniques. They include 
neutron scattering, electron spin resonance and magnetic relaxation,
hyperfine techniques and muon spin resonance experiments.

Whereas the main part of the review concentrates on dipolar effects in
isotropic ferromagnets, chapter 5 concerns  dipolar effects in other 
magnetic systems. These include three--dimensional isotropic
antiferromagnets, bulk uniaxial ferromagnets  and two--dimensional systems.
In the latter case the dipolar interaction leads to long--range order, which
would not be possible for the isotropic Heisenberg model in two dimensions.

Finally, in chapter 6, we discuss alternative derivations of the mode coupling
theory via the generalized Langevin equations of Zwanzig and Mori. These 
methods allow for a systematic derivation of the mode coupling theory in the 
framework of a diagrammatic analysis. Also, some systematic improvements are
possible. We conclude with a summary and an outlook in chapter 7.
Some technical details and important conceptual background material
is collected in the appendices.

%%%%%%%%%%%%%%%%%%%%%%%%%%%%%%%%%%%%%%%%%%%%%%%%%%%%%%%%%%%%%%%%%%%%%%%%%%%%%%%%
\section{ISOTROPIC FERROMAGNETS}
\label{s2}

Our main concern in this review is the immediate vicinity of the critical 
point and the influence of the dipolar interaction on the critical dynamics. 
As a reference and a simpler situation to start with we treat first isotropic
ferromagnets without dipolar interaction. This allows us to introduce the 
basic concepts and results of the different theoretical approaches and 
compare them with the experimental situation.

\subsection{Dynamical Scaling and Hydrodynamics}
\label{s2.1}

In systems where a continuous symmetry is broken, hydrodynamics together with
dynamical scaling allows one to obtain definite conclusions about the 
dynamic critical behavior.

To start with, we remind the reader of the structure of the hydrodynamic 
modes in isotropic ferromagnets. In an isotropic ferromagnet magnetization is 
conserved giving rise to three hydrodynamic equations for the magnetization 
vector ${\bf M} ({\bf x})$. In the paramagnetic phase, i.e. for temperatures 
above the Curie temperature $T_c$ and in zero magnetic field ${\bf H}$, 
($T>T_c, \, {\bf H}=0$) the magnetization obeys a diffusion equation,
\begin{equation}
  {\partial {{\bf M}} \over \partial t} = D \bfnabla^2 {\bf M} \, ,
\label{2.1}
\end{equation}
with a diffusion constant $D$. Hence the long wave length excitations 
are diffusive
\begin{equation}
  \omega (q)  = -i D q^2 \, .
\label{2.2}
\end{equation}

In the ferromagnetic phase the magnetization $\bf M$ is finite and the 
spin fluctuations which are perpendicular to the mean magnetization 
obey spin--wave equations of motion. The spin--wave frequency is according to 
hydrodynamics given by
\begin{equation}
  \omega (q) = {M \over \chi^T (q)} - i q^4 \Lambda \, .
\label{2.3}
\end{equation}
Here $\chi^T (q)$ is the transverse susceptibility, $M$ the magnetization,
and $\Lambda$ a damping 
constant. The real part of the frequency is related to static critical 
quantities. For the imaginary part, the damping, hydrodynamics predicts a 
decay rate proportional to the fourth power of the wave 
number~\footnote{Indeed the microscopic theory of the Heisenberg model 
predicts a spin-wave decay rate of the form $q^4 \left( c_2 (\ln q)^2 + 
c_1 \ln q + c_0 \right)$~\cite{Dyson56,kak61,harris68,vlp68}.}.

{\it Dynamical scaling} states that the critical frequency is of the 
homogeneous form
\begin{equation}
  \omega (q,\xi) = q^z \Omega (q \xi) \, ,
\label{2.4}       
\end{equation}
where $z$ is the dynamic critical exponent and $\xi \propto |T-T_c|^{-\nu}$
the correlation length with the static critical exponent $\nu$.
Using the scaling behavior of the static quantities~\cite{Ma76} 
in the hydrodynamic region 
($ M \sim \xi^{-\beta/\nu}$, and $\chi^T (q) \sim q^{-2} \xi^{-\eta}$) and the 
scaling relations between the static exponents 
($\gamma = \nu (2 - \eta)$, $2 \beta = (d-2+\eta)\nu$) one finds 
for the spin--wave frequency in the hydrodynamic region
\begin{equation}
  {\rm{Re}} \, \left\{ \omega (q,\xi) \right\} = M/\chi^T (q)
  \propto q^2 \xi^{(2-d+\eta)/2}
  = q^{(d+2-\eta)/2} (q \xi)^{(2-d+\eta)/2} \, .
\label{2.5}
\end{equation}
Here $d$ is the spatial dimension of the system. One thus finds as a result 
of the dynamic scaling law, Eq.~(\ref{2.4}) and the hydrodynamic behavior of 
the static quantities an exact relation for the dynamic exponent
\begin{equation}
  z = {d + 2 - \eta \over 2} \, .
\label{2.6}
\end{equation}
Using again Eqs.~(\ref{2.5}) and (\ref{2.6}) the damping coefficient 
$\Lambda$ of the spin--waves and the spin diffusion coefficient $D$ in the 
paramagnetic phase are
\begin{equation}
  \Lambda \sim \xi^{(6-d+\eta)/2}, \quad \hbox{and} \quad
  D \sim \xi^{(2-d+\eta)/2} \, .
\label{2.7}
\end{equation}
The spin diffusion coefficient $D$ goes to zero at the critical 
temperature, a phenomenon known as {\it critical slowing down}.
Thus hydrodynamics and dynamical scaling allow one to
determine the main critical dependencies of the transport
coefficients. The picture emerging from Eqs.~(\ref{2.2})--(\ref{2.7}) is
summarized in Fig.~2.1.

The critical frequency is a function of the wave number and the
inverse correlation length. The hydrodynamic region is given by
$q \ll \xi^{-1}$, whereas the non hydrodynamic critical region by
$q \gg \xi^{-1}$. Of course the critical region is limited to $q
\ll a^{-1}$ and $\xi \gg a$, where $a$ is a microscopic
length scale (e.g. the lattice spacing). 

We close this subsection by giving the hydrodynamic equations of the low
temperature phase~\cite{s70,hh77}
\begin{eqnarray}
  {d \over dt} M_{\bf q}^x &&= {M \over \chi^T (q)} M_{\bf q}^y  - 
                               \Lambda q^4 M_{\bf q}^x \, ,
  \nonumber \\ 
  {d \over dt} M_{\bf q}^y &&= -{M \over \chi^T (q)} M_{\bf q}^x  - 
                                \Lambda q^4 M_{\bf q}^y  \, ,
  \nonumber \\
  {d \over dt} M_{\bf q}^z &&= -\Gamma (q) M_{\bf q}^z \, ,
  \nonumber
\end{eqnarray}
the excitations of which were the basis of our discussion. Here we assumed 
that the magnetization is oriented along the $z$--direction. In addition to 
the transverse equation the magnetization component along the order parameter
obeys a diffusion equation in three dimensions $\Gamma(q) = D_M q^2$.

\subsection{Mode Coupling Theory}
\label{s2.2}

One of the most successful theoretical approaches in critical dynamics 
are mode coupling theories. The first such theory was proposed by Fixman
\cite{f60}, and then put on a more rigorous basis by Kadanoff and
Swift~\cite{ks68}, and especially Kawasaki~\cite{kk67,kk70,kk76}.
In this section we exemplify the mode coupling theory for the critical 
dynamics of an isotropic ferromagnet, where the spins are coupled 
only by short range isotropic exchange interaction. The Hamiltonian 
for such a spin system is given by
\begin{equation}
  H = \int_{\bf q}   J({\bf q}) \, \vS_{\bf q} \cdot \vS_{- {\bf q}} \, ,
\label{2.8}
\end{equation}
Here we have introduced the notation $\int_{\bf q} = v_a \int 
{d^3 q \over (2 \pi)^3}$, where $v_a$ is the volume of the
primitive cell of the Bravais lattice. Upon introducing the cube edge
length $a$ of the corresponding cubic unit cell, the dimensionless
quantity $b = \sqrt{a^3/v_a}$ characterizes the lattice structure. 
The Fourier transforms of the Cartesian components $S^i({\bf x})$ of the 
spin operator are defined by
\begin{equation}
  S^i_{\bf q} = \int d^3x e^{i {\bf q} \cdot {\bf x}} S^i({\bf x}) \, ,
\label{2.9}
\end{equation}                                                               
and $J({\bf q}) = - J_0 + J q^2 a^2$ characterizes the exchange 
interaction. We are retaining only terms up to second order
in the wave vector ${\bf q}$ and have supposed that the exchange interaction
extends up to the second nearest neighbors. For $bcc$ and $fcc$
lattices we have $J = J_1 + J_2$, where $J_1$ and $J_2$ are, 
respectively, the values of the exchange parameters between the nearest and 
next-nearest neighbors. For a simple cubic ($sc$) crystal the relation
is $J = J_1 + 4 J_2$. The parameter $J_0$ does not enter the equations
of motion.
It is convenient to introduce the ladder operators
\begin{equation}
  S^{\pm}_{\bf q} = S^x_{\bf q} \pm iS^y_{\bf q} \, ,
\label{2.10}
\end{equation}                                                               
and $S^z_{\bf q}$ instead of the Cartesian components
of the spin operator.
Then, using the commutation relation for spin operators one finds
for the Hamiltonian (\ref{2.8})
the following set of equations of motion (we take $\hbar =1$)
\begin{equation}
  {d \over dt}S^z_{\bf q} =
  - i \int_{\bf k} 
  \left[ J({\bf k}) - J({\bf q} - {\bf k}) \right]
  S^+_{{\bf q} - {\bf k}} S^-_{\bf k} \, ,
\label{2.11}
\end{equation}                                                               
\begin{equation}
  {d \over dt} S^{\pm}_{\bf q} =
  \pm 2 i  \int_{\bf k} 
  \left[ J({\bf k}) - J({\bf q} - {\bf k}) \right]
  S^{\pm}_{{\bf q} - {\bf k}} S^z_{\bf k} \, .
\label{2.12}
\end{equation}
The equations of motion (\ref{2.11}) and (\ref{2.12}) exhibit explicitly the 
vanishing of $d S^i_{\bf q} / dt$ at ${\bf q}=0$, i.e., the order parameter
itself is a constant of motion. Starting from these microscopic
equations of motion there are a variety of 
different schemes for deriving the mode coupling equations for the spin 
correlation functions~\cite{bm65,w68,rl67,rl69,kk67,h71a,h71b}. 

The quantity of interest is the Kubo
relaxation matrix for a set of dynamical variables $\lbrace
X^{\alpha}({\bf x}) \rbrace$, which is defined by
\begin{equation}
  \Phi^{\alpha \beta}({\bf q},t) =
  i \lim_{\epsilon \to 0} \int \limits_{t}^{\infty} d\tau
  e^{-\epsilon \tau} 
  \langle \lbrack X^{\alpha}({\bf q},\tau),
  X^{\beta}({\bf q},0)^{\dagger} \rbrack \rangle \, ,
\label{2.13}
\end{equation}
where $\langle ... \rangle$ denotes the thermal average. The dynamical 
variables are normalized such that
\begin{equation}
  (X^{\alpha},X^{\beta}) := \Phi^{\alpha \beta}({\bf q},t=0) =
  \delta^{\alpha \beta} \,  ,
\label{2.14}
\end{equation}
i.e., we use normalized spin variables $X^\alpha_{\bf q} (t) =
S^\alpha_{\bf q} (t) / \sqrt{\chi^\alpha ({\bf q})}$, where 
$\chi^\alpha ({\bf q})$ are the static susceptibilities.

One of the most concise ways of deducing mode coupling equations utilizes 
the projection operator technique, originally introduced by Mori~\cite{m65} 
and Zwanzig~\cite{z61}. The main idea is that one can separate the set of
dynamical variables into two classes, one slowly and one fast varying
class. With the aid of this projection operator method the fast
variables are eliminated and one can derive
generalized Langevin equations for the dynamic variables or
equivalently for the correlation functions~\cite{mfs74,mf73,kk73} 
(see also chapter~\ref{s6}, and appendix~\ref{app.b}).
The corresponding equations for the Kubo relaxation function is
\begin{equation}
  {\partial \Phi^{\alpha \beta} ({\bf q},t) \over \partial t} =
  i \omega^{\alpha \gamma} ({\bf q}) \Phi^{\gamma \beta} ({\bf q},t) -
  \int_0^t d \tau \Gamma^{\alpha \gamma} ({\bf q},t-\tau) 
                  \Phi^{\gamma \beta}     ({\bf q},\tau) \, ,
\label{2.15}
\end{equation}
and for its half-sided Fourier transform 
\begin{equation}
  \Phi^{\alpha \beta}({\bf q},\omega) =
  \int \limits_{0}^{\infty} dt \, \Phi^{\alpha \beta}({\bf q},t)
  e^{i \omega t} \, ,
\label{2.16}
\end{equation}   
one obtains
\begin{equation}
  \gvect{\Phi} ({\bf q},\omega) = i
  \left[ 
  \omega  {\bf 1} + \gvect{\omega}  ({\bf q}) + 
  i \gvect{\Gamma} ({\bf q},\omega) 
  \right]^{-1} \, . 
\label{2.17}
\end{equation}
The frequency matrix $\omega^{\alpha \beta}({\bf q})$ is
given by
\begin{equation}
  i \omega^{\alpha \beta}({\bf q}) =
  ({\dot X}^\alpha_{\bf q}, X^\beta_{\bf q}) =
  - i \langle \lbrack X^\alpha_{\bf q},
  X^\beta_{-\bf q} \rbrack \rangle \, ,
\label{2.18}
\end{equation}
where we have used the Kubo identity 
$({\dot A},B) = -i \langle [A,B^\dagger] \rangle$~\cite{ku57}.
The non-linear aspects of the spin dynamics are contained in
the matrix of the transport coefficients (memory matrix) $\gvect{\Gamma}$. 
As a result of the projection operator technique these can be 
written in terms of the Kubo relaxation matrix~\cite{kk70,kk76}
\begin{equation}
  \Gamma^{\alpha \beta}({\bf q},t) =
  (\delta {\dot X}^{\alpha}_{\bf q} (t),
  \delta {\dot X}^{\beta}_{\bf q} (t))
\label{2.19}
\end{equation}

of the non conserved parts of the currents\footnote{Here we have
chosen a linear projection operator ${\cal P} X = (X,X^\alpha)
X^\alpha$. Then the random forces can be written in terms of the
projection operator as $\delta {\dot X}_{\bf q} = \exp [i (1 - {\cal
P}) {\cal L} t] (1-{\cal P}) {\dot X}_{\bf q}$, where ${\cal L}$ is the
Liouville operator. Neglecting the projection operator in the time
development one gets Eq.~(\ref{2.20}); see also Appendix B.}
\begin{equation}
  \delta {\dot X}^{\alpha}_{\bf q} =
  {\dot X}^{\alpha}_{\bf q} - i \omega^{\alpha \beta}({\bf q})
  X^{\beta}_{\bf q} \, .
\label{2.20}
\end{equation}
The simplest approximation which can be made at this stage is to 
consider only two mode decay processes,  which in technical
terms amounts to a factorization of the Kubo formulas
(\ref{2.19}) after insertion of the equations of motion. 
Frequently one makes additionally an approximation for the line shape
(i.e. frequency dependence) of the relaxation matrix, e.g. a Lorentzian 
approximation. In principal, however, 
one can solve directly the set of self consistent equations for the 
shape functions resulting from the decoupling approximation only. For 
the isotropic ferromagnet this was first achieved by Wegner~\cite{w68} 
and Hubbard~\cite{h71a} in the  paramagnetic phase. 
For most practical purposes an 
excellent approximation for the line width can be obtained from the 
mode coupling equations simplified by the Lorentzian
approximation.

\subsubsection {Paramagnetic Phase}
\label{s2.2.1}

In the paramagnetic phase the order parameter is zero 
$<{\bf S}_{\bf q}> \mid_{{\bf q} = 0}=  0$ 
implying that the frequency matrix $\omega^{\alpha \beta}$
vanishes. Upon using the above decoupling procedure one obtains the
following set of coupled integro-differential equations for the Kubo
relaxation function
\begin{equation}
  {\partial \Phi ({\bf q},t) \over \partial t} =
  - \int_0^t d \tau \Gamma({\bf q}, t-\tau)
    \Phi({\bf q}, \tau) \, ,
\label{2.21}
\end{equation}
and the transport coefficients
\begin{equation}
  \Gamma({\bf q}, t) =
  4 k_B T 
  \int_{\bf k} \,  \upsilon({\bf k}, {\bf q}) \,  
  {\chi({\bf q} - {\bf k}) \chi({\bf k}) \over \chi({\bf q}) } \, 
  \Phi({\bf k},t) \Phi({\bf q} - {\bf k},t) \,  ,
\label{2.22}
\end{equation}
with the vertex function ($\hbar =1$)
\begin{equation}
  \upsilon({\bf k}, {\bf q}) =
  \left( J({\bf k}) - J({\bf q} - {\bf k}) \right)^2 = 
  \left(2 J q^2 a^2 \left[ {{\bf q} \cdot {\bf k} \over q^2} - 
  {1 \over 2} \right] \right)^2\, .
\label{2.23}
\end{equation}
Essentially the same equations have been derived by 
numerous authors \cite{bm65,w68,rl67,rl69}, \cite{kk67,kk76},
\cite{h71a,h71b}  using different approaches.
The temperature dependence enters the equations only implicitly via
the correlation length 
\begin{equation}
  \xi = \xi_0 \left( { T - T_c \over T_c } \right)^{- \nu} \, .
\label{2.24}
\end{equation}
The equations for the Kubo relaxation function must in general be
solved numerically. However, in the critical region one can deduce
certain important properties of the solution analytically. 
Upon inserting the static scaling law~\cite{wk74}
\begin{equation}
  \chi(q,\xi) = {1 \over 2 J a^2} \,  q^{-2+\eta} {\hat \chi}(x) \, ,
\label{2.25}
\end{equation}
with the scaling variable $x=1/q\xi$ one can show by inspection that the
solution of Eqs.~(\ref{2.21}) and (\ref{2.22}) fulfills dynamic 
scaling 
\begin{equation}
  \Phi (q,\xi,\omega) = (\Lambda q^z)^{-1} \phi (x,\nu) \, ,
\label{2.26}
\end{equation}
\begin{equation}
  \Gamma (q,\xi,\omega) = \Lambda q^z \gamma (x,\nu) \, ,
\label{2.27}
\end{equation}
with the scaling variable
$\nu = \omega / \Lambda q^z$ and the non universal constant
\begin{equation}
  \Lambda = {a^{5/2} \over b} \sqrt{2 J k_B T_c \over 4 \pi^4} \, ,
\label{2.28}
\end{equation}
where $b$ is a dimensionless parameter which depends on the crystal 
structure (see Table~\ref{tableI}). Note that in
Eqs.~(\ref{2.25})--(\ref{2.27}) 
we have explicitly incorporated the correlation length $\xi$ into the
list of arguments of the correlation functions in order to indicate 
the reduction of arguments accomplished by the dynamic scaling form. 
In order to simplify notation this temperature dependence is in most 
of the remaining text not written out explicitly.

The above mode coupling equations give a dynamic critical exponent
$z = (5+\eta)/2$ instead of the correct expression $z=(5-\eta)/2$
\cite{hh67,mm75,j76,bjw76}.
This inconsistency of the conventional derivation of 
the mode coupling equations is fortunately not a very serious problem
since $\eta$ is very small in the case of 3D ferromagnets, $\eta
\approx 0.05$~\cite{m84}. 
In order to be consistent one has to take for the scaling functions
an Ornstein-Zernike form 
\begin{equation}
 {\hat \chi} (x) = {1 \over 1 + x^2} \, ,
\label{2.29}
\end{equation}      
neglecting the exponent $\eta$. In chapter~\ref{s6} we will give a derivation 
of modified mode coupling equations based on a path integral
formulation of the stochastic equations of motion. There we will
show how the above inconsistency can be resolved by taking into
account certain kinds of vertex corrections which are neglected in
the conventional derivation of mode coupling equations.

The scaling relations (\ref{2.26}) and (\ref{2.27}) for the
Fourier transformed quantities imply for their time dependent
counterparts
\begin{equation}
  \Phi (q,\xi,t) = \phi (x,\tau) \, ,
\label{2.30}
\end{equation}
\begin{equation}
  \Gamma (q,\xi,t) = (\Lambda q^z)^2 \gamma (x,\tau) \, ,
\label{2.31}
\end{equation}
with the scaled time variable
\begin{equation}
  \tau = \Lambda q^z t \, .
\label{2.32}
\end{equation}
The mode coupling equations for the corresponding scaling functions
are
\begin{equation}
  {\partial \phi (x,\tau) \over \partial \tau} =
  - \int_0^\tau d \tau^\prime 
    \gamma(x, \tau-\tau^\prime) \phi(x, \tau^\prime) \, ,
\label{2.33}
\end{equation}
and
\begin{equation}
  \gamma(x, \tau) =
  2 \pi^2 \int_{-1}^{+1} d \eta \int_0^\infty d \rho \rho_-^2 
  {\hat \upsilon} (\rho,\eta) 
  {{\hat \chi} (x/\rho) {\hat \chi} (x/\rho_-) \over {\hat \chi}(x) }
  \phi(x/\rho,\tau \rho^z) \phi(x/\rho_-,\tau \rho_-^z) \, .
\label{2.34}
\end{equation}
The scaled vertex function reads
\begin{equation}
  {\hat \upsilon} (\rho,\eta) =
  2 \left( \rho \eta - 1/2 \right)^2 \, ,
\label{2.35}
\end{equation}
where we have defined $\rho = k/q$, $\rho_- = |{\bf k} - {\bf q}|/q$,
and $\eta = \cos({\bf k},{\bf q})$.

Before turning to the numerical solution of the mode--coupling
equations, let us quote some results which can be obtained analytically.
For temperatures not to close to $T_c$ one can infer from Eqs.~(\ref{2.22}) 
and (\ref{2.23}) that in the limit $q \rightarrow 0$ (hydrodynamic limit) the 
vertex factor $\upsilon ({\bf k},{\bf q})$ and hence the memory kernel 
$\Gamma (q,t)$ becomes small. Hence one could argue that the relaxation
function $\Phi (q,t)$ varies very slowly and the solution of Eq.~(\ref{2.21})
becomes an exponential~\cite{h71a} 
\begin{equation}
  \Phi (q,t) = \exp \left\{ - D q^2 t \right\} \, ,
\label{2.36}
\end{equation}
where the diffusion constant is given by
\begin{equation}
  D = \lim_{q \rightarrow 0} {1 \over q^2} \int_0^\infty
  \Gamma (q,t) dt = 
  \lim_{q \rightarrow 0} {1 \over q^2} \Gamma (q,\omega = 0) \, .
\label{2.37}
\end{equation}
A scaling analysis of the right hand side gives for the temperature
dependence of the diffusion coefficient $D \sim \xi^{-1/2}$
in agreement with the scaling result by Halperin and Hohenberg~\cite{hh67}.

The above argument leading to the spin diffusion behavior has been questioned 
by M\aa nson~\cite{m74}. Starting from a relaxation function which is of spin
diffusion type (see Eq.~(\ref{2.36})) M\aa nson~\cite{m74} showed that the
memory kernel becomes of the form
\begin{equation}
 \Gamma(q,t) \propto t^{-5/2} e^{- D q^2 t / 2 }
\label{2.38}
\end{equation}
for asymptotic times and small $q$. Therefrom M\aa nson~\cite{m74} concludes 
that the spin diffusion type of behavior can not be the correct form of the 
relaxation function at asymptotic times, which would raise some questions 
on the validity of the mode coupling theory (since it invalidates the 
results obtained from a hydrodynamic theory based on the conservation of 
the magnetization). The above argument leading to an exponentially decaying 
relaxation function becomes invalid also close to $T_c$ because the static 
susceptibilities in the expression for the memory matrix diverge and 
therefore the relaxation function no longer varies slowly. A shape crossover
from a Lorentzian to a different critical shape takes place by approaching 
the critical temperature~\cite{h71a}. Nevertheless, a Lorentzian 
approximation for the line shape still gives reasonable an approximation for 
the line width, since the latter is not so sensitive to the precise form of 
the line shape. 

In the Lorentzian approximation the mode coupling equations reduce to a single
integral equation for the line width $\Gamma_{\rm lor} (q) = \Lambda q^z
\gamma_{\rm lor} (x)$
\begin{equation}
  \gamma_{\rm lor} (x) =
  {2 \pi^2 \over {\hat \chi}(x) }
   \int_{-1}^{+1} d \eta \int_0^\infty d \rho \rho_-^2 
  {\hat \upsilon} (\rho,\eta) 
  {{\hat \chi} (x/\rho) {\hat \chi} (x/\rho_-) \over 
  \rho^{5/2} \gamma_{\rm lor} (x/\rho) +   
  \rho_-^{5/2} \gamma_{\rm lor} (x/\rho_-) } \, .
\label{2.39}
\end{equation}
Therefrom one can deduce the asymptotic behavior of the typical
line width analytically
\begin{equation}
  \gamma_{\rm lor} (x) \sim  \cases{
  1 &for $x << 1$ \, , \cr
  \sqrt{x} &for $x >> 1$ \, , \cr}
\label{2.40}
\end{equation}
implying $\Gamma (q) \sim q^{5/2}$ right at $T = T_c$ and
$\Gamma(q) \sim q^2 \xi^{-1/2}$ in the hydrodynamic limit
$q \xi << 1$, i.e the temperature dependence of the diffusion 
constant is given by $D \sim \xi^{-1/2}$~\cite{kk67} 
as we have already deduced from scaling arguments. The full
scaling function resulting from Eq.~(\ref{2.39}) is shown in Fig.~2.2.
It is usually called Resibois--Piette scaling function~\cite{rp70} since
Resibois and Piette did the first numerical solution of the mode
coupling equations in Lorentzian approximation.

In order to find the complete behavior of the relaxation function
$\Phi(q,t)$ one has to solve Eqs.~(\ref{2.33}) and (\ref{2.34}) 
numerically.  This was done by
Wegner~\cite{w68} at $T_c$ and extended to temperatures above $T_c$ by
Hubbard~\cite{h71a,h71b}. The results are shown in Fig.~2.3.
It is found that for small wave vectors (not too close to the
zone boundary) there is a shape crossover 
from a Lorentzian (see Eq.~(\ref{2.36})) to a more Gaussian 
like shape by approaching the critical temperature. The critical 
shape at $T_c$ is essentially the same as the one obtained from 
RG--theory~\cite{d76,bf81} (see also section~\ref{s2.3.1})

Recently, this shape crossover has been reexamined by Aberger and Folk
\cite{af88a} and Frey et al.~\cite{fst89} in detail with emphasis on constant 
energy scans. Their results, shown in Figs.~2.4.a and 2.4.b for the scaling 
function of the spin relaxation function versus time and frequency, 
respectively, confirm the shape crossover from a Lorentzian to a critical 
shape first found by Hubbard~\cite{h71a,h71b}. In addition, strongly
over-damped oscillations in the time-dependent spin relaxation
function at $T_c$ are found. These oscillations, however, do not lead to an
observable structure in the Fourier transform apart from a flatter
decrease at small frequencies. With increasing temperature these
oscillations practically disappear. Presently it is not clear,
whether these oscillations are an artifact of the mode coupling
approximation and go away when higher order terms are included. 
A RG analysis does not show these oscillations~\cite{d76,bf81,i87}.

Hubbard~\cite{h71a} also discusses the shape function for cubic ferromagnets
with nearest-neighbor interaction for wave vectors close to the
Brillouin zone boundary: 
$J({\bf q}) \sim J [cos(q_x a) + cos(q_y a) + cos(q_z a)]$.
He finds that there is a tendency of the shapes to
become more squarer than a Lorentzian and as the wave vectors
come close to the zone boundary one observes the formation of small 
shoulders. Recently Cuccoli et al.~\cite{ctl89} have studied the 
shape of the correlation function at the zone boundary for $EuO$ and $EuS$ 
with a face-centered-cubic lattice taking into account nearest- and
next-nearest-neighbour exchange interaction. The numerical solution of the
mode coupling equations give, as in the simple cubic case
with nearest-neighbor interaction considered by Hubbard~\cite{h71a}, 
inelastic shoulders at the zone boundary, but less intense than 
seen in the experiment~\cite{bkz84,bcs87}.

\subsubsection{Ferromagnetic Phase}
\label{s2.2.2}

In this section we review the mode coupling equations for isotropic 
ferromagnets below the Curie point~\cite{s71,fs88a}. Assuming that the 
spontaneous magnetization points along the $z$--axis the frequency 
matrix is given by                          
\begin{equation}
  \omega^{\alpha \beta}(q) = \omega (q)
  \pmatrix{0&0&0\cr
           0&-1&0\cr
           0&0&+1\cr} \, ,     
\label{2.41}                                      
\end{equation}  
where $\alpha , \beta = z,+,- $. The frequency of the
transverse modes is 
\begin{equation}
  \omega (q) = {M \over \chi^T(q)} \, ,
\label{2.42}
\end{equation}
where $M = \langle S^z_{q=0} \rangle$ denotes the magnetization
and $\chi^T(q)$ the static transverse susceptibility.  Due to the
rotational symmetry of the Hamiltonian the Kubo relaxation matrix
$\Phi^{\alpha \beta}(q,\omega)$ is diagonal          
\begin{equation}
  \Phi^{z z}(q,\omega) =
  {i \over \omega + i \Gamma^{z z}(q,\omega)} \, ,
\label{2.43}
\end{equation}                                                 
\begin{equation}
  \Phi^{\pm \pm}(q,\omega) =
  {2 i \over \omega \mp \omega (q) 
  + i \Gamma^{\pm \pm}(q,\omega)} \, .
\label{2.44}
\end{equation}
The mode coupling approximation for the transport coefficients
\begin{equation}
  \Gamma^{zz}(q,t) = 
  ({\dot S}^z_{{\bf q}}(t),{\dot S}^z_{{\bf q}}(0)) / \chi^L (q)
  \equiv \Gamma(q,t) \, ,
\label{2.45}
\end{equation}
\begin{equation}
  \Gamma^{\pm \pm}(q,t) =
  ({\dot S}^{\pm}_{{\bf q}}(t) \pm 
  i\omega(q) S^{\pm}_{{\bf q}}(t),
  {\dot S}^{\pm}_{{\bf q}}(0) \pm 
  i\omega(q) S^{\pm}_{{\bf q}}(0)) / 2 \chi^T (q) 
  \equiv \Lambda^{\pm}(q,t) \, ,
\label{2.46}
\end{equation}
where $\chi^L(q)$ is the longitudinal susceptibility, results in the 
following set of integral equations~\cite{s71}
\begin{equation}
  \Gamma({\bf q},\omega) =  k_{\rm B} T 
  \int_\nu  \int_{\bf k} 
  \, \upsilon ({\bf k}, {\bf q}) \, 
  {\chi^T ({\bf q} - {\bf k}) \chi^T ({\bf k})  \over \chi^L({\bf q})} \,
  \Phi^{++}({\bf q} - {\bf k}, \omega - \nu)
  \Phi^{--}({\bf k},\nu)  \, ,  
\label{2.47}                                           
\end{equation}             
\begin{equation}
  \Lambda^{\pm}({\bf q},\omega) = 2 k_{\rm B} T
  \int_\nu  \int_{\bf k} 
  \, \upsilon ({\bf k}, {\bf q}) \, 
  {\chi^T ({\bf q} - {\bf k}) \chi^L ({\bf k})  \over \chi^T ({\bf q})} \, 
  \Phi^{\pm \pm}({\bf q} - {\bf k}, \omega - \nu)
  \Phi^{zz}({\bf k},\nu) \, .
\label{2.48}                                          
\end{equation}
Here we have used the notation $\int_\nu = \int {d \nu \over 2 \pi}$.
Eqs.~(\ref{2.43}), (\ref{2.44}) together with
(\ref{2.47}), (\ref{2.48}) constitute 
a complete set of self consistent integral equations for the Kubo 
relaxation functions $\Phi^{\alpha \alpha}(q,\omega)$, which in 
principal could be solved numerically. For $M=0$ and 
$\chi^L = \chi^T$ Eqs.~(\ref{2.47}), (\ref{2.48}) 
reduce to the mode coupling equations for 
the paramagnetic phase, Eqs.~(\ref{2.22}). 

The above mode coupling equations have been analyzed in the 
Lorentzian approximation for the relaxation functions
\begin{equation}
  \Phi^{zz}(q,\omega) = {i  \over
  \omega + i \Gamma(q)}, \quad \quad 
  \Phi^{\pm \pm}(q,\omega) = {2 i  \over
  \omega \mp \omega(q) + i \Lambda^{\pm}(q)} \, ,
\label{2.49}
\end{equation}
with
\begin{equation}
  \Gamma(q) = \Gamma(q,\omega = 0) , \quad        
  \Lambda(q) \equiv \Lambda^+ (q) = {\Lambda^- (q)}^{\ast} 
  = \Lambda^+ (q,\omega(q)) \, .
\label{2.50}
\end{equation}              
The frequency integrals can now be carried out readily and one
finds the following set of coupled integral equations for the
line widths                            
\begin{equation}
  \Gamma({\bf q}) = 
  {4 i k_B T \over \chi^L ({\bf q})}
  \int_{\bf k}
  \, \upsilon ({\bf k}, {\bf q}) \, 
  {\chi^T({\bf q} - {\bf k}) \chi^T ({\bf k}) \over 
  \omega({\bf k}) -\omega({\bf q} - {\bf k}) 
  + i \Lambda({\bf q} - {\bf k}) + i \Lambda^{\ast}({\bf k}) } \, ,
\label{2.51}
\end{equation}
\begin{equation}
  \Lambda({\bf q}) = 
  {4 i k_B T  \over \chi^T ({\bf q})} 
  \int_{\bf k}
  \upsilon ({\bf k}, {\bf q})
  {\chi^T( {\bf q} - {\bf k} ) \chi^L ({\bf k}) \over
   \omega({\bf q}) - \omega({\bf q} - {\bf k})
  + i \Lambda({\bf q} - {\bf k}) + i \Gamma ({\bf k}) } \, .
\label{2.52}
\end{equation}
As is easily seen $\Gamma(q)$ is real, but $\Lambda(q)$ in
general is complex. The imaginary part of the transverse damping
function $\Lambda(q)$ leads to a shift of the frequency of the
transverse spin--waves, which however is a negligible correction
in comparison to the frequency matrix (\ref{2.41}) 
as will be seen later.

In the hydrodynamic regime the Eqs.~(\ref{2.51}), (\ref{2.52}) can be solved
analytically with the result    
\begin{equation}
  \Gamma(q) \propto {q \over \chi^L (q)} \, ,
\quad {\rm and} \quad
  \Lambda(q) \propto 
   q^4 \left[ 
   c_1 \ln  \left( {1 \over q \xi} \right) + c_0 
   \right] \, ,
\label{2.53}
\end{equation}                           
where $c_0$ and $c_1$ are constants~\footnote{The low--temperature spin--wave 
theory of Dyson~\cite{Dyson56} gives in addition a term 
$q^4 \left( \ln (1/q\xi) \right)^2$~\cite{kak61,harris68,vlp68}.
This will probably come out from a mode coupling theory where decays of 
the transverse mode into three transverse modes are included.}.
With the well known scaling properties of the static
susceptibilities neglecting the Fisher exponent $\eta$  
\begin{equation}
  \chi^{L,T} (q) = {1 \over 2 J q^2 a^2} {\hat \chi}^{L,T} (x) \, ,
\label{2.54}
\end{equation}              
Eq.~(\ref{2.41}) gives
\begin{equation}
  \omega(q) = \Lambda q^z {\hat \omega} (x) \, ,
\label{2.55}
\end{equation}          
where the dynamical critical exponent $z$ equals $5/2$ as in
the paramagnetic phase.

The scaling function for the bare frequency of the transverse modes
following from Eq.(\ref{2.42}) can be written as
\begin{equation}
  {\hat \omega}(x) =
  \cases{ f \sqrt{x} &for $T \leq T_c$ \, , \cr
          0                             &for $T \geq T_c$ \, . \cr}
\label{2.56}
\end{equation}
Analyzing the scaling properties of the mode coupling equations and 
combining this with the static and dynamic scaling law it was 
shown~\cite{schinz94a,schinz94b} that the amplitude $f$ for the scaling 
function of the spin--wave frequency is a universal quantity and determined 
by other universal amplitude ratios
\begin{equation}
         f = \left({{\hat c} \over 2}\right)^{1/2}
             \left({R_c \over (R_\xi^+)^d}\right)^{1/2}
             \left({\xi_0 \over \xi_0^T}\right)^{d-2}
             \left({\xi_- \over \xi_+}\right)^{z-2} \, .
\label{2.57}
\end{equation}
Here ${\hat c}$ is an arbitrary normalization constant for the scaling 
functions. If one chooses the value of the scaling functions at criticality 
to be $\gamma(0) = 5.1326$,  ${\hat c}$ becomes 
${\hat c} = 8 \pi^4$~\cite{schinz94b}. The quantities $R_c$ and 
$R_\xi^+$ are universal amplitude ratios as defined in the  review article 
by Privman et al.~\cite{Privman91}, $\xi_0^T$ is a transverse correlation
length below $T_c$~\cite{Privman91}, and $\xi_+$, $\xi_-$ are longitudinal 
correlation lengths above and below $T_c$, respectively.

The amplitude $f$ for the spin--wave frequency can be determined from 
RPA-arguments (see e.g. Ref.~\cite{s71}), which gives $f=\pi^{3/2}$.
This value has been used in consecutive applications of mode coupling theory
on magnets~\cite{fs88a} below $T_c$. Upon using the known values for the 
static amplitude ratios~\cite{Privman91} it is 
found~\cite{schinz94a,schinz94b}
\begin{equation}
f = 9.5 \pm 1.8 \, .
  \label{2.58}
\end{equation}
This amplitude can also be determined from the available experimental data
in Refs.~\cite{bcs87,bkz84,bsbz87,bmt91,Pieper93}, and references cited therein.
The results are summarized in Table~\ref{tableII}~\cite{schinz94a,schinz94b},
where depending on which experiment~\cite{bcs87,bkz84,bsbz87,bmt91,Pieper93}
one analyzes one gets slightly different values for $f$.
                  
Hence Eqs.~(\ref{2.51}), (\ref{2.52}) can be solved by using an dynamic 
scaling Ansatz 
\begin{equation}
  \Gamma(q) = \Lambda q^z \gamma (x), \quad \quad
  \Lambda(q) = \Lambda q^z \lambda (x),
\label{2.59}
\end{equation}          
where the dynamical scaling functions $\gamma (x)$ and $\lambda
(x)$ obey the following set of coupled integral equations
\begin{eqnarray}
  \gamma (x) & = & 2 \pi^2 i \int_{-1}^{+1} d \eta 
  \int \limits_{0}^{ \infty} d \rho \, 
  \rho_{-}^{-2} {\hat \upsilon} (\rho,\eta)
  \, { {\hat \chi}^T ({x / \rho_-})
    {\hat \chi}^T ({x / \rho}) \over {\hat \chi}^L (x) } \nonumber \\
  &\phantom{=} & \times { 1 \over 
    -\rho^z {\hat \omega} ( {x / \rho} ) +
    \rho_{-}^z {\hat \omega} ( {x / \rho_{-}} ) +
    i \rho^z \lambda ( {x / \rho} ) +
    i\rho_{-}^z \lambda^{\ast} ( {x / \rho_{-}} ) } \, ,
  \label{2.60} \\
  \lambda (x) & = & 2 \pi^2 i \int_{-1}^{+1} d \eta 
  \int \limits_{0}^{ \infty} d \rho \, 
  \rho_{-}^{-2} {\hat \upsilon} (\rho,\eta)
  \, { {\hat \chi}^L ({x / \rho_-})
    {\hat \chi}^T ({x / \rho}) \over {\hat \chi}^T (x) } \nonumber \\
  &\phantom{=} & \times { 1 \over {\hat \omega}(x) -
    \rho^z {\hat \omega} ( {x / \rho} ) +
    i \rho^z \lambda ( {x / \rho} ) +
    i \rho_{-}^z \gamma ( {x / \rho_{-}} ) } \, .
\label{2.61}
\end{eqnarray}
Here we have used the same notation as in section~\ref{s2.2.1}.

In order to solve those mode coupling equations one has to know the static
susceptibilities. In the ferromagnetic phase the global continuous rotation 
symmetry is spontaneously broken. Although one of the equivalent directions 
of the order parameter is selected, no free energy is required for an 
infinitesimal quasi-static rotation of the magnetization vector, which in 
turn leads to a diverging transverse correlation length. This physical 
effect is mathematically expressed by the Goldstone 
theorem~\cite{Goldstone61}, stating that there is exactly one massless mode 
for each generator of the broken--symmetry group. In the context of a 
ferromagnet below $T_c$ this implies that the transverse susceptibility is 
given by
\begin{equation}
  \chi^T(q) = {1 \over 2 J q^2 a^2 } \, .
\label{2.62}
\end{equation}
The longitudinal correlation functions entering these integral equations has 
been computed by Mazenko~\cite{m76} to first order in $\epsilon=4-d$ using 
Wilson's matching technique                 
\begin{eqnarray}
  {1 \over \chi^L (q)} &=& 2 J q^2 a^2 \Biggl[ 1 
  - {9 \epsilon \over n + 8} x^2 \left( 1 + \sqrt{1+4x^2} 
  \ln \left( {\sqrt{1+4x^2} - 1 \over 2x} \right) \right) \nonumber 
  \\
  &\phantom{=}& + x^2 \left( { n+8+(5-{n \over 2}) \epsilon \over
  9+(n-1) x^{\epsilon} } \right) \Biggr] \, ,
\label{2.63} 
\end{eqnarray}
where $n$ is the number of spin components. The last term in
$\chi^L(q)$ results from the presence of Goldstone modes below
$T_c$, and it implies that also the longitudinal susceptibility
diverges in the limit $q \rightarrow 0$ for any temperature below
$T_c$.

The resulting numerical solution of the mode coupling equations
(\ref{2.60}), (\ref{2.61}) have been achieved in
Ref.~\cite{fs88a,schinz94a,schinz94b}. The results
are shown in Fig.~2.5 for three different values of the frequency 
amplitude $f$.

One recognizes that the scaling function $Im \lambda (x)$ for the
frequency shift of the transverse modes is very small compared
to ${\hat \omega}(x)$. In the critical region $Im\lambda (x)$
starts at the critical point with infinite slope and is negative
in the hydrodynamical region. The scaling functions for the
longitudinal and transverse line widths split off linearly at the
critical temperature and differ by orders of magnitude in the
hydrodynamic region. This linear split-off of the longitudinal
and transverse widths and the infinite slope of the frequency
shift at the critical temperature below $T_c$ is an immediate
consequence of the presence of Goldstone modes below $T_c$. This
feature can be derived analytically from Eqs.~(\ref{2.51}), (\ref{2.52}). 
The sign of the slope of the longitudinal line width depends on the
magnitude of the amplitude $f$ for the frequency of the spin--waves. 
For values of $f$ close to the RPA-value the slope is positive.
If this value is increased towards the universal value determined in
Ref.~\cite{schinz94a,schinz94b} the slope becomes negative and one gets
a minimum in the longitudinal scaling function $\gamma$. This minimum has been
observed in a recent experiment by B\"oni et al.~\cite{bmt91} (see below). 
Above $T_c$ the scaling function for the line width starts quadratically in
agreement with a renormalization group calculation by Iro
\cite{i87}, but in contrast to the numerically found infinite slope of
Hubbard~\cite{h71a}. It disagrees also with a computation of
Bhattacharjee and  Ferrell~\cite{bf85}, who predict, using Ward identities, 
a linear dependence on ${1 / q \xi}$.

The numerical data can be fitted in the limits $x \gg 1$
(hydrodynamical region) and $x \ll 1$ (critical region) by simple
approximants as summarized in Table~\ref{tableIII} (note that all functions 
are given in units of the value at criticality $\gamma(0) = Re
\lambda(0) \cong 5.1326$)

In unpolarized neutron scattering experiments on $Fe$~\cite{cmnp69}, $Ni$
\cite{mcns69}, and $EuO$~\cite{pdn76} no quasi-elastic peak from spin 
diffusion, as predicted by the mode coupling theory~\cite{fs88a}, was 
discernible. Only the side-peaks originating from the transverse spin 
waves were observed. This is plausible in the light of the mode coupling 
results in Ref.~\cite{fs88a}. In the hydrodynamical region 
($x = {1 / q \xi} \gg 1$) the width of the longitudinal peak is much wider 
than the separation of the transverse peaks~\cite{s71}. Moreover, its 
intensity is smaller than that of the transverse magnons, which
altogether implies that it may be very difficult to distinguish
the longitudinal peak from the background. In the critical region
the line widths are of the same order of magnitude. In this limit
however the frequency of the transverse modes tends to zero.
Using unpolarized neutrons one can only observe a superposition
of the peaks. Lacking a theory for the line width in the critical
region below $T_c$ it was impossible up to recently to resolve the
longitudinal and transverse peaks.

The first observation of the longitudinal peak was reported
by Mitchell et al.~\cite{mcp84} using polarized neutrons.
This study shows in agreement with the theory that the width of the
quasi-elastic longitudinal peak becomes comparable with the spin--wave
energy explaining why this peak was not observed by by neutron
scattering experiments with unpolarized neutrons.
However, there are not enough data as yet to permit a quantitative 
comparison with the theoretical predictions. Furthermore, the material is 
disordered (palladium with $10^{\circ}/_{\circ}$ iron) which makes it not an
ideal system~\cite{bmt91}. Very recently, B\"oni et al.~\cite{bmt91} have
investigated the spin dynamics of a $Ni$ single crystal by means of
polarized neutron scattering. They observe that the longitudinal
fluctuations are quasi-elastic in agreement with our theoretical
predictions~\cite{s71,fs88a} and RG calculations~\cite{mm75}. 
In Fig.~2.6 we show a quantitative comparison of the longitudinal line width, 
obtained from solving the mode coupling equations in
Lorentzian approximation~\cite{zobel,schinz94a,schinz94b}, with the 
experiment~\cite{bmt91}. The light dashed line represents the result of the 
mode coupling equations with an amplitude $f=\pi^{3/2}$, taken from 
RPA-arguments. The solid line and the dot--dashed line represent the results 
of solving the mode coupling equations with an amplitude of $f=5.1326 \times 
1.49$ and $f=5.1326 \times 1.60$, respectively. The agreement 
between theory and experiment is quite well for an appropriate choice of the 
universal amplitude. One should especially note, that the scaling function 
of the longitudinal line width shows a minimum in accord with the 
experimental data.

Finally we note that the above analysis does not take into account effects
from the dipole-dipole interaction. Those effects have up to now not been 
studied quantitatively in the ferromagnetic phase, but, one may expect 
similar effects as above $T_c$, which we are going to describe in the next 
chapter.

\subsection{Renormalization Group Theory}
\label{s2.3}

\subsubsection{Paramagnetic Phase}
\label{s2.3.1}
Renormalization group calculations of the critical dynamics of
ferromagnets start from a stochastic equation of motion for the spin
density $\vS ({\bf x},t)$
\begin{equation}
{\partial \vS ({\bf x},t) \over \partial t} =
  \lambda f \vS \times {\delta {\cal H} \over \delta \vS} +
  \lambda \bfnabla^2  {\delta {\cal H} \over \delta \vS} + \gvect{\zeta} \, ,
\label{2.64}
\end{equation}
where $\gvect{\zeta} ({\bf x},t)$ is a random force with a Gaussian
probability distribution with zero mean and variance
\begin{equation}
  \langle  \zeta^i ({\bf x},t) \zeta^j ({\bf x}^\prime,t^\prime) \rangle =
  2 \Gamma k_B T \delta^{(3)}({\bf x} - {\bf x}^\prime)
  \delta(t - t^\prime) \delta^{ij} \, .
\label{2.65}
\end{equation}
The effective Landau-Ginzburg-Wilson free energy functional is given
by
\begin{equation}
  {\cal H} =
  \int d^d x \left[ { 1\over 2} \left( r {\vS}^2 + 
  (\bfnabla \vS)^2 \right) + {u \over 4!} ({\vS}^2)^2 \right] \, .
\label{2.66}
\end{equation}
These equations can be derived~\cite{mf73,kk73,mfs74} using a 
Mori-Zwanzig projection operator formalism~\cite{m65,z61} (see also 
chapter~\ref{s6}). An exhaustive discussion of these semi-phenomenological
equation of motion can be found in the article by Ma and Mazenko
\cite{mm75}. The first term in Eq.~(\ref{2.64}) describes the
Larmor precession of the spins in the local magnetic field, 
${\delta {\cal H} / \delta \vS}$, and the second term
characterizes the damping of the conserved order parameter.
The precession term of the spins in the local
magnetic field plays a major role in the dynamics. From the RG
analysis in Refs.~\cite{mm75,bjw76} one can infer that its effect 
can be ignored above the upper critical dimension $d_c = 6$, and can be
treated by perturbation theory in $\epsilon = 6-d$\footnote{The upper
critical dimension for the corresponding static problem (see 
Hamiltonian in Eq.~(\ref{2.65})) is $d_c = 4$. Since the
precessional term is of second order in the equations of motion
corresponding to third order in the Lagrangian the upper critical
dimension is shifted to $d_c=6$.}.

The RG--theory proves~\cite{mm75,bjw76} 
the dynamical scaling hypothesis~\cite{fmsss67,hh67} and shows that 
the spin correlation function fulfills 
the dynamical scaling form near the fixed point
\begin{equation}
  C(q,\xi,\omega) = \chi(q,\xi) { 1 \over \omega_c (q,\xi)}
  \phi (x,\nu) \, ,
\label{2.67}
\end{equation}
with the scaling variables $x=1/q\xi$ and 
\begin{equation}
  \nu = {\omega \over \omega_c(q,\xi)} \, ,
\label{2.68}
\end{equation}
where the characteristic frequency has itself the scaling form
\begin{equation}
  \omega_c(q,\xi) = \Lambda q^z \Omega(x) \, .
\label{2.69}
\end{equation}
The dynamic critical exponent $z$ is known exactly from RG theory
\cite{mm75,j76,bjw76}
\begin{equation}
  z = {d+2 -\eta \over 2} \, ,
\label{2.70}
\end{equation}
in accord with the general dynamic scaling considerations of
section~\ref{s2.1}. Here $\eta$ is the Fisher exponent from the static
scaling law
\begin{equation}
  \chi(q,\xi) = q^{-2+\eta} {\hat \chi} (x) \, .
\label{2.71}
\end{equation}
The Fourier transform of the spin correlation function can be
written in the form 
\begin{equation}
  C(q,\xi,\omega) =
  \chi(q,\xi) {1 \over \lambda q^z} Re 
  {2 \over - i {\hat \omega} +
  [{\hat \chi} (x) \Pi (x,{\hat \omega}) ]^{-1} } \, ,
\label{2.72}
\end{equation}
where $\Pi (x,{\hat \omega})$ is the self-energy of the dynamic 
susceptibility and we have defined 
${\hat \omega}  = \omega/ \lambda q^z$.
The asymptotic behavior of the self-energy is known exactly from a
RG--analysis~\cite{bjw76,d76}
\begin{equation}
  \Pi (x,{\hat \omega}) \sim 
  \cases { {\hat \omega}^{(4-z)/z} & for ${\hat \omega} 
           \rightarrow \infty$ \, , \cr
           x^{4-z}       & for $x \rightarrow \infty$ \, . \cr}
\label{2.73}
\end{equation}
One--loop RG--calculations give to order $O(\epsilon)$
\begin{equation}
  \Pi (x,{\hat \omega}) = 1 - \epsilon [ F(x,i{\hat \omega}) + 
                         {1 \over 2} \ln 2 ] + O(\epsilon^2) \, ,
\label{2.74}
\end{equation}
with~\cite{d76,i87}
\begin{equation}
  F(0,i{\hat \omega}) = \cases{ -{1 \over 8} \ln (-i{\hat \omega}) -
                      {3 \over 8} \ln 2 + O({{\hat \omega}}^{-1/2}) 
                      & for ${\hat \omega} \gg 1$ \, , \cr
                      -(\pi + {1 \over 3} )/8 + 3( 2 + 2 \ln 2 - \pi) 
                      {i {\hat \omega} \over 8}
                      & for ${\hat \omega} \ll 1$ \, ,  \cr}
\label{2.75}
\end{equation}
and~\cite{mm75,i87}
\begin{equation}
  F(x,0) = \cases{
           - {1 \over 2} \ln x - (\ln 2 + {3 \over 4} )/2 
           - {13 \over 128} x^{-2}  & for $x \gg 1$ \, , \cr
           -(\pi + {1 \over 3})/8 - (3 \pi - {25 \over 8} - 3 \ln 4)
            {x^2 \over 4} &for $ x \ll 1$ \, . \cr}  
\label{2.76}
\end{equation}
In an $\epsilon$--expansion with respect to the upper critical dimension 
$d_c = 6$ the $\phi^4$--term is irrelevant for $4<d$, hence the static 
critical behavior is classical. This is no more the case for $d<4$, a fact 
which has to be kept in mind if one extends the results of the RG 
$\epsilon$--expansion to $\epsilon = 3$.
The explicit form of $F(x,i{\hat \omega})$ to order $O(\epsilon)$ 
can be found in Ref.~\cite{i87}. The logarithms in the limits 
$(x \rightarrow 0, \nu \rightarrow \infty)$ and $(x \rightarrow
\infty, \nu \rightarrow 0)$ are the $O(\epsilon)$-contributions of
the power law behavior in Eqs.~(\ref{2.73}) which are known exactly 
for any dimension $d$. Exponentiating these logarithms in such a way 
that the exactly known asymptotic behavior Eqs.~(\ref{2.73}) is 
matched one obtains~\cite{bf81,i88a,i88b} the two--parameter 
interpolation formula
\begin{equation}
  \Pi (x,{\hat \omega}) = 
  \left[ \left( 1 + b x^2 \right)^{2 - \epsilon/4} 
         - a i {\hat \omega} 
  \right]^{\epsilon/(8-\epsilon)} \, ,
\label{2.77}
\end{equation}
with $a=0.46$ ($a = {z \over 4} (6 + 6 \ln 2 - 3 \pi)$) and 
$b = 3.16$ for $\epsilon = 3$.
This reasoning for exponentiating the leading logarithms of a first
order $\epsilon$-expansion to match the exactly known asymptotic
results is due to Bhattacharjee and Ferrell~\cite{bf81}. 
One should note,
however, that this exponentiation procedure is not unique. To test 
the validity of the exponentiated expression one would have to 
calculate contributions to order $O(\epsilon^2)$ which would be quite 
cumbersome a task.

The shape function 
\begin{equation}
  \phi(x,{\hat \omega}) = 2 Re
  { 1 \over - i {\hat \omega} + 
   [{\hat \chi} (x) \Pi (x,{\hat \omega}) ]^{-1} }
\label{2.78}
\end{equation}
shows the crossover from the critical shape
at $x=0$ to a Lorentzian at $x=\infty$ in agreement with the mode
coupling results by Hubbard~\cite{h71a} and a more recent reanalysis 
by Aberger and Folk~\cite{af88a}. From Eqs.~(\ref{2.72}) one can also
determine the half width at half maximum $\omega_c$ defined by 
\begin{equation}
  C(q,\xi,\omega_c) = {1 \over 2} C(q,\xi,0) \, .
\label{2.79}
\end{equation}
In Fig.~2.7 the scaling function $\omega_c(x)$ resulting from
Eqs.~(\ref{2.72}) and (\ref{2.79})
is compared with mode coupling results~\cite{rp70,af88a}. Whereas
the MC--result in Lorentzian approximation~\cite{rp70} 
shows large deviations from the RG--result in the hydrodynamic limit,
the complete MC--result abandoning the Lorentzian
approximation~\cite{af88a} follows closely the RG scaling function. 
One should note, however, that if all scaling functions are rescaled 
in such a way that they coincide in the hydrodynamic limit the 
differences between the scaling functions appear much less pronounced.

Let us now compare the theoretical predictions with the experiment.
In early neutron scattering experiments almost all data have
been fitted by a Lorentzian shape function. Recently, however, with
the advances in neutron scattering techniques leading to higher
intensities and better resolution, deviations of the measured
spectra from a Lorentzian have been observed. By comparing the 
RG--result at $T_c$~\cite{d76,bf81} with constant energy scans on $Fe$
\cite{wbs84,l75,l83} it has been shown~\cite{fi85} that theory was
in accord with the data in the observed experimental wave vector and
frequency window. As has been demonstrated in Ref.~\cite{bcs87} the 
peak positions as well as the peak profile in constant--energy scans 
of $EuO$ could be explained on the basis of RG--theory, taking into
account short range exchange interaction only, for wave vectors in 
the range $0.15 \AA^{-1} \leq q \leq 0.3 \AA^{-1}$ and energies 
$0.2 meV \leq \omega \leq 0.4 meV$. 

If a model based on the exchange coupling between neighboring spins
is the correct description of the critical behavior of real
ferromagnets, one would have expected that the results from RG and
MC theories would become even closer to the experimental data as one
comes close to the critical temperature or/and for very large
wavelength. It came as a completely unexpected surprise, when
Mezei~\cite{m86} found in spin echo experiments on $EuO$ that the
observed shape at even smaller wave vectors $q = 0.024 \AA^{-1}$ 
clearly resembled a Lorentzian shape in disagreement with
the predictions of RG theory for the dynamics of an isotropic
Heisenberg ferromagnet, which would give a bell shaped decay.

As we will explain in chapter~\ref{s3} this
ultimate crossover to a Lorentzian can be explained by taking into
account the long--range dipolar interaction.
Further evidence of dipolar effects have been found in $EuS$, where
it was observed that the peak positions (in constant-E scans) do not
scale~\cite{bsbz88}.

Concerning the line width the experimental situation is as follows.
Right at the critical temperature there is almost perfect agreement
of the wave vector dependence of the line width with $\Gamma \sim
q^{5/2}$ in $EuS$~\cite{bsbz87}, $EuO$~\cite{dnp76,bs86,m84}, 
$Fe$~\cite{cmnp69,m82,wbs84} and $Ni$~\cite{mcns69,bmt91}.
In early experimental studies on $Fe$ it seemed that the
experimental data~\cite{pk71,pa72} are in reasonable agreement with 
the theoretically predicted scaling function of Resibois and Piette 
\cite{rp70}. Recent neutron scattering experiments, however, showed large
deviations from the Resibois-Piette scaling function in $Fe$
\cite{m82,m84}, $EuO$~\cite{mfhs89,m88} and $EuS$~\cite{bgkm91}.
This puzzling situation can only be resolved by additionally taking
into account the dipolar interaction, which is the subject of
chapters \ref{s3} and \ref{s4}.

\subsubsection{Ferromagnetic Phase}
\label{s2.3.2}

The critical dynamics below the transition temperature has been
studied also by renormalization group methods. Ma and
Mazenko~\cite{mm75} 
calculated the transport coefficient for the longitudinal
magnetization for small wave vectors in an $\epsilon-$expansion 
$(\epsilon = 6-d)$. Their result was
\begin{equation}
  \Gamma(q) = { {\hat \Gamma}(q) \over \chi^L(q) } q^2 \, ,
\label{2.80}
\end{equation}                                                               
with                                
\begin{equation}
  {\hat \Gamma} \propto q^{d-6 \over 6} \, .
\label{2.81}
\end{equation}
With $\chi^L(q) \propto {1 \over q}$ in $d = 3$ dimensions this
would give $\Gamma(q) \propto q^{5 \over 2}$ in contradiction to
the mode coupling result for small q (i.e. in the hydrodynamical
limit~\cite{rp70}). However, Sasv\'ari's~\cite{sa77} reanalysis of 
Ma and Mazenko's~\cite{mm75} exponentiation method showed, 
that taking into account
the regular parts of ${\hat \Gamma}(q)$ results in
\begin{equation}
  {\hat \Gamma}(q) \propto q^{d-6 \over 3} \, .
\label{2.82}
\end{equation}
This leads to $\Gamma(q) \propto q^2$ for $d=3$ dimensions, in
agreement with the mode coupling result in Table~\ref{tableIII}. The
$q^4-$dependence of the transverse transport coefficient in the
hydrodynamical limit is also confirmed by the renormalization
group calculations~\cite{mm75}.  

A thorough renormalization group study of the critical dynamics of
a Heisenberg ferromagnet below $T_c$ is still lacking. Such a study 
would have to take into account all the peculiarities resulting from the
presence of the Goldstone modes below $T_c$. As a first step towards this
end, there is a recent study~\cite{Taeuber92} of the critical dynamics of 
the $O(n)$--symmetric relaxational models with either non-conserved (model A) 
or conserved order parameter (model B) below the transition temperature (see
also chapter~\ref{s5}).

\newpage

\begin{table}
\setdec 0.00
\caption{Crystal structure dependent parameters of cubic Bravais lattices. 
$c$ counts the number of next nearest neighbors to a given lattice site. The
parameter $b$ is defined as $b = ({a^3/v_a})^{1/2}$, and characterizes the 
lattice structure. $\delta$ is the distance between nearest neighbors ions 
and $v_a$ the  volume of the Bravais lattice primitive cell. $a$ is the cube 
edge.}
\bigskip\bigskip
\begin{tabular}{lccc}
&sc&bcc&fcc\cr
\tableline
$c$           & $6$     & $8$              & $12$             \\
$b$           & $1$     & $\sqrt{2}$       & $2$              \\
$\delta / a$  & $1$     & $\sqrt{3}/2$     & $\sqrt{2}/2$     \\
$v_a / a^3$  & $1$     & $1/2$            & $1/4$            \\
\end{tabular}
\label{tableI}
\end{table}

\vskip 1truecm

\begin{table}
\setdec 0.00
\caption{Experimental values for the spin--wave amplitude ${\hat b}=f/5.1326$.}
\bigskip\bigskip
\begin{tabular}{lccccc}
{ }&Fe&Ni&Co&EuO&EuS\\
\tableline
${\hat b}$&1.5(1)&1.5(1)&1.6(2)&1.3(2)&1.4(2)\\ 
$$&1.8(1)&2.1(1)&&&1.9(3)\\ 
\end{tabular}
\bigskip
\noindent {\small Data are collected from
Refs.~\cite{bcs87,bkz84,bsbz87,bmt91,Pieper93}, and references cited therein.}
\label{tableII}
\end{table}

\vskip 1truecm
\begin{table}
\setdec 0.00
\caption {Asymptotic behavior of the scaling functions
below the critical temperature in units of the value at
criticality $\gamma (0)$.} 
\bigskip\bigskip
\begin{tabular}{lccccc}
 &${\gamma(x) / \gamma(0)}$&$\Gamma(q)$&${Re \lambda(x) \over
\gamma(0)}$&${Im \lambda(x) \over \gamma(0)}$&$Im \Lambda(q); \, \,
Re \Lambda(q)$ \\
\noalign{\smallskip \hrule \smallskip}
$x \gg 1$&$1.37 + 0.37{x^{-1/2} \over {\hat \chi}^L (x)}$&
$q^2 \xi^{-1/2}$&$0.16 x^{-3/2} \ln(x)$&$-0.07 x^{-3/2} \ln(x)$&
$q^4 \xi^{3/2} \ln \left( {1 \over q \xi} \right)$ \cr
$x \ll 1$&$1.0 + 0.55 x$&
$q^{5/2}$&$1.0 -1.34 x$&$0.77 x^{1/2}$&
$q^2 \xi^{-1/2}; \, \, q^{5/2}$ \\
\end{tabular}
\label{tableIII}
\end{table}

\vfill

\eject

\newpage

\centerline{\bf Figure Captions:}
\bigskip

\noindent {\bf Figure 2.1:} \par
\noindent {The macroscopic domain of wave vector $q$ and correlation
length $\xi$. In the three shaded regions the correlation
functions have different characteristic behaviors. (a)
Hydrodynamic regions: $q \xi \ll 1$, $T > T_c$, and $T< T_c$, (b)
critical region: $q \xi \gg 1$, $T \approx T_c$. There is a
change--over from under-damped spin--waves to spin diffusion when the
temperature is raised from below $T_c$ to above $T_c$, as
schematically indicated in the diagram.}
\vskip 1 truecm

\noindent {\bf Figure 2.2:} \par
\noindent {The Resibois--Piette scaling function versus $x=1/q \xi$, 
resulting from the numerical solution~\cite{rp70} of the mode coupling 
equations in Lorentzian approximation for a Heisenberg ferromagnet in the 
paramagnetic phase.}
\vskip 1 truecm

\noindent {\bf Figure 2.3:} \par
\noindent {The universal function $\phi(\omega/\sigma q^{5/2},1/q \xi)$
for several values of $1/q \xi$: A) $1/q \xi = 0.$, B) $1/q \xi = 0.2$,
C) $1 / q \xi = 1.04$. Taken from Ref.~\cite{h71a} . The scale $\sigma$
is defined in Ref.~\cite{h71a}. }
\vskip 1truecm

\noindent {\bf Figure 2.4a:} \par
\noindent {The spin relaxation function 
$\phi(\Lambda q^{5/2} t,1/q \xi)$ for several values of $1/q\xi$
indicated in the graph.}
\vskip 1truecm

\noindent {\bf Figure 2.4b:} \par
\noindent {The scaling function $\phi(\omega/\Lambda q^{5/2},1/q \xi)$ for 
the spin relaxation function $\Phi(q,\omega) = 
\phi(\omega/\Lambda q^{5/2},1/q \xi)/\Lambda q^{5/2}$ for several values of 
$1/q\xi$ indicated in the graph, showing the shape crossover to a Lorentzian 
shape for $1/q \xi \geq 1$.}
\vskip 1truecm

\noindent {\bf Figure 2.5:} \par
\noindent {Dynamic scaling function for ferromagnets  with
short range exchange interaction only versus $1 / q \xi$  below $T_c$
for three different amplitudes of the spin-wave frequency:
$f = \pi^{3/2}$ (solid), $f = 7.65$ (dashed), and $f = 8.20$ (point--dashed).
Taken from Ref.~\cite{schinz94b,schinz94c}.}
\vskip 1truecm

\noindent {\bf Figure 2.6:} \par
\noindent {Comparison of the longitudinal line width with the polarized 
neutron scattering experiments on $Ni$~\cite{bmt91}. All line widths are 
normalized to $1$ at criticality. The result of mode coupling theory is 
shown for two different values of the spin-wave frequency amplitude 
$f=\pi^{3/2}$ (solid),  $f = 7.65$ (dashed), and $f = 8.20$ (point-dashed).
Taken from Ref.~\cite{schinz94a,schinz94b}.}
\vskip 1truecm

\noindent {\bf Figure 2.7:} \par
\noindent {Comparison of the line width obtained from mode coupling and 
renormalization group analysis. Taken from Ref.~\cite{i88b}.}
\vskip 1truecm
\vfill

\newpage
\vfill
\begin{figure}[h]
  \centerline{{\epsfysize=5in \epsffile{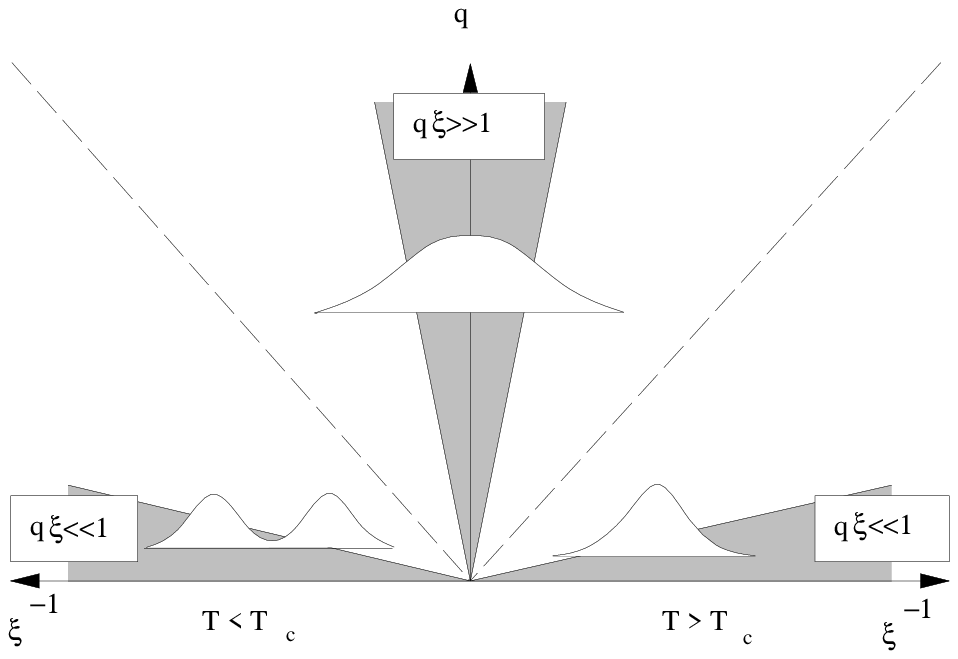}}}
\bigskip \bigskip
\noindent {\bf Figure 2.1:}   {The macroscopic domain of wave vector $q$ and 
           correlation length $\xi$. In the three shaded regions the 
           correlation functions have different characteristic behaviors. (a)
           Hydrodynamic region: $q \xi \ll 1$, $T > T_c$, and $T< T_c$, (b)
           critical region: $q \xi \gg 1$, $T \approx T_c$.  There is a
           change--over from under-damped spin--waves to spin diffusion when 
           the temperature is raised from below $T_c$ to above $T_c$, as
           schematically indicated in the diagram.}
  \label{fig21}
\end{figure}
 
\newpage
\vfill

\begin{figure}[h]
 \centerline{{\epsfysize=5in \epsffile{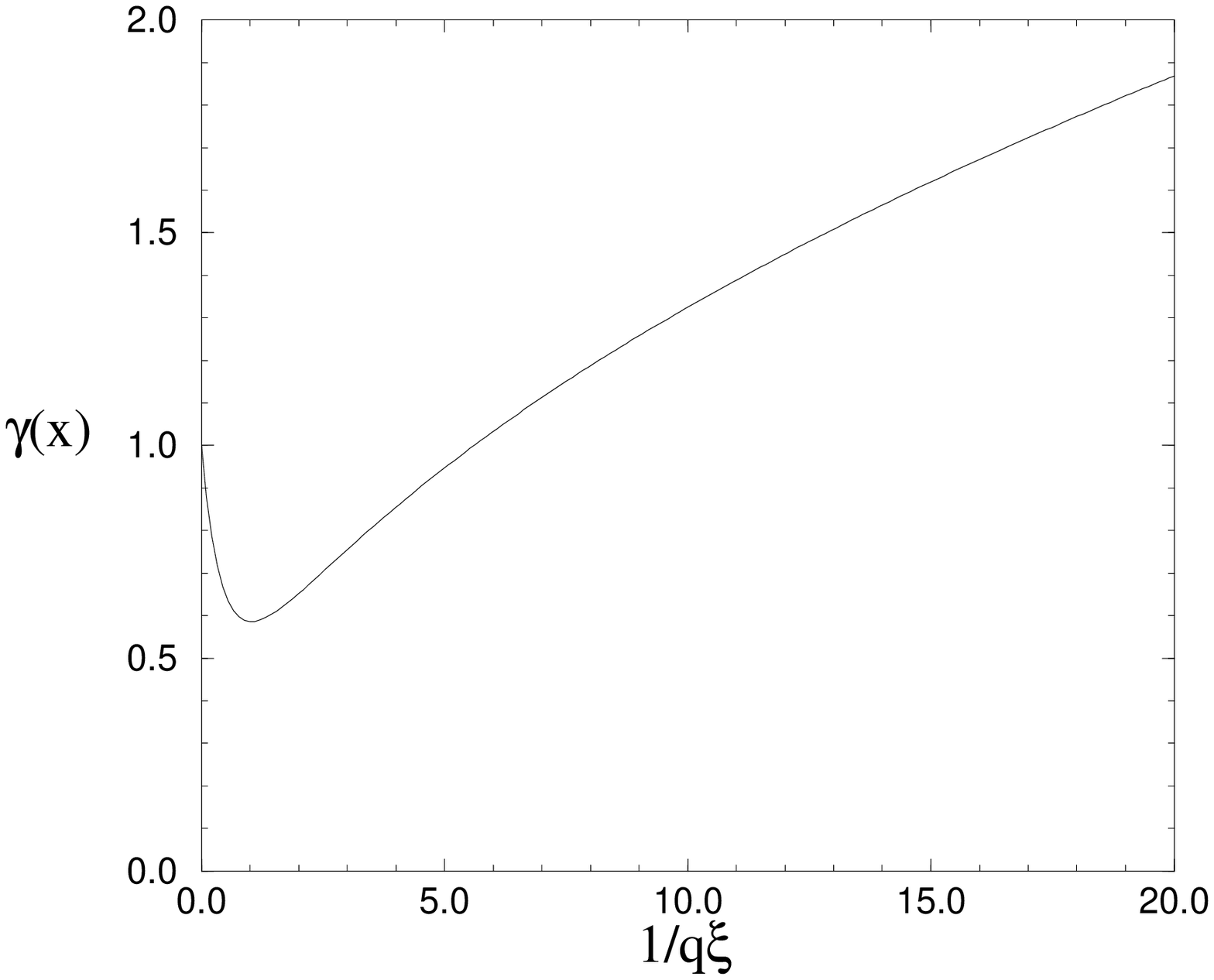}}}
\bigskip \bigskip
\noindent {\bf Figure 2.2:}  {The Resibois--Piette scaling function versus 
           $x=1/q \xi$, resulting from the numerical solution~\cite{rp70}
           of the mode 
           coupling equations in Lorentzian approximation for a Heisenberg 
           ferromagnet in the paramagnetic phase.}
  \label{fig22}
\end{figure}
 
\newpage
\vfill

\begin{figure}[h]
   \centerline{\rotate[r]{\epsfysize=5in \epsffile{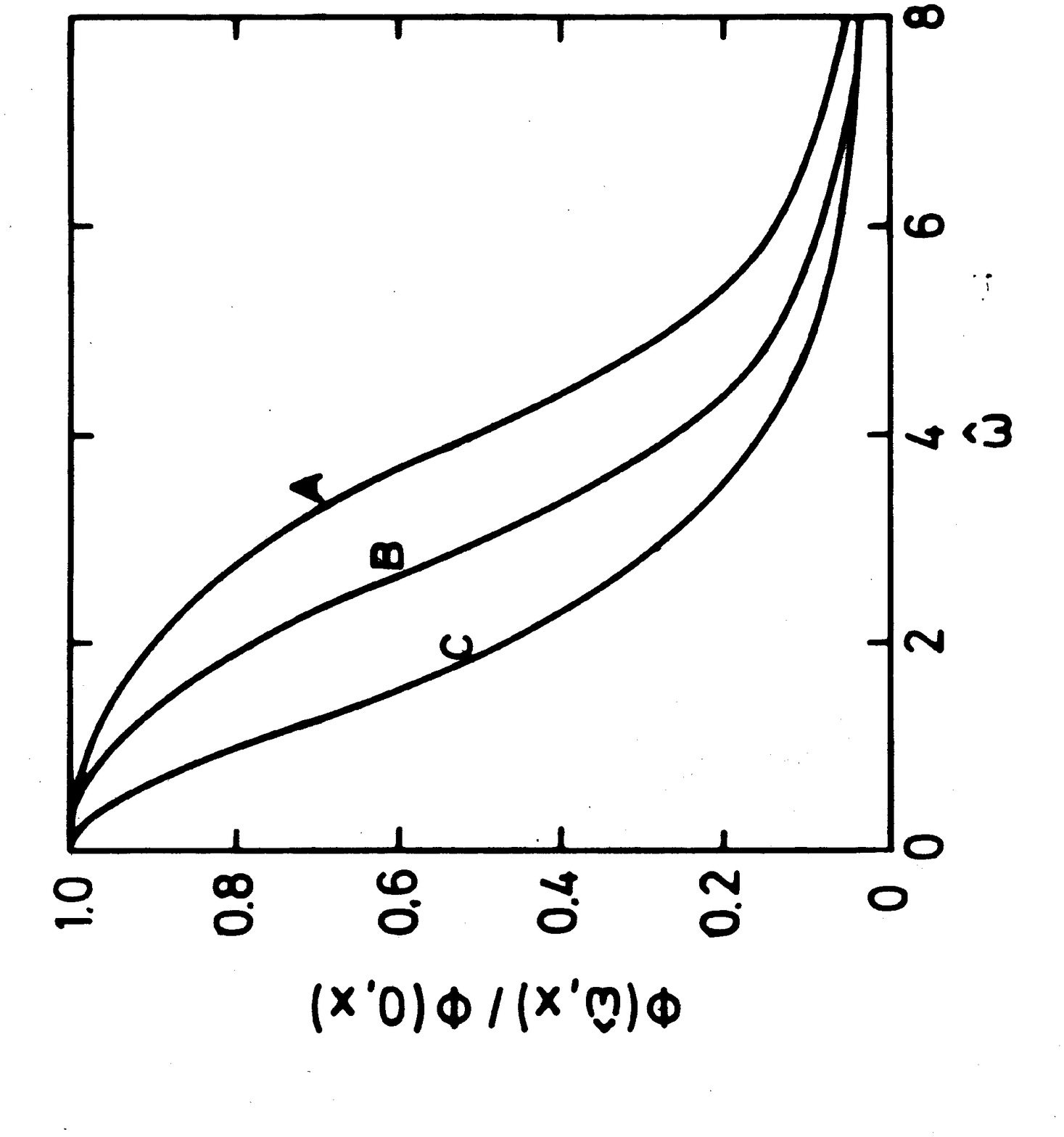}}}
\bigskip \bigskip
\noindent {\bf Figure 2.3:}   {The universal function $\phi(\omega/\sigma 
           q^{5/2},1/q \xi)$ for several values of 
           $1/q \xi$: A) $1/q \xi = 0.$, B) $1/q \xi = 0.2$,
           C) $1 / q \xi = 1.04$. Taken from Ref.~\cite{h71a} . 
           The scale $\sigma$ is defined in Ref.~\cite{h71a}.}
  \label{fig23}
\end{figure}

\newpage

\vfill

\begin{figure}[h]
   \centerline{\rotate[r]{\epsfysize=5in \epsffile{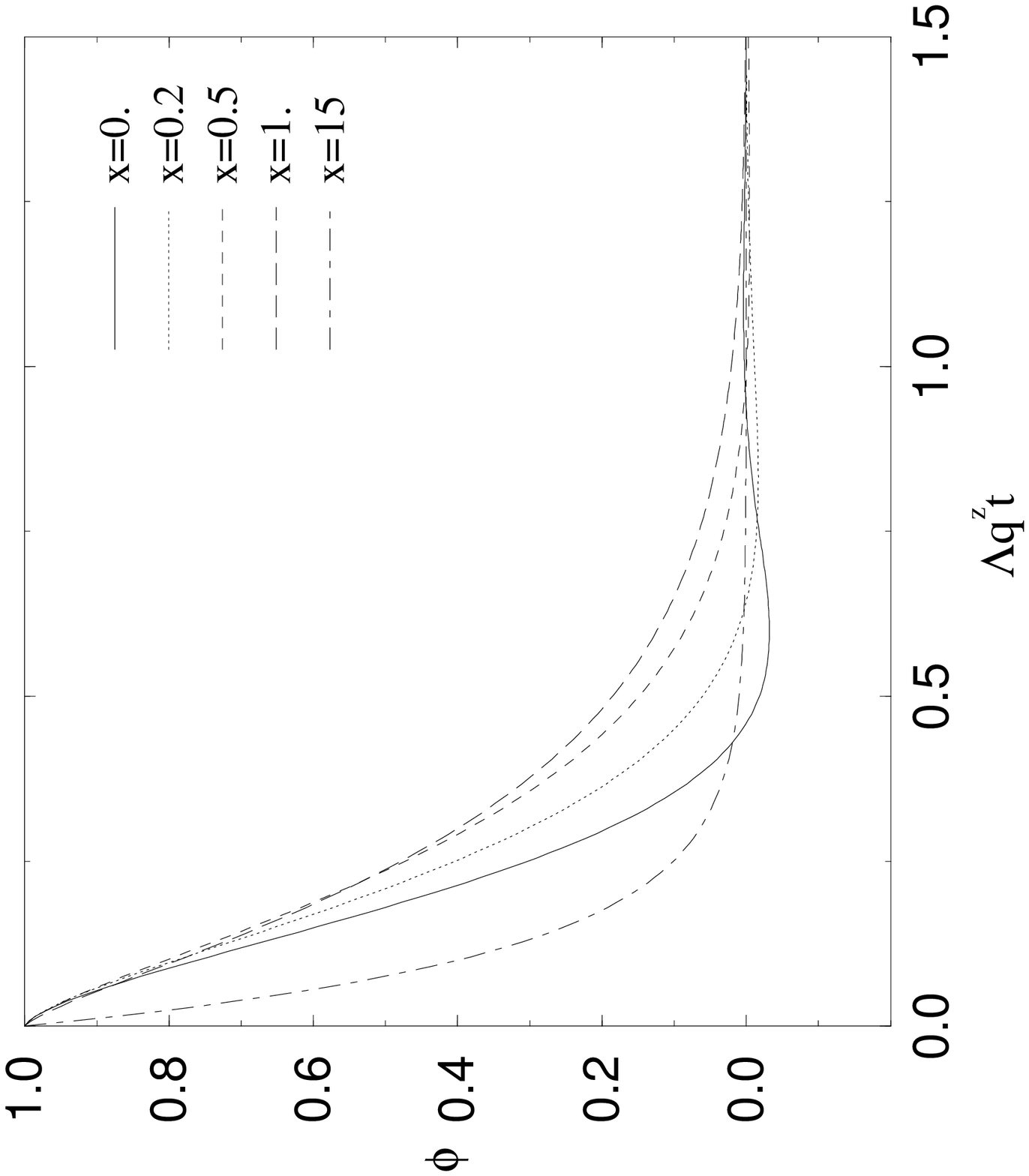}}}
\bigskip \bigskip
\noindent {\bf Figure 2.4.a:}   {The spin relaxation function 
           $\phi(\Lambda q^{5/2} t,1/q \xi)$ for several values of $1/q\xi$
           indicated in the graph.}
  \label{fig24a}
\end{figure}

\newpage

\vfill

\begin{figure}[h]
   \centerline{\rotate[r]{\epsfysize=5in \epsffile{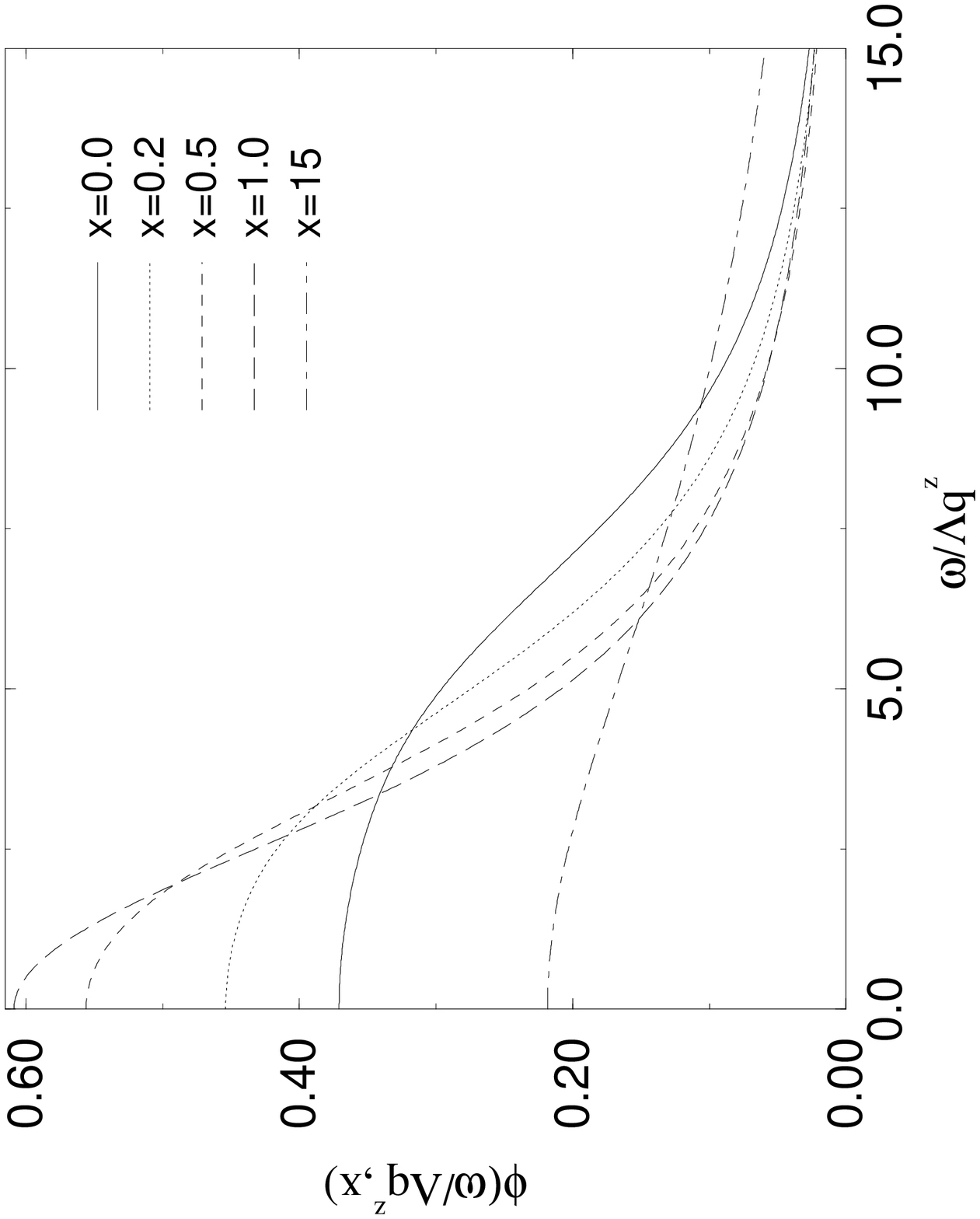}}}
\bigskip \bigskip
\noindent {\bf Figure 2.4.b:}   {The scaling function 
           $\phi(\omega/\Lambda q^{5/2},1/q \xi)$ for the spin relaxation 
           function $\Phi(q,\omega) = \phi(\omega/\Lambda 
           q^{5/2},1/q \xi)/\Lambda q^{5/2}$ for several values of $1/q\xi$ 
           indicated in the graph, showing the shape crossover to a 
           Lorentzian shape for $1/q \xi \geq 1$.}
  \label{fig24b}
\end{figure}

\newpage

\vfill

\begin{figure}[h]
  \centerline{\rotate[r]{\epsfysize=5.5in \epsffile{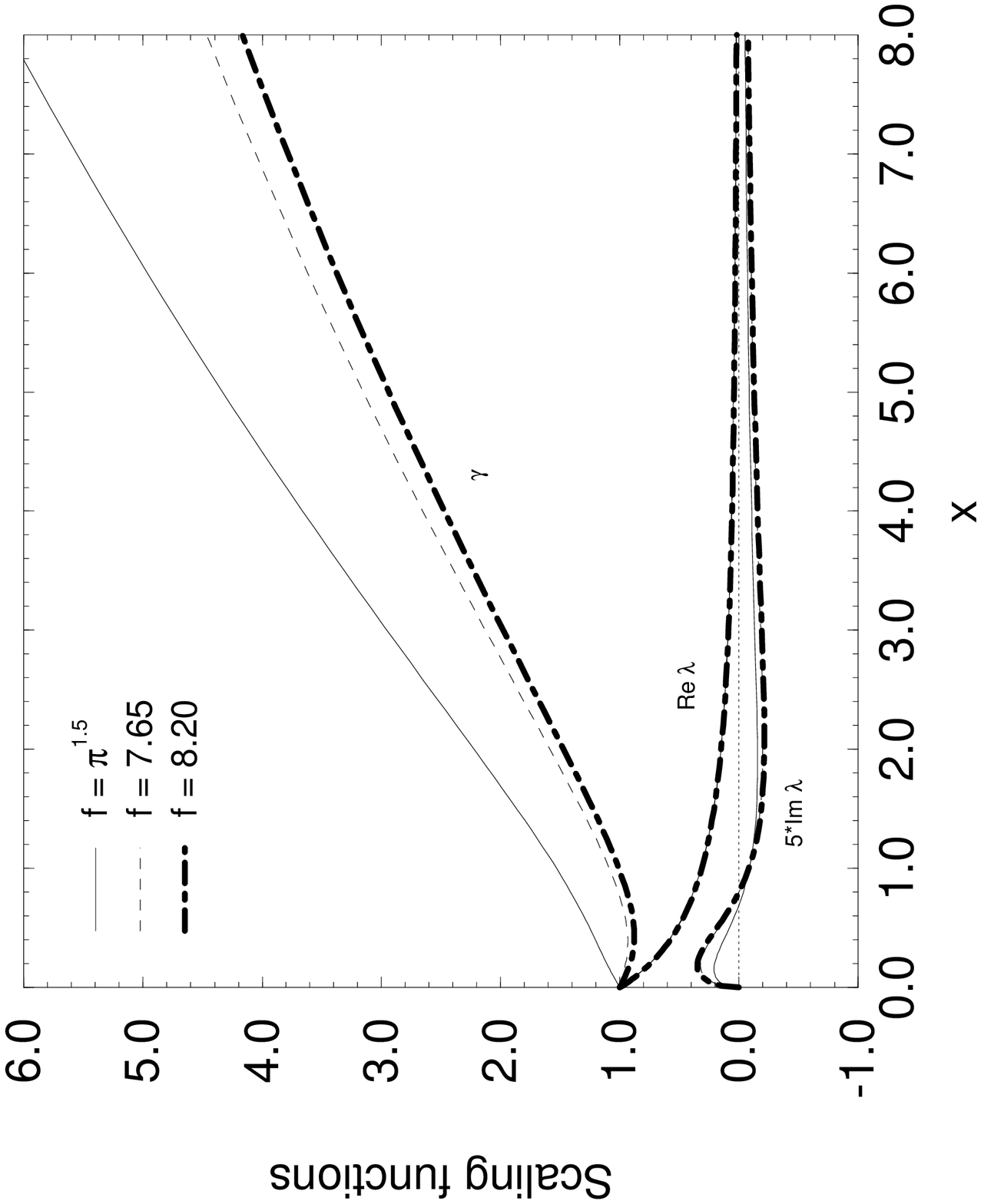}}}
\bigskip \bigskip
\noindent {\bf Figure 2.5:}   {Dynamic scaling function for ferromagnets  with
           short range exchange interaction only versus $1 / q \xi$  below 
           $T_c$ for three different amplitudes of the spin-wave frequency:
           $f = \pi^{3/2}$ (solid), $f = 7.65$ (dashed), and $f = 8.20$ 
           (point-dashed). Taken from Ref.~\cite{schinz94b,schinz94c}.}
  \label{fig25}
\end{figure}

\newpage

\vfill

\begin{figure}[h]
  \centerline{{\epsfysize=4in \epsffile{fig26.ps.bb}}}
\bigskip 
\noindent {\bf Figure 2.6:}   {Comparison of the longitudinal line width with 
           the polarized neutron scattering experiments on $Ni$~\cite{bmt91}.
           All line widths are normalized to $1$ at criticality. The result 
           of mode coupling theory is shown for two different values of the 
           spin-wave frequency amplitude $f=\pi^{3/2}$ (solid), $f = 7.65$ 
           (dashed), and $f = 8.20$ (point--dashed). 
           Taken from Ref.~\cite{schinz94a,schinz94b}.}
  \label{fig26}
\end{figure}

\newpage

\vfill

\begin{figure}[h]
  \centerline{\rotate[r]{\epsfysize=6in \epsffile{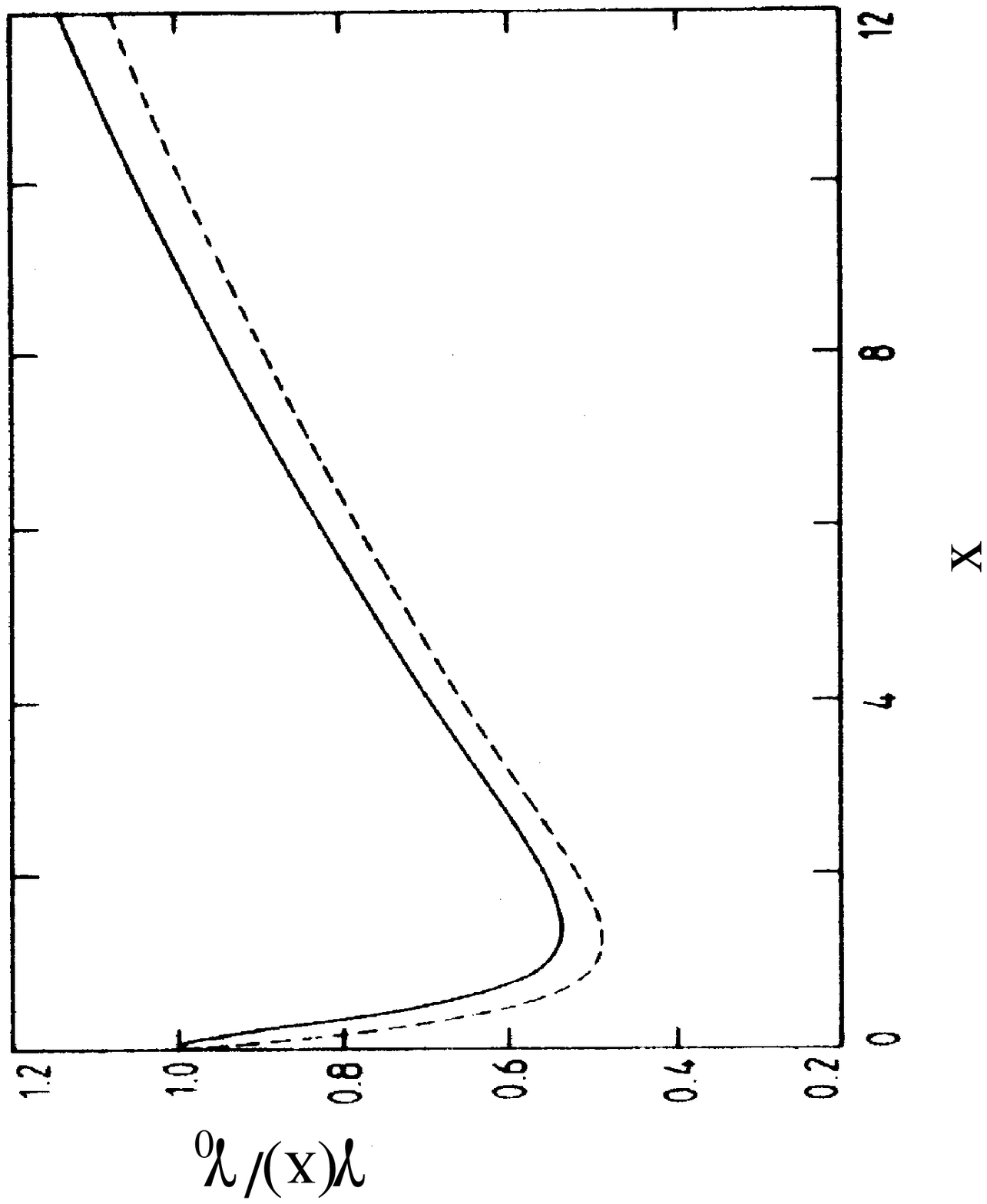}}}
\bigskip \bigskip
\noindent {\bf Figure 2.7:}   {Comparison of the line width obtained form mode
           coupling and renormalization group analysis. Taken from 
           Ref.~\cite{i88b}.}
  \label{fig27}
\end{figure}

\section{DIPOLAR FERROMAGNETS}
\label{s3}

In this chapter we review the static and dynamic critical behavior of dipolar
ferromagnets, i.e., spin systems with both short--range exchange and 
long--range dipole-dipole interaction. Special emphasis is put on the 
discussion of the mode coupling theory in the paramagnetic phase.

But, before turning to the detailed analysis we would like to emphasize the 
following characteristic features of the dipole-dipole interaction, which 
have important implications on the critical dynamics. (1) In contrast to the 
short range exchange interaction the dipolar interaction is
long-ranged and thus dominates the asymptotic critical behavior of
ferromagnets. (2) It introduces an anisotropy of the spin
fluctuations longitudinal and transverse to the wave vector $\bf q$. 
This implies that the longitudinal static susceptibility
remains finite for ${\bf q} \rightarrow 0$ and $T \rightarrow
T_c$ \cite{af73}. (3) The order parameter no longer is conserved as
can be inferred from the equations of motion. (4) The
strength of the dipolar interaction introduces, besides the
correlation length $\xi$, a second length scale $q^{-1}_D$, where
$q_{_{D}}$ is the so called dipolar wave vector defined below. This
leads to generalized scaling laws for the relaxation functions
and the line widths.

\subsection{The Model--Hamiltonian}
\label{s3.1}

Our starting point is a Hamiltonian for a spin system including both 
isotropic short--range exchange and long--range dipolar interactions 
\begin{equation}
  H = - \sum_{l \neq l^\prime} 
        \left[ 
        J_{l l^\prime} \delta^{\alpha \beta} + A_{l l^\prime}^{\alpha \beta}
        \right] S_l^\alpha S_{l^\prime}^\beta \, ,
\label{3.1}
\end{equation}  
where $J_{l l^\prime}$ denotes the short range exchange interaction, usually 
restricted to nearest or next nearest neighbors, and 
$A_{l l^\prime}^{\alpha \beta}$ is the dipolar interaction tensor given by
\begin{equation}
 A_{l l^\prime}^{\alpha \beta} = - {1 \over 2} (g_L \mu_B)^2 
 \left( {\delta^{\alpha \beta} \over |{\bf x}_l - {\bf x}_{l^\prime}|^3}
      - {3 (x_l - x_{l^\prime})^\alpha (x_l - x_{l^\prime})^\beta 
        \over |{\bf x}_l - {\bf x}_{l^\prime}|^5} 
 \right) \, .
\label{3.2}
\end{equation}
Here $g_L$ is the Land\'e factor, $\mu_B$ the Bohr magneton.  
As shown by Cohen and Keffer~\cite{Cohen55} the lattice sums
\begin{equation}
 A_{\bf q}^{\alpha \beta} = \sum_{l \neq 0}
                            A_{l 0}^{\alpha \beta}
                            e^{i {\bf q} \cdot {\bf x}_l }
\label{3.3}
\end{equation}
can be evaluated by using Ewald's method~\cite{Ewald21}, and one finds for
infinite three--dimensional cubic lattices~\cite{Cohen55,af73,a73}
\begin{eqnarray}
 A_{\bf q}^{\alpha \beta} v_a = {1 \over 2} (g_L \mu_B)^2
 \Biggl\{
 &&{4 \pi \over 3} 
 \left( \delta^{\alpha \beta} - {3 q^\alpha q^\beta \over q^2} \right)
 \nonumber \\
 + &&\alpha_1 q^\alpha q^\beta + 
 \left[\alpha_2 q^2 - \alpha_3 (q^\alpha)^2 \right] \delta^{\alpha \beta} 
 + {\cal O} (q^4, (q^\alpha)^4, (q^\alpha)^2 (q^\beta)^2) \Biggr\}  \, ,
\label{3.4}
\end{eqnarray}
where $v_a$ is the volume of the primitive unit cell with lattice constant $a$, 
and $\alpha_i$ are constants, which depend on the lattice structure 
(see Table~\ref{tableIV}). Upon expanding the exchange interaction,
\begin{equation}
J_{\bf q} = {\sum_l}^\prime J_{l0}
            e^{i {\bf q} \cdot {\bf x}_l }
            \approx J_0 - J q^2 a^2 + {\cal O}(q^4) \, ,
\label{3.5}
\end{equation}
and keeping only those
terms, which are relevant in the sense of renormalization--group theory
this results in the following effective Hamiltonian for dipolar ferromagnets
\begin{equation}
  H = \sum_{\bf q} 
      \left[ -J_0 + J q^2 a^2  + J g  {q^\alpha q^\beta \over q^2}    
      \right] 
      S^\alpha_{- \bf q} S^\beta_{\bf q} \, ,
\label{3.6}
\end{equation}
where the Fourier--transform of the spin variables is defined by
\begin{equation} 
 S_{\bf q}^\alpha = {1 \over \sqrt{N}} \sum_l S_l^\alpha 
 e^{i {\bf q} \cdot {\bf x}_l } \, .
\label{3.7}
\end{equation}
Here we have defined a dimensionless quantity $g$ as the ratio of the
dipolar energy $(g_L \mu_B)^2/a^3$ and exchange energy $2J$, multiplied by a
factor ${4 \pi a^3 / v_a}$, which depends on the lattice structure.
\begin{equation}
 g = {4 \pi a^3 \over v_a} \, {(g_L \mu_B)^2 / a^3  \over 2 J} 
 \propto {{\rm Dipolar \, Energy} \over {\rm Exchange \, Energy}} \, .
\label{3.8}
\end{equation}
Strictly speaking there are dipolar corrections of order ${\cal O}(q^2)$ to the
exchange coupling. But, those can be neglected, since the strength of the
dipolar interaction is small compared with the exchange interaction.

The dipolar interaction induces an anisotropy of the static as well
as the dynamic spin-spin correlation function with respect to the
wave vector ${\bf q}$. It was shown by Aharony and Fisher\cite{af73}, that the 
static transverse susceptibility diverges 
with the dipolar critical exponent $\gamma$ as the critical
temperature is approached, whereas the longitudinal susceptibility 
remains finite. The finite value is inversely proportional to the 
strength of the dipolar interaction $g$. The dipolar
anisotropy becomes substantial when $q^2 + \xi^{-2} \leq q_{_{D}}^2$, where the
dipolar wave vector is defined as $g = (q_{_{D}} a)^2$. The
strong suppression of the longitudinal fluctuations have  
been observed in $EuS$ and $EuO$ by polarized neutron scattering
experiments \cite{kmgf86}.

The matrix of the static susceptibility is given by
\begin{equation}   
  \chi^{\alpha \beta} (g,{\bf q}) = 
  \chi^T(g,q)  
  \left(\delta^{\alpha \beta} - {q^\alpha q^\beta \over q^2} \right) 
  + \chi^L(g,q)  
  {q^\alpha q^\beta \over q^2} \, ,
\label{3.9}
\end{equation}   
where one often uses the Ornstein-Zernike forms for the longitudinal
and transverse susceptibilities
\begin{eqnarray}
  \chi^T(g,q) &=& {\Upsilon \over q^2 + \xi^{-2}} \, , 
  \label{3.10} \\
  \chi^L(g,q) &=& {\Upsilon \over q^2 + q_{_{D}}^2 + \xi^{-2}} \, ,
  \label{3.11}
\end{eqnarray}
with the non--universal static amplitude
\begin{equation}
  \Upsilon = {(g_L \mu_B)^2 \over 2 J a^2 } \, .
\label{3.12}
\end{equation}
Conventional mode coupling theory does not account for effects of the
critical exponent $\eta$, which will be neglected in the
following\footnote{For a refined version of mode 
coupling theory
which allows for a consistent treatment of the Fisher exponent
$\eta$ see chapter VI.}. Here $\xi = \xi_0 \left( {T - T_c \over T_c}
\right)^{- \nu}$ is the correlation length. The static crossover
from Heisenberg to dipolar critical behavior is partly contained
in $\xi$ through the effective temperature dependent 
exponent~\cite{bkn76,fs88c,fs91} $\nu \cong {\gamma_{\rm eff} / 2}$.
The full dipolar crossover in the static critical behavior has been 
studied in Refs.~\cite{Nattermann76,bkn76,Bruce77,fs88c,fs91}.

The tensorial structure of the static susceptibility suggests a
decomposition of the spin operator $ {\bf S}_{\bf q}$ into one
longitudinal and two transverse components with respect to the
wave vector ${\bf q}$, i.e.,
\begin{equation}   
  {\bf S}_{\bf q} =
  S^L_{\bf q} {\hat {\bf q}} + S^{T_1}_{\bf q} {\hat  {\bf t}}^1({\hat {\bf q}})
  + S^{T_2}_{\bf q} {\hat {\bf t}}^2({\hat {\bf q}}) \, ,
\label{3.13}
\end{equation}   
where the orthonormal set of unit vectors is defined by          
\begin{equation}   
  {\hat {\bf q}} = {{\bf q} \over q}, \quad
  {\hat  {\bf t}}^1({\hat {\bf q}}) 
  = {{\bf q} \times {\bf e}_3 \over 
    \left( {q_1^2 + q_2^2} \right)^{1/2} } 
  , \quad
  {\hat {\bf t}}^2({\hat {\bf q}}) = {\hat {\bf q}} \times {\hat {\bf t}}^1
  ({\hat {\bf q}}) \, .
\label{3.14}
\end{equation}   
For vanishing components of ${\bf q}$ the limits are taken in the
order of increasing Cartesian components.

From the Hamiltonian, Eq.~(\ref{3.6}), one deduces the following 
microscopic Heisenberg equations of motion~\cite{fe86,fs87}                           
\begin{eqnarray}
  {d \over dt} S_{{\bf q}}^L =
  &&J a^2 \int {v_a d^3k \over (2 \pi)^3} {1 \over k}
  \Biggl\{ {\bf q} \cdot (2 {\bf k} - {\bf q})
  \left( \sqrt{k_1^2+k_2^2} \lbrace S_{{\bf q} - {\bf k}}^{T_1},
  S_{{\bf k}}^L \rbrace + k_3 \lbrace S_{{\bf q} - {\bf k}}^{T_1},
  S_{{\bf k}}^{T_2} \rbrace \right) 
\nonumber \\
  &&+g \sqrt{k_1^2+k_2^2} \lbrace S_{{\bf q} - {\bf k}}^{T_1},
  S_{{\bf k}}^L \rbrace
  \Biggr\} \, ,
\label{3.15}
\end{eqnarray}
\begin{eqnarray}
  {d \over dt} S_{{\bf q}}^{T_1} =
  &&-J a^2 \int {v_a d^3k \over (2 \pi)^3} {1 \over k}
  \Biggl\{ {\bf q} \cdot (2 {\bf k} - {\bf q})
  \Biggl[ 
  {-k_1 k_3 \over \sqrt{k_1^2+k_2^2}}
  \lbrace S_{{\bf q} - {\bf k}}^{T_1},S_{{\bf k}}^L \rbrace 
  +k_1\lbrace S_{{\bf q} - {\bf k}}^{T_1},S_{{\bf k}}^{T_2}\rbrace  
  \nonumber \\
  &&+ {k_2 (k_3 q - k^2) \over \mid {\bf q} - {\bf k} \mid 
  \sqrt{k_1^2+k_2^2}}
  \lbrace S_{{\bf q} - {\bf k}}^{T_2},S_{{\bf k}}^L \rbrace 
  + {1 \over 2} {k_2 q \over \mid {\bf q} - {\bf k} \mid} 
  \left( \lbrace S_{{\bf q} - {\bf k}}^L,S_{{\bf k}}^L \rbrace
  - \lbrace S_{{\bf q} - {\bf k}}^{T_2},S_{{\bf k}}^{T_2} \rbrace
  \right) \Biggr] \nonumber \\
  &&+g \Biggl[ { -k_1 k_3 \over \sqrt{k_1^2+k_2^2} } 
  \lbrace S_{{\bf q} - {\bf k}}^{T_1},S_{{\bf k}}^L \rbrace
  + { k_2 (k_3 q - k^2) \over \mid {\bf q} - {\bf k} \mid 
  \sqrt{k_1^2+k_2^2} }
  \lbrace S_{{\bf q} - {\bf k}}^{T_2},S_{{\bf k}}^L \rbrace 
  \Biggr] \Biggr\} \, ,
\label{3.16}
\end{eqnarray}
and for $S_{{\bf q}}^{T_2}$ correspondingly, where $\lbrace ,
\rbrace$ denotes the anticommutator. The terms proportional to
$g$, resulting from the dipolar term in the Hamiltonian remain
finite as the wave vector ${\bf q}$ tends to zero, whereas all the other
terms vanish in this limit. This is responsible for the fact that the
dipolar forces lead to a relaxational dynamics in the limit of
long wavelengths, i.e., the order parameter is no longer a conserved
quantity.

\subsection{Mode Coupling Theory for the Paramagnetic Phase}

\subsubsection{General Mode Coupling Equations}

As we have explained in chapter II, mode coupling theory amounts in an
factorization approximation for the transport coefficients
\begin{equation}   
  \Gamma^L(q,g,t) =
  {(g_L \mu_B)^2 \over \chi^L (q,g)} 
  ({\dot S}^L ({\bf q},t),{\dot S}^L ({\bf q},0)) \, ,
\label{3.17}
\end{equation}               
\begin{eqnarray}
  \Gamma^T(q,g,t) 
  &&={(g_L \mu_B)^2  \over \chi^T (q,g)} 
  ({\dot S}^{T_1} ({\bf q},t),
  {\dot S}^{T_1} ({\bf q},0)) = \nonumber \\
  &&={(g_L \mu_B)^2  \over \chi^T (q,g)} 
  ({\dot S}^{T_2} ({\bf q},t),
  {\dot S}^{T_2} ({\bf q},0)) \, . 
\label{3.18}
\end{eqnarray}
The mode coupling equations resulting from considering two--mode decay processes
have been derived in Refs.~\cite{fe86,fs87} ($\alpha = L,T$) and are given by
\begin{eqnarray}
  \Gamma^{\alpha} (q,g,t) =
  &&2 (2 J a^2)^2 {k_B T \over (g_L \mu_B)^2}
  \int  {v_a d^3 k \over (2 \pi)^3} 
  \sum_{\beta, \sigma} 
  \upsilon_{\beta \sigma}^{\alpha} (k,q,g,\theta) 
  \left( \delta^{\sigma ,T} + \delta^{\alpha ,T} 
        \delta^{\beta ,L} \delta^{\sigma ,L} \right) 
  \nonumber \\
  &&\times 
  {\chi^{\beta} ({\bf k},g) 
   \chi^{\sigma} ({\bf q} - {\bf k},g) \over 
   \chi^{\alpha} ({\bf q},g)} 
  \Phi^{\beta} ({\bf k},g,t) 
  \Phi^{\sigma} ({\bf q} - {\bf k},g,t) \, .
\label{3.19}
\end{eqnarray}
Here the $k$-integration runs over the first Brillouin zone (BZ).
The vertex functions  $\upsilon_{\beta \sigma}^{\alpha}$ for
the decay of the mode $\alpha$ into the modes $\beta$ and
$\sigma$ are given by  \cite{fs87}
\begin{equation}   
  \upsilon_{TT}^L (k,q,g,\eta) =
  2 q^4 \cos^2 \theta {\faka} \, ,
\label{3.20}
\end{equation}   
\begin{equation}   
  \upsilon_{LT}^L (k,q,g,\eta) =
  2 q^4 \sin^2 \theta {\fakb} \, ,
\label{3.21}
\end{equation}   
\begin{equation}   
  \upsilon_{TT}^T (k,q,g,\eta) =
  q^4 \sin^2 \theta {\faka}
  \left( 1 + { q^2 \over 2 {\mid {\bf q} - {\bf k} \mid}^2 }  
  \right) \, ,
\label{3.22}
\end{equation}   
\begin{equation}   
  \upsilon_{LT}^T (k,q,g,\eta) =
  q^4 {\fakb}
  \left( 2 - \left( 1 + { q^2 \over 
  {\mid {\bf q} - {\bf k} \mid}^2  } \right) \sin^2 \theta    
  \right) \, ,
\label{3.23}
\end{equation}   
\begin{equation}   
  \upsilon_{LL}^T (k,q,g,\eta) =
  q^4 \sin^2 \theta {\faka}  
  {q^2 \over 2 {\mid {\bf q} - {\bf k} \mid}^2 } \, ,
\label{3.24}
\end{equation}         
with $\eta = \cos \theta$. In passing we note that there were certain attempts to develop 
a mode coupling theory already twenty years ago, although nobody succeeded
to derive the appropriate mode coupling equations. A type of mode coupling calculation
was used by Huber~\cite{hu71} to determine the uniform spin relaxation for 
temperatures larger than the dipolar crossover temperature. This analysis was
extended by Finger~\cite{fi77b}, who put forward certain scaling estimates and computed 
the uniform spin relaxation in the strong dipolar region.
An attempt to construct a mode coupling theory was launched by Borckmans et al. 
\cite{bwd77}, using an incomplete basis and ending up with equations containing 
undetermined vertices. The theoretical and experimental situation was rather
controversial and no explanation was available for the apparent contradictions
~\cite{k86,m87}. Only in 1986 a complete self--consistent mode coupling 
theory was developed by the authors~\cite{fe86,fs87} and its various properties 
and consequences were studied in detail~\cite{fs88a,fst88,fs88c,fst89}. 
Some of the results were confirmed numerically by Aberger and Folk~\cite{af88b,af89},
and by Kalashnikov and Tret'jakov~\cite{kt90a,kt90b} using analytic approximants.

The mode coupling result for the transport coefficients,
Eqs.~(\ref{3.19}), together with the relation
\begin{equation}   
  {\partial \over \partial t} \Phi^{\alpha} (q,g,t) =
  - \int \limits_{0}^{t} d \tau \Gamma^{\alpha} (q,g,t - \tau) 
    \Phi^{\alpha} (q,g,\tau)
\label{3.25}
\end{equation}   
for the Kubo relaxation functions constitute, as in chapter II, a
complete set of self consistent equations.

As emphasized before, the dipolar interaction introduces a second 
length scale $q_{_{D}}^{-1}$ besides the correlation length $\xi$. This 
entails the following extension of the static scaling law for the
spin susceptibilities
\begin{equation}   
  \chi^{\alpha}(q,\xi,g) = 
  \Upsilon q^{-2} {\hat \chi}^{\alpha} (x,y) \, ,
\label{3.26}
\end{equation}   
with the scaling variables
\begin{equation}   
  x={1 \over q \xi} \hfil \quad {\rm and} \quad \hfil 
  y={q_{_{D}} \over q} \, .
\label{3.27}
\end{equation}
Note, that here and in the remainder of this section we have 
explicitly indicated the dependence of the susceptibility on the 
correlation length. In all other parts of this review we suppress
this dependence for notational convenience.   
Since the vertex functions $\upsilon^{\alpha}_{\beta \sigma}$ 
are proportional to the fourth power of the wave number,  
$\upsilon^{\alpha}_{\beta \sigma} \propto q^4$ and because of the
homogeneity of Eq.~(\ref{3.26}), the relaxation functions and
line widths derived from Eqs.~(\ref{3.19})-(\ref{3.25})
obey the dynamical scaling laws
\begin{equation}   
  \Phi^{\alpha}(ql,gl^2,\omega l^z) =
  l^{-z} \Phi^{\alpha} (q,\xi,g,\omega) \, ,
\label{3.28}
\end{equation}   
and
\begin{equation}   
  \Gamma^{\alpha} (ql,gl^2,\omega l^z) = 
  l^z \Gamma^{\alpha} (q,\xi,g,\omega) \, ,
\label{3.29}
\end{equation}   
with $z={5 / 2}$ and a scaling parameter $l$. We emphasize
that despite of $z$ assuming the isotropic value ${5 / 2}$,
there is a crossover to dipolar critical behavior contained in
the functional form of the correlation functions, as will become
clear below.

An immediate consequence of Eq.~(\ref{3.29}) is the following scaling
property of the characteristic longitudinal and transverse
frequencies $\omega_c^{\alpha}(q,\xi,g)$
\begin{equation}   
  \omega_c^{\alpha}(q,\xi,g) =
  \Lambda q^z \Omega^{\alpha} (x,y) \, ,
\label{3.30}
\end{equation}   
where $\Lambda$ is a non universal coefficient.

Now there are various ways to rewrite the scaling laws Eqs.~(\ref{3.28})
and (\ref{3.29}) by appropriate choices of the scaling parameter $l$. 
If one sets $l=q^{-1}$ one finds
\begin{equation}   
  \Phi^{\alpha}(q,\xi,g,\omega) =
  q^{-z} {\hat \Phi}^{\alpha} 
  ({1 \over q \xi},{g \over q^2},{\omega \over q^z}) \, ,
\label{3.31}
\end{equation}   
and
\begin{equation}   
  \Gamma^{\alpha}(q,\xi,g,\omega) =
  q^{z} {\hat \Gamma}^{\alpha} 
  ({1 \over q \xi},{g \over q^2},{\omega \over q^z}) \, .
\label{3.32}
\end{equation}   
A disadvantage of the representation
given in Eqs.~(\ref{3.31},\ref{3.32}) is that both the
crossover of the time scales and of the shapes of the correlation
functions are intermixed in ${\hat \Phi}^{\alpha}$. Since the time
scales for the isotropic and dipolar critical and hydrodynamic
behavior differ quite drastically, it is more natural to measure
frequencies in units of the characteristic frequencies. Hence we
fix the scaling parameter by the condition
\begin{equation}   
  l^z =
  {1 \over \Lambda q^z \Omega^{\alpha}(x,y)} \, ,
\label{3.33}
\end{equation}   
and find from Eqs.~(\ref{3.31},\ref{3.32}) the scaling forms
\begin{equation}   
  \Phi^{\alpha} (q,\xi,g,\omega) =
  {1 \over \Lambda q^z \Omega^{\alpha} (x,y)}
  \phi^{\alpha} (x,y,\nu_{\alpha} ) \, ,
\label{3.34}
\end{equation}   
and
\begin{equation}   
  \Gamma^{\alpha} (q,\xi,g,\omega) =
  \Lambda q^z \Omega^{\alpha} (x,y) \gamma^{\alpha}
  (x,y,\nu_{\alpha} ) \, ,
\label{3.35}
\end{equation}   
with  the scaling variable for the frequency
\begin{equation}   
   \nu_{\alpha} = 
  {\omega \over \Lambda q^z \Omega^{\alpha} (x,y)} \, . 
\label{3.36}
\end{equation}   
With Eq.~(\ref{3.33}) one has separated the crossover of the frequency 
scales and the crossover of the shapes of the correlation functions. 
The former mainly is contained in $\Omega^{\alpha} (x,y)$ the latter
in $\phi^{\alpha}(x,y,\nu_{\alpha})$.

There is still some freedom in the choice of the
characteristic frequencies $\omega_c^{\alpha}$
in Eq.~(\ref{3.33}); for instance, one could take the half width at
half maximum (HWHM) of the frequency dependent Kubo functions.
This, however, would require to solve Eqs.~(\ref{3.19}) and 
(\ref{3.25}) simultaneously for the time scales and the shapes of the
correlation functions. Therefore, in the following we will use as
characteristic frequencies the half widths resulting from the
Lorentzian approximation for the line shape (see section 3.3).
The Lorentzian line widths qualitatively obey the same scaling laws as
the HWHM and have the same asymptotic (hydrodynamic, dipolar,
isotropic) properties. Thus this choice for the characteristic
frequencies solely is a matter of numerical convenience and does
not introduce any approximations in the final result. From the final 
result one can obtain the HWHM and rewrite the scaling functions in 
terms of these new variables. 

Eqs.~(\ref{3.34}), (\ref{3.35}) and (\ref{3.36}) imply for the Laplace
transformed quantities the scaling laws
\begin{equation}   
  \Phi^{\alpha}(q,\xi,g,t) =
  \phi^{\alpha}(x,y,\tau_{\alpha}) \, ,
\label{3.37}
\end{equation}   
and
\begin{equation}   
  \Gamma^{\alpha}(q,\xi,g,t) =
  \left[ \Lambda q^z \Omega^{\alpha}(x,y) \right]^2
  \gamma^{\alpha}(x,y,\tau_{\alpha}) \, ,
\label{3.38}
\end{equation}   
where the scaled time variables $\tau_{\alpha}$ are
given by
\begin{equation}   
   \tau_{\alpha} =
  \Lambda q^z \Omega^{\alpha} (x,y) t \, .
\label{3.39}
\end{equation}   
One should note that the characteristic time scales $1/[\Lambda
q^z \Omega^{\alpha} (x,y)]$ are different for the longitudinal
and transverse modes. This is mainly due to the non critical 
longitudinal static susceptibility implying that the
longitudinal characteristic frequency $\Lambda q^z \Omega^L
(x,y)$ shows no critical slowing down asymptotically. In other
words for $T=T_c$ and $q \rightarrow 0$ the longitudinal
characteristic frequency does not tend to zero, which
implies an effective longitudinal dynamical critical exponent 
$z_{\rm eff}^L = 0$ for the wave vector dependence
in the limit $q \rightarrow 0$. In comparison, the effective exponent
for the transverse characteristic frequency at $T_c$ shows a 
crossover from $z_{\rm eff}^T = 5/2$ to $z_{\rm eff}^T = 2$ (see also
the following section). This mode coupling result disagrees with
a calculation based on nonlinear spin wave theory, where $z_{\rm eff}^T = 1$
is found in the dipolar region~\cite{Maleev74}.

Inserting Eqs.~(\ref{3.37},\ref{3.38}) together with the static scaling
law (\ref{3.26}) into Eqs.~(\ref{3.19}) and (\ref{3.25}) we find the 
following coupled integro-differential equations
\begin{eqnarray}
  \gamma^{\alpha}  (x,y,\tau_{\alpha}) = 
  &&2 \left( { \pi \over \Omega^{\alpha} (x,y) } \right)^2
  \int \limits_{-1}^{+1} d \eta
  \int \limits_{0}^{\rho_{\rm cut}} d \rho \rho_{-}^{-2}
  \sum_{\beta,\sigma} 
  {\hat \upsilon}_{\beta \sigma}^{\alpha} (y,\rho,\eta)
  \left( \delta^{\sigma ,T} + \delta^{\alpha ,T} 
        \delta^{\beta ,L} \delta^{\sigma ,L} \right) \nonumber \\
  &&\times { {\hat \chi}^{\beta} \left( {x \over \rho},
                               {y \over \rho} \right)
  {\hat \chi}^{\sigma} \left( {x \over \rho_-},
                               {y \over \rho_-} \right)
  \over {\hat \chi}^{\alpha} (x,y) }
  \phi^{\beta} \left( {x \over \rho}, {y \over \rho},
                      \tau_{\alpha \beta}(x,y,\rho) \right)
  \phi^{\sigma} \left( {x \over \rho_-}, {y \over \rho_-},
                      \tau_{\alpha \sigma}(x,y,\rho_-) \right) \, ,
\label{3.40}
\end{eqnarray}
and
\begin{equation}   
  {\partial \over \partial \tau_{\alpha}} 
   \phi^{\alpha} (x,y,\tau_{\alpha}) =
  - \int \limits_{0}^{\tau_{\alpha}} d \tau 
  \gamma^{\alpha} (x,y,\tau_{\alpha} - \tau) 
  \phi^{\alpha} (x,y,\tau) \, ,
\label{3.41}
\end{equation}   
connecting the scaling functions for the transport
coefficients with the scaling functions for the Kubo relaxation
functions. In Eq.~(\ref{3.40}) we introduced the notations $\rho = {k
/ q}$, $\rho_- = {\mid {\bf q} - {\bf k} \mid / q}$, $\eta
= \cos ({\bf q},{\bf k})$ and 
$\tau_{\alpha \beta} (x,y,\mu) = \tau_{\alpha} \mu^z
{\Omega^{\beta} \left( {x / \mu}, {y / \mu} \right) /
\Omega^{\alpha} (x,y)}$.

The non universal frequency scale resulting from the
transformation of Eq.~(\ref{3.19}) in Eq.~(\ref{3.40}) is
\begin{equation}   
  \Lambda = {a^{5 / 2} \over b} \sqrt{ 2 J k_B T  \over 4 \pi^4 }  \, .
\label{3.42}
\end{equation}            
The apparent critical dynamic exponent contained in  Eq.~(\ref{3.40}) equals 
$z = {5 / 2}$. However, as noted before, the crossover to
dipolar behavior is contained in the scaling functions for the
transport coefficients $\gamma^{\alpha}(x,y,\tau_{\alpha})$, the
scaling functions for the Kubo relaxation functions
$\phi^{\alpha} (x,y,\tau_{\alpha})$ and the scaling functions of
the characteristic frequencies $\Omega^{\alpha} (x,y)$. 

The scaled vertex functions appearing in Eq.~(\ref{3.40}) 
read~\cite{fs87}
\begin{equation}   
{\hat \upsilon}_{\beta \beta}^{\alpha} =
  \left[ 2 \eta^2 \delta^{\alpha ,L}
  + \left( 1 - \eta^2 \right)                                    
    \left( \delta^{\beta ,T} + {1 \over 2 \rho_{-}^2} \right)
    \delta^{\alpha ,T} \right]
  \left( \rho \eta - {1 \over 2} \right)^2 \, ,
\label{3.43}
\end{equation}   
\begin{equation}   
  {\hat \upsilon}_{LT}^{\alpha} =
  \left[ 2 \left(1 - \eta^2 \delta^{\alpha ,L} \right)
  - \left( 1 - \eta^2 \right)                                    
    \left( 1 + {1 \over \rho_{-}^2} \right)
    \delta^{\alpha ,T} \right]
  \left( \rho \eta - {1 \over 2} + {y^2 \over 2} \right)^2 \, ,
\label{3.44}
\end{equation}                            
which are related to the vertex functions $\upsilon_{\beta
\sigma}^{\alpha}$ of Eqs.~(\ref{3.20}-\ref{3.24}) by $\upsilon_{\beta
\sigma}^{\alpha} = q^4 {\hat \upsilon}_{\beta \sigma}^{\alpha}$.
For both longitudinal and transverse modes, the dipolar
interaction enters only in vertices involving decays into a
longitudinal and a transverse mode, since the dipolar interaction
enters the Hamiltonian only through the longitudinal modes.

Because the $k$-integration is restricted to the Brillouin zone the
$\rho$-integration of Eq.~(\ref{3.40}) contains the cutoff
\begin{equation}   
  \rho_{\rm cut} = {q_{BZ} \over q} = {q_{BZ} \over q_{_{D}}} y  \, ,
\label{3.45}
\end{equation}       
where $q_{BZ}$ denotes the boundary of the first Brillouin zone.
All other material dependent parameters are contained in the
frequency scale $\Lambda$. The cutoff is
important for small times, because the integrand of Eq.~(\ref{3.40})
is of order 1 for $t=0$ and $\rho \gg 1$. Hence for small
times also wave vectors near the zone boundary contribute to the
relaxation mechanism.

As explained before, in the numerical solution of the MC--equations
one has taken~\cite{fst88,fst89} for the characteristic frequencies
the line widths resulting from the Lorentzian approximation 
of the MC equations, i.e.,
\begin{equation}   
  \Omega^{\alpha} (x,y) = \gamma^{\alpha}_{\rm lor} (x,y) \, .
\label{3.46}
\end{equation}   
Using as input the solution of the mode coupling equations in the
Lorentzian approximation one can  solve the
complete set of MC--equations for different
values of $q_{BZ}$. Because there are three scaling variables
($x$, $y$ and $\nu_{\alpha}$) it is impossible to present here
all numerical results. Instead, in chapter IV, we will
concentrate on a limited number of temperatures and wave vectors
motivated by the available experiments on the substances of
primary interest $EuS$, $EuO$ and $Fe$.

\subsubsection{Lorentzian Approximation}

For later reference  we want to close this section with quoting the 
results from the so called
Lorentzian approximation \cite{fs87,fs88a}. 
The results of the numerical solution of
the full mode coupling equations will be discussed in chapter IV
in conjunction with the experimental data.

If the transport coefficients
vary only slowly with $\omega$ we may approximate the relaxation
functions by Lorentzians, i.e., we replace the transport
coefficients by their values at $\omega= 0$    
\begin{equation}   
\Gamma^L_{\rm lor} (q,\xi,g) = \Gamma^L (q,\xi,g,\omega =0)
  ,\quad \Gamma^T_{\rm lor} (q,\xi,g) = \Gamma^T (q,\xi,g,\omega =0).
\label{3.47}
\end{equation}   
Despite a Lorentzian being not the correct shape of the correlation
function for all values of the scaling variables, the resulting line 
widths obtained in the Lorentzian approximation already capture most of 
the crossover in the time scale.

The Lorentzian line widths obey the scaling law
\begin{equation}   
  \Gamma^{\alpha}_{\rm lor} (q,\xi,g) =
  \Lambda q^z \gamma^{\alpha}_{\rm lor} (x,y).
\label{3.48}
\end{equation}   
From Eq.~(\ref{3.20}) it is then easily inferred that the 
scaling functions of the Lorentzian line widths
$\gamma_{\rm lor}^{\alpha} (x,y)$ are determined by the coupled
integral equations 
\begin{eqnarray}
  \gamma_{\rm lor}^{\alpha} (x,y) =
  &&{ 2 \pi^2 \over {\hat \chi}^{\alpha} (x,y) }
  \int \limits_{-1}^{+1} d \eta
  \int \limits_{0}^{\infty} d \rho \rho_{-}^{-2}
  \sum_{\beta} \sum_{\sigma} 
  {\hat \upsilon}_{\beta \sigma}^{\alpha} (y,\rho,\eta)
\nonumber \\
  &&\left( \delta^{\sigma ,T} + \delta^{\alpha ,T} 
        \delta^{\beta ,L} \delta^{\sigma ,L} \right)
  { {\hat \chi}^{\beta} \left( {x \over \rho},
                               {y \over \rho} \right)
  {\hat \chi}^{\sigma} \left( {x \over \rho_-},
                               {y \over \rho_-} \right)
  \over \rho^{5 / 2} \gamma_{\rm lor}^{\beta}
  \left( {x \over \rho}, {y \over \rho} \right)
  + \rho_{-}^{5 / 2} \gamma_{\rm lor}^{\sigma}
  \left( {x \over \rho_-}, {y \over \rho_-} \right) }.
\label{3.49}
\end{eqnarray}
As summarized in Table~3.1 the mode coupling equations 
(\ref{3.49}) can
be solved analytically in the dipolar (D) and isotropic (I)
critical (C) and hydrodynamic (H) limiting regions. These are
defined by  DC: $y \gg1$, $x \ll 1$; IC: $y \ll 1$, $x \ll 1$;
DH: $y \gg x$, $x \gg 1$; IH: $y \ll x$, $x \gg 1$.

Concerning the critical dynamical exponent one finds for the
longitudinal line width a crossover from $z = {5 / 2}$ in the
isotropic critical region to $z = 0$ in the dipolar critical
region, whereas for the transverse line width the crossover is
from $z = {5 / 2}$ to $z = 2$. The precise position of this
crossover can only be determined numerically. 

The numerical solution \cite{fe86,fs87} of Eq.~(\ref{3.49}) shows that 
the dynamic crossover for the transverse width is shifted to smaller wave vectors by 
almost one order of magnitude with respect to the static crossover, 
whereas the crossover for the longitudinal width occurs at the static
crossover. For the numerical solution of the mode coupling equations it is
convenient to introduce polar coordinates
$$ 
  r = \sqrt{x^2+y^2}, \quad {\rm and} \quad \varphi = \arctan (y/x) \, .
$$
The transverse and longitudinal scaling functions $\gamma^T(r, \varphi)$ and 
$\gamma^L (r, \varphi)$ are shown in Figs.~3.1 and 3.2 as a function of the 
radial scaling variables $r$ and $\varphi$. A different representation
of the results can be given by plotting the linewidth versus the single
scaling variable $x$ for several values of $\varphi$. This is shown in 
Figs.~3.3-5, where we have drawn the two-parameter scaling functions 
$\gamma_{\rm lor}^{L,T}(x,y)$ in units of the value at criticality 
$\gamma_0 = \gamma_{\rm lor}^{L(T)} (0,0) \cong 5.1326$. The physical content 
of the two parameter scaling surfaces is illustrated best by considering
cuts for fixed $q_{_{D}}$ and various temperatures. In Figs.~3.3,4 the
scaling functions versus $x = 1/q\xi$ are displayed for different
values of $\varphi = \arctan (q_{_{D}} \xi) = N \pi / 20$ with $N =
0,1,...,9$. For $\varphi = 0$, corresponding to vanishing dipolar coupling $g$, 
the scaling functions coincide with the Resibois-Piette scaling function.
If the strength of the dipolar
interaction $q_{_{D}}$ is finite, the curves approach the Resibois-Piette
scaling function for small values of the scaling variable $x$ and
deviate therefrom with increasing $x$. For a given material, $q_{_{D}}$ 
is fixed and the parametrisation by $\varphi$ corresponds to a
parametrisation in terms of the reduced temperature $(T-T_c)/T_c$.

To examine the dipolar crossover precisely at the Curie point,
Fig.~3.5 displays the scaling function for the transverse and
longitudinal width at $T=T_c$ versus the wave number, i.e., $y^{-1} =
q/q_{_{D}}$. This graph shows that the crossover from the isotropic
Heisenberg to dipolar critical dynamics in the transverse line width
occurs at a wave number, which is almost one order
of magnitude smaller than the static crossover wave vector $q_{_{D}}$.
The crossover of the longitudinal width, from $z=2.5$ to $z=0$, is
more pronounced and occurs in the intermediate vicinity of $q_{_{D}}$.
The reason for the different location of the dynamic crossover is
mainly due to the fact that it is primarily the longitudinal static
susceptibility which shows a crossover due to the dipolar 
interaction. Since the change in the static critical exponents is
numerically small the transverse static susceptibility is nearly the
same as for ferromagnets without dipolar interaction \cite{af73}. 
Hence the crossover in the transverse width is purely a dynamical 
crossover,
whereas the crossover of the longitudinal width being proportional
to the inverse static longitudinal susceptibility is enhanced by the
static crossover. In order to substantiate these arguments we have
plotted in the inset of Fig.~3.5 the scaling functions of the Onsager
coefficients ${\hat \chi}^\alpha \gamma^\alpha$ at the critical
temperature versus $q/q_{_{D}}$, showing only the dynamical crossover.

\subsubsection{Selected Results of the Complete Mode Coupling Equations}

All the scaling functions for the dipolar ferromagnet depend on the 
three scaling variables $x = 1 / q \xi$, $y = q_{_{D}}/ q$ and 
$\nu_\alpha = \omega / \Lambda q^z  \Omega^\alpha (x,y)$. Therefore it is
impossible to present all the results obtained from the numerical 
solution of the complete mode coupling equations. In this section we 
intend to review the most important features of the shape and line width 
crossover reported in Refs.~\cite{fst88,fst89}. Further results and 
discussion will be presented in the next chapter in conjunction with a 
comparison with experimental data.

The results from the Lorentzian approximation show that the
dipolar line width crossover in the transverse line width sets in at a
wave vector almost one order of magnitude smaller than the dipolar wave
vector $q_{_{D}}$. In order to get information about the line shape,
one has to dismiss this approximation and solve the complete set of mode 
coupling equations, Eqs.~(\ref{3.40}) and (\ref{3.41}). This has been 
achieved in Ref.~\cite{fst88,fst89}, and one finds the following
crossover scenario. First of all, the crossover in the line shape sets in
at wave vectors of the order of the dipolar wave vector.

Let us first consider the case of temperatures very close to $T=T_c$.
Fig.~3.6 and 3.7 show the transverse and longitudinal scaling functions
$\phi^\alpha (r,\varphi,\tau_\alpha)$ versus the scaling variables $r$
and $\tau_\alpha$ for $\varphi=1.49$. Referring to $EuO$, characterized
by the (non universal) parameters $q_{_{D}} = 0.147 \AA^{-1}$, $T_c =
69.1 K$ and $\xi_0 = 1.57 \AA$, this corresponds to a temperature 
$T \approx (1 + 0.003) T_c $. The line shapes of the longitudinal and transverse
relaxation function agree in the isotropic Heisenberg limit, i.e., for
$r \rightarrow 0$ corresponding to large values of the wave vector $q$ 
($q \gg q_{_{D}}$).
In this limit the dipolar interaction becomes negligible and the shape is of 
the Hubbard-Wegner type as discussed in section II. Upon increasing the value of the
scaling variable $r$ the line shape of the transverse and longitudinal relaxation
function become drastically different. Whereas the transverse relaxation function shows
a nearly exponential decay, pronounced over-damped oscillations show up for the
transverse relaxation function. The shape crossover is also shown in Fig.~3.8, where
the transverse and longitudinal relaxation functions are plotted versus the scaling
variable $\tau_\alpha$ for three different values of the scaling variable
$r=\sqrt{x^2 + y^2}$ ($r=0$, $r=1$, and $r=10$).

For temperatures well separated from $T_c$, i.e. $q_{_{D}} \xi \ll 1$, the dipolar
interaction becomes negligible. Hence, the difference in the shape crossover of the 
longitudinal and transverse relaxation function diminishes with decreasing 
$q_{_{D}} \xi$. For $q_{_{D}} \xi \ll 1$ the shape crossover as a function
of $r$ corresponds to the crossover from the critical (Hubbard-Wegner) shape to the 
hydrodynamic shape as discussed in section II (Note that for $q_{_{D}} \xi \ll 1$, the 
scaling variable $r$ reduces to $x=1/q\xi$.). Figs.~3.9 and 3.10 show the shape
crossover for $q_{_{D}} \xi = 3.52$ ($\varphi=1.294$) for the transverse and 
longitudinal relaxation function, respectively. Figs.~3.11 and 3.12 display the 
corresponding crossover for $q_{_{D}} \xi = 1.46$ ($\varphi=0.97$). In the latter case
the shape crossover of the transverse and longitudinal relaxation function are already
quite similar and almost identical to the shape crossover found for the isotropic
Heisenberg ferromagnet without dipolar interaction. The various crossover scenarios
can also be read off from Fig.~3.8.

Finally let us add a comment on how the line shape crossover affects the line width
crossover. It has been shown~\cite{fst88,fst89} (see Section IV) that there are only 
slight changes of the line width crossover, when the line width is determined from the 
full solution of the mode coupling equations, as compared to the widths obtained from 
the Lorentzian approximation. Roughly speaking, the overall effect of taking into 
account the correct line shape is approximately a shift by a constant factor. For 
details we refer the reader to section IV.

\subsection{Spin Wave Theory in the Ferromagnetic Phase}

Holstein and Primakoff~\cite{Holstein40} have investigated the dynamics of dipolar
ferromagnets far below $T_c$ using linear spin--wave theory. Upon neglecting 
fluctuations in the longitudinal component, $S^z_l \approx S$, they get the 
following equations of motion for the transverse spin fluctuations (see also
Ref.~\cite{Lovesey91})
\begin{eqnarray}
 i {d \over dt} S_{\bf k}^+ = &&A_{\bf k} S_{\bf k}^+ + 
                               B^\star_{\bf k} S_{-\bf q}^- \, ,
 \label{3.50} \\
 i {d \over dt} S_{\bf k}^- = &&A^\star_{\bf k} S_{-\bf q}^- + 
                               B_{\bf k} S_{\bf k}^+ \, ,
\label{3.51}
\end{eqnarray}
in terms of the raising and lowering operators $S^{\pm}_l = S_l^x \pm i S_l^y$. 
From these equations of motion the spin--wave dispersion follows (see also
Ref.~\cite{Keffer66})
\begin{equation}
  \epsilon_{\bf k} = \sqrt{A_{\bf k} + |B_{\bf k}|} \, 
                     \sqrt{A_{\bf k} - |B_{\bf k}|} \, .
\label{3.52}
\end{equation}
The coefficients $A_{\bf k}$ and $B_{\bf k}$ are given in terms of the exchange
interaction $J_{\bf k} = 2 S \sum_m J_{lm} e^{i {\bf k} \cdot ({\bf x}_l - {\bf x}_m)}$
and the dipolar interaction tensor $A^{\alpha \beta}_{\bf k}$ ( see Eq.~(\ref{3.3}) )
\begin{eqnarray}
  A_{\bf k} &&= g_L \mu_B H_0 + (J_0 - J_{\bf k}) + (2E_0 + E_{\bf k}) \, ,
  \label{3.53} \\
  E_{\bf k} &&=  S A^{zz}_{\bf k} \, , 
  \label{3.54} \\
  B_{\bf k} &&=- S 
  \left( A^{xx}_{\bf k} -  A^{yy}_{\bf k} -2 i  A^{xy}_{\bf k} \right)  
  = S \sqrt{ (A^{xx}_{\bf k} -  A^{yy}_{\bf k})^2 + 4 {A^{xy}_{\bf k}}^2 } 
    e^{-2i \varphi_{\bf k}} \, .
\label{3.55}
\end{eqnarray}
For crystals with cubic symmetry the dipolar tensor $A^{\alpha \beta}_{\bf k}$ is given 
by  $A^{\alpha \beta}_{\bf k} \approx {1 \over 2} (g_L \mu_B)^2 {4 \pi \over 3} 
\left(\delta^{\alpha \beta} - {k^\alpha k^\beta \over k^2} \right)$ near the zone
center $ka \ll 1$, plus small structure dependent terms proportional to $k^2$ 
(see Section III.1)). Note however that the dipolar tensor becomes severely structure 
dependent (requiring numerical evaluation, see Ref.~\cite{Cohen55}) as ${\bf k}$ 
approaches the zone boundary. 

Hence, for $ka \ll 1$ one finds for crystals with cubic symmetry (for more 
details we refer to the review by Keffer~\cite{Keffer66})
\begin{eqnarray}
  A_{\bf k} = &&g_L \mu_B H_0 + (J_0 - J_{\bf k}) + 
              {1 \over 2} g_L \mu_B M_0 
              \left[ 4 \pi - 4 \pi (k^z / k)^2 \right] \, ,
 \label{3.56} \\
  B_{\bf k} =&&{1 \over 2} g_L \mu_B M_0 \left[ 4 \pi { (k^x)^2 - i (k^y)^2 \over k^2} 
               \right] 
            = 2 \pi g_L \mu_B M_0 \sin^2 \Theta_{\bf k} \exp (-2 i \varphi_{\bf k}) \, ,
 \label{3.57}
\end{eqnarray}
where $\Theta_{\bf k}$ and $\varphi_{\bf k}$ define the orientation of the wave vector
${\bf k}$ with respect to the $z$--axis; $M_0$ is the saturation magnetization.
Note, that the latter equations are strictly valid for long thin samples only. 
Otherwise, one has to take into account demagnetization effects, which amount in an 
additional magnetic field $H_0 \rightarrow H_0 - N^z M_0$, where $N^z$ is a 
demagnetization factor (~\cite{Keffer66}, see also Ref.~\cite{Anderson55}).

Using the formalism of Holstein and Primakoff~\cite{Holstein40} the influence of the 
dipole-dipole interaction on the inelastic scattering of neutrons has been investigated
by Elliott and Lowde~\cite{Elliott55} (see also \cite{Lowde65}).

The neutron scattering cross section is proportional to the correlation function 
corresponding to the spin fluctuations (see e.g. Ref.~\cite{Lovesey84,Collins89}). 
If one considers $N$ identical atoms with fixed positions one finds for the 
magnetic partial differential cross section for unpolarized neutrons in forward
direction 
\begin{eqnarray}
  {d^2 \sigma \over d \Omega d E^\prime} 
  =&&{k^\prime \over k} {N \over \hbar} r_0^2 |F ({\bf Q}) |^2 
  {1 \over N} \sum_{l l^\prime} e^{ i {\bf Q} \cdot ({\bf x}_l - {\bf x}_{l^\prime}) }
  \int_{- \infty}^{+ \infty} {dt \over 2 \pi} e^{ -i \omega t} 
  \langle S^T_l (t) S^T_{l^\prime} (0) \rangle
  \nonumber \\
  =&&{k^\prime \over k} {N \over \hbar} r_0^2 |F ({\bf Q}) |^2  
    S^T({\bf Q},\omega) \, ,
\label{3.58}
\end{eqnarray}
where $S^\perp_l (t)$ is the component of the spin density perpendicular to the momentum 
transfer (scattering vector) ${\bf Q} = {\bf k} - {\bf k}^\prime$, and 
$\hbar \omega$ is the neutron energy transfer. The length $r_0 = - 5.391 fm$ is
analogous to the nuclear scattering length (see also chapter IV), and $F ({\bf Q})$
the magnetic form factor. We have also defined the transverse magnetic scattering function 
$S^T({\bf Q},\omega)$ by 
\begin{equation}
  S^T({\bf Q},\omega) = 
  \left(\delta^{\alpha \beta} - {Q^\alpha Q^\beta \over Q^2} \right)
  \int_{- \infty}^{+ \infty} {dt \over 2 \pi} e^{ -i \omega t} 
  \langle S^\alpha_{-\bf q} (t) S^\beta_{\bf q} (0) \rangle \, ,
\label{3.59}
\end{equation} 
where ${\bf Q} = {\bf q} + {\mbox{\boldmath $\tau$}}$ with ${\bf q}$ the wave
vector of the magnon and ${\mbox{\boldmath $\tau$}}$ a reciprocal lattice vector. 

The equations of motion, Eqs.~\ref{3.50} and \ref{3.51}, 
are diagonalized in terms of the Bose operators $a$ and
$a^\dagger$ that satisfy
\begin{equation}
  [a_{\bf q}, a^\dagger_{{\bf q}^\prime}] = \delta_{ {\bf q} {\bf q}^\prime }
\label{3.60}
\end{equation}
with 
\begin{equation}
  S^+_{\bf q} = u_{\bf q} a_{\bf q} + v_{\bf q} a^\dagger_{-\bf q} \, .
\label{3.61}
\end{equation}
A convenient choice for the coefficients $u_{\bf q}$ and $v_{\bf q}$ is
\cite{Keffer66,Lovesey91}\footnote{Holstein and Primakoff~\cite{Holstein40} have 
ignored the phase  relationship between $u_{\bf q}$ and $v_{\bf q}$, without which an
incorrect expression for the scattering cross section between spin--waves and 
neutrons would be obtained. This error has been corrected by 
Lowde~\cite{Elliott55,Lowde65,Keffer66}.}.
\begin{eqnarray}
  u_{\bf q}^2 &= (2SN) (A_{\bf q} + \epsilon_{\bf q})/ 2 \epsilon_{\bf q} \, 
  &\label{3.62} \\
  v_{\bf q}   &= -u_{\bf q}B^\star_{\bf q}/(A_{\bf q}+ \epsilon_{\bf q}) \, .
  &\label{3.63}
\end{eqnarray}

The contributions from single--spin--wave events to the dynamic structure factor was 
calculated by Elliot and Lowde~\cite{Elliott55,Lowde65} 
\begin{equation}
  S^T({\bf Q},\omega) = {S \over 2} (n_{\bf q} + 1)
  \left[ 
  \left(1+{(Q^z)^2 \over Q^2} \right) {A_{\bf q} \over \epsilon_{\bf q}} +
  \left(1-{(Q^z)^2 \over Q^2} \right) {|B_{\bf q}| \over \epsilon_{\bf q}} 
  \cos[2(\varphi_{\bf Q} - \varphi_{\bf q})] 
  \right] \delta(\omega + \epsilon_{\bf q})\, ,
\label{3.64}
\end{equation}
where $n_{\bf q}$ is the occupation number of the magnon oscillator with wave vector 
${\bf q}$. The second term in the dynamic structure factor 
is due to the dipolar interaction. For zero dipolar interaction or for wave vectors 
larger than the dipolar wave vector $q_{_{D}}$ the dynamic structure factor reduces to 
a simple angular distribution proportional to $1+(Q^z / Q)^2$. When dipolar effects 
become of importance, the angular dependence of the scattered intensity gets quite 
complicated.

More recently Lovesey et al.~\cite{Lovesey91,Trohidou93} have extended Lowde's analysis 
to scattering of polarized neutrons, and a discussion of the longitudinal cross section
$\langle S_l^z S_{l^\prime}^z \rangle$, which contains two--spin--wave scattering 
events. Furthermore, they analyze the static susceptibilities in the framework of 
linear spin--wave theory, correcting work by Toh and Gehring~\cite{Toh90} who used the 
incorrect expressions for the coefficients $u_{\bf q}$ and $v_{\bf q}$ from 
Ref.~\cite{Holstein40}.

As a first step beyond the spin-wave theories described above, Toperverg, and 
Yashenkin~\cite{Toperverg93} have used the perturbation approach developed by Vaks, 
Larkin and Pikin~\cite{vlp68}, to investigate the frequency dependence of the 
uniform transverse and longitudinal susceptibilities~\footnote{An excellent 
review on the theoretical and experimental work prior to 1984 was given by 
Maleev \cite{maleev87}, with a particular emphasis on Green's function 
techniques.}. 
The applicability of their 
results is mainly restricted to low and intermediate temperatures. Their perturbation
approach for the dipolar interaction breaks down not only close to the critical
temperature, but also at any temperature for low frequencies. For this parameter 
regime a non perturbative approach like mode coupling theory is needed to account for 
the strong fluctuations. A first attempt towards such a theory has been made some
time ago by Raghavan and Huber~\cite{rh76}. There are, however, several shortcomings in
their approach. First of all, the static susceptibilities used in their analysis do not 
account for the coexistence anomalies and dipolar crossover properly. Instead the 
longitudinal susceptibility is taken of Ornstein-Zernike form $\chi_L \propto 1/(q^2 +
\xi^{-2})$, which neglects dipolar crossover effects as well as the by now well known 
coexistence anomaly  $\chi_L \propto 1/q$ in the limit $q \rightarrow 0$. The 
expression used for the transverse susceptibility is valid in certain limits only. 
Furthermore, in the presence of dipolar interaction, for a general angle between the 
wave vector and the direction of the spontaneous magnetization, there are three and not
just two non degenerate eigenvalues for the static susceptibility matrix. This fact has 
been neglected completely. Also, they didn't evaluate the complete functional form of
the relaxation functions, but used instead a parametrisation, which is strictly valid
in the hydrodynamic regime only, and calculated the corresponding parameters.
Nevertheless, the theory seems to give a quite reasonable description
of the data obtained by a neutron scattering study on $EuO$~\cite{Passell72} in the
range $q \xi \leq 1$ and $q_{_{D}} \xi \leq 1$ for not too small values of the wave
vector $q$. Beyond this range the approximate treatment of the dipolar interaction in 
Ref.~\cite{rh76} breaks down. 

More recently Lovesey~\cite{Lovesey93} reported on an approximative mode coupling 
approach for dipolar ferromagnets below $T_c$. Similar to Raghavan and Huber the spin 
wave dispersion relation used by Lovesey is applicable for not too small wave vectors 
only. The analysis in Ref.~\cite{Lovesey93} is restricted to the exchange region 
$q_{_{D}}\xi \leq 1$. This is due to the assumption made in Ref.~\cite{Lovesey93}, that
the only non vanishing relaxation kernels are those for the spin fluctuations 
longitudinal and transverse to the direction of the spontaneous magnetization. 
Neglecting off-diagonal matrix elements excludes the applicability of the theory to the
dipolar region. Note, that in the dipolar region close to $T_c$ the memory function and
the relaxation functions become -- similar to the situation above $T_c$ -- diagonal in 
terms of the spin fluctuations longitudinal and transverse with respect to the 
wave vector and not to the spontaneous magnetization. Furthermore, the analysis in  
Ref.~\cite{Lovesey93} is restricted to very small wave vectors. In this limit,  
however, the expressions for the static susceptibility and the spin--wave dispersion 
relation used in Ref.~\cite{Lovesey93} are not valid, since they do not account for the
presence of the subtle combined effects of Goldstone modes and dipolar anisotropy. 
Also, no self-consistent solution but only a first iteration of the mode coupling
equations, based on neglecting the damping, is performed.
In summary, a thorough mode coupling analysis of the effects of the dipolar
interactions in the ferromagnetic phase is still a very challenging theoretical problem
and the topic of ongoing research~\cite{schinz94b}.

\subsection{Renormalization Group Theory of Time--Dependent Ginzburg--Landau Models in
the Ferromagnetic Phase}

The spin--wave theory, reviewed in the preceding section, is only a first step towards a
more rigorous theory of the critical dynamics of dipolar ferromagnets below the Curie
temperature. For a thorough understanding of the dynamics in the ferromagnetic phase
a renormalization group theory or a mode coupling approach analogous to section III.B,
which takes into account the effects of the critical fluctuations,
would be necessary. A detailed analysis requires a treatment of a modified model 
$J$~\cite{hh77} appropriate for the dynamics of isotropic ferromagnets, where dipolar
forces are included. 

Recently, the effects of the dipolar interaction on the critical dynamics of the
$n$--component time--dependent Ginzburg--Landau models (model A~\cite{hh77}) 
below the critical temperature have been studied within a generalized minimal
subtraction scheme~\cite{Taeuber93}. The corresponding Langevin equation of motion 
reads (Model A: $a=0$; Model B: $a=2$)
\begin{equation}
  {\partial {\bf S} ({\bf x},t) \over \partial t} =
  - \lambda (i \bfnabla)^a {\delta H [ \{ S^\alpha \} ] \over 
                            \delta {\bf S} ({\bf x},t) } 
  + {\gvect {\zeta}} ({\bf x},t) \, ,
\label{3.65}
\end{equation}
where the stochastic forces are characterized by a Gaussian probability distribution 
function with zero mean and variance
\begin{equation}
  \langle  \zeta^\alpha ({\bf x},t) \zeta^\beta ({\bf x}^\prime,t^\prime) \rangle =
  2 \lambda k_B T (i \bfnabla)^a \delta^{(3)}({\bf x} - {\bf x}^\prime)
  \delta(t - t^\prime)  \delta^{\alpha \beta}\, .
\label{3.66}
\end{equation}
The Ginzburg--Landau effective free energy functional reads
\begin{eqnarray}
  H [ \{ S^\alpha \} ] =
  {1 \over 2} \int_q 
  \Biggl[ 
  &&\sum_{\alpha,\beta=1}^{min(d,n)} \left[ (r+q^2)   P^T_{\alpha \beta} +
                                          (r+g+q^2) P^L_{\alpha \beta} 
                                     \right]
    S^\alpha ({\bf q}) S^\beta (-{\bf q})  \nonumber \\
  && + \sum_{\alpha=min(d,n)+1}^{n} S^\alpha ({\bf q}) S^\alpha (-{\bf q}) 
  \Biggr] \, ,
\label{3.67}
\end{eqnarray}
where the general situation $n \neq d$ is considered. Here $P^T_{\alpha \beta} =
\delta^{\alpha \beta} - q^\alpha q^\beta / q^2$ and $P^L_{\alpha \beta} =
q^\alpha q^\beta / q^2$ denote the transverse and longitudinal projectio operator.

Those relaxational models neglect mode coupling
terms resulting from the reversible motion of the spins in the local magnetic field.
It is expected~\cite{Taeuber93}, however, that most of the conclusions based on the 
relaxational models will also hold for models with mode coupling terms.

As a consequence of the spontaneously broken symmetry there are $n-1$ massless
Goldstone modes in the ordered phase of ideally isotropic systems. These massless
modes lead to infrared singularities (coexistence singularities) in certain 
correlation functions for {\it all} temperatures below $T_c$. Based on the analysis
of the effects of the critical and Goldstone fluctuations for the $O(n)$--symmetric
time--dependent Ginzburg--Landau models~\cite{Taeuber92}, it has been 
investigated~\cite{Taeuber93} how the coexistence anomalies are modified when dipolar 
forces or weak anisotropies are included. For later reference the coexistence anomalies
of the isotropic relaxational models are collected in Table~VI.

The analysis in Ref.~\cite{Taeuber93} shows that the influence of the dipolar 
interaction on the
coexistence singularities is quite subtle. Although the model explicitly breaks the
$O(n)$--symmetry, not all transverse modes lose their Goldstone character, but only 
their effective number is reduced by one. Hence, while for $n=2$ a crossover to an
asymptotically uncritical theory takes place, for $n \geq 3$ coexistence anomalies
persist, governed by a dipolar coexistence fixed point~\cite{Taeuber93}.

Below $T_c$ there are two preferred axes, the axis defined by the direction of the
spontaneous magnetization and the axis defined by the wave vector ${\bf q}$. This
leads to a complex structure of the correlation functions already on the harmonic level.
It is quite remarkable, however, that a one--loop theory for the two--point cummulants
becomes an exact representation in the ordered phase in the coexistence limit
(${\bf q} \rightarrow 0$ and $\omega \rightarrow 0$)~\cite{Taeuber93}.

For $n=d \geq 3$ it
is found~\cite{Taeuber93}, that the power laws characteristic of the coexistence
anomalies are {\it not} changed by the presence of the dipolar interaction. Hence the
same power laws as in Table VI apply to the dipolar case also. It is also interesting 
to note that there is the following exact amplitude ratio of the longitudinal
response function in the dipolar and the isotropic case:
\begin{equation}
 {\chi_L ({\bf q}, \omega)_{\rm dipolar} \over 
  \chi_L ({\bf q}, \omega)_{\rm isotropic}} =
 {n-2 \over n-1} \, .
\label{3.68}
\end{equation}
For $n=3$ this universal amplitude ratio is $1/2$, in accord with the results of
Pokrovsky\cite{Pokrovsky79} and Toh and Gehring~\cite{Toh90}, obtained in the framework
of a spin--wave theory.

As is apparent from the value of the dipolar coexistence fixed point $u_{CD}^\star =
{6 (4-d) / (n-2)}$, the situation $n=2$ requires a separate discussion. In this case
there are no massless modes left, since the dipolar interaction reduces the number of
Goldstone modes from $n-1$ to $n-2$. As a consequence the crossover below $T_c$ is from 
a critical theory at $T_c$ to a Gaussian theory, i.e. the fluctuations die out on
leaving the critical region. A qualitative summary of the various crossover scenarios
is given in Table VII~\cite{Taeuber93}.

\newpage

%%%%%%%%%%%%%%%%%%%%%%%%%%%%%%%%%%%%%%%%%%%%%%%%%%%%%%%%%%%%%%%%%%%%

% the tables follow here

\begin{table}
\setdec 0.00
\caption{Coefficients in the Taylor expansion of $A_{\bf q}^{\alpha \beta}$ for
three--dimensional cubic lattices, taken from~[Aharony73a,Aharony73b]. $c$ denotes 
the coordination number, $v_a$ the volume of the primitive unit cell and $\alpha_i$
are lengths characterizing the dipolar interaction.}
\bigskip \bigskip
\begin{tabular}{lccc}
{\rm Lattice}     &  {\rm sc} &  {\rm bcc}  & {\rm fcc}  \\ 
\tableline
$c$           & $6$      & $8$             & $12$        \\
$v_a [a^3]$     & $1$      & $4/3\sqrt{3}$   & $1/\sqrt{2}$   \\
$\alpha_1 [a]$    & $1.2755$ & $1.7420$        & $2.8313$      \\
$\alpha_2 [a]$    & $0.1649$ & $-0.321$        & $-0.335$       \\
$\alpha_3 [a]$    & $1.7700$ & $0.8210$        & $1.823$         \\
\end{tabular}
\label{tableIV}
\end{table}

\vskip 1truecm

\begin{table}
\setdec 0.00
\caption{Asymptotic behavior of
the scaling functions for the longitudinal and transverse Lorentzian
line width in the paramagnetic phase. The different regions dipolar
critical (DC), isotropic critical (IC), dipolar hydrodynamic (DH),
and isotropic hydrodynamic (IH) are defined by  DC: 
$y \gg1$, $x \ll 1$; IC: $y \ll 1$, $x \ll 1$;
DH: $y \gg x$, $x \gg 1$; IH: $y \ll x$, $x \gg 1$.}
\bigskip \bigskip
\begin{tabular}{lcc}
   &$\gamma^T$    &$\gamma^L$ \\
\tableline
DC &$y^{1/2}$     &$y^{5/2}$  \\
IC &1             &1          \\
DH &$y^{1/2} x^2$ &$y^{5/2}$  \\
IH &$x^{1/2}$     &$x^{1/2}$  \\
\end{tabular}
\label{tableV}
\end{table}

\vskip 1truecm

\begin{table}
\setdec 0.00
\caption{Coexistence anomalies of the isotropic relaxational models for the
longitudinal dynamic susceptibility $\chi_L ({\bf q},\omega)$ and correlation function 
$G_L ({\bf q},\omega)$.}
\bigskip \bigskip
\begin{tabular}{lll}
   &Model A    &Model B \\
\tableline
$a$ & $0$ & $2$ \\
$\chi_L ({\bf q}, 0)$                       &  $\propto q^{d-4}$           &  $\propto q^{d-4}$                 \\
$Re\left[\chi_L (0, {\omega/q^a})\right]$   &  $\propto \omega^{(d-4)/2}$  &  $\propto (\omega/q^2)^{(d-4)/2}$  \\
$G_L (0, {\omega/q^a})$                     &  $\propto \omega^{d/2-3}$    &  $\propto (\omega/q^2)^{d-4}$      \\
\end{tabular}
\label{tableVI}
\end{table}

\vskip 1truecm

\begin{table}
\setdec 0.00
\caption{The influence of the dipolar interaction on the coexistence anomalies.
The table summarizes the various crossover scenarios possible if the number of
components $n$ and the dimensionality $d$ of space is varied.}
\bigskip \bigskip
\begin{tabular}{lcc}
   &$d=2$ $d=3$    &$d=4$ \\
\tableline
$n = 1$    & Crossover to a Gaussian theory & as for $g=0$ \\
$n = 2$    & Crossover to a Gaussian theory & no anomalies \\
$n \geq3$  & $u_{CD}^\star = {6 (4-d) \over n-2} \rightarrow {\rm anomalies}$
           & logarithmic corrections \\
\end{tabular}
\label{tableVII}
\end{table}

\newpage

\centerline{\bf Figures captions:}
\vskip 1.5truecm
\noindent{\bf Figure 3.1:}
           {Scaling function $\gamma^T$ for the transverse width 
           in Lorentzian approximation versus 
           $r={1 / q \xi} (1+(q_{_D}\xi)^2)^{1/2}$ and 
           $\varphi=\arctan (q_{_D}\xi)$.
\vskip0.5truecm
\noindent{\bf Figure 3.2:}
           Scaling function $\gamma^L$ for the transverse width 
           in Lorentzian approximation versus 
           $r={1 / q \xi} (1+(q_{_D}\xi)^2)^{1/2}$ and 
           $\varphi=\arctan (q_{_D}\xi)$.}
\vskip 0.5truecm
\noindent{\bf Figure 3.3:}
           {Scaling function $\gamma^T$ for the transverse width 
           in Lorentzian approximation versus 
           ${1 / q \xi}$ for values of $\varphi={N \pi / 20}$ 
           with N indicated in the graph.}
\vskip0.5truecm
\noindent{\bf Figure 3.4:}
          {Scaling function $\gamma^L$ for the longitudinal width 
           in Lorentzian approximation 
           versus ${1 / q \xi}$ for values of $\varphi={N \pi/ 20}$ 
           with N indicated in the graph.}
\vskip 0.5truecm
\noindent{\bf Figure 3.5:}
           {Scaling functions for the transverse (solid) and the 
           longitudinal (point-dashed) widths versus ${q / q_{_{D}}}$
           in Lorentzian approximation at the
           critical temperature. {\bf Inset:} Scaling functions for the
           transverse (solid) and the longitudinal (point-dashed) 
           Onsager coefficients versus ${q / q_{_{D}}}$ at the critical 
           temperature.}
\vskip 0.5truecm
\noindent{\bf Figure 3.6:}
           {Scaling function of the transverse Kubo relaxation function 
            $\phi^T (r,\varphi,\tau_T)$ at $\varphi = 1.49$ (close to the critical
            temperature) versus $\tau_T$ and $r = \sqrt{(q_{_{D}}/q)^2 + (1/q\xi)^2}$.}
\vskip 0.5truecm
\noindent{\bf Figure 3.7:}
           {Scaling function of the longitudinal Kubo relaxation function 
            $\phi^L (r,\varphi,\tau_L)$ at $\varphi = 1.49$ (close to the critical
            temperature) versus $\tau_L$ and $r = \sqrt{(q_{_{D}}/q)^2 + (1/q\xi)^2}$.}
\vskip 0.5truecm
\noindent{\bf Figure 3.8:}
           {Scaling function of the longitudinal and transverse Kubo relaxation 
            function $\phi^\alpha (r,\varphi,\tau_T)$ versus $\tau_\alpha$ 
            for three different values of $\varphi$ 
            ($\varphi = 1.49, \, 1.294, \, 0.97$). In each graph the scaling function
            is shown for $r=0$ (solid), $r=1$ (dashed), and $r=10$ (dot-dashed). }
\vskip 0.5truecm
\noindent{\bf Figure 3.9:}
           {Scaling function of the transverse Kubo relaxation function 
            $\phi^T (r,\varphi,\tau_T)$ at $\varphi = 1.294$ 
            versus $\tau_T$ and $r = \sqrt{(q_{_{D}}/q)^2 + (1/q\xi)^2}$.}
\vskip 0.5truecm
\noindent{\bf Figure 3.10:}
           {Scaling function of the longitudinal Kubo relaxation function 
            $\phi^L (r,\varphi,\tau_L)$ at $\varphi = 1.294$ 
            versus $\tau_L$ and $r = \sqrt{(q_{_{D}}/q)^2 + (1/q\xi)^2}$.}
\vskip 0.5truecm
\noindent{\bf Figure 3.11:}
           {Scaling function of the transverse Kubo relaxation function 
            $\phi^T (r,\varphi,\tau_T)$ at $\varphi = 0.97$ 
            versus $\tau_T$ and $r = \sqrt{(q_{_{D}}/q)^2 + (1/q\xi)^2}$.}
\vskip 0.5truecm
\noindent{\bf Figure 3.12:}
           {Scaling function of the longitudinal Kubo relaxation function 
            $\phi^L (r,\varphi,\tau_L)$ at $\varphi = 0.97$ 
            versus $\tau_L$ and $r = \sqrt{(q_{_{D}}/q)^2 + (1/q\xi)^2}$.}

\newpage
\newpage

\begin{figure}
%  \centerline{\rotate[r]{\epsfysize=5in \epsffile{???.eps}}}
\phantom{.}
\vfill
\bigskip \bigskip\bigskip
\noindent{\bf Figure 3.1:}
           {Scaling function $\gamma^T$ for the transverse width 
           in Lorentzian approximation versus 
           $r={1 / q \xi} (1+(q_{_D}\xi)^2)^{1/2}$ and 
           $\varphi=\arctan (q_{_D}\xi)$.}
  \label{fig31}
\end{figure}
\newpage

\begin{figure}
%  \centerline{\rotate[r]{\epsfysize=5in \epsffile{???.eps}}}
\phantom{.}
\vfill
\bigskip \bigskip\bigskip
\noindent{\bf Figure 3.2:}
           {Scaling function $\gamma^L$ for the transverse width 
           in Lorentzian approximation versus 
           $r={1 / q \xi} (1+(q_{_D}\xi)^2)^{1/2}$ and 
           $\varphi=\arctan (q_{_D}\xi)$.}
  \label{fig32}
\end{figure}
\newpage

\begin{figure}
  \centerline{\rotate[r]{\epsfysize=5in \epsffile{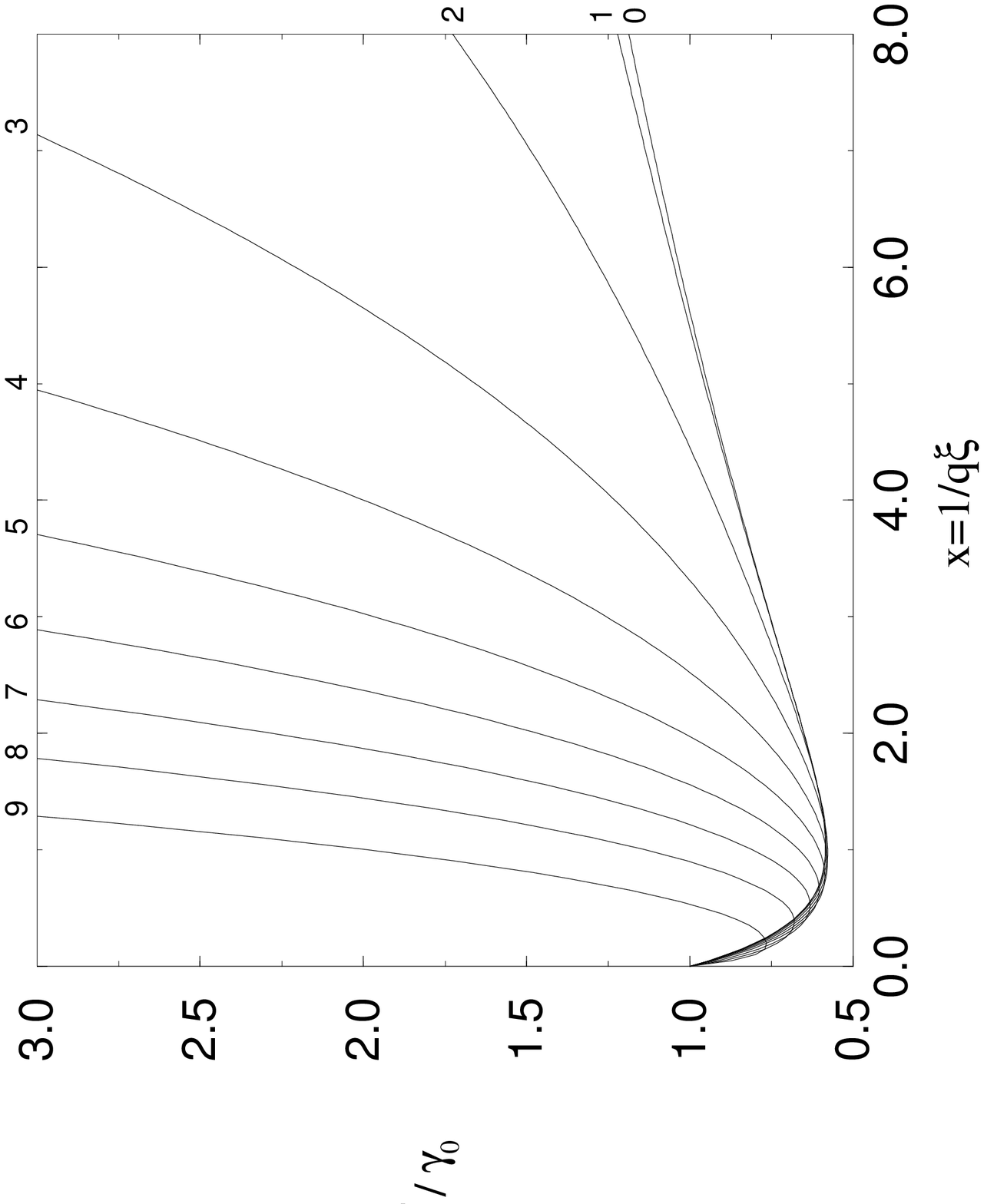}}}
\bigskip \bigskip\bigskip
\noindent{\bf Figure 3.3:} {Scaling function $\gamma^T$ for the transverse width in 
           Lorentzian approximation versus 
           ${1 / q \xi}$ for values of $\varphi={N \pi / 20}$ 
           with N indicated in the graph.}
  \label{fig33}
\end{figure}

\newpage

\begin{figure}
  \centerline{\rotate[r]{\epsfysize=5in \epsffile{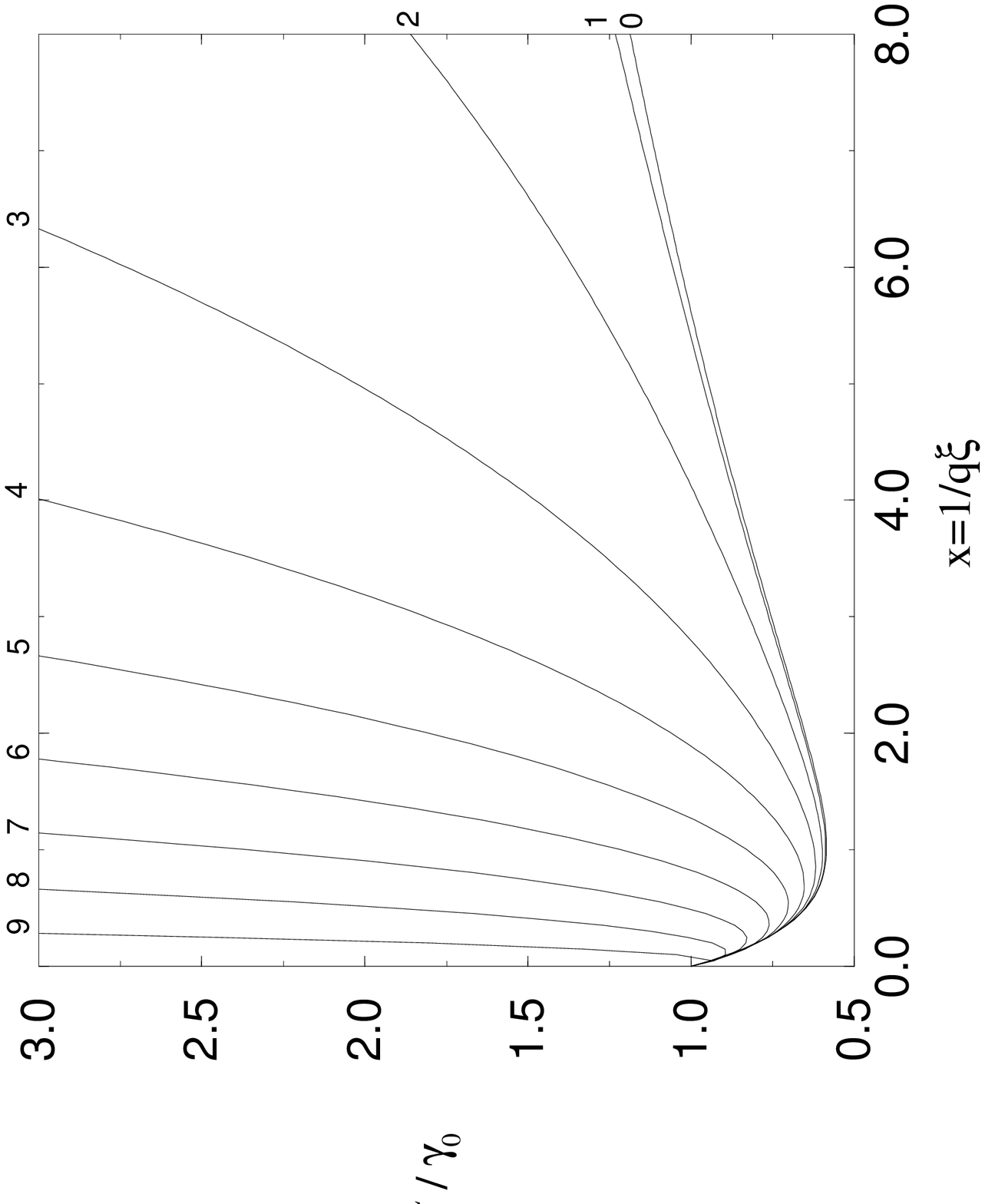}}}
\bigskip \bigskip\bigskip
\noindent{\bf Figure 3.4:} {Scaling function $\gamma^L$ for the longitudinal width 
           in Lorentzian approximation versus ${1 / q \xi}$ for values of 
           $\varphi={N \pi/ 20}$ 
           with N indicated in the graph.}
  \label{fig34}
\end{figure}

\newpage

\begin{figure}
  \centerline{\rotate[r]{\epsfysize=5in \epsffile{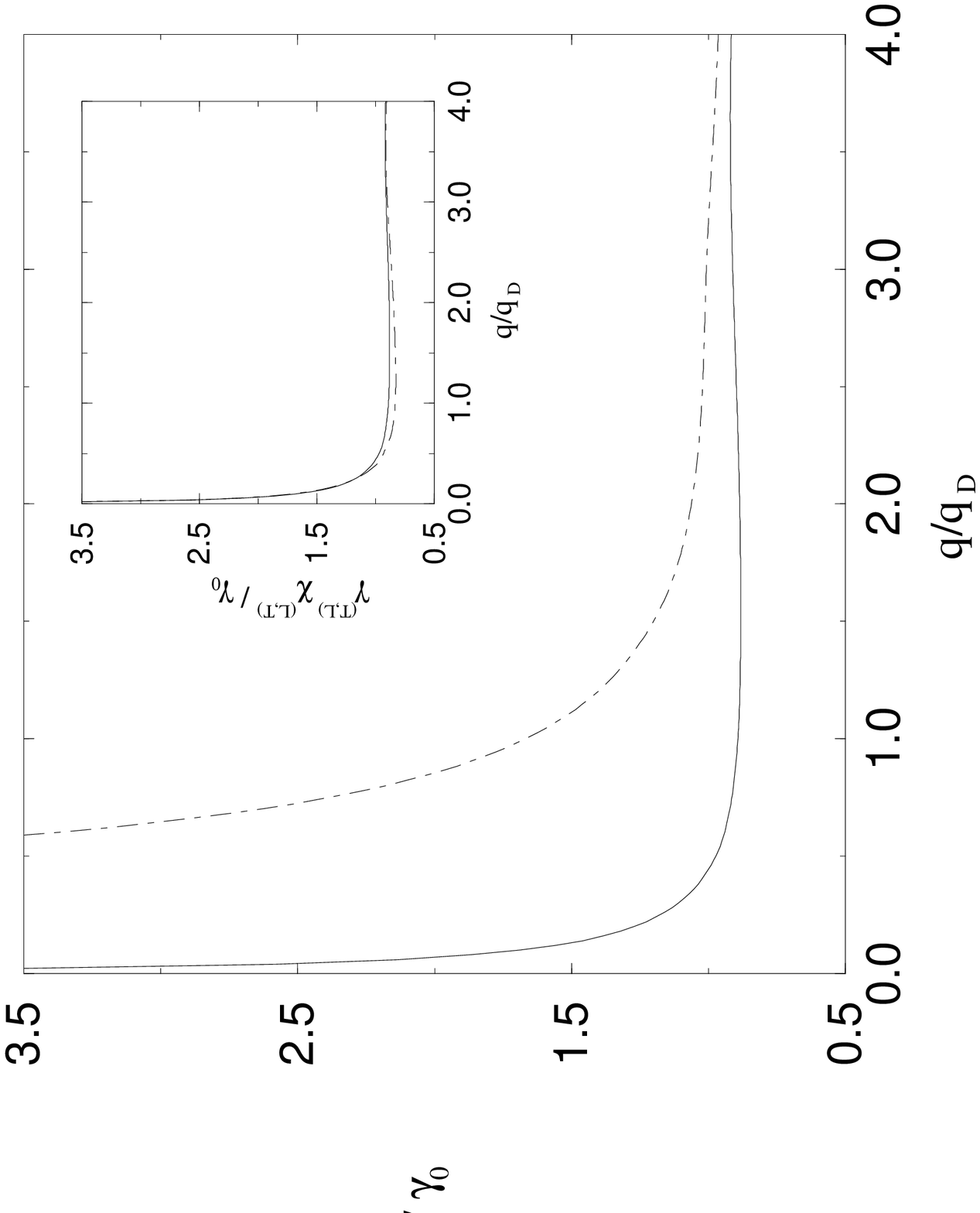}}}
\bigskip \bigskip\bigskip
\noindent{\bf Figure 3.5:} {Scaling functions for the transverse (solid) and the 
           longitudinal (point-dashed) widths versus ${q / q_{_{D}}}$ 
           in Lorentzian approximation at the
           critical temperature. {\bf Inset:} Scaling functions for the
           transverse (solid) and the longitudinal (point-dashed) 
           Onsager coefficients versus ${ q / q_{_{D}} }$ at the critical 
           temperature.}
  \label{fig35}
\end{figure}

\begin{figure}
  \centerline{\rotate[r]{\epsfysize=8in \epsffile{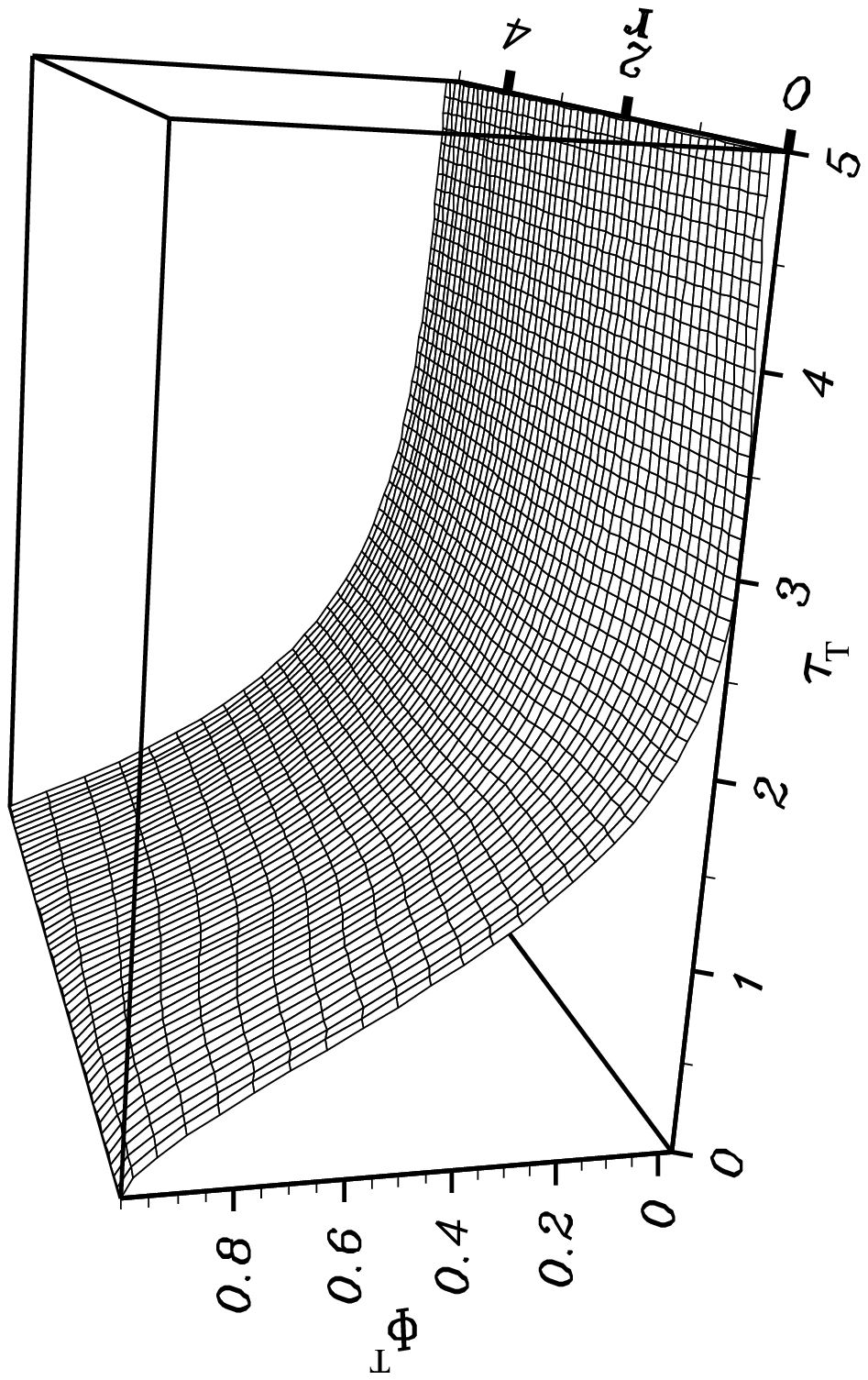}}}
\bigskip \bigskip\bigskip
\noindent{\bf Figure 3.6:} {Scaling function of the transverse Kubo relaxation
          function $\phi^T (r,\varphi,\tau_T)$ at $\varphi = 1.49$ (close to the 
          critical temperature) versus $\tau_T$ and $r = \sqrt{(q_{_{D}}/q)^2 + 
          (1/q\xi)^2}$.}
  \label{fig36}
\end{figure}

\begin{figure}
  \centerline{\rotate[r]{\epsfysize=8in \epsffile{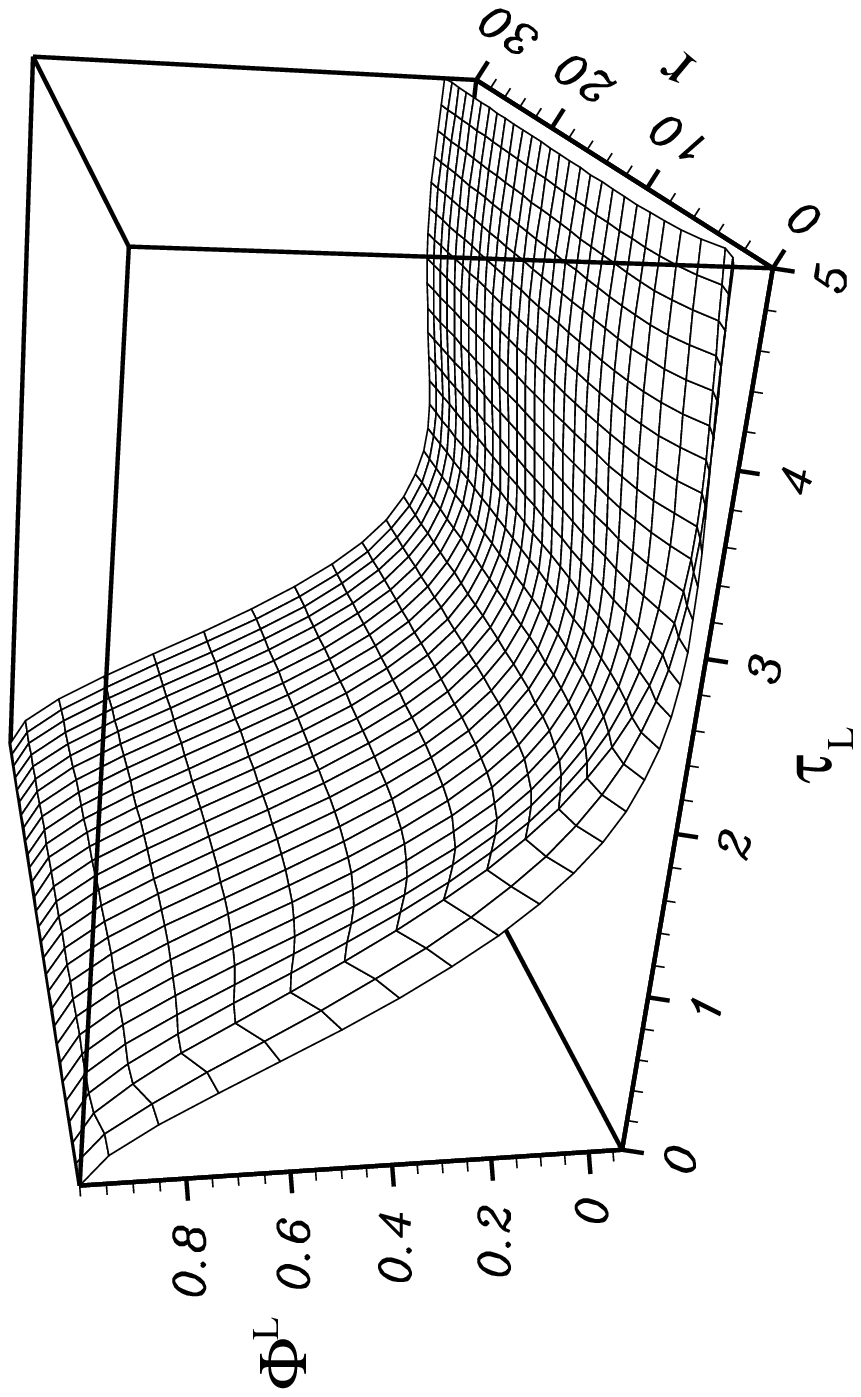}}}
\bigskip \bigskip\bigskip
\noindent{\bf Figure 3.7:} {Scaling function of the longitudinal Kubo relaxation
          function $\phi^L (r,\varphi,\tau_L)$ at $\varphi = 1.49$ (close to the 
          critical temperature) versus $\tau_L$ and $r = \sqrt{(q_{_{D}}/q)^2 + 
          (1/q\xi)^2}$.}
  \label{fig37}
\end{figure}   

\begin{figure}
  \centerline{{\epsfysize=6in \epsffile{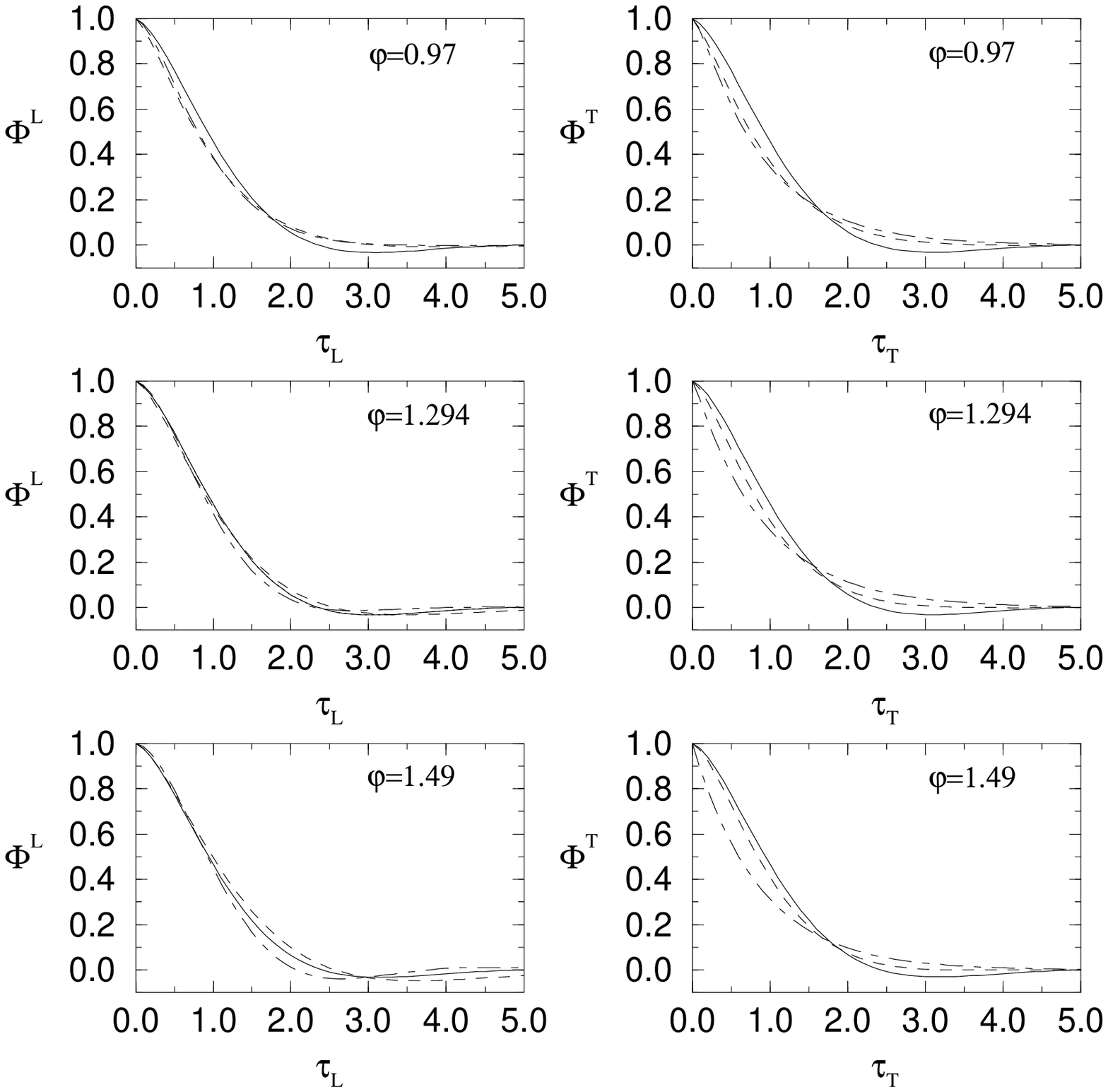}}}
\bigskip \bigskip\bigskip
\noindent{\bf Figure 3.8:} 
            {Scaling function of the longitudinal and transverse Kubo relaxation 
            function $\phi^\alpha (r,\varphi,\tau_T)$ versus $\tau_\alpha$ 
            for three different values of $\varphi$ 
            ($\varphi = 1.49, \, 1.294, \, 0.97$). In each graph the scaling function
            is shown for $r=0$ (solid), $r=1$ (dashed), and $r=10$ (dot-dashed). }
  \label{fig38}
\end{figure}

\begin{figure}
  \centerline{\rotate[r]{\epsfysize=8in \epsffile{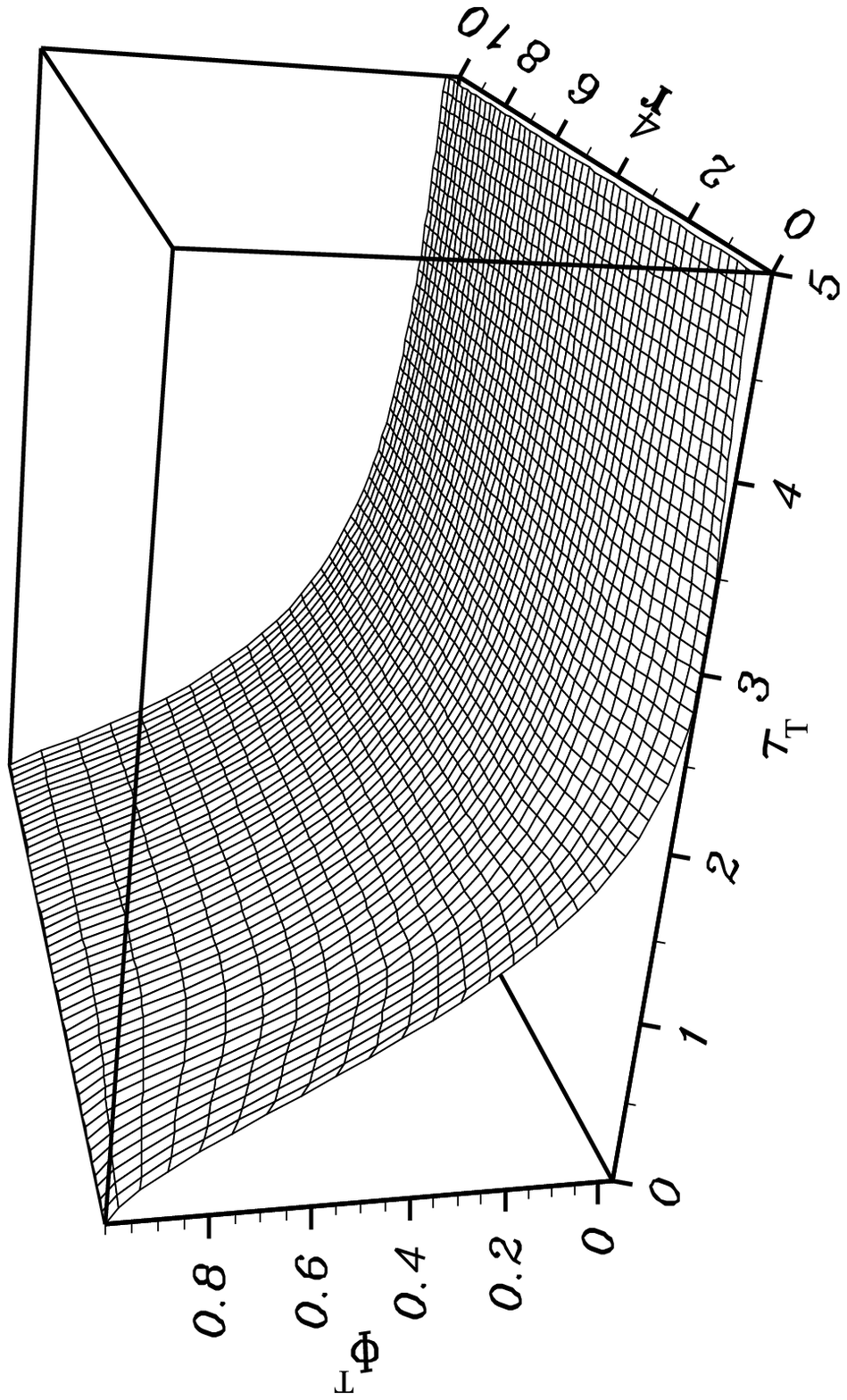}}}
\bigskip \bigskip\bigskip
\noindent{\bf Figure 3.9:} {Scaling function of the transverse Kubo relaxation
          function $\phi^T (r,\varphi,\tau_T)$ at $\varphi = 1.294$
          versus $\tau_T$ and $r = \sqrt{(q_{_{D}}/q)^2 + 
          (1/q\xi)^2}$.}
  \label{fig39}
\end{figure}

\begin{figure}
  \centerline{\rotate[r]{\epsfysize=8in \epsffile{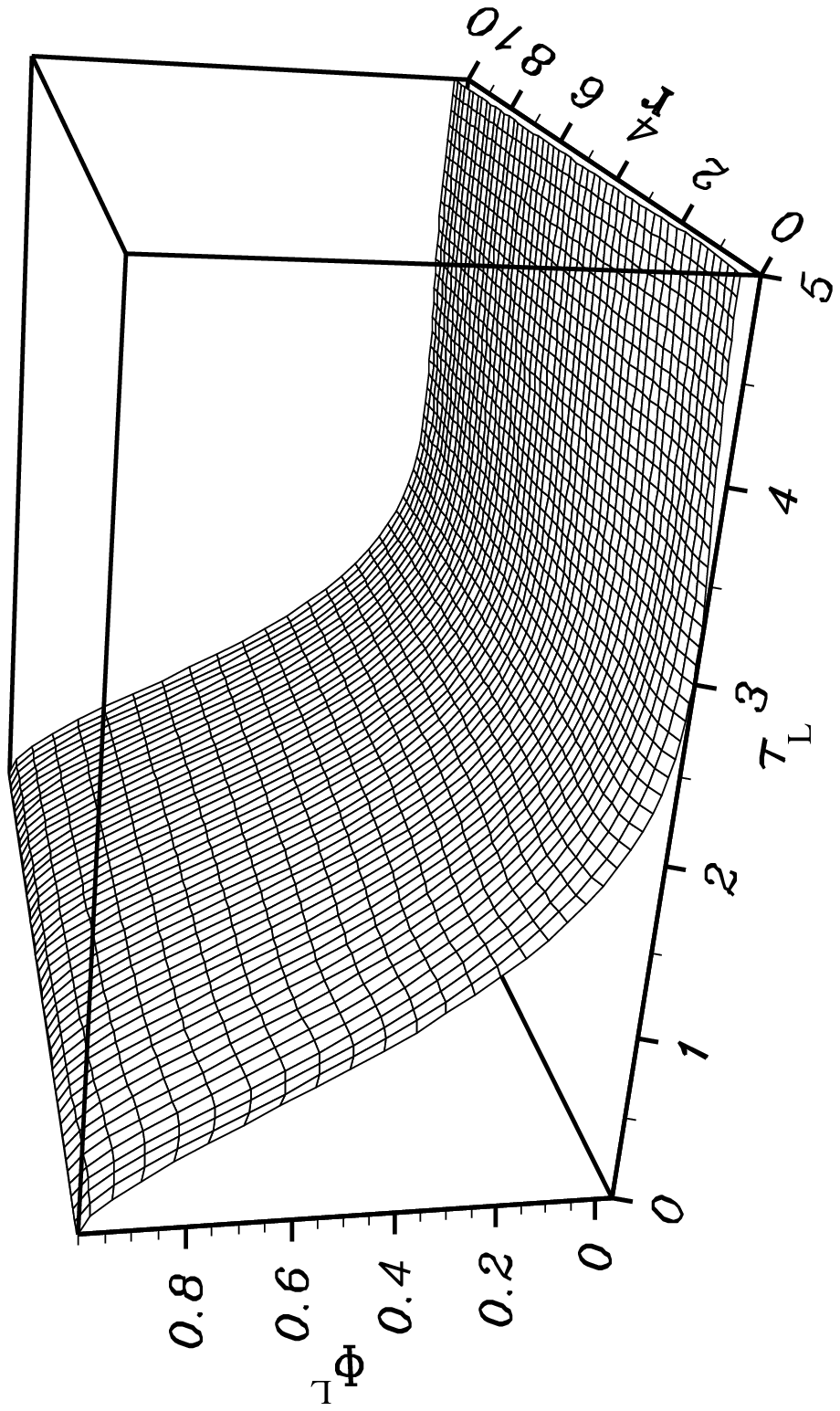}}}
\bigskip \bigskip\bigskip
\noindent{\bf Figure 3.10:} {Scaling function of the longitudinal Kubo relaxation
          function $\phi^L (r,\varphi,\tau_L)$ at $\varphi = 1.294$
          versus $\tau_L$ and $r = \sqrt{(q_{_{D}}/q)^2 + 
          (1/q\xi)^2}$.}
  \label{fig310}
\end{figure}

\begin{figure}
  \centerline{\rotate[r]{\epsfysize=8in \epsffile{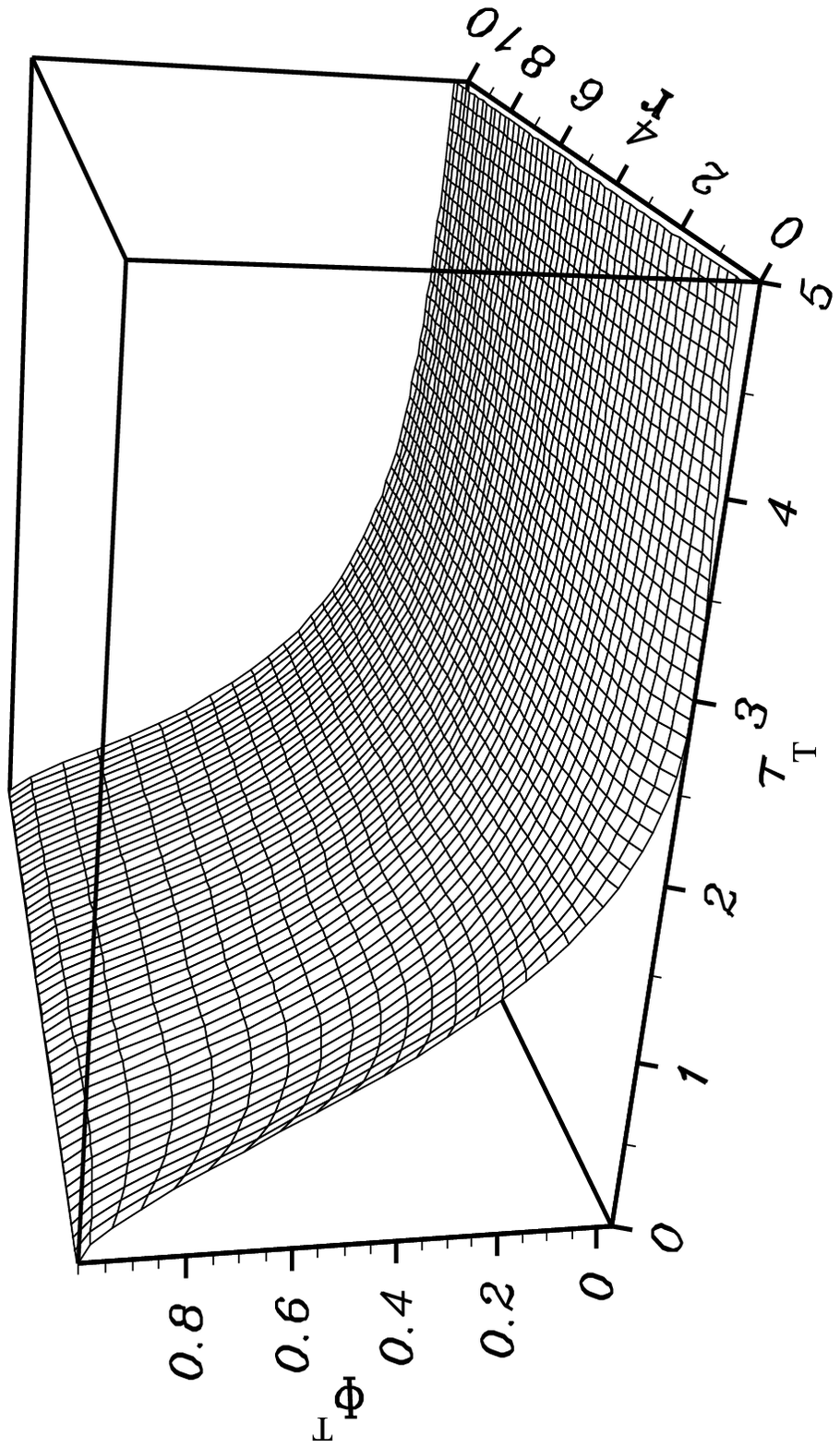}}}
\bigskip \bigskip\bigskip
\noindent{\bf Figure 3.11:} {Scaling function of the transverse Kubo relaxation
          function $\phi^T (r,\varphi,\tau_T)$ at $\varphi = 0.97$
          versus $\tau_T$ and $r = \sqrt{(q_{_{D}}/q)^2 + 
          (1/q\xi)^2}$.}
  \label{fig311}
\end{figure}

\begin{figure}
  \centerline{\rotate[r]{\epsfysize=8in \epsffile{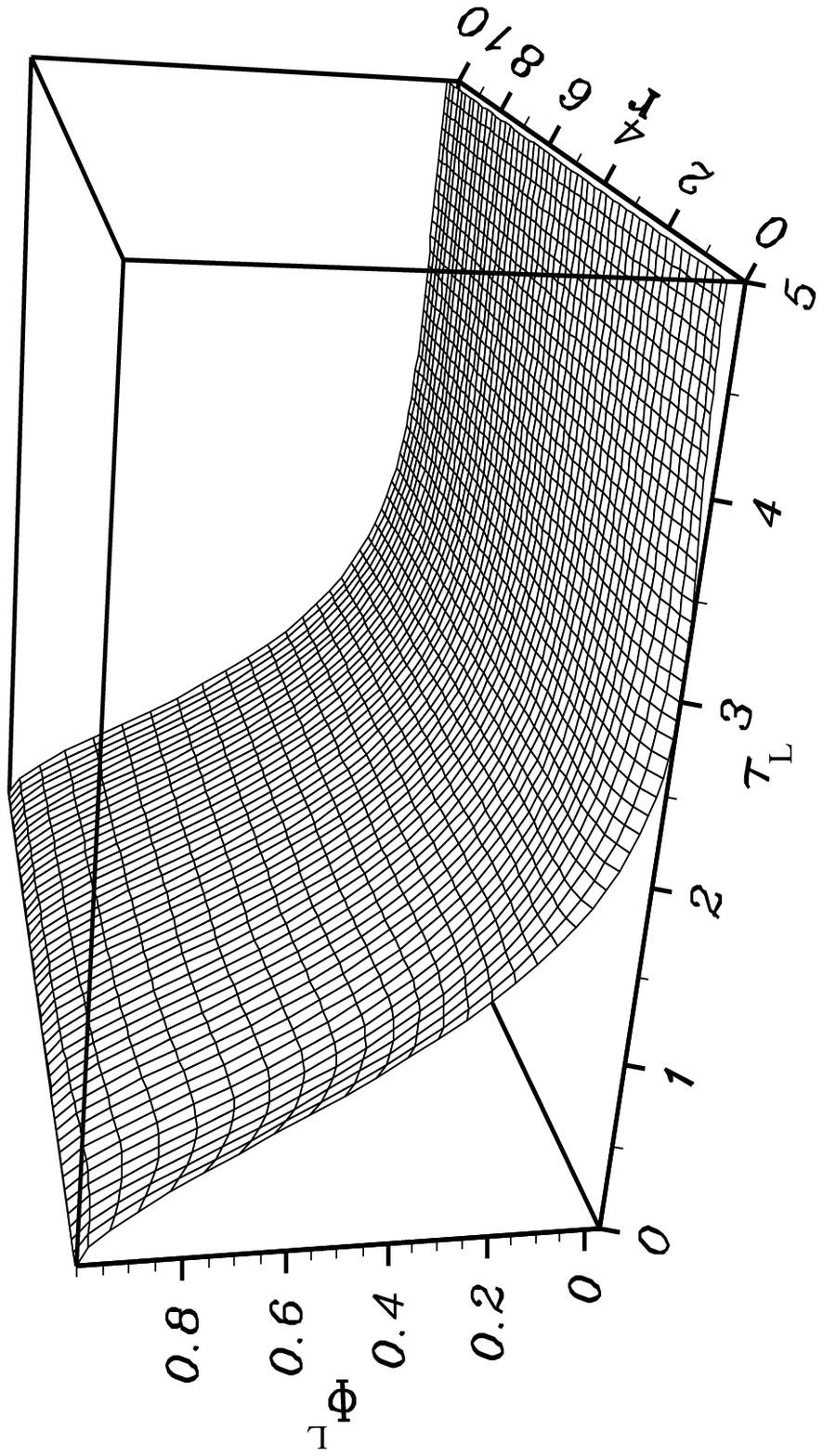}}}
\bigskip \bigskip\bigskip
\noindent{\bf Figure 3.12:} {Scaling function of the longitudinal Kubo relaxation
          function $\phi^L (r,\varphi,\tau_L)$ at $\varphi = 0.97$
          versus $\tau_L$ and $r = \sqrt{(q_{_{D}}/q)^2 + 
          (1/q\xi)^2}$.}
  \label{fig312}
\end{figure}

\newpage

%%%%%%%%%%%%%%%%%%%%%%%%%%%%%%%%%%%%%%%%%%%%%%%%%%%%%%%%%%%%%%%%%%%%%%%%%%%%%%%%
\section{APPLICATION TO EXPERIMENTS}
\label{s4}
In this chapter we review the results from the numerical solution of
the mode coupling equations for dipolar ferromagnets above the
transition temperature and compare them with experiments. 
Furthermore we give a short account of the experimental data recorded
below $T_c$ and future theoretical developments.

There are several experimental techniques like
neutron scattering (NS), electron spin resonance (ESR), magnetic
relaxation (MR), hyperfine interaction (HFI) and muon spin
resonance ($\mu$SR) with various complementary characteristics: 
different wave vector range, short-range point-like probes in real space 
(HFI, $\mu$SR) vs. probes  in reciprocal space (NS, ESR, MR). The critical 
dynamics of isotropic ferromagnets such as $EuO$, $EuS$, $Fe$, $Ni$ and
many other materials have been studied by one or several of the above 
methods. 

To emphasize the decisive role of the dipolar interaction we note that 
the experimental situation prior to the theoretical work in
Refs.~\cite{fe86,fs87,fs88a} was, however, puzzling in many ways.
On the one hand, in hyperfine interaction experiments on $Fe$ and 
$Ni$ one observed a crossover in the dynamical critical exponent from 
$z={5/2}$ to $z=2$~\cite{rh72,gh73,hcs82,hrk89}, i.e., a crossover to a 
dynamics with a non conserved order parameter. This was confirmed by 
electron spin resonance and magnetic relaxation
experiments \cite{kkw76,ksbk78,kp78,ks78,dg80,k88}, 
where a non vanishing Onsager coefficient at zero wave vector was
found. These results indicated that the critical dynamics of isotropic
ferromagnets cannot be explained solely on the basis of the short range
exchange interaction, which would lead to an exponent $z={5/2}$ for the
whole wave vector range. However, on the other hand a
critical exponent $z={5/2}$ was deduced from the wave vector
dependence of the line width observed in neutron scattering
experiments right at the critical temperature by three different experimental 
groups in Refs.~\cite{cmnp69,dnp76}, \cite{m82,m84,m86,m87,m88}
and \cite{wbs84,bs86,bcs87,bsbz87}.
Nevertheless and even more puzzling, the data for the line widths above 
the transition temperature~\cite{m82,m84,bsbz88} could not be described by 
the Resibois-Piette scaling function resulting from a mode coupling (MC) 
theory~\cite{rp70} and a renormalization group (RG)
theory~\cite{mm75,i87}, which take into account the short range exchange 
interaction only (see chapter~\ref{s2}). 

Those apparent experimental discrepancies could be resolved by 
the mode coupling theory \cite{fe86,fs87,fs88a} described in
chapter~\ref{s3}, which on top of the exchange interaction takes into
account the dipole-dipole interaction present in all real
ferromagnets. The results of this theory are reviewed in the following
and compared with the experimental data.

\subsection{Neutron scattering}
\label{s4.1}

Inelastic neutron scattering has been used since decades to investigate 
the spin dynamics of magnetic systems providing information about the spin 
correlation function\cite{cmnp69,mcns69,pk71,pa72,pdn76,dnp76}. It has become
one of the most important experimental techniques for studying the properties
of condensed matter systems.

The cross section for magnetic inelastic scattering of polarized neutrons
(with polarization ${\bf P}$) is
given by~\cite{Marshall71,Lovesey84}
\begin{eqnarray}
  {d^2 \sigma \over d \Omega d E^\prime} =
   &&r_0^2 \, {k^\prime \over k} \, \sum_{l,d} \sum_{l^\prime,d^\prime}
     e^{i {\bf Q} \cdot ({\bf R}_{l,d}-{\bf R}_{l^\prime,d^\prime})} 
     {1 \over 2} g_d F_d^\star ({\bf Q})
     {1 \over 2} g_{d^\prime} F_{d^\prime}^\star ({\bf Q}) \nonumber \\
   &&\times {1 \over 2 \pi \hbar} \int_{-\infty}^{\infty} dt
     e^{- i \omega t} 
     \left[
     \langle
     {\bf S}_{l,d}^{\perp} \cdot {\bf S}_{l^\prime,d^\prime}^{\perp} (t)
     \rangle +
     i {\bf P} \cdot 
     \langle
     {\bf S}_{l,d}^{\perp} \times {\bf S}_{l^\prime,d^\prime}^{\perp} (t)
     \rangle 
     \right] \, ,
\label{4.1}
\end{eqnarray}
where the spin fluctuations transverse to the neutron wave vector transfer 
(scattering vector)
${\bf Q} = {\bf k} - {\bf k}^\prime$ are defined by  ${\bf S}_{l,d}^{\perp} = 
{\bf Q} \times \left[ {\bf S}_{l,d} \times {\bf Q} \right] /Q^2$.
${\bf R}_{l,d} = {\bf l} + {\bf d}$ denotes the position vector of a
nucleus in a rigid lattice, where ${\bf d}$ is the site of the nucleus
(with a gyromagnetic ratio $g_d$ and a form factor $F_d ({\bf Q}$) 
within the unit cell ${\bf l}$). The quantity $r_0 = -0.5391 \, fm$ is a 
useful unit for the magnetic scattering length.

In neutron scattering experiments using unpolarized neutrons one measures 
the transverse scattering function $S^T({\bf Q},\omega)$
\begin{equation}
  S^T ({\bf Q},\omega) \propto 
  \left( \delta^{\alpha \beta} - {Q^\alpha Q^\beta \over Q^2} \right)
  \langle
  S_{\bf q}^\alpha (\omega) S_{-{\bf q}}^\beta (-\omega) 
  \rangle \, ,
\label{4.2}
\end{equation}
where ${\bf q}$ denotes the neutron wave vector transfer with respect to
the nearest reciprocal lattice vector $\gvect{\tau}$, i.e. the 
scattering vector reads ${\bf Q} = {\bf q} + \gvect{\tau}$.
Eq.~(\ref{4.2}) implies that only spin fluctuations transverse to the neutron 
wave vector transfer ${\bf q}$ can be measured in the forward direction, where
$\gvect{\tau} = {\bf 0}$. In order to measure the longitudinal fluctuations 
one has to go to a finite Bragg peak $\gvect{\tau}$ with ${\bf q}$ 
perpendicular to ${\bf Q} \approx \gvect{\tau}$. The separation of the
longitudinal and transverse spin fluctuations can be achieved by using a
polarized neutron beam. A typical scattering geometry is shown in Fig.~4.1,
where the polarization ${\bf P}$ is defined by a small guide field 
${\bf H}_{\rm v}$. For more details on neutron scattering using polarized 
neutrons we refer the reader to the books by 
Marshall and Lovesey~\cite{Marshall71}, Williams~\cite{Williams88} and 
Lovesey~\cite{Lovesey84}.

Recent advances in high resolution neutron scattering techniques, like 
e.g. the Neutron Spin Echo (NSE) technique~\cite{m72}, have 
allowed to extend the early studies 
to critical fluctuations of substantially larger wavelengths and even
made it possible to distinguish between different line shapes~\cite{m86}.

In this subsection we compare the results of mode coupling theory with neutron
scattering data on $EuO$, $EuS$ and $Fe$. For a quantitative comparison we
need several material parameters (non--universal quantities) like the
magnitude of the Brillouin zone boundary, exchange constant, etc. (see
Table~\ref{table4.1}). 

\subsubsection{Shape crossover}
\label{s4.1.1}

Due to the recent advances in high resolution neutron scattering techniques it
became possible to resolve not only the width but also the shape of the spin
correlation functions. This progress made it possible to test the predictions 
for the line shape from MC--theory~\cite{w68,h71a} and 
RG--theory~\cite{d76,bf85}, which take into account solely the short--range 
exchange interaction. The latter theories for the critical dynamics were 
expected to fit the experimental data in an increasing quantitative way as 
one was able to measure the correlation functions at smaller and smaller 
wave vectors, since the influence of additional irrelevant interactions 
should diminish as one moves closer to criticality. It came as quite a 
surprise when Mezei found a nearly exponential decay of the Kubo relaxation 
function by spin  echo experiments on $EuO$ at $q=0.024 \AA^{-1}$ and 
$T=T_c$~\cite{m86}, which was in drastic disagreement with the bell--like shape
predicted~\cite{w68,h71a,d76,bf85}. 

Actually -- in contrast to the expectation -- only at larger wave 
vectors the RG-- and MC--calculations, taking into account exchange 
interaction only, were in  accord with the experiment. This was shown by a 
comparison~\cite{fi85} of the shape of the RG--theory~\cite{d76,bf85} with 
the heuristic shape of Ref.~\cite{fi85}. Also, a discrete version of the 
MC--theory gives quite reasonable an agreement with experiment for large 
wave vectors close to the Brillouin zone boundary~\cite{ctlv88,ctl89}. 

In order to explain the anomalous exponential decay found by 
Mezei~\cite{m86}, it was asserted in Refs.~\cite{lw86,bpclt87} that neither 
MC-- nor RG--theory is valid in the time regime probed by Mezei's~\cite{m86} 
experiments. To fit the  experimental results a 
``hybrid theory''~\cite{bpclt87} was proposed, which is a phenomenological 
interpolation scheme between the short and long time limits. In contrast it 
was shown on the basis of a mode coupling theory~\cite{fst88,fst89}, which 
take into account the effects of the long--range dipolar interaction between 
the spins, that the experimental data can be explained quite naturally 
without any need of a special treatment of the short--time behavior. The 
results of this mode coupling analysis will be reviewed in this subsection.

We have seen in chapter~\ref{s3} that the dynamic crossover in the 
{\it line width} sets in at wave vectors one order of magnitude smaller than 
found in the static susceptibility. Here we consider the {\it shape} 
crossover, i.e., changes in the functional form of the spin correlation 
function. From the solution of the dipolar MC--equations one finds the 
following scenario for the shape crossover. 

\begin{itemize}
\item{} In the immediate vicinity of the 
critical temperature the {\it transverse} relaxation function shows a nearly 
{\it exponential decay} in time for wave vectors smaller than the dipolar wave
vector $q_{_{\rm D}}$. Passing to wave vectors larger than $q_{_{\rm D}}$ 
there is a shape crossover to a Hubbard--Wegner shape \cite{h71a,w68} 
corresponding to the isotropic Heisenberg fixed point (see chapter~\ref{s2}). 
This shape crossover takes place close to the dipolar wave vector 
$q_{_{\rm D}}$ in remarkable contrast to the crossover in the line width 
which sets in at wave vectors almost one order of magnitude smaller.

\item{} For wave vectors smaller than $q_{_{\rm D}}$ and for temperatures in 
the intermediate vicinity of the critical temperature the {\it longitudinal} 
Kubo relaxation function exhibits a Gaussian decay at small times and damped 
oscillations for larger times. This shape is quite different from the 
exponential decay found for the transverse scaling function. Passing to 
larger wave vectors the shapes of the transverse and longitudinal relaxation 
functions become identical (Hubbard--Wegner shape).

\item{} For temperatures well separated from the critical temperature (i.e.
for temperatures where the correlation length becomes less than the typical
dipolar length scale $q_{_{\rm D}}^{-1}$: $q_{_{\rm D}} \xi \leq 1$) the 
shape of the transverse and longitudinal relaxation becomes identical. Upon 
passing from small to large wave vectors both relaxation functions show a 
crossover from a hydrodynamic shape, which is nearly exponential, to the bell
like Hubbard--Wegner shape.

\end{itemize}

In order to be specific and in view of spin echo experiments on {\it EuO} 
above $T_c$~\cite{m88,mfhs89} we have displayed the Kubo relaxation function 
versus time for a set of wave vectors ($q\, = \, 0.018 \AA^{-1}$, $0.025 
\AA^{-1}$, $0.036 \AA^{-1}$, $0.071 \AA^{-1}$, $0.150 \AA^{-1}$) in 
Figs.~4.2-5. In Figs.~4.2 and 4.3 these are plotted versus the scaled time 
variables $\tau_{\alpha}$ for a temperature in the immediate vicinity of 
$T_c$ ($T=T_c+0.25K$) and in Figs.~4.4 and 4.5 for a temperature 
$T=T_c + 8K$ well separated from $T_c$. For wave vectors 
$q \ll q_{_{\rm D}}=0.147\,\AA^{-1}$ one can infer from Fig.~4.2 that the 
transverse scaling function decays nearly exponentially in time. This implies 
a  Lorentzian like shape for the frequency dependent relaxation function for 
$q \ll q_{_{\rm D}}$ (solid line in Fig.~4.2). For larger wave vectors, 
$q \geq q_{_{\rm D}}$, the curves look similar to Gaussians for small times 
and oscillate for larger times, $\tau_T \geq 3$ (see point--dashed curve in
Fig.~4.2). The oscillations of the longitudinal relaxation function
(Fig.~4.3) leads to the side peaks in Fig.~4.9, where the
longitudinal correlation function is plotted versus frequency for 
a fixed wave vector.

Further away from $T_c$ the line shape crossover of the
transverse relaxation function is much less pronounced and the
shape resembles more a Gaussian even for wave vectors much smaller
than $q_{_{\rm D}}$. In Figs.~4.4 and 4.5 the transverse and longitudinal
Kubo functions are shown for the same wave vectors as in Figs.~4.2 and 4.3 at 
$T=T_c+8K$. For this temperature the shape of the
longitudinal and transverse Kubo function is nearly the same (compare 
Figs.~4.4 and 4.5). This is precisely what one would have expected, because 
the influence of the dipolar forces decreases with separation from the 
critical point and then there is no difference any more between longitudinal 
and transverse modes.

To single out the line shape crossover the appropriate time
scales are $\tau_{\alpha}$ of Eq.~\ref{3.39} used in Figs.~4.2-5.
On the other hand, for comparison with experiments it may be more
convenient to present the transverse Kubo relaxation function
versus the time scale $\tau^{\prime} = \Lambda_{\rm lor} q^{5/2} t$.
Such plots are exhibited in Fig.~4.6 for $T = T_c$ and
the wave vectors $q = 0.018 \AA^{-1}, \, 0.036 \AA^{-1}, \,
0.150 \AA^{-1}, \, 0.3\AA^{-1}$ and in Fig.~4.7 for $T = T_c +
0.5K$ and $q = 0.018 \AA^{-1}, \, 0.025 \AA^{-1}, \, 0.036
\AA^{-1}, \, 0.071 \AA^{-1}$.

At $T=T_c$ triple-axis spectrometer scans at $q \geq 0.15 \AA^{-1}$
\cite{mfhs89} show non-Lorentzian line shapes in agreement with
predictions based on models with short range exchange interaction
only and in agreement with the above MC result. Neutron Spin Echo (NSE)
studies at much smaller wave vectors \cite{m86,mfhs89} lead to 
Lorentzian line shapes in agreement with the above crossover scenario
for dipolar ferromagnets.
In Fig.~4.8 data from NSE studies \cite{m86} on $EuO$ right at
the critical temperature are shown for
the transverse Kubo relaxation function at the
wave vector $q = 0.024 \AA^{-1}$ (solid line) versus time in $nsec$.
We have used the theoretical value for the non universal scale
$\Lambda = 7.1/5.1326 \, meV \, \AA^{5/2}$ (see Table~\ref{table4.1}). 
There is an excellent
agreement with the experimental data for $t \leq 1 \, nsec$. The
experimental data are above the theoretical curve for $t \geq 1
\, nsec$. This may be due to finite collimation effects in this
time domain, as noted by Mezei~\cite{m86}. To substantiate this point
we have also plotted in Fig.~4.8 the relaxation function at $q =
0.028 \AA^{-1}$ (point-dashed curve), which is significantly
higher than the curve for $q = 0.024 \AA^{-1}$ for $t \geq 1 \,
nsec$. The fairly large difference of the curves with $q = 0.024
\, \AA^{-1}$ and $q = 0.028 \, \AA^{-1}$ is implied by the 
close proximity to the crossover region.

In order to exhibit the difference from the MC theory including
only short range exchange interaction the dashed curve in Fig.~4.8
shows the solution of Eqs.~\ref{3.40} and \ref{3.41} for this special case, 
i.e., $y=0$, $x=0$ and
$\rho_{\rm cut} = {q_{_{\rm BZ}} / q}$ with $q=0.024 \, \AA^{-1}$. The
result differs drastically
from the actual line shape including the long--range dipolar interaction.
It is important to realize, that the crossover in the line shape
starts nearly at $q_{_{\rm D}}$, whereas the line width still scales with
the isotropic critical dynamic exponent $z={5 / 2}$ in this
wave vector region. 

Very recently the theoretical predictions concerning the line shape
of the longitudinal relaxation function~\cite{fst88,fst89} have been verified
experimentally~\cite{bgkm91,Goerlitz92a}. Fig.~4.9 shows the longitudinal
Kubo relaxation function for $EuS$ at $T= 1.006 T_c$ and $q = 0.19
\AA^{-1}$. The normalized experimental data de-convoluted by  a maximum 
entropy method agree quite well with the theoretical prediction (solid line). 
Especially the double peak structure corresponding to the over-damped 
oscillations in time are observed.

\subsubsection{Line width crossover}
\label{s4.1.2}

From the frequency and wave vector dependent Kubo relaxation
function one can define a characteristic line width by the half
width at half maximum (HWHM)
\begin{equation}
  Re \left[ \Phi^\alpha (\omega = \Gamma^\alpha_{_{\rm HW}}) \right] 
  = {1 \over 2} Re \left[ \Phi(\omega = 0) \right] \, ,
\label{4.3}
\end{equation}
which obeys the scaling law
\begin{equation}
  \Gamma^\alpha_{_{\rm HW}} (q) = 
  \Lambda_{_{\rm HW}} q^z \gamma^\alpha_{_{\rm HW}} (x,y) \, . 
\label{4.4}
\end{equation}
Let us start by comparing the theoretical and experimental line
widths for $Fe$, $EuO$ and $EuS$ precisely at the critical
temperature. Fig.~4.10 shows the scaling functions for the 
transverse and longitudinal line width in the Lorentzian approximation,
normalized such that the line width approaches $1$ for $q \gg q_{_{D}}$. 
As already noted earlier, a remarkable result of mode coupling
theory is that the dynamic dipolar crossover of the longitudinal
line width starts near the dipolar wave vector $q \approx q_{_{D}}$, whereas
for the transverse line width it is shifted to a wave vector which is 
about one order of magnitude smaller than $q_{_{D}}$. This explains why no 
appreciable deviation from the exchange scaling prediction 
$\Gamma^T (q) \propto q^{5/2}$ in the wave vector range accessible up to now  
was found in Refs.~\cite{cmnp69,dnp76}, \cite{m82,m84,m86,m87,m88} and
\cite{wbs84,bs86,bcs87,bsbz87} in neutron scattering experiments right at the 
critical temperature. The experimental data in Fig.~4.10 are collected from 
Refs.~\cite{cmnp69,dnp76}, \cite{m82,m84,m86,m87,m88} and
\cite{wbs84,bs86,bcs87,bsbz87}
and are normalized with respect to the theoretical value of the
non--universal frequency scale $\Lambda_{\rm lor} = 5.1326 \times
\Lambda$. For a comparison the solid line in Fig.~4.10 shows the scaling
function of the transverse line width as it has been obtained from the
solution of the complete mode coupling equations~\cite{fst88,fst89}. Whereas 
the general functional form is quite similar to the scaling function obtained 
in the Lorentzian approximation, their asymptotic values at large values
differ by a factor of approximately $1.2$, which leads to an even better
quantitative agreement between theory and experiment. It still needs
experiments at smaller momentum transfer to detect the increasing of $\Gamma$
because of the crossover to the dipolar regime. 

In contrast to the data on the wave vector dependence right at the critical 
temperature, experiments on the scaling function above 
the critical temperature revealed huge deviations from the Resibois-Piette 
scaling function, as is evident from the neutron scattering data on 
$Fe$ \cite{m82} shown in the scaling plot in Fig.~4.11, which become the more
pronounced the closer to $T_c$. 

This figure also shows a comparison of the HWHM (dashed-dotted line)
of the transverse Kubo functions resulting from Eqs.~\ref{3.40} and \ref{3.41} 
and the Lorentzian line width $\gamma_{\rm lor}^{T}$ (solid line) 
determined by Eq.~\ref{3.49}. In the case of $Fe$
curves a, b, c and d correspond to the temperatures $T-T_c \, =
\, 1.4K, \, 5.8K, \, 21.0K, \, {\rm and} \, 51.0K$. 
All curves are normalized with respect to their value at criticality.
This implies that the non
universal frequency scale is then found to be $\Lambda_{\rm lor} =
5.1326 \, \Lambda = 107.2 \, meV \, \AA^{5/2}$ in the Lorentzian
approximation and $\Lambda_{\rm HW} = 1.37 \times \Lambda_{\rm lor} = 147 meV 
\, \AA^{5/2}$ for the HWHM. The latter value is in quite
reasonable agreement with the
experimental value $\Lambda_{\rm exp} = 130 \, meV \, \AA^{5/2}$.
The Lorentzian line width and the HWHM have nearly
the same $(q\xi)^{-1}$-dependence. Especially for small wave
vectors the difference comes about only because of the different
non universal frequency scales. In summary, the line width is 
affected by the line shape crossover only to a minor extent, since 
the former is only a coarse feature of the relaxation function.
At small wave vectors only the non-universal scale is modified. This
allows one to use the Lorentzian approximation for some later 
applications of the theory to NS, ESR, MR, and HFI experiments.

As can be inferred from Fig.~4.11 the complete solution of the mode
coupling theory is in reasonable agreement with the experiment
close to $T_c$ ($T-T_c \, = \, 1.4K, \, 5.8K$) and gives an
improvement over the Lorentzian approximation. The minor
differences may be due to the following reasons. (i) As mentioned
above the measured scaling functions of the transverse relaxation
function were fitted to an exponential line, which is not the
correct shape. (ii) Because the dipolar crossover temperature
$T_{_{\rm D}}$ of $Fe$ is $8.6K$, static crossover effects not taken into
account in the experiments may cause some changes. Furthermore
the non universal scale of the correlation length $\xi_{_{0}}$ is
affected by experimental uncertainties. A change in $\xi_{_{0}}$ would
lead to a horizontal shift of the data points in Fig.~4.11.  

Larger differences show up for temperatures further away from
$T_c$ ($T-T_c=21K,\,51K$), which cannot be accounted for by the
shape crossover or static crossover effects. Here the measured
line widths are larger than the theoretical. In order to
explain this, it is necessary to take into account the van Hove
terms and further relaxation mechanisms due to irrelevant
interactions. Such irrelevant interactions are unimportant as concerns the
critical behavior, but nevertheless may play an important role for
temperatures well separated from $T_c$ (see Ref.~\cite{fst89}).   
Pseudo-dipolar forces have been studied in Refs.~\cite{fst89} and 
Ref.~\cite{af89}.
Those additional interactions (crystal field interactions, the spin orbit
interaction leading to pseudo-dipolar terms) are presumably less important in
magnets with localized moments such as $EuO$ and $EuS$. This can be seen from 
the comparison of the line width with the experimental data. Fig.~4.12
(Fig.~1 from Ref.~\cite{mfhs89}, and Fig.~1 from Ref.~\cite{m88})
shows a scaling plot of 
the transverse line width data for temperatures 
$T-T_c = 0.5K$, $2K$, and $4K$ in $EuO$, which are in quantitative
agreement with the mode coupling theory (solid lines). 

Recently the transverse and for the first time the longitudinal spin 
fluctuations have been measured on $EuS$ using inelastic scattering 
of polarized neutrons \cite{bgkm91,Goerlitz92a}. The
observed relaxation rates follow rather precisely the {\it
transverse} and {\it longitudinal} line width
calculated from the dipolar mode coupling theory (see Figs.~4.13 and
4.14). Especially, these
measurements confirm the behavior of the longitudinal line width
predicted by mode coupling theory \cite{fs87,fs88a}.

\subsubsection{Constant energy scans}
\label{s4.1.3}

In this subsection we return to an analysis of the line shape and review
results obtained by scans at constant frequency as opposed to constant wave
vector in the previous sections. As will become clear in the
following, certain features of the line shape of the correlation function can 
be accentuated by constant energy scans for the scattering function 
$S^T(q,\omega)$. Characteristic  quantities in these scans are the 
peak position $q_{_{0}}$ and the HWHM $\Delta q$. For an isotropic 
Heisenberg ferromagnet RG--theory predicts a flat curve for 
the reduced peak positions $q_{_{0}}(\Lambda/\omega)^{2/5}$
plotted versus the scaled frequency $\omega \xi^{5/2}/\Lambda$
\cite{i87,fi86}. This theoretical result has been confirmed
experimentally in certain energy and wave vector regions\cite{bcs87}.
However, more recently B\"oni et al.~\cite{bsbz88} have found pronounced 
departures from the isotropic scaling law in $EuS$ in a region, where 
according to Refs.~\cite{fs87,fs88a} the dipolar interaction should have a 
considerable effect on the dynamics. This is quite similar to the situation 
in constant wave vector scans, where one finds deviations from the 
Resibois-Piette scaling function \cite{m82,m84}.

One of the most striking new features introduced by the dipolar
interaction is the generalized dynamical scaling Eq.~(\ref{3.31}) with
the additional scaling variable $y=q_{_{\rm D}}/q$. In conventional
constant $q$ scans plotted versus the scaling variable $x=1/ q \xi$ 
the line widths are not represented by a single scaling
function but by a series of curves as exhibited in Figs.~3.1 and 3.2., 
where each curve corresponds to a fixed temperature.
The failure of the isotropic scaling law in constant energy scans 
can also be attributed to the influence of the dipolar 
forces\cite{fst88,fst89}.

The generalized scaling law for the Kubo functions and characteristic
frequencies leads to the following scaling law for the peak position 
in constant energy scans~\cite{fst89}
\begin{equation}
  q_{_{0}} \left( \Lambda \over \omega \right)^{2/5} =
  {\cal Q} \left( \varphi, 
  { \omega \over \Lambda \left( \xi^{-2} + q_{_{\rm D}}^2 \right)^{z/2} } 
           \right) \, .
\label{4.5}
\end{equation}
From the above generalized scaling laws it becomes obvious that 
one obtains a set of curves parameterized by the scaling variable
$ \varphi = \arctan ( q_{_{D}} \xi ) $, 
if Eq.~(\ref{4.5}) containing the two parameter scaling 
function $\cal Q$ is plotted versus $\omega \xi^{5/2} / \Lambda$. 
In Fig.~4.15 the scaled peak position 
$q_{_{0}} (\Lambda_{\rm lor}/\omega)^{2/5}$ is plotted versus 
${\hat \omega} = \omega \xi^{5/2} / \Lambda_{\rm lor}$
for the following set of scaling variables $\varphi=$ a) 1.490,
b) 1.294, c) 0.970 and d) $\varphi=0$. Case d) corresponds to an
isotropic ferromagnet, i.e., 
$q_{_{\rm D}}=0$. $\Lambda_{\rm lor}$ is related
to $\Lambda$ by $\Lambda_{\rm lor} = 5.1326 \Lambda$. 
The above values for $\varphi$ correspond in the
case of $EuS$ ($q_{_{\rm D}} = 0.27 \AA^{-1}$, $\xi_{_{0}}=1.81 \AA$) to the
temperatures $T=$ a) $1.01 T_c$, b) $1.06 T_c$, and c)$1.21 T_c$.
Due to the generalized scaling law Eq.~(\ref{4.5}) the curves coincide
with the isotropic theory for high frequency but deviate for
small frequency. The frequency, where the deviation from the isotropic result 
sets in, increases upon approaching the critical temperature.

The steep drop off of the scaled peak positions at particular
$\varphi$ dependent values of $\hat \omega$ has been explained as follows
~\cite{fst89}. The dipolar forces imply that the order parameter is no
more conserved and the uniform relaxation rate $\Gamma_{_{0}}$ becomes finite. 
Hence the scattering function $S^T(q,\omega)$ remains finite for vanishing 
wave vector
\begin{equation} 
   S^T(q=0,\omega) \propto 
   {\Gamma_{_{0}} \xi^2 \over \omega^2 +
   \Gamma_{_{0}}^2} \quad {\rm for} \quad T \geq T_c \quad. 
\label{4.6}
\end{equation}     
Since $\Gamma_{_{0}}$ is proportional to $\xi^{-2}$ Eq.~(\ref{4.6}) reduces to 
\begin{equation} 
   S^T(q=0,\omega) \propto  {1 \over \omega^2} 
   \quad {\rm for} \quad T=T_c \quad. 
\label{4.7}
\end{equation}
Therefore, the maximum of the constant energy scan at $q=0$
increases strongly with decreasing frequency $\omega$, whereas
the local maximum at finite $q$ is shifted to smaller $q$
(see also the inset in Fig.~4.15). As a result of this competition
only the maximum at $q=0$ survives for low frequencies. In order
to substantiate this, typical constant energy scans are shown in
the inset of Fig.~4.15, where $S^T(q,\omega)/S^T(0,\omega)$ (in
arbitrary units) is plotted versus $1/r$ for $\varphi = 1.294$
and for a set of scaled frequencies ( ${\hat \omega} = 10^{L/10}$
with $L=8, \, 10, \, 12, \, 14, \, 16$ indicated in the graph).
The corresponding scaled peak positions are indicated in the
scaling plot by crosses. This behavior explains, why the scaled
peak positions for dipolar ferromagnets show such a steep drop
off at small frequencies. 

Finally, it is important to note that the characteristic deviations from the
isotropic theory exhibited in Fig.~4.15 result from the crossover in
the time scale and not so much from the crossover in the shape
function. This can be inferred from Fig.~4.16 where the scaled peak
positions $q_{_{0}} \left({\Lambda \over \omega}\right)^{2/5}$ in
Lorentzian approximation is shown. Compared to Fig.~4.15 the maxima are
overemphasized in Fig.~4.16, but the portions with the steep slope
are nearly identical. The differences of the exact mode coupling
theory and the Lorentzian approximation can also be seen in the
insets of Figs.~4.15 and 4.16.

\subsection{Electron spin resonance and magnetic relaxation}
\label{s4.2}

In electron spin resonance (ESR) and magnetic relaxation (MR) experiments 
one measures the electronic response function at zero wave vector and finite
frequency
\begin{equation}
  \chi^\alpha ({\bf q}={\bf 0},\omega) =
  {\Gamma^\alpha ({\bf q},\omega) \chi^\alpha ({\bf q}) \over
  i \omega + \Gamma^\alpha ({\bf q},\omega)} \mid_{{\bf q} \rightarrow 0} \, ,
\label{4.8}
\end{equation}
where $\Gamma^\alpha ({\bf q},\omega)$ is the frequency dependent relaxation 
function in Eq.~(\ref{3.19}). Therefrom one determines the kinetic coefficient
\begin{equation}
  L(\omega) = 
  \Gamma^\alpha ({\bf q},\omega) 
  \chi^\alpha ({\bf q}) \mid_{{\bf q} \rightarrow 0}
\label{4.9}
\end{equation}
for the homogeneous magnetization dynamics. In Lorentzian approximation 
one finds for the kinetic coefficient at zero frequency
\begin{equation}
  L_0 (q_{_{\rm D}},\xi) =
  { g^2 J^2 \over 3 \pi^2} k_{_{\rm B}} T
  \int \limits_{0}^{\infty} dk k^2
  { \chi^L (k,q_{_{\rm D}}) \chi^T (k,q_{_{\rm D}}) \over
    \Gamma^L (k,q_{_{\rm D}}) + \Gamma^T (k,q_{_{\rm D}}) } \, ,
\label{4.10}
\end{equation}
where the dependence of the Onsager coefficient on the dipolar wave vector and 
the temperature is now indicated explicitly. 
Note that the kinetic coefficient for the homogeneous relaxation is the same
for the longitudinal and transverse spin fluctuations. Using the generalized 
dynamic scaling and introducing polar coordinates ($r=\sqrt{x^2 + y^2}$, 
${y / x}= \tan \varphi$) one finds~\cite{fs88a}
\begin{equation}
  L_0 (q_{_{\rm D}},\xi) = B F(1/q_{_{D}} \xi) \, ,
\label{4.11}
\end{equation}
with the universal crossover function 
\begin{equation}
  F(1/q_{_{D}} \xi) = 
  \left( 1 + {1 \over q_{_{\rm D}}^2 \xi^2} \right)^{-7 / 4}
  \int \limits_{0}^{\infty} dr r^{5 / 2}
  { {\hat \chi}^L (r,\varphi) {\hat \chi}^T (r,\varphi) \over
    \gamma^L (r,\varphi) + \gamma^T (r,\varphi) } \, ,
\label{4.12}
\end{equation}                                                               
and the non--universal constant 
\begin{equation}
  B = {4 \pi^2 \over 3} \Lambda q_{_{\rm D}}^{5 / 2} \, . 
\label{4.13}
\end{equation}

If there were no dipolar interaction ($g = 0$), one
would simple find a vanishing Onsager coefficient due to the
factor $g^2$ in equation (\ref{4.10}). With regard to the crossover
function $F$ of Eq.~(\ref{4.12}) it is natural to define the reduced crossover
temperature by
\begin{equation}
  \tau_{_{\rm D}} = {T_{_{\rm D}} - T_c \over T_c}  
               = (q_{_{\rm D}}^2 \xi_0^2)^{1 / \phi} \, ,
\label{4.14}
\end{equation}
where the crossover exponent $\phi$ equals the susceptibility exponent 
$\gamma$ for a ferromagnet without dipolar interaction~\cite{af73}. 
If one neglects the dipolar crossover of the correlation length, the 
scaling variable $q_{_{\rm D}}^2 \xi^2$ can be written as 
$q_{_{\rm D}}^2 \xi^2 = \left( {\tau_{\rm cross}  / \tau} \right)^{\gamma}$ 
and the crossover temperature in terms of the dipolar wave vector is given by
\begin{equation}
  q_{_{\rm D}} \xi_{_{\rm D}} = 1 \, .
\label{4.15}
\end{equation}                                                               
The crossover temperatures $T_{_{\rm D}}$ resulting from Eq.~(\ref{4.15}) can 
be found in Table~\ref{table4.2}. 

The Onsager coefficient $L = \Gamma^{\alpha} \chi^{\alpha}$ does not depend 
on the sample shape~\cite{fi77a} and is the same for the transverse and the 
longitudinal mode. The universal crossover function $F(\rho)$ with $\rho =
1/q_{_{D}} \xi$ is plotted in Fig.~4.17 in units of its value at criticality
$F(0)$. For temperatures larger than the dipolar crossover temperature 
$F(\rho)$ shows a $\xi^{7/2}$ power law behavior, which was first shown in
Refs.~\cite{hu71,rh76}. In the strong dipolar limit, i.e., very close
to the transition temperature, the Onsager coefficient approaches a 
constant \cite{fi77b} reflecting the non conserved
nature of the order parameter due to the presence of the dipolar
interaction. 

The kinetic Onsager coefficient at zero wave vector has been
measured by magnetic relaxation
experiments~\cite{kkw76,deHaas77,ksbk78,dg80} and ESR 
experiments~\cite{ksbk78,kkw76,kp78,ks78} for the Onsager
coefficient in $EuS$, $EuO$, $CdCr_2S_4$, $CdCr_2Se_4$ and
$Ni$~\cite{sa67,sb75}. In order to extract the contribution of the
critical fluctuations one has to subtract a non-critical background $L_{\rm
bg}$ which in the critical region of all ferromagnets (except $Ni$) is
very small, $L_{\rm bg} \ll L_{\rm cr}$ (see Table~\ref{table4.2}). 
The critical part of the Onsager coefficient 
\begin{equation}
  \hbar L_0 (q_{_{\rm D}},\xi) = 
  L_{\rm d} { F(\rho) \over F(0) } = B F(\rho)
\label{4.16}
\end{equation}
is compared with the crossover function $F$ in Fig.~4.18 
(Fig.~1 of Ref.~\cite{k88}). The non--universal parameter 
$\hbar L_{\rm d} = B F(0)$ was fixed by the data for the kinetic 
coefficient in the center of the critical region. There is an
excellent agreement (except $Ni$ and $EuO$) of the observed crossover 
with the results of the mode coupling theory. Especially, the data
for $EuS$ follow the theoretical function quite closely. 
The ESR--data on $Ni$~\cite{sa67}, shown in Fig.~4.18, deviate from
the mode coupling result. Sp\"orel and Biller~\cite{sb75}, however,
find an increase in the electron paramagnetic resonance line width
of $Ni$ near $T_c$, in accord with mode coupling theory but 
opposite to what was found in Ref.~\cite{sa67}. It seems that the
observation of the peaklike broadening depends critically on the 
quality of the sample. Recently Li et al.~\cite{Li90} have
demonstrated that it is possible to observe the critical
broadening of the linewidth in ultrathin $Ni$ films.
As further possible sources for the deviations in $Ni$ and $EuO$ from the
dipolar crossover sample imperfections (e.g. oxygen vacancies in
$EuO$, and internal stress in $Ni$) have been suggested~\cite{k88}.

The values of the fit-parameter $L_{\rm d}$ are listed in 
Table~\ref{table4.2}. But, even these values agree with the theoretical 
prediction as can be inferred from the inset in Fig.~4.18. The minor 
differences can be attributed to some uncertainties in the dipolar 
wave vector~\cite{k88}.

Because ESR experiments are performed in a magnetic field $B$ there
should be an effect on the relaxation rate starting at $\gamma B
= \Gamma(q=\xi^{-1})$ \cite{kk76}. Those data points are not given in
Fig.~4.18. The influence of the magnetic field on the relaxation
rate at zero wave vector has recently been studied by K\"otzler et
al. \cite{kkkw91}.
In analyzing the effect of the magnetic field in terms of the
internal isothermal susceptibility
$\chi_{\rm int} = \delta M / \delta H_{\rm int}$ ($H_{\rm int}$ is
the internal magnetic field related to the external magnetic field
by demagnetization corrections) the data were found to collapse on a
single curve. Hence, empirically, the field effect on the kinetic coefficient 
can be accounted for by using $q_{_{\rm D}}^2 \chi_{\rm int}$ instead of
$q_{_{\rm D}}^2 \xi^2$ as the scaling variable and the same scaling function
$F$ as for the case of zero magnetic field.

Up to now we have studied the behavior of the uniform relaxation at zero
frequency only. Actually, the kinetic coefficient is frequency dependent, 
which was observed first by De Haas and Verstelle \cite{deHaas77}. Measuring 
the uniform dynamic susceptibility in $EuO$ they have found a deviation from 
a Debye spectrum (Lorentzian form), which according to Eqs.~(\ref{4.8}) and 
(\ref{4.9}) is equivalent to a frequency dependent kinetic coefficient. 
Similar deviation from a Lorentzian have been found quite recently in $EuS$ 
\cite{Grahl91,Dombrowski94}.
In order to study the frequency dependence of the kinetic coefficient
one would have to solve the full frequency dependent mode coupling 
equations and therefrom deduce $L(q_{_{\rm D}}, \xi, \omega)$. As a
first approximation one can use~\cite{Koetzler94,Dombrowski94}
\begin{equation}
  L(q_{_{\rm D}}, \xi, \omega) =
  {g^2 J^2 \over 3 \pi^2} k_{\rm B} T 
  \int_0^\infty dk k^2 
  {\chi^L(k,q_{_{\rm D}}) \chi^T(k,q_{_{\rm D}}) \over
  \Gamma^L (k,q_{_{\rm D}})  + \Gamma^T (k,q_{_{\rm D}})  + i \omega} \, ,
\label{4.17}
\end{equation}
which is obtained as the first iteration step in a self consistent
determination of the frequency dependent kinetic coefficient, starting
with the damping coefficients $\Gamma^{L,T}(q,q_{_{\rm D}}, \xi)$ from the 
Lorentzian 
approximation. A scaling analysis of the frequency dependent kinetic 
coefficient gives 
\begin{equation}
 L(q_{_{\rm D}}, \xi, \omega) = B  F (1/q_{_{D}}\xi,\omega \xi^z /\Lambda) \, ,
\label{4.18}
\end{equation}
with the scaling function  
\begin{equation}      
  F (1/q_{_{D}}\xi,\omega/\Lambda q_{_{D}}^z) =                       
  \left( 1 + {1 \over q_{_{\rm D}}^2 \xi^2} \right)^{-7 / 4}
  \int \limits_{0}^{\infty} dr r^{5 / 2}
  { {\hat \chi}^L (r,\varphi) {\hat \chi}^T (r,\varphi) \over
    \gamma^L (r,\varphi) + \gamma^T (r,\varphi) +
    i (\omega/\Lambda q_{_{D}}^z)  r^z \sin^z \varphi} \, .
\label{4.19}
\end{equation}    
The real and imaginary part of $F$ is shown in Figs.~4.19a and 4.19b
versus the scaled frequency ${\hat \omega} = \omega/\Lambda q_{_{D}}^z$
for several values of $\varphi = \arctan (q_{_{D}}\xi)$, indicated in the
graphs.                                          
For small and large values of ${\hat \omega}$ the real part of the scaling 
function behaves asymptotically as
\begin{equation}    
  {\rm Re} F (1/q_{_{D}}\xi,\omega/\Lambda q_{_{D}}^z) \approx
  F (1/q_{_{D}}\xi) 
  \cases{  
         1-({\hat \omega}/{\hat \omega_{c1}  (1/q_{_{D}}\xi) })^2      
         & for  ${\hat \omega} \ll 1$ \, ,  \cr   
         ({\hat \omega}/{\hat \omega_{c2} (1/q_{_{D}}\xi) })^{-1-1/z}  
         & for ${\hat \omega} \gg 1$ \, .   
        }
\label{4.20}
\end{equation}    
The corresponding scaling functions for the Onsager coefficient at zero 
frequency $F (1/q_{_{D}}\xi)$ and the scaling functions $\omega_{c2}  
(1/q_{_{D}}\xi)$ and $\omega_{c1}  (1/q_{_{D}}\xi)$ characterizing the 
large and low frequency behavior, respectively, are shown in 
Figs.~4.20a,b,c. As can be inferred from these figures, the scaling function 
$\omega_{c1}  (1/q_{_{D}}\xi)$ is nearly constant. This finding is in accord 
with experiments by Dombrowski et al.~\cite{Dombrowski94}. The experiments 
have been analyzed assuming a Lorentzian shape for the kinetic 
coefficient~\cite{Dombrowski94}
\begin{equation}
  L(q_{_{\rm D}}, \xi, \omega) \approx 
  { L(q_{_{\rm D}}, \xi) \over 1 + i \omega / \omega_c } \, .
\label{4.21}
\end{equation}
This shape differs from the theoretical results presented above, especially 
in the large frequency limit, i.e. for $\omega \gg \Lambda q_{_{D}}^z$. 
At low frequencies the theoretical result has the same expansion as the 
Lorentzian approximation used in analyzing the experiments. It is found 
experimentally that the characteristic frequency  $\omega_c$ is nearly 
independent on temperature. This is in accord with the above theoretical 
result, that the scaling function $\omega_{c1}  (1/q_{_{D}}\xi)$ is nearly 
constant over the whole temperature range. For a more detailed comparison 
with the experiment further analysis of the data on the basis of the 
results of the mode coupling theory for the frequency dependence of the 
Onsager coefficient seems to be necessary. It would be interesting to see, 
whether the experiment confirms the predicted large frequency behavior 
$L(\omega) \propto \omega^{-1-1/z}$.

Recently, the homogeneous magnetization dynamics has also been investigated
in the ferromagnetic phase~\cite{Dombrowski94}. It is found that the scaling
function for the kinetic coefficient below $T_c$ agrees exactly with that 
observed earlier above $T_c$. Since a complete theory of the critical dynamics
below $T_c$ is still lacking, these results are not explained yet. However,
those experimental findings are a clear indication of the importance of the
dipolar interaction below the critical temperature.

\subsection{Hyperfine Interactions}
\label{s4.3}

There are several nuclear, i.e., hyperfine interaction (HFI), methods commonly
used to study critical fluctuations in magnets. These are nuclear magnetic 
resonance (NMR)~\cite{sbew80}, M\"ossbauer effect (ME)~\cite{ksgh76,kh78}, 
perturbed angular correlations (PAC)~\cite{rh72,gh73,csh80,hcs82} of gamma 
rays, and muon spin rotation ($\mu$SR) (see also section~\ref{s4.3}). The 
application of nuclear techniques to study critical phenomena in magnets has 
recently been reviewed by Hohenemser et al.~\cite{hrk89}. For recent work on 
the application of $\mu$-SR for the investigation of spin dynamics the reader 
may consult Refs.~\cite{sc85,co87,ha89}. All of the hyperfine interaction
methods are local probes which are related to a wave vector integral of the 
spin correlation functions. As such they offer a complement to neutron
scattering. Dynamic studies, using hyperfine interaction probes, utilize
the process of nuclear relaxation produced by time--dependent hyperfine
interaction fields, which reflect fluctuations of the surrounding electronic 
magnetic moments. 

The hyperfine interaction stems from the magnetic interaction of the electrons
with the magnetic field produced by the nucleus. The hyperfine interaction
of a nucleus with spin ${\bf I}$, $g$--factor $g_N$ and mass $m_N$ with one
of the surrounding electrons with spin ${\bf S}$ and orbital momentum 
${\bf L}$ can be written in the form~\cite{Bethe57,Schwabl91}
\begin{equation}
  H_{\rm hyp} = 
  {Z e_0^2 g_N \over 2 m_N m c^2} 
  \left[
  {1 \over r^3} {\bf I} \cdot {\bf L} +
  {8 \pi \over 3} \delta^{(3)} ({\bf x}) {\bf I} \cdot {\bf S} -
  {1 \over r^3} {\bf I} \cdot {\bf S} + 
  {3 ({\bf I}\cdot{\bf x}) ({\bf S}\cdot{\bf x}) \over r^5}
  \right] \, .
\label{4.x1}
\end{equation}
The first term represents the interaction of the orbital momentum of the 
electron with the nuclear magnetic moment of the nucleus. The second term is
the Fermi contact interaction and the last two terms represent the dipolar
interaction~\footnote{Actually, also the Fermi contact interaction comes from 
the magnetic dipole interaction. In the evaluation of the matrix elements of
the (residual) dipolar interaction, represented by the last two terms, an
infinitesimal sphere around the origin has to be excluded. The singular part
of the dipolar interaction coming from the interior of the sphere is
represented by the Fermi contact term.}.
The Fermi contact interaction is finite only for electrons having a finite
probability density at the nucleus, i.e. bound $s$-electrons or itinerant
electrons. 

For a substitutional nucleus the main contribution to the
hyperfine field comes from the bound $s$-electrons~\cite{Goldanski68}, 
which are polarized by the
magnetic ions. The large (and often negative) hyperfine field for 
transition--element ions comes from the polarization of the core $s$ electrons
by the spin density of the unpaired $3d$ electrons, which then contribute to 
the hyperfine field through the Fermi contact interaction~\cite{Watson61}. 
The residual dipolar interaction is negligible in this case. Note also that 
this term is zero for cubic symmetry.
The Hamiltonian, Eq.~(\ref{4.x1}), can also be used for the analysis of spin 
resonance experiments with muons ($\mu$SR). However, these do not have bound
electrons and hence the Fermi contact term involves only conduction electrons
and is of the same order of magnitude as the (residual) dipolar 
interaction~\cite{Denison79,Meier81}
(see also section~\ref{s4.4}). For instance in $Pd$, there is a dipolar 
contribution by the $Pd$ $f$--electrons which also polarize the 
$s$--electrons which in turn contribute via the Fermi contact term to the 
hyperfine field at the muon site.

After the above general remarks, let us now return to those hyperfine 
interaction probes (ME, PAC, NMR), where the Fermi contact term gives the 
dominant contribution to the hyperfine field at the nucleus. Then the 
corresponding interaction Hamiltonian reduces to 
\begin{equation}
  {\cal H}(t) = {{\it A}_{\rm contact}} \, {\bf I} \cdot {\bf S}(t) \, .
\label{4.22}
\end{equation} 
If the spin auto--correlation time, $\tau_c$, is much shorter than the Larmor
period, $1/\omega_L$, and the nuclear lifetime, $\tau_N$, (motional narrowing 
regime) the nuclear relaxation rate $\tau_R^{-1}$ is directly proportional to 
the (averaged) spin autocorrelation time $\tau_c$ 
\begin{equation}
  \tau_{_R}^{-1} = C_{\rm hf} \tau_c \, ,
\label{4.23}
\end{equation}
with a hyperfine coupling constant $C_{\rm hf}$~\cite{hcs82}. The nuclear 
relaxation rate equals the spin-spin relaxation rate $\tau_R^{-1} = T_2^{-1}$ 
in the case of NMR, is proportional to the relaxation induced excess line 
width $\Delta \Gamma$ for ME, and for PAC $\tau_R$ is given by 
the relaxation time of the perturbation factor.

The (averaged) spin autocorrelation time $\tau_c$ is defined by a time-integral
\begin{equation}
  \tau_c = {1 \over 2} \int \limits_{-\infty}^{+\infty} dt
  {1 \over 3} \sum_{\alpha} G^{\alpha \alpha} ({\bf r} = 0,t)
  = {1 \over 6} \sum_{\alpha} \int_{\bf q} G^{\alpha \alpha} ({\bf q}, 
  \omega =0) \, .
\label{4.24}
\end{equation} 
over the spin autocorrelation function
$G^{\alpha \alpha} ({\bf r},t) = {1 \over 2} \langle \lbrace
S^{\alpha} ({\bf r},t),S^{\alpha} (0,0) \rbrace \rangle$. Hence
hyperfine interaction methods provide an integral property of the
spin-spin correlation function. Upon using the 
fluctuation dissipation theorem (FDT), which in the special
case $\omega = 0$ reduces to 
\begin{equation}
  G^{\alpha \alpha} (q,\omega = 0) = 
  2 k_{_{\rm B}} T {\chi^{\alpha} ({\bf q},g) \over 
                    \Gamma^{\alpha} ({\bf q},g)} \, ,
\label{4.25}
\end{equation}
one finds for the autocorrelation time
\begin{equation}
  \tau_c = {k_{_{\rm B}} T \over V_q} \int_{\rm BZ} d^3q {1 \over 3} 
  \sum_{\alpha} {\chi^{\alpha} ({\bf q},g) \over \Gamma^{\alpha} 
  ({\bf q},g)} \, .
\label{4.26}
\end{equation}
The ${\bf q}$-integration extends over the Brillouin zone (BZ), the
volume of which is $V_q$. 

Important information about the behavior of the auto-correlation
time can be gained from a scaling analysis. Upon
using the static and dynamic scaling laws Eq.~(\ref{4.26}) can be written as
\begin{equation}
  \tau_c \propto 4 \pi \int dq q^{-z} {1 \over 3} \sum_{\alpha}
  { {\hat \chi}^{\alpha} \left( q \xi,{q / q_{_{\rm D}}} \right)
    \over \gamma^{\alpha} \left( q \xi,{q / q_{_{\rm D}}}
    \right)  } \, .
\label{4.27}
\end{equation}
If there were no dipolar interaction, one could extract the
temperature dependence from the integral in Eq.~(\ref{4.27}) 
with the result $\tau_c \propto \xi^{z-1}$. This expression can be
used to define an effective dynamical exponent $z_{\rm eff} (\tau)$, 
which depends on the correlation length by 
\begin{equation}
  \tau_c \propto \xi^{z_{\rm eff}-1} \propto 
  \left( {T- T_c \over T_c} \right)^w \, . 
\label{4.28}
\end{equation}
In the presence of dipolar forces one finds
after introducing polar coordinates as in section~\ref{s4.2}   
\begin{equation}
  \tau_c = H {\left( 1 + {1 \over q_{_{\rm D}}^2 \xi^2} \right)}^{(1-z)/2}
  \int \limits_{r_0}^{\infty} dr r^{z-2} {1 \over 3}  
  \sum_{\alpha}
  { {\hat \chi}^{\alpha} (r,\varphi) \over
    \gamma^{\alpha} (r,\varphi)} \, ,
\label{4.29}
\end{equation}     
where $z = {5 / 2}$ and the non universal constant $H$ is given by~\cite{fs89}
\begin{equation}
  H = { (k_{_{\rm B}} T)^2 \over 
        32 \pi^6 \left( \Lambda a^{-5 / 2} \right)^3 }
      \left( q_{_{\rm D}} a \right)^{3 / 2} \, .
\label{4.30}
\end{equation}                                                 
The lower cutoff $r_0$ is 
\begin{equation}
  r_0 = {q_{_{\rm D}} \over q_{_{\rm BZ}}} 
  \sqrt{ 1 + {1 \over q_{_{\rm D}}^2 \xi^2} } \, ,  
\label{4.31}                  
\end{equation}                  
where $q_{_{\rm BZ}}$ is the boundary of the Brillouin zone. In the
critical region it can be disregarded and replaced by $r_0=0$, 
since $q_{_{\rm BZ}} \gg q_{_{\rm D}}$ and the integrand in Eq.~(\ref{4.29}) 
is proportional to $\sqrt{r}$ for small $r$. For very small $\xi$ (outside 
the critical region) the cutoff reduces the autocorrelation time with respect 
to the critical value.  

One should note, that the dominant wave vectors contributing to the
relaxation time $\tau_c$ in Eq.~(\ref{4.26}) are close to the zone center.
Quantitative estimates for the wave vector range probed by $\mu$-SR
experiments on $Fe$ and $Ni$, which also apply for other hyperfine experiments,
are given in Ref.~\cite{Reotier94b}.

The averaged autocorrelation time is a sum of two parts
\begin{equation}
  \tau_c = {1 \over 3} (\tau_L + 2 \tau_T ) \, ,
\label{4.32}
\end{equation}
which we call longitudinal and transverse relaxation times    
\begin{equation}
  \tau_{\alpha} = H I_{\alpha} (\varphi)  \, ,
\label{4.33}
\end{equation}                                                      
where we have defined
\begin{equation}
  I_{\alpha} (\varphi)= {\left( 1 + 
  {1 \over q_{_{\rm D}}^2 \xi^2} \right)}^{-3 / 4}
  \int \limits_{r_0}^{\infty} dr \sqrt{r} \quad 
  { {\hat \chi}^{\alpha} (r,\varphi) \over
    \gamma^{\alpha} (r,\varphi) } \, , \quad \alpha = L,T \, .
\label{4.34}
\end{equation}                                                      
These two relaxation times are shown in Fig.~4.21 as functions of
the scaling variable ${1 / q_{_{\rm D}} \xi}$, where we have
neglected the lower cutoff $r_0$. Since the static
longitudinal susceptibility does not diverge one finds that the 
longitudinal relaxation time is non-critical, whereas the transverse 
relaxation time diverges like $\tau_T \propto \xi$.  This corresponds 
to an effective dynamical exponent $z_{\rm eff} = 2$. When leaving the
dipolar critical region there is a crossover to the isotropic
Heisenberg region, where both curves join and the relaxation times
$\tau_L=\tau_T=\tau_c$ are characterized  by another  power
law $\tau_c \propto \xi^{3 / 2}$ corresponding to an effective
dynamical exponent $z_{\rm eff}= {5 / 2}$.

Let us now compare with hyperfine experiments on $Fe$ and $Ni$.
Early PAC~\cite{rh72,gh73} and ME~\cite{ksgh76,kh78} were performed
well in the dipolar region and an effective exponent $z=2$ was found.
The crossover in the dynamic exponent from $z=2.5$ to $z=2$ was
first observed by Chow et al.~\cite{csh80,hcs82}.
The autocorrelation time $\tau_c$ is shown in Fig.~4.22 in units of
the non universal constant $H$ (see Table~\ref{table4.2}) versus the scaling
variable ${1 / q_{_{\rm D}} \xi}$. The data points in Fig.~4.22 are results 
of HFI experiments on $Fe$ and $Ni$~\cite{rh72,gh73,csh80,hcs82} 
for the autocorrelation time $\tau_c$ (in units of their non universal 
frequency scale $H$). As before there
is no fit parameter for the scaling variable ${1 / q_{_{\rm D}} \xi}$.
So we conclude that our MC--theory accounts well for the
experimental data demonstrating the universal crossover behavior
from $z_{\rm eff}={5 / 2}$ to $z_{\rm eff}=2$ as the critical
temperature is approached.                                  

Finally we would like to note that ME~\cite{cch84,cch86a} and 
PAC~\cite{cch86b} experiments on the dynamics of $Gd$ show neither 
$z=2.5$ nor $z=2.0$ but $1.74 < z < 1.82$ depending on whether the 
system is assumed to have Heisenberg or Ising static exponents. Recent 
$\mu$-SR experiments give similarly depressed values of the critical 
dynamic exponent $z$~\cite{whkw86,whkw89} (see section~\ref{s4.4}). 
In Ref.~\cite{Reotier94a} the $\mu$-SR data~\cite{whkw86,hwkk90} 
have been compared with the results from MC--theory for isotropic dipolar 
ferromagnets~\cite{fs88a,fs89}, and it is found that the available 
data are found to be in agreement with the MC--theory.
$Gd$ is supposed to have an uniaxial exchange anisotropy along the $c$-axis. 
Hence one expects that in the asymptotic region there is one critical mode 
along the $c$-axis with $z_{\rm eff} = 2$ and two uncritical modes with 
$z_{\rm eff} = 0$ perpendicular to the $c$-axis~\cite{fi77b}. Since for 
polycrystalline probes $\tau_c$ is an average over the relaxation rates of 
these modes, it is possible that the experimental data in the crossover region 
yield effective exponents less than $z=2$. Here more experimental work
with single crystals and a more detailed mode coupling analysis
beyond the scaling analysis in Ref.~\cite{fi77b} would be needed to
clarify the situation. 

\subsection{Muon Spin Relaxation ($\mu$SR)}
\label{s4.4}

In muon spin relaxation ($\mu$SR) experiments one observes the muon
precession in an applied magnetic field or in a local magnetic field
inside the sample\footnote{Muon spin relaxation experiments are
often performed in zero applied external field}. For recent work on the
application of $\mu$SR to the investigation of spin dynamics we
refer the interested reader to Refs.~\cite{sc85,co87,ha89}.

In analogy with other implanted probes\footnote{The positive muon
ends up in interstitial sites in solids, and in metals the actual
site is in most cases identical to that preferred by hydrogen.}, 
like those discussed in section~\ref{s4.3}, the muon spin will relax through 
interactions with fluctuating magnetic fields in its surroundings. 
However, in contrast the local magnetic field at the interstitial site 
of the muon has comparable contributions from the Fermi contact field 
{\it and} the dipolar field. 
In rare earth materials the residual dipolar field is high, and often the
dominant field \cite{ka82,hkwc86,ka90}. It has a reduced symmetry as compared 
to the Fermi contact coupling. Whereas the Fermi contact coupling is isotropic 
in space, the residual dipolar field has an anisotropy with respect to the wave 
vector. In the case of a hyperfine field dominated by the residual dipolar field 
this implies that the spin fluctuations transverse and longitudinal with 
respect to the wave vector contribute with different weights to the $\mu$SR 
relaxation rate. Hence, one has a situation quite different from 
e.g. PAC and ME measurements where the interaction of the substitutional 
probe with the host atoms is mainly a contact hyperfine coupling.
This feature of $\mu$-SR offers the possibility to design
experiments which allow to distinguish between spin fluctuations
longitudinal and transverse to the wave vector 
${\bf q}$~\cite{yrf93a,yrf93b,Reotier94a}. 

Among the basic ferromagnets muon relaxation has been measured in
$Fe$ \cite{hffs86} and $Ni$ \cite{nyim84}, which have cubic crystal structure,
and in $Gd$ \cite{whkw86,whkw89} which crystallizes in a hexagonal lattice.
The data for the muon relaxation rate in $Fe$ were interpreted by a dynamic 
exponent $z=2.0$~\cite{hffs86} which would indicate that the data are taken in
the dipolar region. Actually, the data cover both the dipolar and the 
isotropic region -- the crossover temperature for iron is 
$T_{_{\rm D}} - T_c = 8.6 K$ (see Table~\ref{table4.1}) -- with fairly
large where the data error bars in the isotropic region, so that a crossover 
from $z=2.5$ to $z=2.0$ is not excluded by the data \cite{fa91}. The 
observations in $Ni$ \cite{nyim84} seemed to be conflicting with 
PAC~\cite{csh80} measurements, if one assumes that in both cases the contact 
field is the dominant contribution to the local magnetic field at the muon 
site. However, one has to  take into account that the interaction of the 
muon with the surrounding is more complicated than for the other hyperfine 
probes, especially as already pointed out by Yushankhai~\cite{yu89} one has 
to take into account the classical dipolar interaction between the muon 
magnetic moment and the lattice ion magnetic moments. This has recently 
been done in Ref.~\cite{yrf93a,yrf93b}, the results of which we will 
review next.

A zero--field $\mu$SR measurement consists of measuring the depolarization 
function $P_z(t)$~\cite{sc85}, 
\begin{equation}
  P_z(t) = {1 \over 2} Tr 
  \left[ \rho_m \sigma_z \sigma_z (t) \right] \, ,
\nonumber 
\end{equation}
where $\rho_m$ is the density operator of the magnet and $\sigma_z$ the
projection of the Pauli spin operator of the muon spin on the $z$-axis.
$Tr[A]$ stands for the trace of $A$ over the muon and magnet quantum states.
Since the magnetic fluctuations are sufficiently rapid, the depolarization
they induce can be treated in the motional narrowing limit. One 
finds~\cite{yrf93a}
\begin{equation}
  P_z(t) = \exp \left[ - \lambda_z t \right] \, ,
\label{4.35}
\end{equation}
with the damping rate
\begin{equation}
 \lambda_z = {\pi {\cal D} \over V}
 \int {d^3 q \over (2 \pi)^3}
 \left\{ \left[ {\bf G} ({\bf q}) {\bf C} ({\bf q})  {\bf G} (-{\bf q}) 
 \right]^{xx} +
         \left[ {\bf G} ({\bf q}) {\bf C} ({\bf q})  {\bf G} (-{\bf q}) 
 \right]^{yy}
 \right\} \, ,
\label{4.36}
\end{equation}
where 
\begin{equation}
 C^{\alpha \beta} ({\bf q}) =
 {1 \over 2} \left[ \langle S^\alpha_{\bf q} (\omega) S^\beta_{-{\bf q}} 
                    (-\omega) \rangle + 
                    \langle S^\beta_{\bf q} (\omega) S^\alpha_{-{\bf q}} 
                    (-\omega) \rangle 
             \right] \mid_{\omega = 0}
\label{4.37}
\end{equation}
is the symmetrized correlation function at zero frequency, and
\begin{equation}
  {\cal D} = \left( {\mu_0 \over 4 \pi} \right)^2 
             \gamma_\mu^2 (g_L \mu_B)^2 \, .
\label{4.38}
\end{equation}
$\gamma_\mu = 851.6 \, Mrad \, s^{-1} \, T^{-1}$ is the gyro-magnetic ratio 
of the muon. The tensor ${\bf G}$ characterizes the type of interactions by
which the muon spin is coupled to the magnetic moments of the
material considered. It is given by 
\begin{equation}
  G^{\alpha \beta} ({\bf q}) = r_\mu H \delta^{\alpha \beta} + 
                               D^{\alpha \beta} ({\bf q}) \, .
\label{4.39}
\end{equation}
The first term describes the Fermi contact coupling, where $r_\mu$ is the 
number of nearest--neighbor magnetic ions to the muon localization site and
$H$ measures the strength of the hyperfine coupling (see
Table~\ref{table4.1}). The 
simple form for the contact coupling is valid only if the muon site is a 
center of symmetry. The second term characterizes the dipolar coupling and 
can be calculated via Ewald's method~\cite{yrf93a}. To lowest order in the 
wave vector ${\bf q}$ one finds
\begin{equation}
 D^{\alpha \beta} ({\bf q}) = 4 \pi \left[ B^{\alpha \beta} ({\bf q}=0) - 
                                           {q^\alpha q^\beta \over q^2} 
                                    \right] \, .
\label{4.40}
\end{equation}
The values of the tensor components 
$B^{\alpha \beta} ({\bf q}=0)$ depend on the muon 
site. For muons in a tetrahedral or octahedral interstitial site in an $fcc$
lattice one finds $B^{\alpha \beta} ({\bf q}=0) = {1 \over 3} 
\delta^{\alpha \beta}$~\cite{Kronmueller79}. The situation is more complex 
for a $bcc$ lattice~\cite{Kronmueller79}.

For simplicity we confirm ourselves to the case of an $fcc$ lattice and refer 
the reader to  Refs.~\cite{yrf93a,yrf93b,Reotier94a} for a discussion of the 
more complicated cases of an $bcc$ cubic crystal or a hexagonal lattice 
structure. For an $fcc$ lattice the damping rate is found to be~\cite{yrf93a}
\begin{equation}
  \lambda_z = {8 \over 3} {\mu_0 \over 4 \pi} \gamma_\mu^2 {k_B T \over v_a}
  \int_0^{q_{BZ}} dq q^2 
  \left[
  2 p^2 {\chi^T (q) \over \Gamma^T (q)} + (1-p)^2  
        {\chi^L (q) \over \Gamma^L (q)} 
  \right] \, ,
\label{4.41}
\end{equation}
where we have defined 
\begin{equation}
  p = {1 \over 3} + {r_\mu H \over 4 \pi} \, .
\label{4.42}
\end{equation}
When the Fermi contact interaction is large compared to the classical 
dipolar interaction, i.e., $\vert p \vert \gg 1 $, Eq.~\ref{4.41} reduces to 
the result obtained in section ~\ref{s4.3}
\begin{equation}  
  \lambda_z^{\rm contact}  = 
                {1 \over 6 \pi^2} {\mu_ 0 \over 4\pi} 
                (\gamma_\mu r_\mu H)^2 
                {k_BT \over v_a} \int^{ q_{BZ}}_0dq \, q^2 
                \left[ 2{\chi^ T(q) \over \Gamma^T(q)}+
                        {\chi^ L(q) \over \Gamma^ L(q)} 
                 \right] \, .
\label{4.43}
\end{equation}
On the other hand when the Fermi interaction is negligible, i.e. $ p\simeq 1/3
$, we get from Eq.~\ref{4.41}
\begin{equation}
  \lambda_z^{dipolar}  = {16 \over 27} {\mu_ 0 \over 4\pi} \gamma^2_\mu 
                {k_BT \over v_a} \int^{ q_{BZ}}_0dq \, q^2 
                \left[ {\chi^ T(q) \over \Gamma^T(q)}+
                      2{\chi^ L(q) \over \Gamma^ L(q)} 
                 \right] \, .
\label{4.44}
\end{equation}
While Eq.~(\ref{4.43}) has been derived under the hypothesis that the muon spin
interacts with the lattice spins through the isotropic hyperfine interaction,
Eq.~(\ref{4.44}) has been obtained supposing that the coupling is only due to 
the classical dipolar interaction. One should note that the relative weight 
of the transverse and the longitudinal modes is reversed in going from a pure 
contact to a pure dipolar interaction of the muon with the lattice spins. 
Hence in combing PAC with $\mu$-SR measurements it should be possible to 
distinguish between transverse and longitudinal relaxation rates.

Upon introducing polar coordinates as in section~\ref{s4.3} one can write the 
damping rate in scaling form
\begin{equation}
 \lambda_ z  = {\cal
 W} \left[ 2 p^2 I_T (\varphi) + (1-p)^2 I_L (\varphi) 
                        \right] \, ,
\label{4.45}
\end{equation}
where the non universal constant ${\cal W}$ is given by
\begin{equation}
 {\cal W}  ={8\pi^{ 3/2} \over 3P} \sqrt{{ \mu_ 0 \over 4\pi}} \
 {\gamma^ 2_\mu \hbar q^{3/2}_{_{D}} \over g_L\mu_ B} \sqrt{ k_BT_C} \, .
\label{4.46}
\end{equation}
The theoretical values of ${\cal W}$ for the four magnets
considered primarily in this section are listed in Table~\ref{table4.1}. 
The scaling functions $I^{L,T}(\varphi )$ are defined in section~\ref{s4.3}.
Fig.~4.21 indicates that while $I_T(\varphi)$ exhibits a
strong temperature dependence in the whole temperature range of the critical
region, $I_L(\varphi)$ is practically temperature independent for 
$q_{_{\rm D}}\xi >1$. Therefore, the temperature dependence of the $\mu$SR 
damping rate $\lambda_ z$ depends on the relative weight of $I_T(\varphi)$ 
and $I_L(\varphi)$ which is controlled by the parameter $p$. As first noticed 
by Yushankhai~\cite{yu89}, the transverse fluctuations do not contribute to 
$\lambda_z$ if $r_\mu H/4\pi =-1/3$. In this case $\lambda_z$ becomes 
temperature independent near $T_c$. 

In Figs.~4.23 and 4.24 experimental data for $Ni$~\cite{nyim84} and 
$Fe$~\cite{hffs86} are compared with the results from MC--theory. One 
notices that most of the data have been recorded in the crossover temperature
region where $\lambda_ z$ is not predominantly due to the transverse 
fluctuations. Again the MC--theory gives a quantitative explanation of the 
observed relaxation rates.

\newpage

\begin{table}
\setdec 0.0
\caption{Material parameters for $Ni$, $Fe$, $EuO$ and $EuS$. The parameter 
are defined in the main text.}
\bigskip\bigskip
\begin{tabular}{lcccc}
 &$Ni$ &$EuO$ &$EuS$ &$Fe$\\
\tableline
$a (\AA)$ &$3.52$$^{(1)}$ &$5.12$$^{(2)}$ &$5.95$$^{(2)}$ &$2.87$$^{(1)}$\\
$T_C\ (K)$&$627$$^{(1)}$&$69.1$$^{(2)}$&$16.6$$^{(2)}$&$1043$$^{(1)}$\\
$q_{_{D}}\ \left(\AA^{ -1}
\right)$&$0.013$$^{(3)}$&$0.147^{(3,4)}$&$0.245^{(1,35)}$&$0.045$$^{(4)}$$(0.033)$$^{(3)}$\\
$\xi_ 0\ (\AA)$&$1.23$$^{(5)}$&$1.57$$^{(6)}$&$1.81$$^{(6)}$&$0.95$$^{(7)}$$(0.82)$ $^{(8)}$ \\
$r_\mu H/4\pi \ (T=0\ K)$&$-0.11$$^{(9)}$&$0\ (?)$&$0\ (?)$&$-0.51$$^{(9)}$\\
$\Upsilon \ (\AA)$&$0.00184$&$0.725$&$3.16$&$0.0239\ (0.0129)$\\
$\Lambda_{ {\rm exp}}\ \left(meV \AA^{ 5/2} \right)$&$350$$^{(5)}$&$8.7$$^{(4)}$$(8.3)$$^{(10)}$&
$2.1$$^{(11)}$$(2.25)$$^{(12)}$&$130$$^{(13)}$\\
$\Lambda_{ {\rm lor}}\ \left(meV \AA^{ 5/2} \right)$&$241$&$7.09$&$2.08$&$90.0\ (123)$\\
${\cal W}_{th}\ (MHz)$&$0.358$&$5.35\ (5.60)$&$6.87\ (6.41)$&$2.98\ (2.56)$\\
\end{tabular}
\bigskip
\noindent {\small \baselineskip = 9pt The values have been taken from the references 
(1)~\cite{Kittel71}, (2)~\cite{pdn76}, (3)~\cite{k88}, (4)~\cite{m86}, (5)~\cite{bmt91}, 
(6)~\cite{n76b}, (7)~\cite{wbs84}, (8)~\cite{n76}, (9)~\cite{Denison79}, (10)~\cite{bs86}, 
(11)~\cite{bsbz87}, (12)~\cite{bsbz88}, and (13)~\cite{m82}. When no reference is indicated, 
the parameter has been computed from a formula given in the main text with parameters given 
in tables III.1 and IV.1.}
\label{table4.1}
\end{table}

\newpage

\newpage

\begin{table}
\caption{The dipolar wave vector $q_{_{\rm D}}$, the Curie
temperature $T_c$, the experimental non universal constant for
the Onsager coefficient $L_{\rm d}$, $L_{\rm bg}$ and for the 
autocorrelation time $H_{\rm exp}$.}
\bigskip\bigskip
\begin{tabular}{lcccccc}
&$EuS$&$EuO$&$CdCr_2S_4$&$CdCr_2Se_4$&$Fe$&$Ni$ \\
\tableline
$q_{_{\rm D}} [{\AA}^{-1}]$ & $0.245$ & $0.147$ & $0.058$ & $0.034$ &
$0.045(0.033)$ & $0.013$ \\
$T_c [K]$ & $16.6$ & $69.1$ & $84.4$ & $127.8$ & $1043$ & $627$ \\
$\hbar L_{\rm d} (\mu eV)$ & $38$ & $24$ & $5.9$ & $4.4$ & $-$ &
$3.0$ \\
$\hbar L_{\rm bg} (\mu eV)$ & $1.8$ & $0.64$ & $0.01$ & $0.01$ & $-$ & $-$ \\
$H_{\rm exp}[10^{-13} sec]$ & $-$ & $-$ & $-$ & $-$ & $6.0$ &
$9.2$ \\
\end{tabular}
\bigskip 
{\small  The values are taken from Refs.~\cite{k88} and the references in 
Table~\ref{table4.1}.}
\label{table4.2}
\end{table}

\centerline{\bf Captions to the figures:}
\bigskip
\bigskip

\noindent {\bf Figure 4.1:} Typical scattering geometry in neutron scattering 
   experiments. The magnetic guide field ${\bf H}_V$ is used to define the
   beam polarization ${\bf P}$. In order to measure the longitudinal spin 
   fluctuations ${\bf S}^L$ one has to go to a finite Bragg peak
   $\gvect {\tau}$ with the wave vector ${\bf q}$ perpendicular to the 
   scattering vector ${\bf Q} = \gvect{\tau} + {\bf q}$. For unpolarized
   neutrons only the transverse fluctuations can be measured in the forward
   direction, $\gvect{\tau} = {\bf 0}$.
\bigskip

\noindent {\bf Figure 4.2: } Scaling function of the transverse Kubo
relaxation function versus the scaled time variable $\tau_T$ at
$T=T_c+0.25K$ for $q=0.018 \AA^{-1}$ (solid), $q=0.025 \AA^{-1}$ (dotted), 
$q = 0.036 \AA^{-1}$ (dashed), $q=0.071 \AA^{-1}$ (long dashed) and 
$q=0.150 \AA^{-1}$ (dot-dashed).
\bigskip

\noindent {\bf Figure 4.3: } The same as in Fig.~4.2 for the
longitudinal Kubo relaxation function.
\bigskip

\noindent {\bf Figure 4.4: } Scaling function of the transverse Kubo
relaxation function versus the scaled time variable $\tau_T$ at
$T=T_c+8.0K$ for the same set of wave vectors $q$ as in Figs.~4.2 and 4.3.
\bigskip

\noindent {\bf Figure 4.5: } The same as in Fig. 4.4 for the
longitudinal Kubo relaxation function.
\bigskip

\noindent {\bf Figure 4.6: } Scaling function of the transverse Kubo
relaxation function versus $\tau^{\prime}=\Lambda_{\rm lor} q^z t$
at $T=T_c$ for $q=0.025 \AA^{-1}$ (solid), $q=0.036 \AA^{-1}$
(dotted), $q=0.150 \AA^{-1}$ (dashed) and $q=0.300 \AA^{-1}$ 
(long dashed).
\bigskip

\noindent {\bf Figure 4.7: } Scaling function of the transverse Kubo
relaxation function versus $\tau^{\prime}= \Lambda_{\rm lor}
q^z t$ at $T=T_c+0.5K$ for $q=0.018 \AA^{-1}$ (solid), $q=0.025
\AA^{-1}$ (dotted), $q=0.036 \AA^{-1}$ (dashed) and $q=0.071 \AA^{-1}$ 
(long dashed).
\bigskip

\noindent {\bf Figure 4.8: } Transverse Kubo relaxation function
$\Phi^T(q,g,t)$ at $q=0.024 \AA^{-1}$ (solid line) and $q=0.028
\AA^{-1}$ (dashed-dotted line) for dipolar ferromagnets versus
time in $nsec$. The dashed line is the Kubo relaxation function
for short range exchange interaction only at $q=0.024 \AA^{-1}$.
Data points for $q = 0.024 \AA^{-1}$ from Fig.~1 of Ref.~\cite{m86}.
\bigskip

\noindent {\bf Figure 4.9: } Normalized experimental data deconvoluted by a
maximum entropy method. The solid line shows the spectral shape function
predicted by mode coupling theory. The full circles (dashed line)
indicate(s) the deconvoluted data. Figure taken from Ref.~\cite{Goerlitz92a}.
\bigskip

\noindent {\bf Figure 4.10: } Scaling functions versus $y^{-1}={q /
q_{_{\rm D}} }$ at $T_c$ (in units of the theoretical non universal
constant $\Lambda$) for (i) the HWHM of the complete solution of
the MC equations for the transverse Kubo relaxation
function (solid), (ii) transverse (point-dashed) and (iii)
longitudinal line (dashed) width in Lorentzian approximation. Experimental
data of the transverse line width for 
$EuO$ ( $\sqcup$ Ref.~\cite{bs86}, $\diamondsuit$ Ref.~\cite{m84} )
and $Fe$ ( $\triangle$ Ref.~\cite{m82}). 
\bigskip

\noindent {\bf Figure 4.11: } Scaling function of the transverse line
width in Lorentzian approximation (solid lines) and of the HWHM
of the complete solution (dashed lines) versus the scaling
variable $x=(q \xi)^{-1}$ for a set of temperatures              
($T-T_c=$ a) 1.4K, b) 5.8K, c) 21K, d) 51K). The Resibois-Piette
function (solid) and the HWHM from the complete solution of the
MC equations (dashed) without dipolar interaction is also plotted
e). Experimental results for $Fe$ from Refs.~\cite{m82,m84} ( $T-T_c =$ 
$(\sqcup) \, 1.4K$, $(\diamondsuit) \, 5.8K$, $(\triangle) \,
21.0K$, $(\circ) \, 51.0K$ ).
\bigskip

\noindent {\bf Figure 4.12: } 
Scaling plot of the temperature dependence of the relaxation rate of
critical fluctuations in $EuO$ above the Curie temperature $T_c = 69.3 K$.
The solid lines represent the results from mode coupling theory~\cite{fs87}.
The data were taken between $q = 0.018 \AA^{-1}$ and $q = 0.15 \AA^{-1}$.
Taken from Ref.~\cite{m88,mfhs89} ($\kappa_1 = 1/\xi$).
\bigskip

\noindent {\bf Figure 4.13: } 
Comparison of the longitudinal linewidth of Lorentzian fits normalized to 
$\Lambda_{\rm exp} q^{5/2}$ with $\Lambda_{\rm exp} = 2.1 meV \AA^{5/2}$
to the dipolar dynamic scaling function predicted by mode coupling theory 
for $q_{_{D}} = 0.245 \AA^{-1}$ for $EuS$. Figure taken from Ref.~\cite{bgkm91}
($\kappa = 1/\xi$).
\bigskip

\noindent {\bf Figure 4.14: } 
The same as Fig.~4.13 for the transverse widths. Figure taken from 
Ref.~\cite{bgkm91}.
\bigskip

\noindent  {\bf Figure 4.15: } Scaled peak positions
$q_0(\Lambda_{\rm lor}/\omega)^{2/5}$ for constant energy scans of
the scattering function for the complete solution of the MC
equations versus the scaling variable ${\hat \omega}$ for
$\varphi=$ a) 1.490, b) 1.294, c) 0.970 and d) for the isotropic
case, i.e., $\varphi=0$. {\bf Inset:}
$S^T(q,\omega)/S^T(0,\omega)$ in arbitrary units versus ${1 \over
r}$ for $\varphi=1.294$ for some typical values of the scaled
frequency (${\hat \omega} = 10^{L/10}$ with $L=$ 8 (solid), 10
(dashed), 12 (point dashed), 14 ($- \cdot \cdot - \cdot \cdot -
$) and 16 ($- \cdot \cdot \cdot - \cdot \cdot \cdot - $)
indicated in the graph by crosses).
\bigskip

\noindent {\bf Figure 4.16: } The same as in Fig.~4.15 for the scattering
function resulting from the Lorentzian approximation.
\bigskip

\noindent {\bf Figure 4.17: } Universal crossover function $F(1/q_{_{\rm D}} 
\xi)/F(0)$ with $F(0) = 0.1956$ for the Onsager kinetic coefficient at zero 
frequency and zero wave vector versus the scaling variable $1/q_{_{\rm D}} 
\xi$.
\bigskip

\noindent {\bf Figure 4.18: } 
Critical part of the Onsager kinetic coefficient for the homogeneous spin 
dynamics above $T_c$ of $CdCr_2Se_4$ (Ref.~\cite{kp78}), $CdCr_2S_4$ 
(Ref.~\cite{ks78}), $EuO$ (Ref.~\cite{ksbk78}), $EuS$ (filled circles: 
Ref.~\cite{kkw76}, open circles: Ref.~\cite{k88}), and $Ni$ 
(Ref.~\cite{sa67}). The solid lines represent the result from mode coupling 
theory~\cite{fs88a}. {\bf Inset:} Nonuniversal amplitudes 
$\hbar L_d q_{_{\rm D}}^{-3/2}$ for the kinetic coefficient and the relaxation
rate at the critical point $\hbar L_q q^{-1} q_{_{\rm D}}^{-1/2} = 
\Gamma^T (q,T=T_c) / q^{5/2} \approx 5.1326 \Lambda$ for ferromagnets 
including $Fe$ and $Co$. Figure taken from Ref.~\cite{k88}.
\bigskip

\noindent {\bf Figure 4.19: }
The real and imaginary parts of $F$ are shown in a) and b)
versus the scaled frequency ${\hat \omega} = \omega/\Lambda q_{_{D}}^z$
for several values of $\varphi = \arctan (q_{_{D}}\xi)$: $\varphi_1 = 
0.99 {\pi \over 2}$ (top curve), $\varphi_2 = 0.90 {\pi \over 2}$, 
$\varphi_3 =  {\pi \over 2} (1-10^{-0.5})$, $\varphi_4 =  {\pi \over 2} (1-10^{-0.25})$,
$\varphi_5 =  {\pi \over 2} (1-10^{-0.1})$, $\varphi_6 =  {\pi \over 2} (1-10^{-0.025})$
(bottom curve).
\bigskip

\noindent {\bf Figure 4.20: }
The scaling functions for the a) Onsager coefficient at zero frequency
$F (1/q_{_{D}}\xi)$ and b) the scaling functions $\omega_{c2}  
(1/q_{_{D}}\xi)$ and c) $\omega_{c1}  (1/q_{_{D}}\xi)$ characterizing the 
large and low frequency behaviour, respectively. 
\bigskip

\noindent {\bf Figure 4.21: } 
Scaling functions $I_{L,T} (\varphi)$ for the transverse and longitudinal 
auto-correlation-time $\tau_{L,T}$ versus ${1 / q_{_{\rm D}} \xi}$.
\bigskip

\noindent {\bf Figure 4.22: } 
Auto correlation time $\tau_c /H$ (in units of 
the non universal constant H) versus the scaling variable ${1 \over
q_{_{\rm D}} \xi}$ (solid line). Experimental results for the auto
correlation time in units of $H_{exp}$ for $Fe$ (Ref.~\cite{hcs82}: $\sqcup$)
and $Ni$ (Ref.~\cite{rh72}: $\diamondsuit$, Ref.~\cite{gh73}: 
$\triangle$, Ref.~\cite{hcs82}: $\circ$). 
\bigskip

\noindent {\bf Figure 4.23: } 
Temperature dependence of the $\mu$SR damping rate for metallic $Ni$. The 
points are the experimental data of Nishiyama et al.~\cite{nyim84}. The full 
line is the result of the model which takes the muon dipolar interaction into 
account. The dashed line gives the prediction when this latter interaction is 
neglected.
\bigskip

\noindent {\bf Figure 4.24: } 
Temperature dependence of the $\mu$SR damping 
rate for metallic $ Fe $. The points are the experimental data of Herlach et 
al.~\cite{hffs86}. The curves are the predictions of mode-coupling theory for 
different sets of material parameters. In Fig.~4.24a (Fig.~4.24b) the muon is 
supposed to diffuse between tetrahedral (octahedral) sites. The full and 
dashed lines are present the results obtained with $\left(q_{_{D}},\xi_ 
0 \right)$ equal to $\left(0.033\ \AA^{ -1}, 0.82\ \AA \right)$ and to 
$\left(0.045\ \AA^{-1},0.95\ \AA \right)$ respectively; see 
Table~\ref{table4.1}. 
\vfill

\newpage
\vfill
\begin{figure}[h]
  \centerline{\rotate[r]{\epsfysize=5in \epsffile{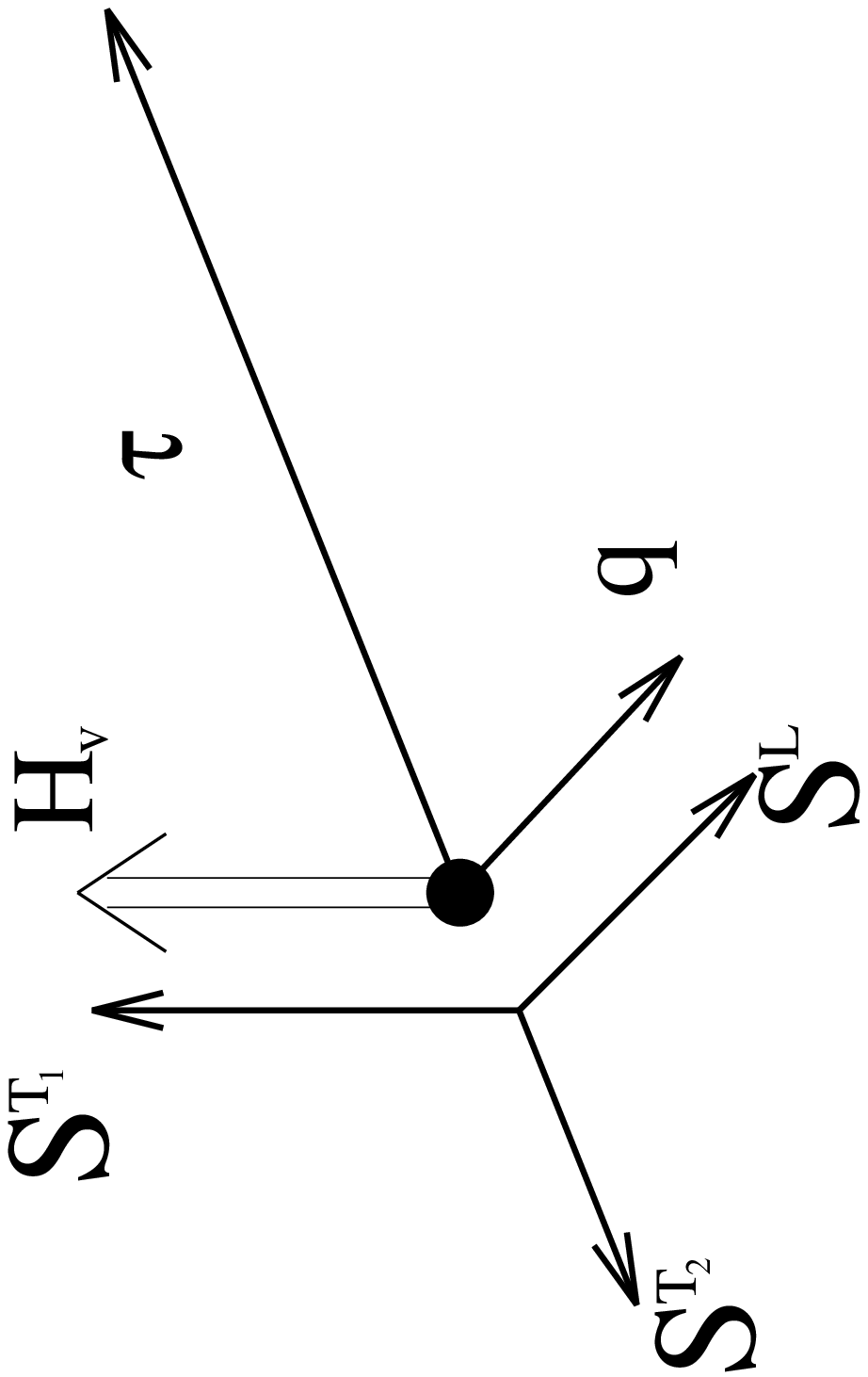}}}
\bigskip \bigskip
  {{\bf Figure 4.1:} Typical scattering geometry in neutron scattering 
   experiments. The magnetic guide field ${\bf H}_V$ is used to define the
   beam polarization ${\bf P}$. In order to measure the longitudinal spin 
   fluctuations ${\bf S}^L$ one has to go to a finite Bragg peak
   $\gvect {\tau}$ with the wave vector ${\bf q}$ perpendicular to the 
   scattering vector ${\bf Q} = \gvect{\tau} + {\bf q}$. For unpolarized
   neutrons only the transverse fluctuations can be measured in the forward
   direction, $\gvect{\tau} = 0$.}
\label{fig4.1}
\end{figure}
\newpage
\vfill
\begin{figure}[h]
  \centerline{\rotate[r]{\epsfysize=5in \epsffile{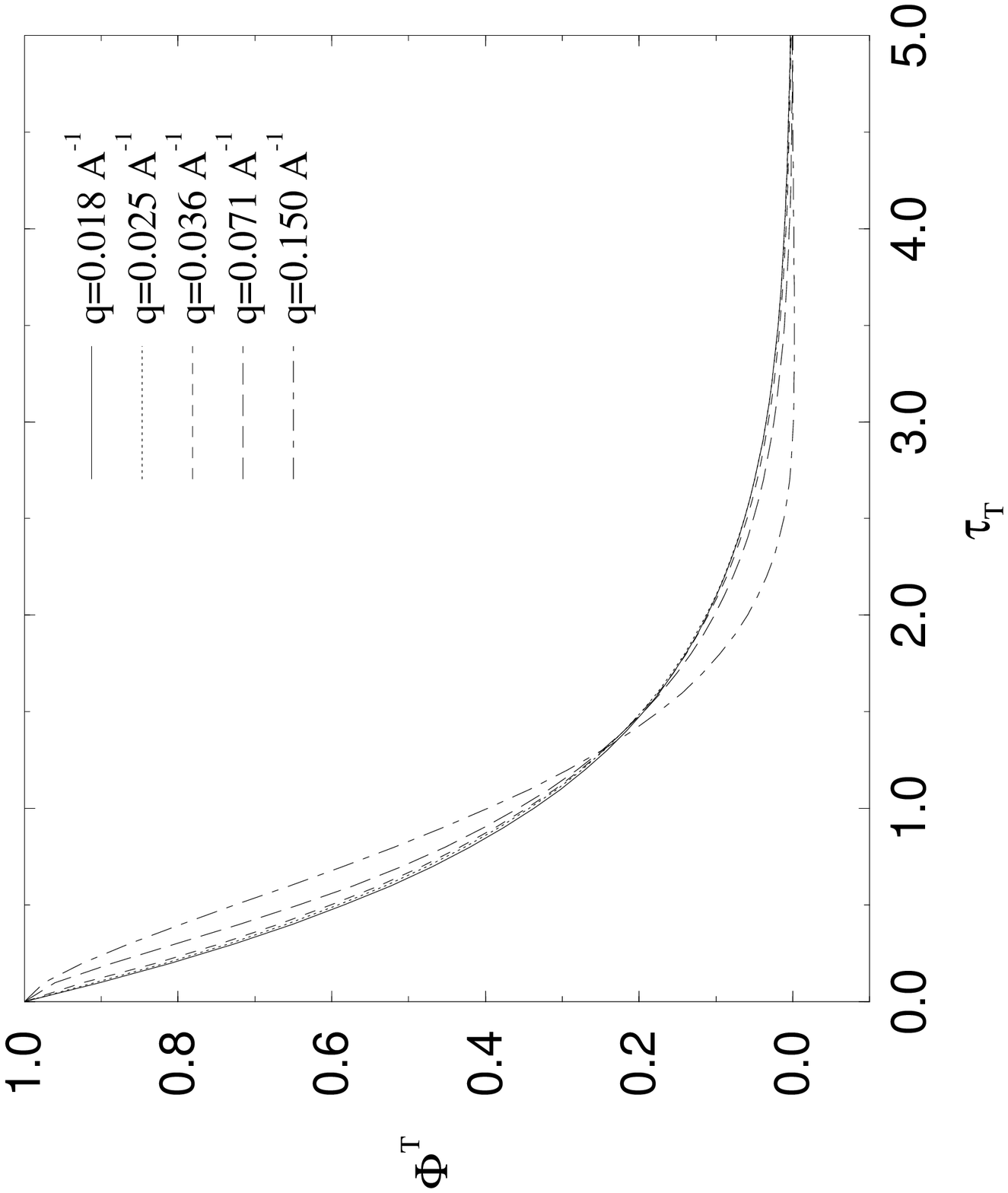}}}
\bigskip \bigskip
  {{\bf Figure 4.2:} Scaling function of the transverse Kubo
   relaxation function versus the scaled time variable $\tau_T$ at
   $T=T_c+0.25K$ for $q=0.018 \AA^{-1}$ (solid), $q=0.025
  \AA^{-1}$ (dotted), $q = 0.036 \AA^{-1}$ (dashed), $q=0.071 \AA^{-1}$ 
  (long dashed) and $q=0.150 \AA^{-1}$ (dot-dashed).}
\label{fig4.2}
\end{figure}
\newpage
\vfill
\begin{figure}[h]
  \centerline{\rotate[r]{\epsfysize=5in \epsffile{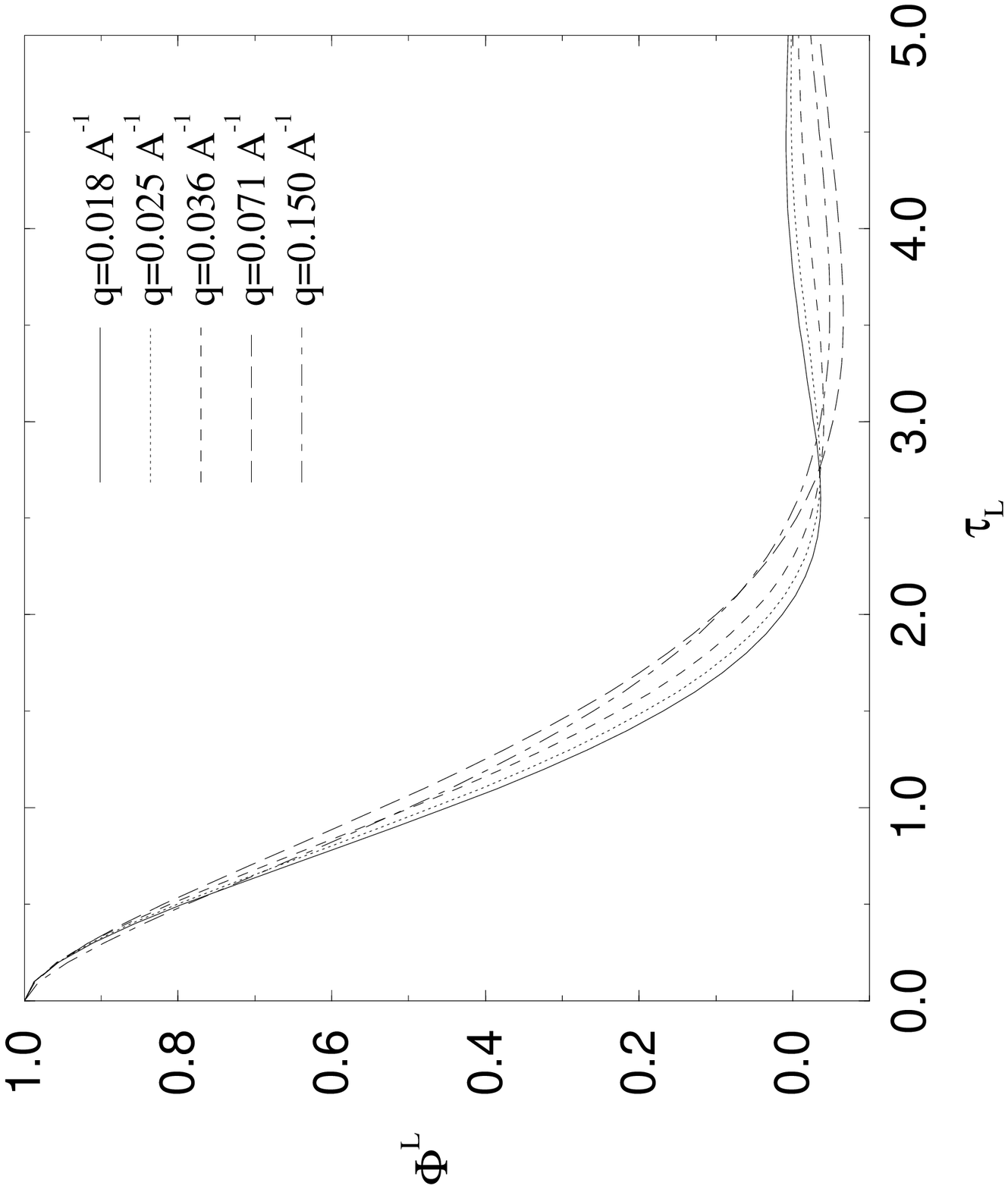}}}
\bigskip \bigskip
  {{\bf Figure 4.3:} The same as in Fig.~4.2 for the
   longitudinal Kubo relaxation function.}
\label{fig4.3}
\end{figure}
\newpage
\vfill
\begin{figure}[h]
 \centerline{\rotate[r]{\epsfysize=5in \epsffile{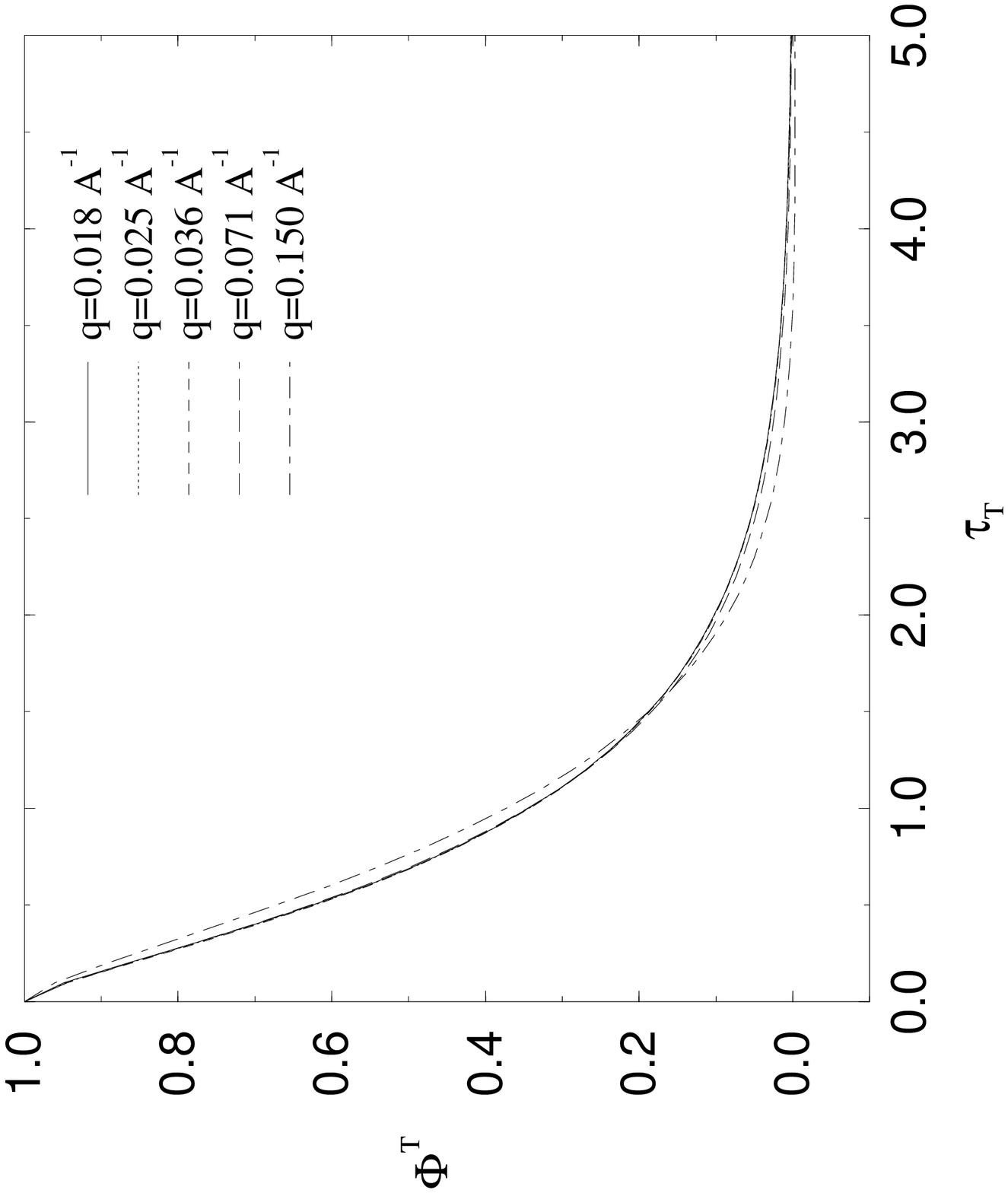}}}
\bigskip \bigskip
  {{\bf Figure 4.4:} Scaling function of the transverse Kubo
   relaxation function versus the scaled time variable $\tau_T$ at
   $T=T_c+8.0K$ for the same set of wave vectors $q$ as in Figs.~4.2
   and 4.3.}
\label{fig4.4}
\end{figure}
\newpage
\vfill
\begin{figure}[h]
  \centerline{\rotate[r]{\epsfysize=5in \epsffile{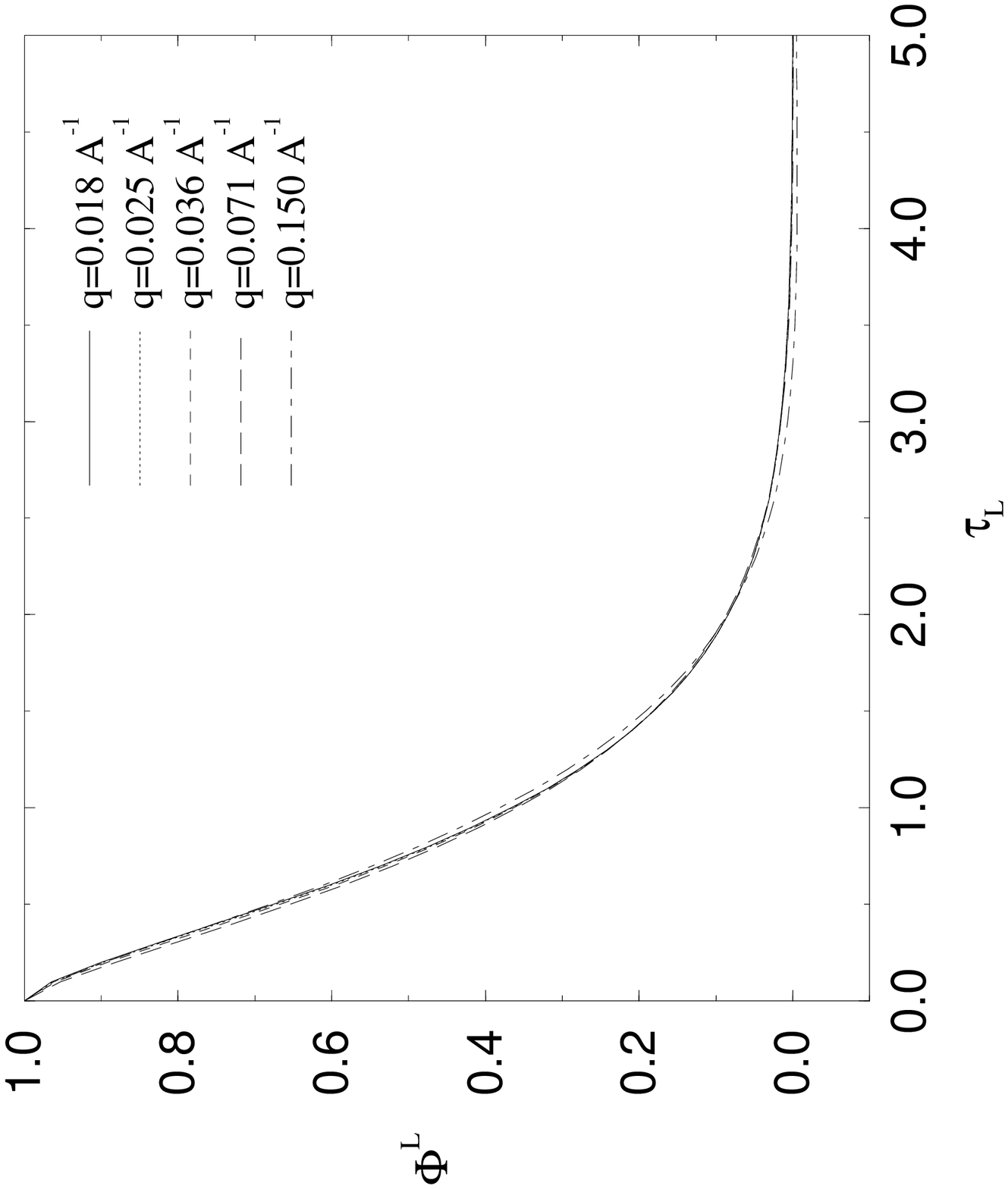}}}
\bigskip \bigskip
  {{\bf Figure 4.5:} The same as in Fig.~4.4 for the
   longitudinal Kubo relaxation function.}
\label{fig4.5}
\end{figure}
\newpage
\vfill
\begin{figure}[h]
  \centerline{\rotate[r]{\epsfysize=6in \epsffile{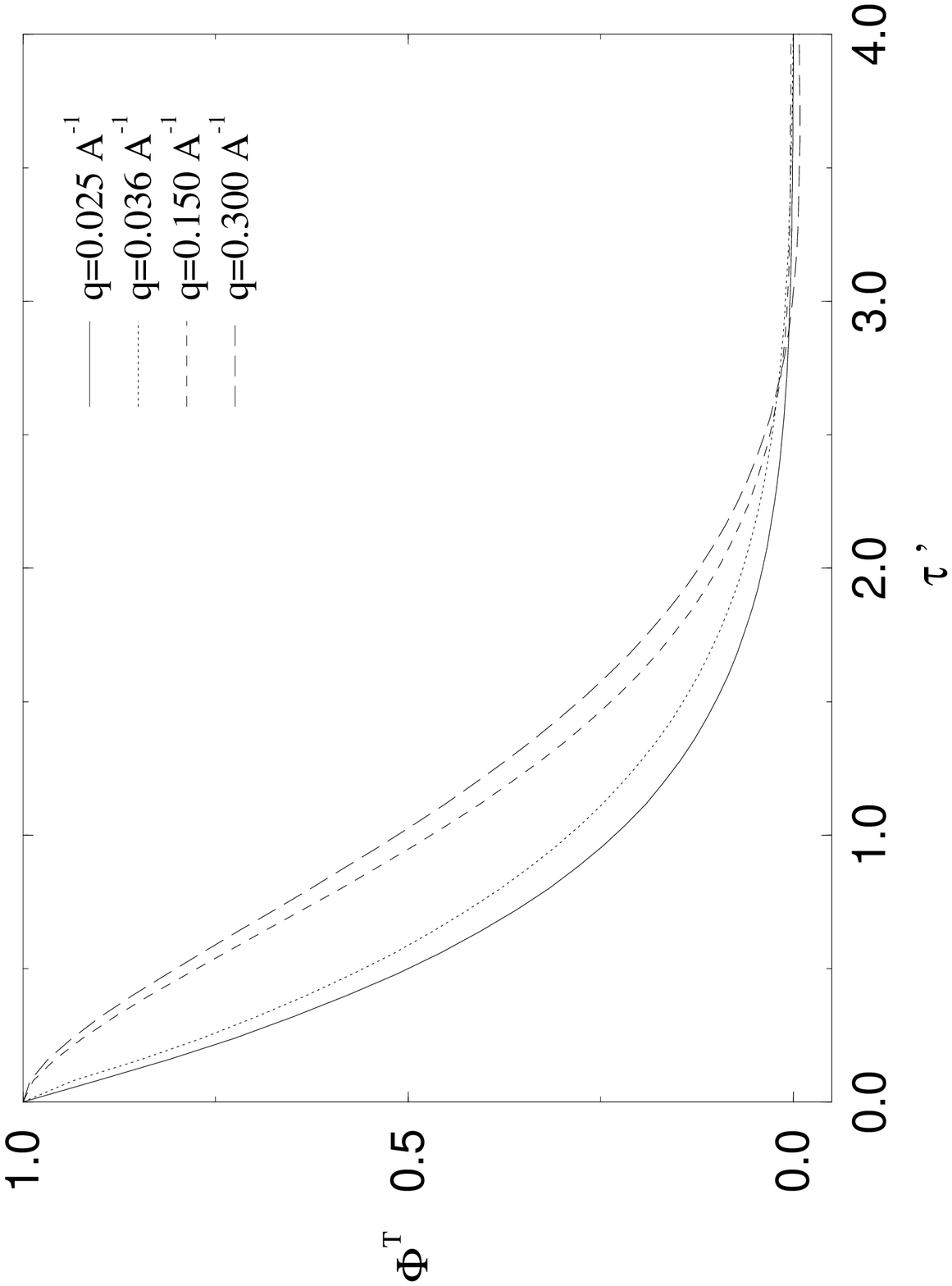}}}
\bigskip \bigskip
  {{\bf Figure 4.6:} Scaling function of the transverse Kubo
   relaxation function versus $\tau^{\prime}=\Lambda_{\rm lor} q^z t$
   at $T=T_c$ for $q=0.025 \AA^{-1}$ (solid), $q=0.036 \AA^{-1}$
   (dotted), $q=0.150 \AA^{-1}$ (dashed) and
   $q=0.300 \AA^{-1}$ (long dashed).  }
  \label{fig4.6}
\end{figure}
 \newpage
\vfill
\begin{figure}[h]
  \centerline{\rotate[r]{\epsfysize=5in \epsffile{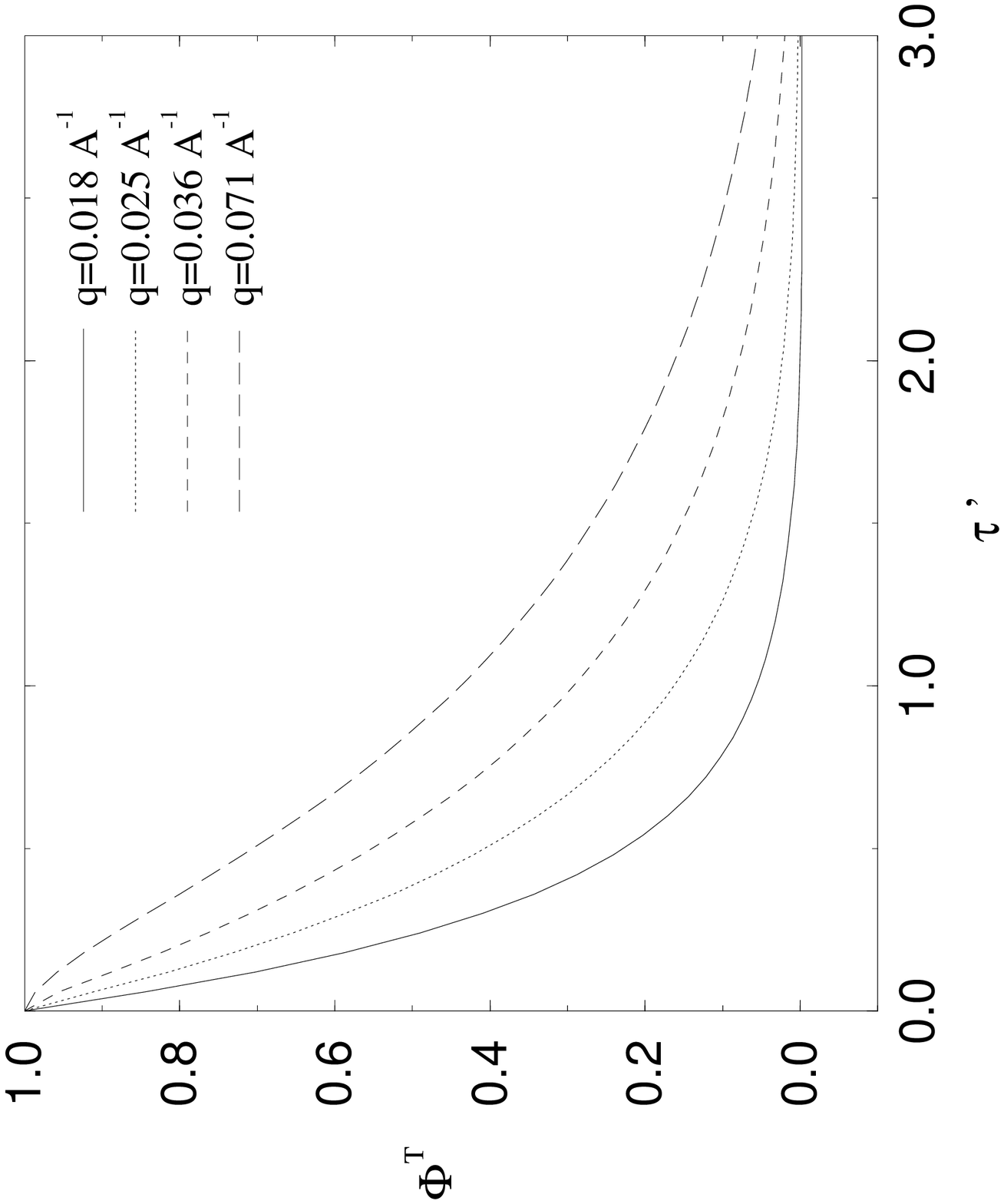}}}
\bigskip \bigskip
  {{\bf Figure 4.7:} Scaling function of the transverse Kubo
   relaxation function versus $\tau^{\prime}= \Lambda_{\rm lor}
   q^z t$ at $T=T_c+0.5K$ for $q=0.018 \AA^{-1}$ (solid), $q=0.025
   \AA^{-1}$ (dotted), $q=0.036 \AA^{-1}$ (dashed) and $q=0.071 \AA^{-1}$ 
  (long dashed).}
\label{fig4.7}
\end{figure}
\newpage
\vfill
\begin{figure}[h]
  \centerline{\rotate[r]{\epsfysize=6in \epsffile{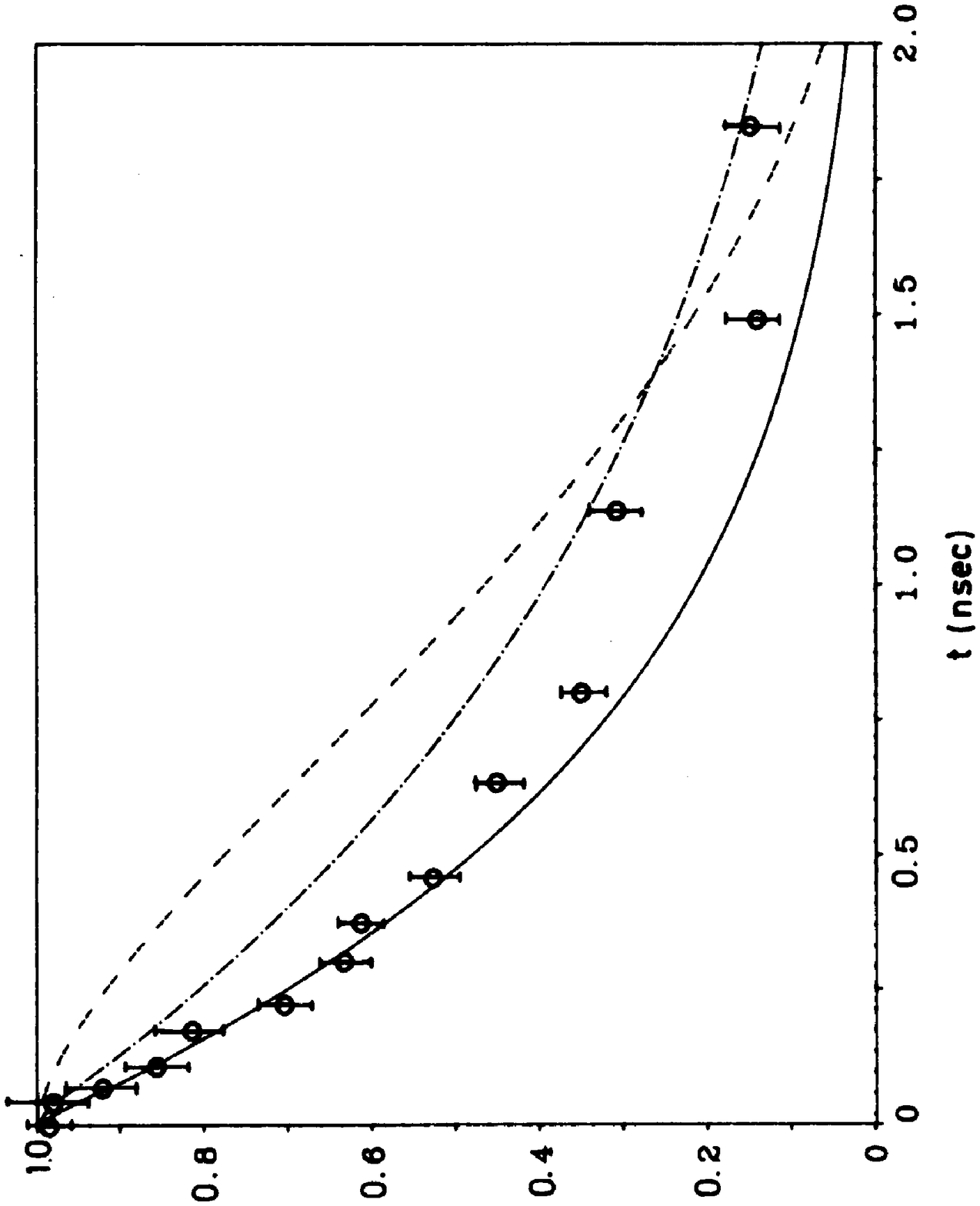}}}
\bigskip \bigskip
  {{\bf Figure 4.8:} Transverse Kubo relaxation function
   $\Phi^T(q,g,t)$ at $q=0.024 \AA^{-1}$ (solid line) and $q=0.028
   \AA^{-1}$ (dashed-dotted line) for dipolar ferromagnets versus
   time in $nsec$. The dashed line is the Kubo relaxation function
   for short range exchange interaction only at $q=0.024 \AA^{-1}$.
   Data points $q = 0.024 \AA^{-1}$ from Fig.~1 of Ref.~\cite{m86}.}
\label{fig4.8}
\end{figure}
\newpage
\vfill
\begin{figure}[h]
  \centerline{\rotate[r]{\epsfysize=5in \epsffile{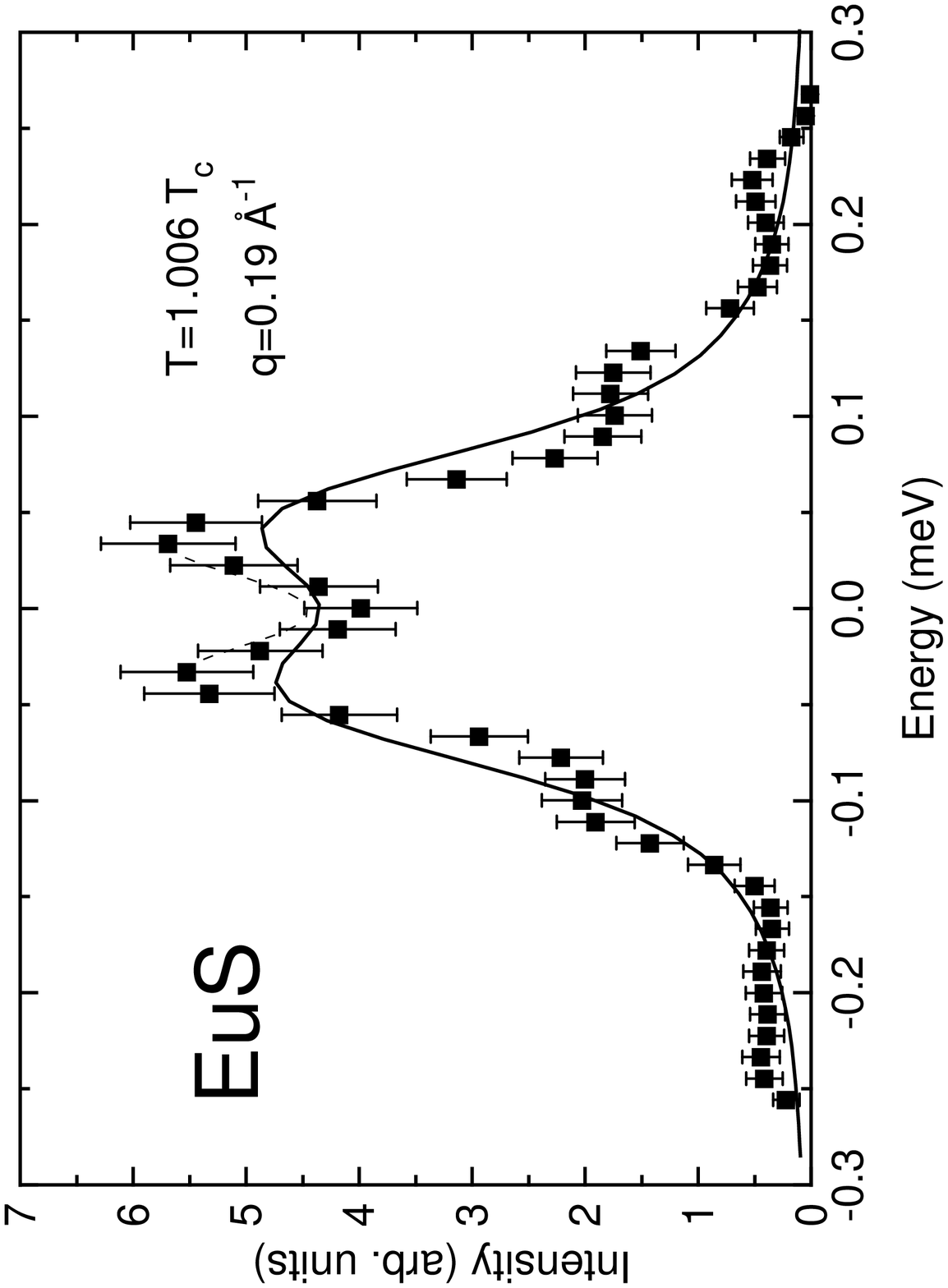}}}
\bigskip \bigskip
  {{\bf Figure 4.9:} Normalized experimental data deconvoluted by a
   maximum entropy method. The solid line shows the spectral shape function
   predicted by mode coupling theory. The full circles (dashed line)
   indicate(s) the deconvoluted data. Taken from Ref.~\cite{Goerlitz92a}.}
\label{fig4.9}
\end{figure}
\newpage
\vfill
\begin{figure}[h]
  \centerline{\rotate[r]{\epsfysize=5in \epsffile{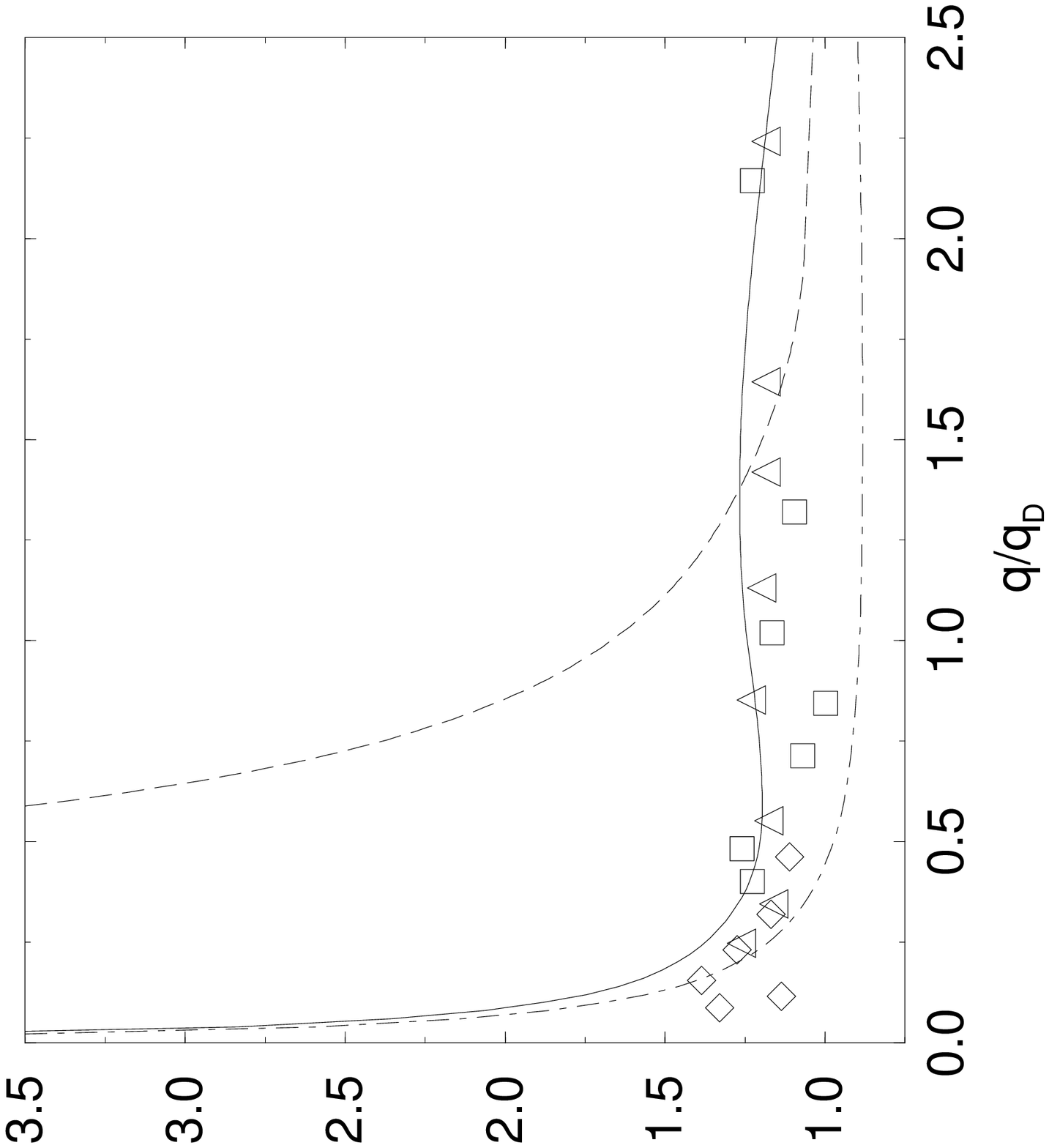}}}
\bigskip \bigskip
  {{\bf Figure 4.10:} Scaling functions versus $y^{-1}={q /
   q_{_{\rm D}} }$ at $T_c$ (in units of the theoretical non universal
   constant $\Lambda$) for (i) the HWHM of the complete solution of
   the MC equations for the transverse Kubo relaxation
   function (solid), (ii) transverse (point-dashed) and (iii)
   longitudinal line (dashed) width in Lorentzian approximation. Experimental
   data of the transverse line width 
   for $EuO$ ($\sqcup$ Ref.~\cite{bs86}, $\diamondsuit$ Ref.~\cite{m84})
   and $Fe$ ($\triangle$ Ref.~\cite{m82}).}
\label{fig4.10}
\end{figure}
\newpage
\vfill
\begin{figure}[h]
  \centerline{\rotate[r]{\epsfysize=5in \epsffile{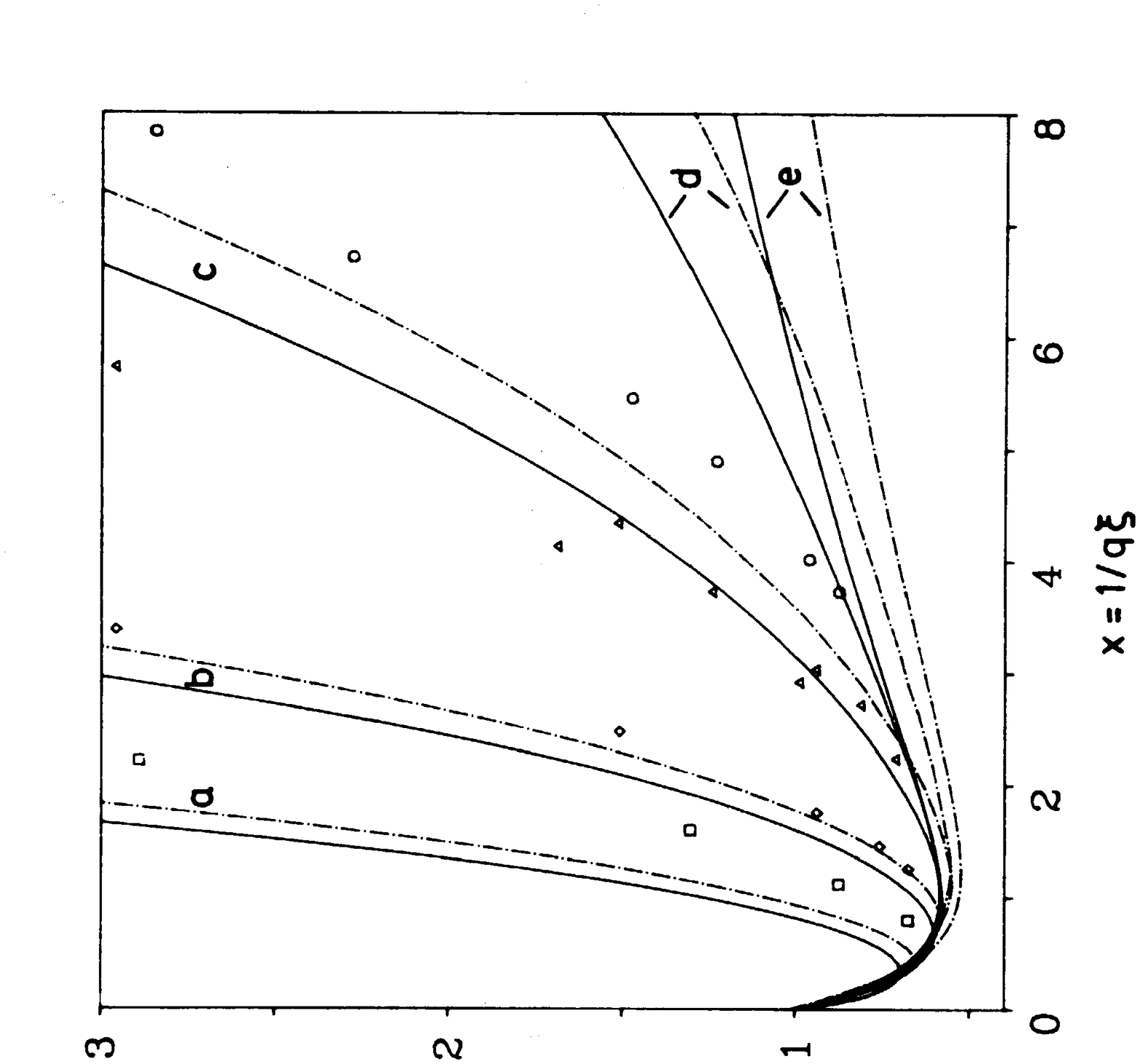}}}
\bigskip \bigskip
  {{\bf Figure 4.11:} Scaling function of the transverse line
   width in Lorentzian approximation (solid lines) and of the HWHM
   of the complete solution (dashed lines) versus the scaling
   variable $x=(q \xi)^{-1}$ for a set of temperatures              
   ($T-T_c=$ a) 1.4K, b) 5.8K, c) 21K, d) 51K). The Resibois-Piette
   function (solid) and the HWHM from the complete solution of the
   MC equations (dashed) without dipolar interaction is also plotted
   e). Experimental results for $Fe$ from Refs.~\cite{m82,m84} ( $T-T_c =$ 
   $(\sqcup) \, 1.4K$, $(\diamondsuit) \, 5.8K$, $(\triangle) \,
   21.0K$, $(\circ) \, 51.0K$ ).}
\label{fig4.11}
\end{figure}
\newpage
\vfill
\begin{figure}[h]
  \centerline{{\epsfysize=5in \epsffile{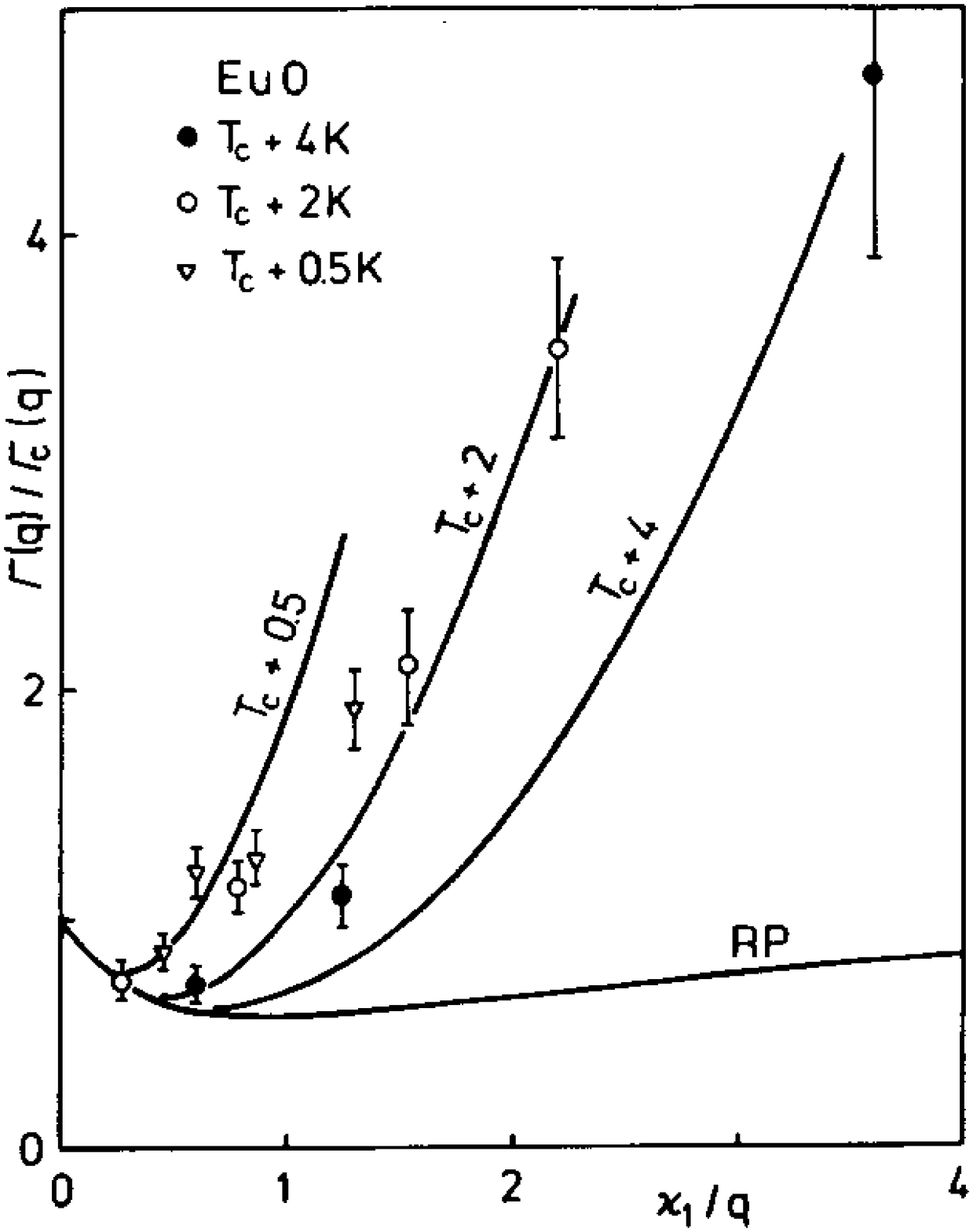}}}
\bigskip \bigskip
  {{\bf Figure 4.12:} Scaling plot of the temperature dependence of the 
   relaxation rate of critical fluctuations in $EuO$ above the Curie 
   temperature $T_c = 69.3 K$. The solid lines represent the results from 
   mode coupling theory~\cite{fs87}. The data were taken between $q = 
   0.018 \AA^{-1}$ and $q = 0.15 \AA^{-1}$. Taken from Ref.~\cite{m88,mfhs89}
  ($\kappa_1 = 1/\xi$).}
\label{fig4.12}
\end{figure}
\newpage
\vfill
\begin{figure}[h]
  \centerline{\rotate[r]{\epsfysize=5in \epsffile{fig413.eps}}}
\bigskip \bigskip
  {{\bf Figure 4.13:} Comparison of the longitudinal linewidth of Lorentzian 
   fits normalized to $\Lambda_{\rm exp} q^{5/2}$ with $\Lambda_{\rm exp} = 
   2.1 meV \AA^{5/2}$ to the dipolar dynamic scaling function predicted by 
   mode coupling theory for $q_{_{D}} = 0.245 \AA^{-1}$ for $EuS$. Figure taken 
   from Ref.~\cite{bgkm91} ($\kappa = 1/\xi$).}
\label{fig4.13}
\end{figure}
\newpage
\vfill
\begin{figure}[h]
  \centerline{\rotate[r]{\epsfysize=6in \epsffile{fig412.eps}}}
\bigskip \bigskip
  {{\bf Figure 4.14:} The same as Fig.~4.13 for the transverse widths. Figure 
   taken from Ref.~\cite{bgkm91}.}
\label{fig4.14}
\end{figure}
\newpage
\vfill
\begin{figure}[h]
  \centerline{\rotate[r]{\epsfysize=6in \epsffile{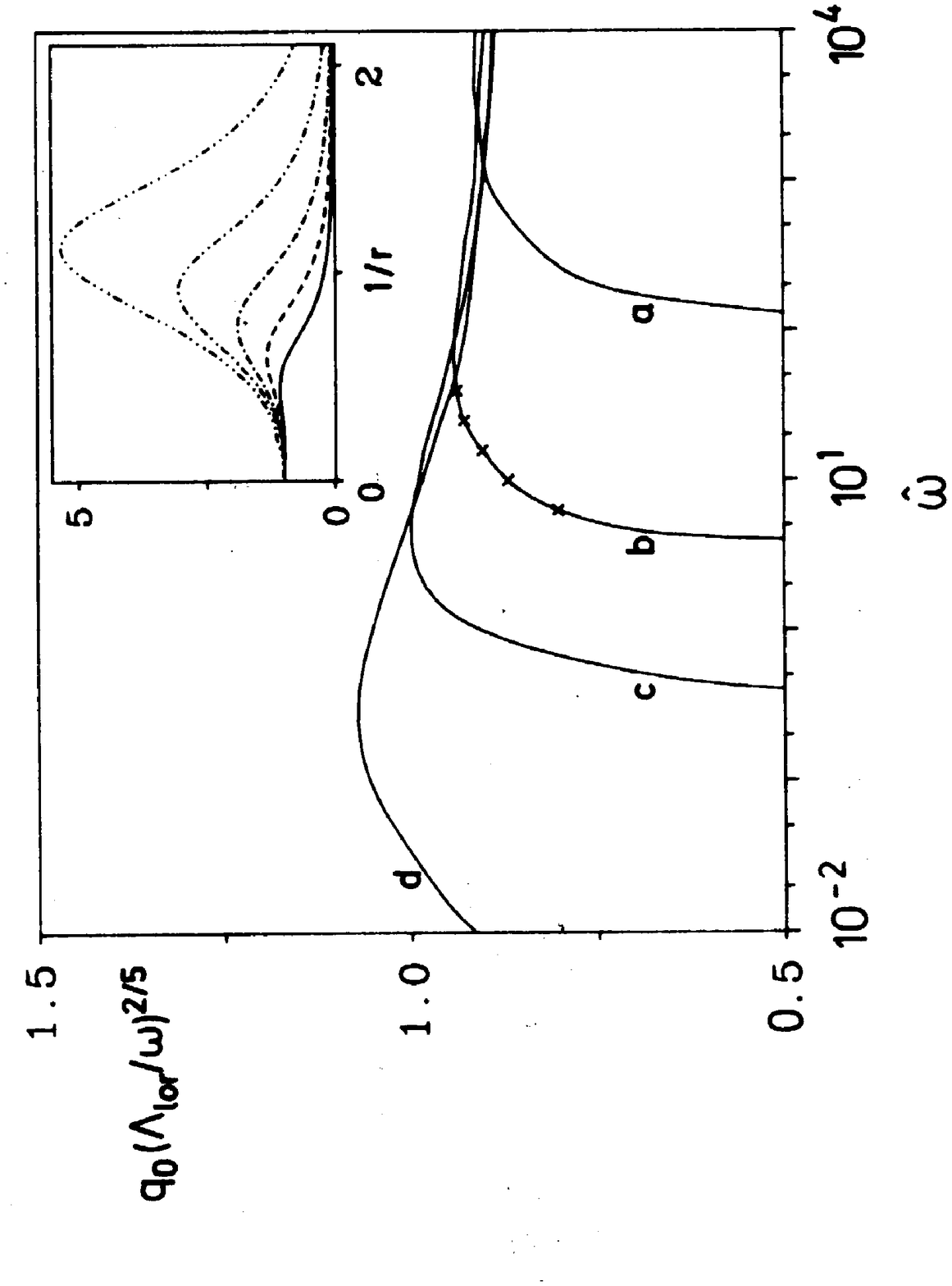}}}
\bigskip \bigskip
  {{\bf Figure 4.15:} Scaled peak positions 
   $q_0(\Lambda_{\rm lor}/\omega)^{2/5}$ for constant energy scans of
   the scattering function for the complete solution of the MC
   equations versus the scaling variable ${\hat \omega}$ for
   $\varphi=$ a) 1.490, b) 1.294, c) 0.970 and d) for the isotropic
   case, i.e., $\varphi=0$. {\bf Inset:}
   $S^T(q,\omega)/S^T(0,\omega)$ in arbitrary units versus ${1 \over r}$ 
   for $\varphi=1.294$ for some typical values of the scaled frequency 
   (${\hat \omega} = 10^{L/10}$ with $L=$ 8 (solid), 10 (dashed), 12 
   (point dashed), 14 ($- \cdot \cdot - \cdot \cdot - $) and 16 
   ($- \cdot \cdot \cdot - \cdot \cdot \cdot - $) indicated in the graph by 
   crosses).}
\label{fig4.15}
\end{figure}
\newpage
\vfill
\begin{figure}[h]
  \centerline{\rotate[r]{\epsfysize=6in \epsffile{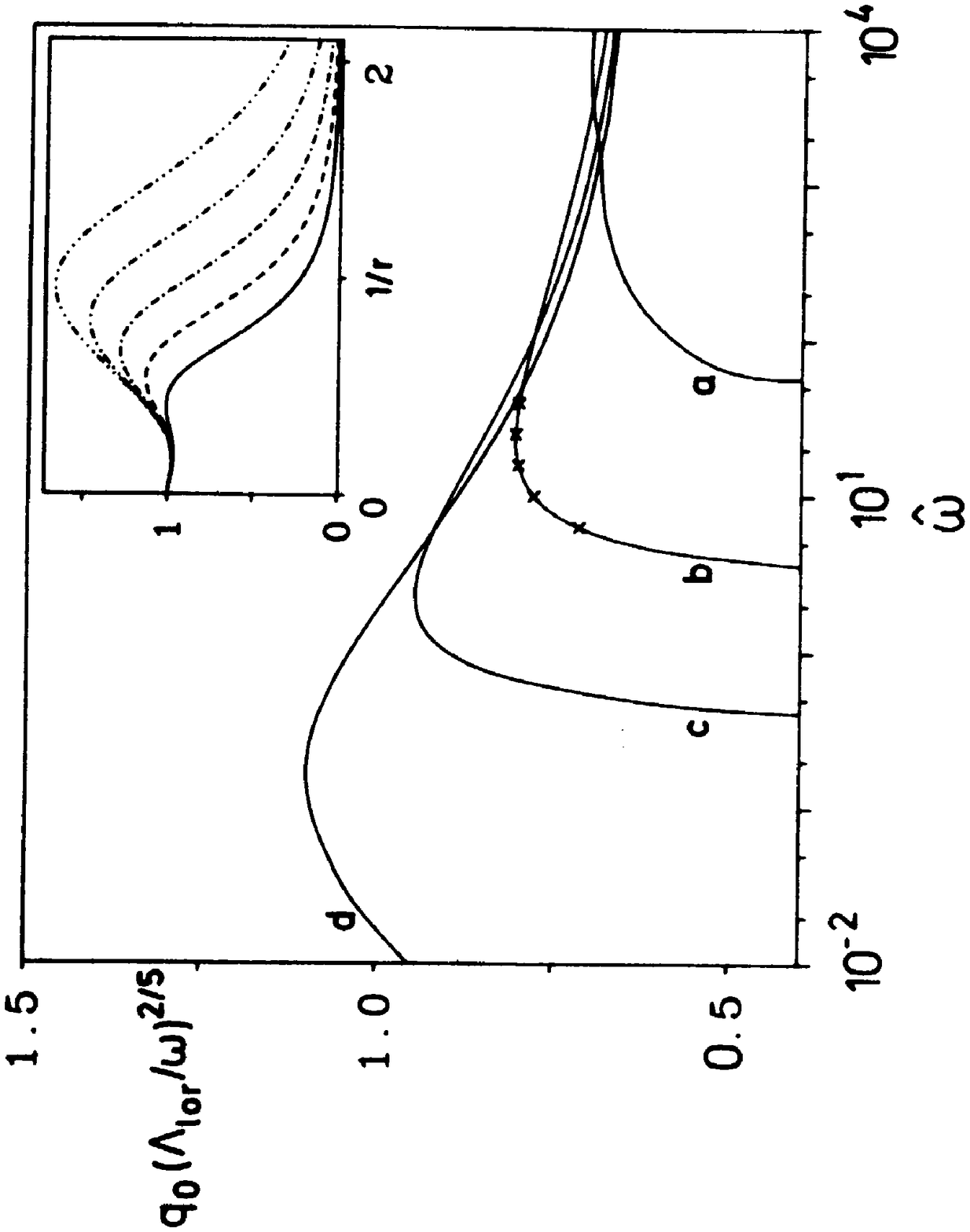}}}
\bigskip \bigskip
  {{\bf Figure 4.16:} The same as in Fig.~4.15 for the scattering
   function resulting from the Lorentzian approximation.}
\label{fig4.16}
\end{figure}
\newpage
\vfill
\begin{figure}[h]
  \centerline{\rotate[r]{\epsfysize=5in \epsffile{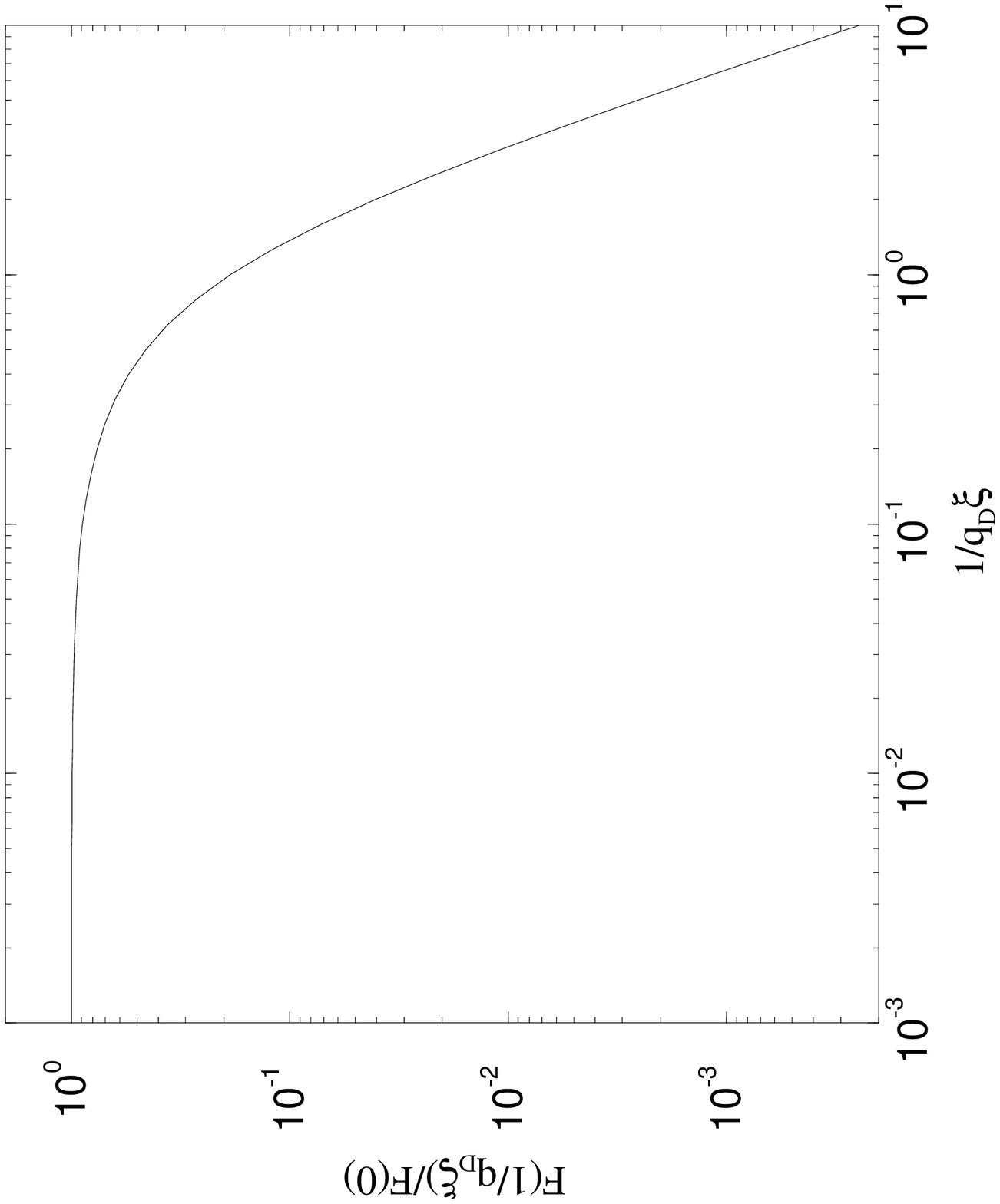}}}
\bigskip \bigskip
  {{\bf Figure 4.17:} Universal crossover function 
   $F(1/q_{_{\rm D}} \xi)/F(0)$ with $F(0) = 0.1956$ for the Onsager kinetic 
   coefficient at zero frequency and zero wave vector versus the scaling 
   variable $1/q_{_{\rm D}} \xi$.}
\label{fig4.17}
\end{figure} 
\newpage
\vfill
\begin{figure}[h]
 \centerline{\rotate[r]{\epsfysize=5in \epsffile{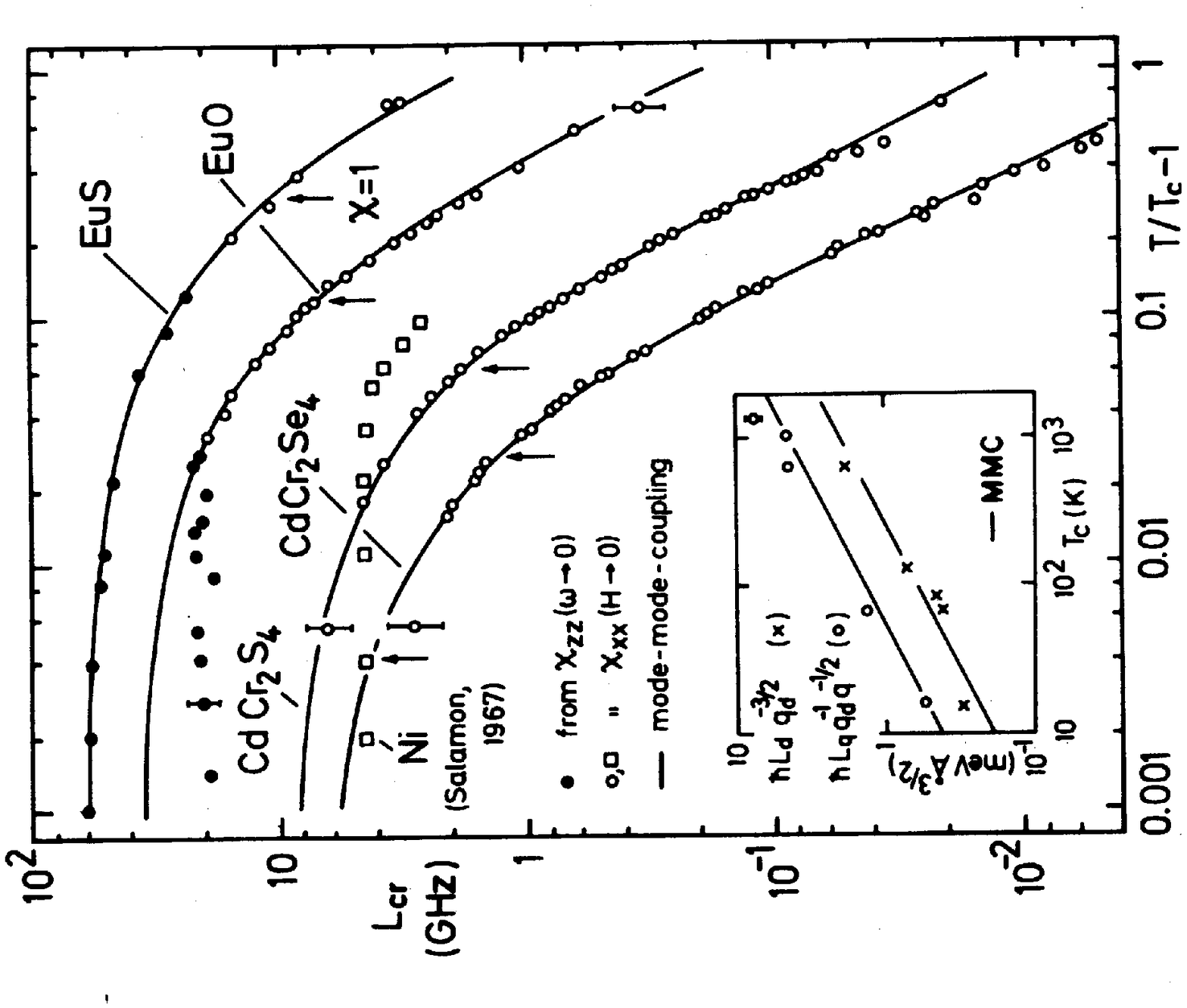}}}
\bigskip \bigskip
  {{\bf Figure 4.18:} Critical part of the Onsager kinetic coefficient for the 
   homogeneous spin dynamics above $T_c$ of $CdCr_2Se_4$ (Ref.~\cite{kp78}), 
   $CdCr_2S_4$ (Ref.~\cite{ks78}), $EuO$ (Ref.~\cite{ksbk78}), $EuS$ (filled 
   circles: Ref.~\cite{kkw76}, open circles: Ref.~\cite{k88}), and $Ni$ 
   (Ref.~\cite{sa67}). The solid lines represent the result from mode coupling
   theory~\cite{fs88a}. {\bf Inset:} Nonuniversal amplitudes 
   $\hbar L_d q_{_{\rm D}}^{-3/2}$ for the kinetic coefficient and the
   relaxation rate at the critical point $\hbar L_q q^{-1} q_{_{\rm D}}^{-1/2}
   = \Gamma^T (q,T=T_c) / q^{5/2} \approx 5.1326 \Lambda$ for ferromagnets 
   including $Fe$ and $Co$. Figure taken from Ref.~\cite{k88}.}
\label{fig4.18}
\end{figure}
\newpage
\vfill
\begin{figure}[h]
 \centerline{\rotate[r]{\epsfysize=7in \epsffile{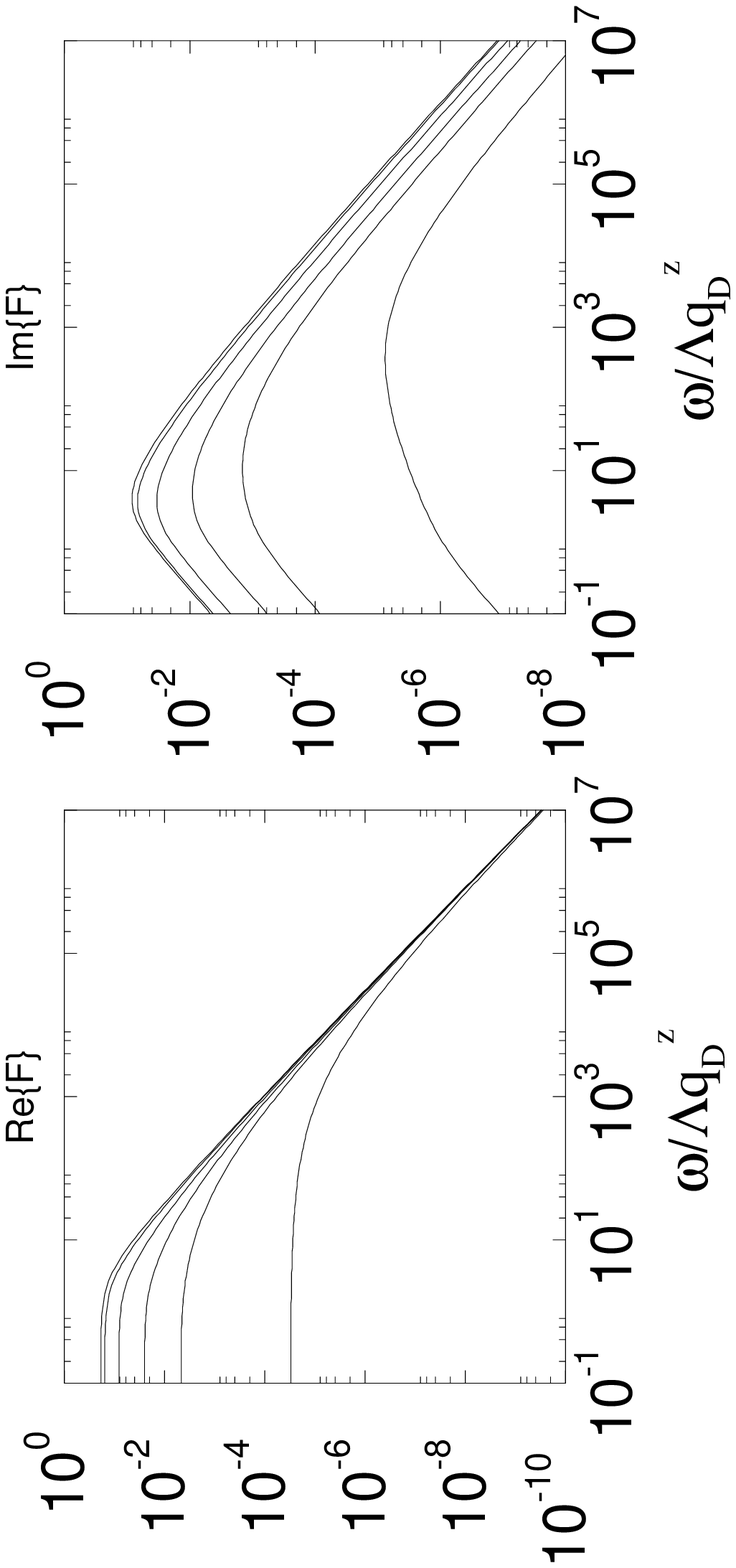}}}
\bigskip \bigskip
  {{\bf Figure 4.19:} The real and imaginary parts of $F$ are shown in a) and 
   b) versus the scaled frequency ${\hat \omega} = \omega/\Lambda q_{_{D}}^z$
   for several values of $\varphi = \arctan (q_{_{D}}\xi)$: $\varphi_1 = 
   0.99 {\pi \over 2}$ (top curve), $\varphi_2 = 0.90 {\pi \over 2}$, 
   $\varphi_3 =  {\pi \over 2} (1-10^{-0.5})$, $\varphi_4 =  {\pi \over 2} (1-10^{-0.25})$,
   $\varphi_5 =  {\pi \over 2} (1-10^{-0.1})$, $\varphi_6 =  {\pi \over 2} (1-10^{-0.025})$
   (bottom curve).}
\label{fig4.19}
\end{figure}
\newpage
\vfill
\begin{figure}[h]
 \centerline{\rotate[r]{\epsfysize=6in \epsffile{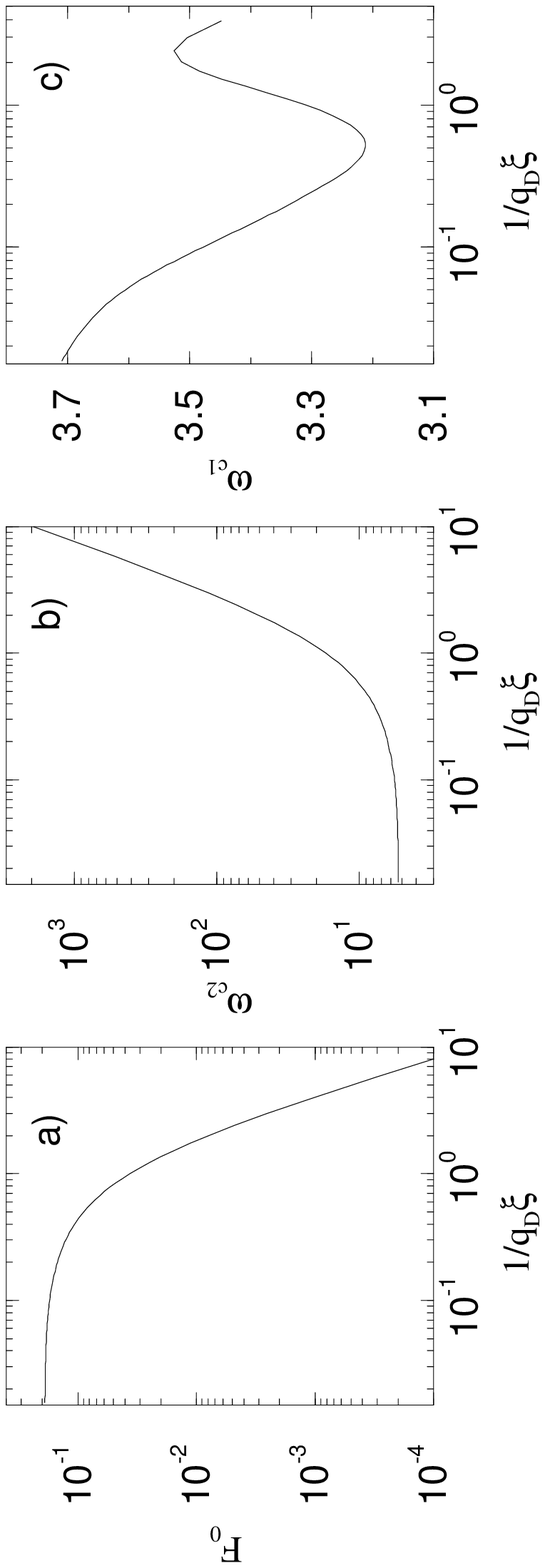}}}
\bigskip \bigskip
  {{\bf Figure 4.20:} The scaling functions for the a) Onsager coefficient at 
  zero frequency $F (1/q_{_{D}}\xi)$ and b) the scaling functions 
  $\omega_{c2}  (1/q_{_{D}}\xi)$ and c) $\omega_{c1}  (1/q_{_{D}}\xi)$ 
  characterizing the large and low frequency behaviour, respectively.}
\label{fig4.20}
\end{figure}
\newpage
\vfill
\begin{figure}[h]
 \centerline{{\epsfxsize=5.0truein \epsffile{hfi.ps.bb}}}
\bigskip\bigskip
 {{\bf Figure 4.21:} Scaling functions $I_{L,T} (\varphi)$ for the transverse 
  and longitudinal auto-correlation-time $\tau_{L,T}$ versus 
  ${1 / q_{_{\rm D}} \xi}$.}
\label{fig4.21}
\end{figure}
\newpage
\vfill
\begin{figure}[h]
 \centerline{\rotate[r]{\epsfysize=5in \epsffile{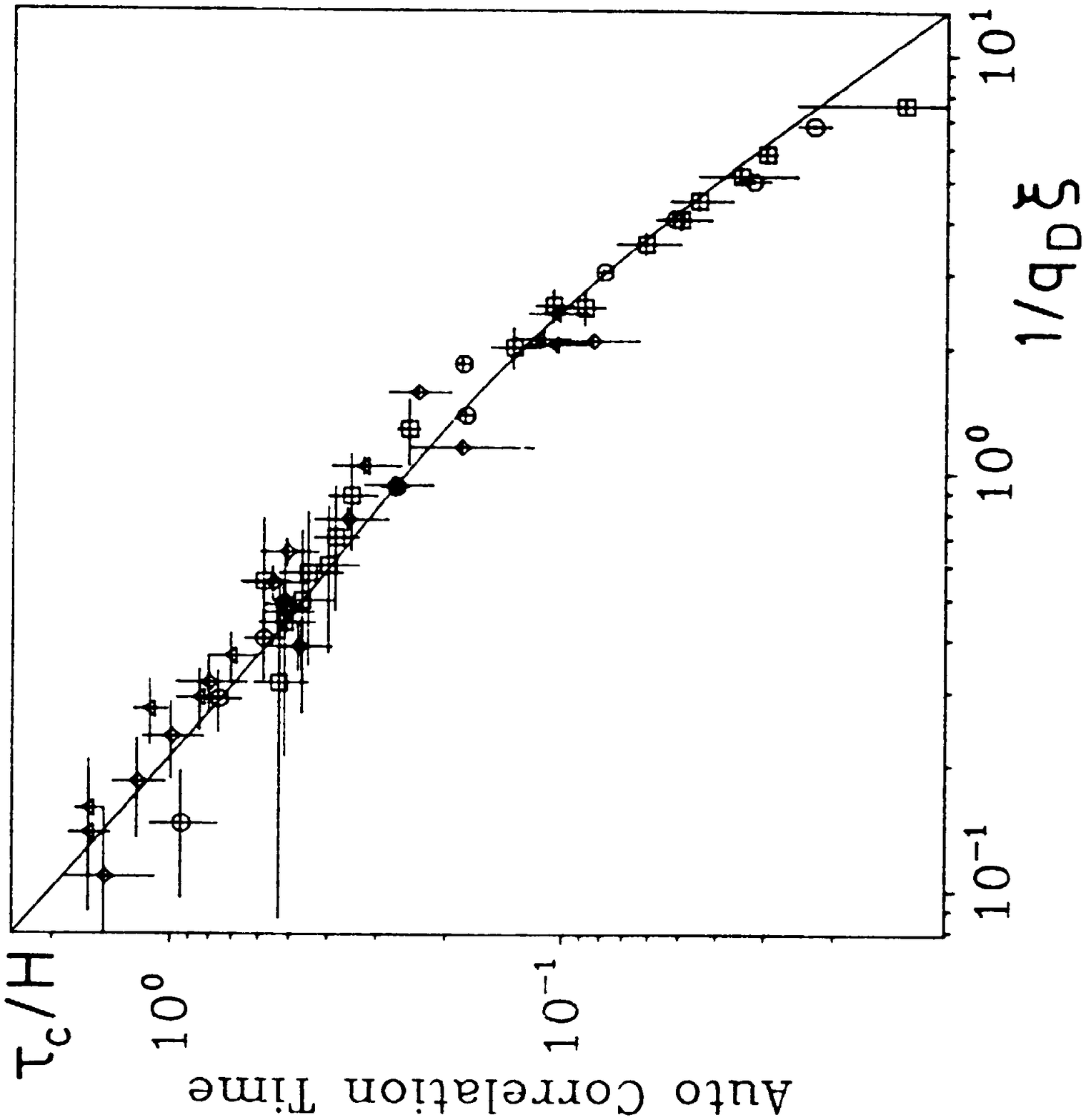}}}
\bigskip \bigskip
  {{\bf Figure 4.22:} Auto correlation time $\tau_c /H$ (in units of the
   non universal constant H) versus the scaling variable ${1 \over
   q_{_{\rm D}} \xi}$ (solid line). Experimental results for the auto
   correlation time in units of $H_{exp}$ for $Fe$ (Ref.~\cite{hcs82}: 
   $\sqcup$) and $Ni$ (Ref.~\cite{rh72}: $\diamondsuit$, Ref.~\cite{gh73}: 
   $\triangle$, Ref.~\cite{hcs82}: $\circ$).}
\label{fig4.22}
\end{figure}
\newpage
\vfill
\begin{figure}[h]
  \centerline{\rotate[r]{\epsfysize=5in \epsffile{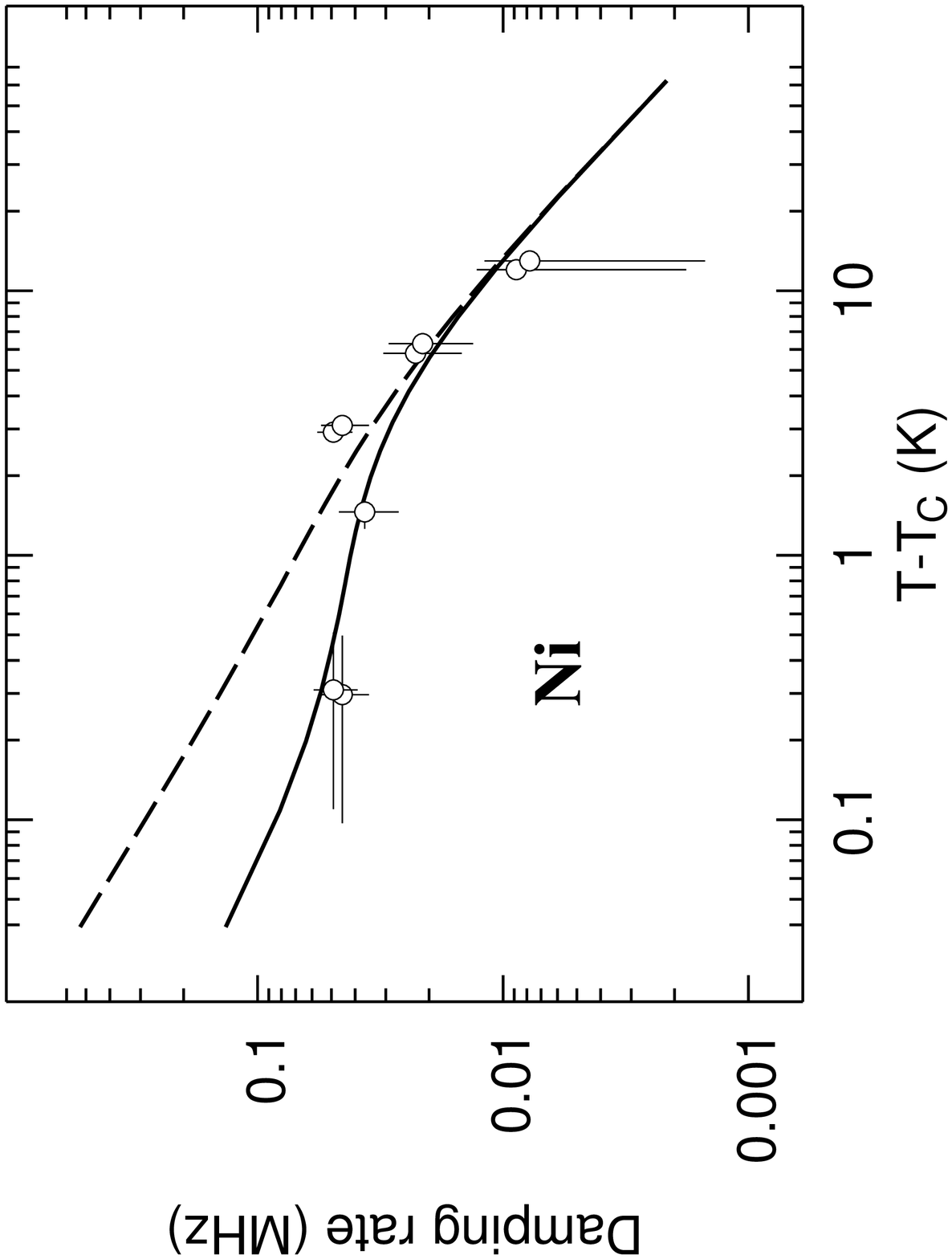}}}
\bigskip \bigskip
  {{\bf Figure 4.23:} Temperature dependence of the $\mu$SR damping rate for
   metallic $Ni$. The points are the experimental data of Nishiyama et 
   al.~\cite{nyim84}. The full line is the result of the model which takes 
   the muon dipolar interaction into account. The dashed line gives the 
   prediction when this latter interaction is neglected.}
\label{fig4.23}
\end{figure}
\newpage
\vfill
\begin{figure}[h]
  \centerline{\rotate[r]{\epsfysize=4in \epsffile{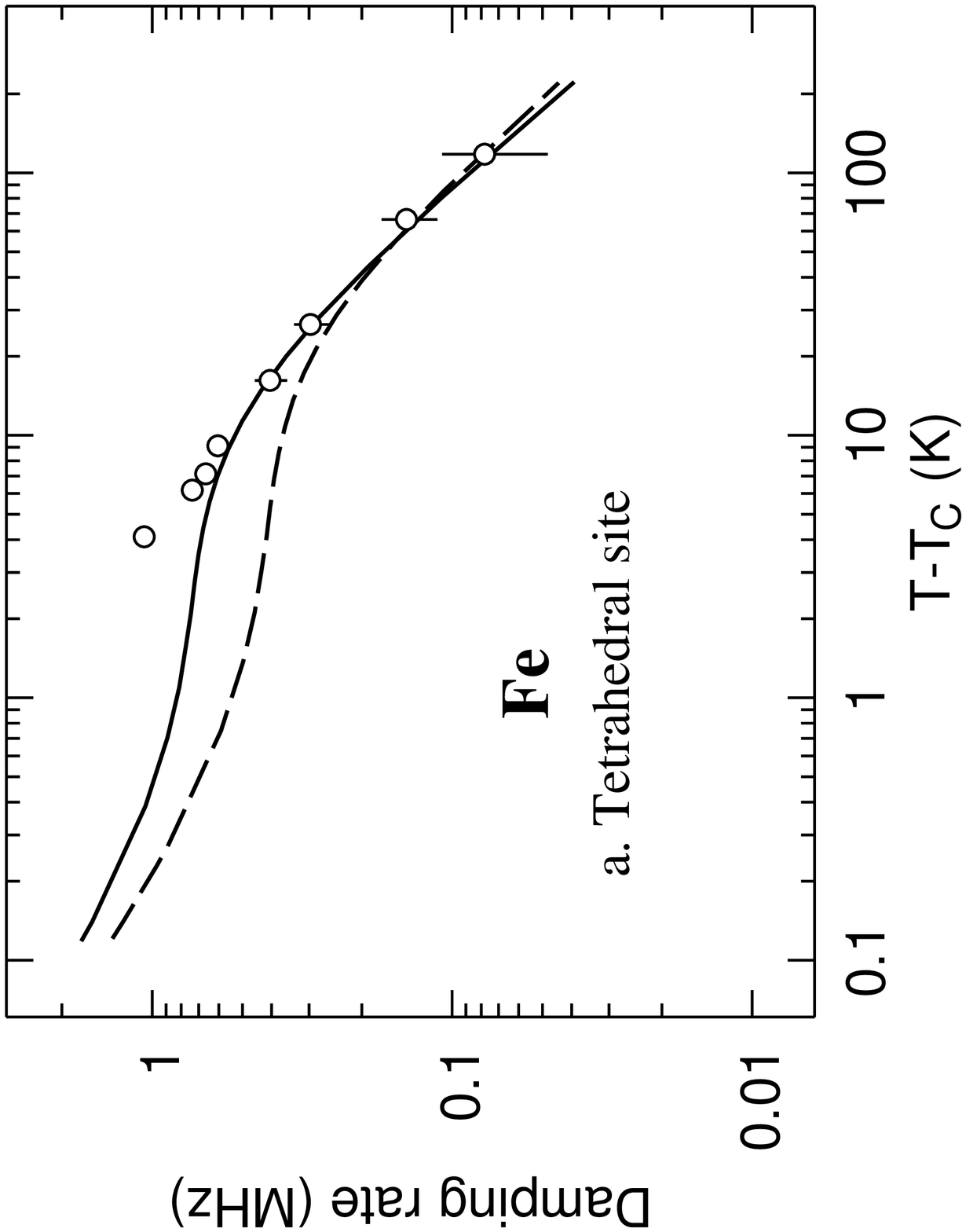}}}
\bigskip
  \centerline{\rotate[r]{\epsfysize=4in \epsffile{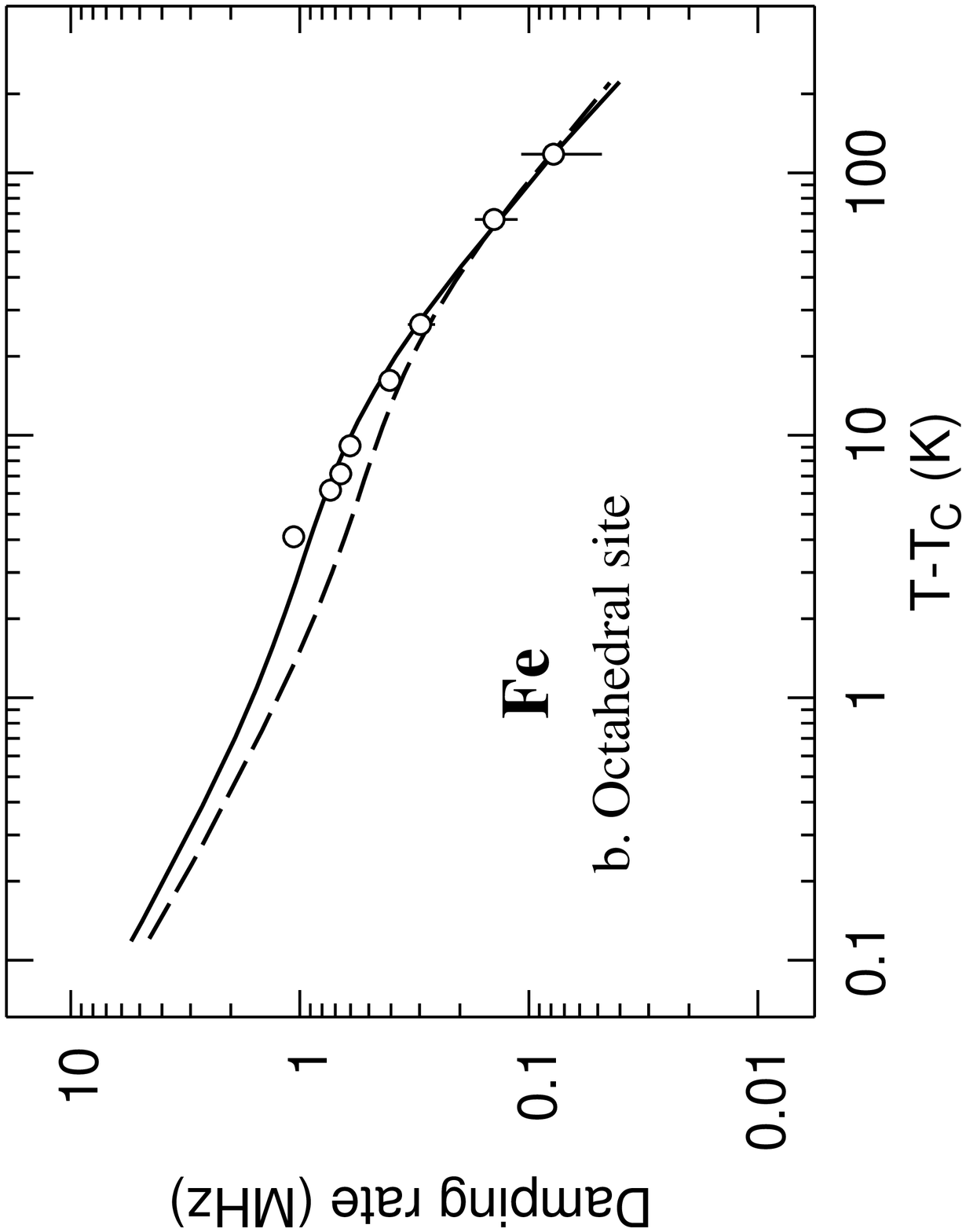}}}
\bigskip \bigskip
  {{\bf Figure 4.24:} Temperature dependence of the $\mu$SR damping rate for
   metallic $ Fe $. The points are the experimental data of Herlach et 
   al.~\cite{hffs86}. The curves are the predictions of mode-coupling theory 
   for different sets of material parameters. In Fig.~4.24a (Fig.~4.24b) the 
   muon is supposed to diffuse between tetrahedral (octahedral) sites. The
   full and dashed lines  present the results obtained with 
   $\left(q_{_{D}},\xi_ 0 \right)$ equal to $\left(0.033\ \AA^{ -1}, 0.82\ 
   \AA \right)$ and to $\left(0.045\ \AA^{-1},0.95\ \AA \right)$ respectively;
   see Table~\ref{table4.1}.}
\label{fig4.24}
\end{figure}

%%%%%%%%%%%%%%%%%%%%%%%%%%%%%%%%%%%%%%%%%%%%%%%%%%%%%%%%%%%%%%%%%%%%%%%%%%%%%%%%
\section{OTHER DIPOLAR SYSTEMS}
\label{s5}

Dipolar interactions are of importance not only for the critical behavior
of three--dimensional isotropic ferromagnets, but also for other magnetic 
materials such as antiferromagnets, ferromagnets with an uniaxial or planar
exchange anisotropy. Furthermore, dipolar interactions play an important 
role in two--dimensional systems, especially with regard to the existence 
of long--range order. 

\subsection{Dipolar Antiferromagnets}
\label{s5.1}
\subsubsection{Hamiltonian and equation of Motion}
\label{s5.1.1}

The order parameter of an antiferromagnet is the staggered
magnetization, i.e., in the case of two sublattices the difference of
the magnetization on these. Due to the alternating nature of the
order parameter one supposes that the effect of the dipolar interaction
on the long--range order parameter fluctuations is averaged out and thus
one expects no influence on the static critical behavior. Indeed in Ref.~\cite{a73}
it was shown that in contrast to the ferromagnetic case, the main 
parameters characterizing the dipolar interaction become irrelevant 
for the critical statics of antiferromagnets. Yet there is an
important effect on the dynamics, since the dipolar forces
lead to new damping processes in the long wave length dynamics of the
magnetization, because they break conservation of the total
magnetization. Since the magnetization and staggered
magnetization are coupled dynamically, there should be a
crossover in the dynamic critical exponents and related features in
the dynamical scaling functions.
This effect has been studied theoretically for temperatures above 
and at the N\'eel temperature $T_N$ by means of mode--coupling 
theory~\cite{Fischer90}.

The Hamiltonian of a dipolar antiferromagnet reads
\begin{equation}
  H=-\sum_{l\ne l'}\sum_{\alpha\beta}
     \left( J_{ll'}\delta^{\alpha\beta} + A^{\alpha\beta}_{ll'} \right)
     S_l^{\alpha}S_{l'}^{\beta} \, ,
\label{5.1}
\end{equation}
with spins ${\bf S}_l$ at lattice sites ${\bf x}_l$. The first term in 
brackets is the exchange interaction $J_{ll'}$ and the second the 
dipole-dipole interaction with the dipolar tensor given by
\begin{equation}
  A^{\alpha\beta}_{ll'}  = 
  -{1\over 2}(g_L\mu_B)^2
  \left( {\delta^{\alpha\beta} \over 
  |{\bf x}_l - {\bf x}_{l'}|^3}-
  {3({\bf x}_l-{\bf x}_{l'})^\alpha({\bf x}_l-{\bf x}_{l'})^\beta 
  \over |{\bf x}_l-{\bf x}_{l'}|^5}
  \right) \, .
\label{5.2}
\end{equation}

We define Fourier transformed quantities by
\begin{eqnarray}
  S_l^{\alpha} = 
        &&{1 \over \sqrt{N}} \, \sum_{\bf q} e^{i {\bf q} \cdot {\bf x}}
          S_{\bf q}^\alpha \, , \label{5.3} \\
  J_{l l^\prime} = {1 \over N} 
  &&\sum_{\bf q} e^{i {\bf q} \cdot ({\bf x}_l-{\bf x}_{l^\prime})}
          J_{\bf q} \, ,  \label{5.4} \\
  A^{\alpha \beta}_{l l^\prime} = {1 \over N} 
  &&\sum_{\bf q} e^{i {\bf q} \cdot ({\bf x}_l-{\bf x}_{l^\prime})}
               A_{\bf q}^{\alpha \beta} \, , 
\label{5.5}
\end{eqnarray}
and get 
\begin{equation}
  H = - \sum_{\bf q} 
      \left(J_{\bf q} \delta^{\alpha \beta} +  A_{\bf q}^{\alpha \beta} 
      \right) 
      S_{\bf q}^\alpha S_{- \bf q}^\beta
\label{5.6}
\end{equation}
for the Hamiltonian in terms of Fourier components.
In order to study the long--wavelength behavior of the model one needs to know
the behavior of the dipole--dipole interaction tensor $A^{\alpha \beta}_{\bf q}$ at
small wave vectors. The result depends on the lattice structure and the
dimensionality of the system. For a three-dimensional system the expansion has
been given in section III (see Eqs.~(\ref{3.4})). In two dimensions the
dipole-interaction tensor has been analyzed by Maleev [Maleev76]. He finds
for small wave vectors
\begin{equation}
  A^{\alpha \beta}_{\bf q} = 
  A^{(0)} \left( {1 \over 3} \delta^{\alpha \beta} - {\hat z}^\alpha
                                                   {\hat z}^\beta \right)
 +A^{(1)} \, q \, \left( {\hat z}^\alpha {\hat z}^\beta - {q^\alpha q^\beta 
  \over q^2} \right) \, ,
\label{5.7}
\end{equation}
where ${\bf q} = (q_x,q_y,0)$ is an in--plane wave vector and $z$ denotes the 
direction perpendicular to the plane ${\bf r} = (r_x,r_y,0)$. 
The constants $A^{(0)}$ and $A^{(1)}$ depend on 
the lattice structure
\begin{eqnarray}
   A^{(0)} = &&{3 \over 4} (g_L \mu_B)^2 \sum_{{\bf r} \neq 0} { 1 \over r^3}\, , 
\label{5.8} \\
   A^{(1)} = && {\pi \over v_2} (g_L \mu_B)^2 \, ,
\label{5.9}
\end{eqnarray}
where $v_2$ is the volume of the 2D unit cell.

With the standard commutation relation for spin operators 
$[S_{\bf k}^\alpha,S^\beta_{\bf q}]=i \hbar \epsilon_{\alpha \beta \gamma}
S^\gamma_{{\bf k} + {\bf q}}$ one gets the following equations of
motion~\cite{Pich94b},
\begin{eqnarray}
  {d \over dt} S_{\bf5. q}^x = 
  \sum_{\bf k} 
  \Biggl[
  &&\left( A_{\bf k}^{yy} + J_{\bf k} \right) \, 
                   \left( 
                   S_{{\bf k}+{\bf q}}^z S_{-{\bf k}}^y + 
                   S_{\bf k}^y S_{{\bf q}-{\bf k}}^z
                   \right) \nonumber \\
 -&&\left( A_{\bf k}^{zz} + J_{\bf k} \right) \, 
                   \left( 
                   S_{{\bf k}}^z S_{{\bf q}-{\bf k}}^y + 
                   S_{{\bf q}+{\bf k}}^y S_{-{\bf k}}^z
                   \right) \nonumber \\
 +&&A_{\bf k}^{xy} \left( 
                   S_{{\bf k}}^x S_{{\bf q}-{\bf k}}^z + 
                   S_{{\bf q}+{\bf k}}^z S_{-{\bf k}}^x
                   \right) \nonumber \\
 -&&A_{\bf k}^{xz} \left( 
                   S_{{\bf k}}^x S_{{\bf q}-{\bf k}}^y + 
                   S_{{\bf q}+{\bf k}}^y S_{-{\bf k}}^x
                   \right) \nonumber \\
 +&&A_{\bf k}^{yz} \left( 
                   S_{{\bf q}+{\bf k}}^z S_{-{\bf k}}^z + 
                   S_{{\bf k}}^z S_{{\bf q}-{\bf k}}^z -
                   S_{{\bf k}}^y S_{{\bf q}-{\bf k}}^y - 
                   S_{{\bf q}+{\bf k}}^y S_{-{\bf k}}^y
                   \right)
  \Biggr]
\label{5.10}
\end{eqnarray}

\subsubsection{Critical Behavior of 3D Dipolar Antiferromagnets}
\label{s5.1.2}

Now we turn to the critical dynamics of dipolar antiferromagnets. 
Specializing to the three-dimensional isotropic case and retaining only those 
terms which are dominant in the long wavelength limit the Hamiltonian for a
simple cubic lattice reduces to~\cite{Fischer90} 
\begin{eqnarray}
  H = &&J \int_{\bf q} 
      \left[ 
            \left((qa)^2 -6\right) \delta^{\alpha \beta} - 
            g {q^\alpha q^\beta \over q^2}
      \right]
      M_{\bf q}^\alpha M_{- \bf q}^\beta \nonumber \\ 
      +&&J \int_{\bf q} 
      \left( 6-(qa)^2 \right)  \delta^{\alpha \beta}
      N_{\bf q}^\alpha N_{- \bf q}^\beta \, ,
\label{5.11}
\end{eqnarray}
where $M_{\bf q}^\alpha = S_{\bf q}^\alpha$ denotes the magnetization and 
      $N_{\bf q}^\alpha = S_{{\bf q}+{\bf q}_0}^\alpha$ the staggered
magnetization, and we have introduced the abbreviation $\int_{\bf q} = 
\int v_a {d^3q \over (2 \pi)^3}$. The wave vector 
${\bf q}_0 = {\pi \over a} (1,1,1)$ characterizes the antiferromagnetic 
modulation.

Starting from this Hamiltonian mode coupling equations for the
correlation functions of the magnetization and staggered magnetization have 
been derived and analyzed in Ref.~\cite{Fischer90}.
For instance for the longitudinal magnetization mode one finds
\begin{eqnarray}
\Gamma^L_M&&({\bf q},t) = 4\Lambda_{\rm af}^2
\int\limits_0^{q_{BZ}} k^2 dk
\int\limits_{-1}^1 d(\cos \vartheta) \biggl[  
({g\over 12})^2\sin ^2\vartheta
\Phi^T_M({\bf k}_+,\xi,t)
\Phi^L_M({\bf k},t) \nonumber \\
&&+{ ( 2 {\bf q}\cdot{\bf k}+{\bf q}^2)^2 \over
({\bf k}^2+\xi^{-2})({\bf k}_+^2+\xi^{-2})}
\left(\sin ^2 \vartheta 
\Phi^T_N({\bf k}_+,t)
\Phi^L_N({\bf k},t)
+\cos ^2\vartheta
\Phi^T_N({\bf k}_+,t)
\Phi^T_N({\bf k},t)\right)
\biggr] \, ,
\label{5.12}
\end{eqnarray}
where we have used the abbreviations ${\bf k}_+ = {\bf k} + \bf q$,
$\Lambda_{\rm af} = a^{3/2} \sqrt{J k_{B} T \over 8 \pi^2}$ and
$q_{BZ}={2\pi\over 2a}\root 3 \of{3\over4\pi}$.

It is found that the memory kernels obey generalized scaling laws of the form
\begin{eqnarray}
  \Gamma_{M,N}^{L,T} (q, g,\xi,t) = 
  &&\Lambda_{\rm af}^2 q^{2z} \gamma_{M,N}^{L,T} (x,y,\tau) \, , 
\label{5.13} \\
  \Phi_{M,N}^{L,T} (q, g,\xi,t) = 
  &&\phi_{M,N}^{L,T} (x,y,\tau) \, ,
\label{5.14}
\end{eqnarray}
where the scaling variables are defined by $x=1/q \xi$, $y = q_A/q$, and 
$\tau = \Lambda_{\rm af} q^z t$ with the dynamic critical exponent $z = 3/2$. 
The characteristic wave vector $q_A$ is related to the dipolar wave vector 
$q_{_{D}} = \sqrt{g}/a$ by
\begin{equation}
  q_A = \left({1 \over 12}\right)^{2/3} (q_{_{D}} a)^{4/3} q_{BZ} \, .
\label{5.15}
\end{equation}
The value $3/2$ is the dynamical critical exponent for
isotropic antiferromagnets and any crossover to dipolar behavior is
contained in the dynamic scaling function.
In evaluating the transport coefficients $\Gamma_{M,N}^{L,T}$ only two mode
decay processes have been considered (see Eq.~(4.12)). Without 
dipolar interaction, i.e., for the isotropic exchange antiferromagnet, the
magnetization modes can decay into staggered modes 
only~\cite{kk76,w69,rp70,jp73}. 

The dipolar interaction leads in addition to
a decay into two magnetization modes, dominating for small wave vectors. In 
the strong dipolar limit ($T \rightarrow T_N$ and $q \rightarrow 0$) the mode
coupling equations for the magnetization modes can be solved exactly with the 
result
\begin{equation}
  \Gamma^\lambda_M (q,\xi,t) = 
  {\Lambda_{\rm af} g \over 12 \sqrt{{14 \over 9} + {\pi^2 \over 8}}} 
  q_{BZ}^{3/2} 
  \left[ {16 \over 9} \delta^{\lambda L} + 
         \left( {\pi^2 \over 8} - {2 \over 9} \right) \delta^{\lambda T} 
  \right] \, \delta(t) 
\label{5.16}
\end{equation}
corresponding to a relaxation of the magnetization modes enforced by the 
dipolar interaction. The phase space for this decay is the full Brillouin 
zone, whereas the decay into staggered modes is weighted by the critical 
static susceptibilities. Therefore, with separation from the critical 
temperature and for larger wave vectors the decay of the magnetization mode 
into two staggered modes dominates the decay into two magnetization modes.

The mode coupling equations for dipolar antiferromagnets have been solved in 
the Lorentzian approximation in Ref.~\cite{Fischer90}. The results for the 
transverse scaling function for the magnetization and staggered magnetization
$\gamma^T_{M,N} (x,\varphi)$ versus $x$ for different values of $\varphi = 
N\pi/30$ with $N=0,1,...,14$ are displayed in Figs.~5.1 and 5.2. 
The corresponding plots for the longitudinal scaling functions
are very similar~\cite{Fischer90}. The curves with $N=0$ correspond to the
isotropic case. It is important to note, that the effect of the dipolar 
interaction on the scaling functions of the magnetization and staggered 
magnetization modes is rather different. The scaled line width for the 
magnetization modes exceeds the isotropic curve by an amount which increases 
with $T$ on approaching $T_c$ (i.e. larger values of $\varphi = 
\arctan (q_{_{D}} \xi)$). This reflects the crossover from a diffusive 
behavior $\Gamma_M \propto q^2 q_A^{-1/2}$ to a relaxational behavior 
$\Gamma_M \propto q_A^{3/2}$. The line width of the staggered magnetization 
at fixed scaling variable $x$ decreases with $T$ approaching $T_c$ since the 
magnetization modes become uncritical. But, this change is much less
pronounced than for the magnetization modes and more important,
the asymptotic hydrodynamic dependence on the wave vector and
correlation length is unmodified, i.e., the hydrodynamic behavior
is always relaxational.

In the dipolar (D) and isotropic (I) critical (C) and
hydrodynamical (H) limiting regions the mode coupling equations
can be solved analytically. These regions are defined by DC: $y
\gg 1$,$x \ll 1$; IC: $y \ll 1$, $x \ll 1$; DH: $y \gg x$, $x \gg
1$;IH: $y \ll x$, $x \gg 1$. The results are summarized in 
Table~\ref{table51}, where we note that these asymptotic power laws are the 
same for the longitudinal and transverse fluctuations. In the immediate 
vicinity of $q=q_{_A}$ there is a dynamic crossover from $z_{\rm eff}=3/2$ to 
$z_{\rm eff}=2$ for the staggered magnetization and from $z_{\rm eff}=3/2$ 
to an uncritical value $z_{\rm eff}=0$ for the magnetization modes.

$RbMnF_3$ is one of the most thoroughly studied isotropic antiferromagnets
~\cite{Tucciarone71,Corliss70,Lau69}. Taking the values for the exchange
coupling and the static susceptibility of the magnetization from
Refs.~\cite{Tucciarone71,Windsor66,Huber70}, the characteristic wave vector
$q_{_A}$ is about $0.02 \AA^{-1}$. Unfortunately
the neutron scattering data \cite{Tucciarone71,Corliss70,Lau69} for the line 
width of the staggered magnetization modes are limited to the isotropic 
region and to our knowledge there are no data for the line widths of the 
magnetization modes.

The dipolar effects could be observed more
readily in $EuSe$ and $EuTe$ because of their smaller transition
temperatures $4.8K$ and $9.7K$ implying larger $q_{_A}$. Experiments
in the appropriate wave vector and temperature region are, however,
still lacking.

\subsection{Uniaxial Dipolar Ferromagnets}
\label{s5.2}

We have mainly concentrated on the influence of dipolar forces on otherwise
isotropic ferromagnets. But, even in systems which are anisotropic to start
with such as uniaxial ferromagnets, the dipolar interaction has a significant
effect, because of its long--range nature.

The influence of the dipolar interactions on the critical statics of 
isotropic and uniaxial ferromagnets is quite different. Whereas for 
isotropic ferromagnets the dipolar interaction leads only to a slight 
modification of the critical exponents, Larkin and 
Khmelnitskii~\cite{Larkin69} have discovered that uniaxial dipolar
ferromagnets show classical behavior with logarithmic corrections in three 
dimensions. This system was then studied by Aharony~\cite{Aharony73c} by 
means of the renormalization group method. Aharony employed an isotropic 
elimination procedure and calculated the critical exponents to first order in 
$(3-d)$ as well as logarithmic corrections~\cite{Aharony75}. Note, that the 
dipolar interaction leads to a shift in the upper critical dimension from $4$
to $d_c = 3$. 

The existence of logarithmic corrections was verified experimentally for a 
number of uniaxial ferromagnetic substances~\cite{Ahlers75,Frowein82}. 
However, these experiments were performed in regions of the reduced 
temperature, where departures from the asymptotic behavior are expected and 
are observed indeed. In particular, a maximum in  the effective exponent of 
the susceptibility~\cite{Frowein82} has been found. On the basis of a 
generalized minimal subtraction scheme the latter crossover from Ising 
behavior with non classical exponents to asymptotic uniaxial dipolar
behavior,  which is characterized by classical exponents with logarithmic 
corrections, has been analyzed in Ref.~\cite{fs88b,fs90}. The 
theoretical results for the specific heat and susceptibility are in
quantitative agreement with measurements on $LiTbF_4$, where solely the 
strength of the dipolar coupling constant entered as an adjustable parameter
~\cite{fs88b,fs90}.

The Landau-Ginzburg free energy functional for a n-component uniaxial spin 
system with an isotropic exchange coupling and dipolar interaction is given 
by~\cite{Larkin69,Aharony73c} 
\begin{eqnarray}
{\cal H} = &&-{1 \over 2} \inq{k}
  \left[ r + k^2 + g^2 {q^2 \over k^2} \right] 
  S^{\alpha} ({\bf k}) S^{\alpha} (- {\bf k}) \nonumber \\
  &&-{u_0 \over 4!} \inq{ \{ k_i \} } 
  S^{\alpha} ({\bf k}_1) S^{\alpha} ({\bf k}_2)
  S^{\beta} ({\bf k}_3) 
  S^{\beta} (- {\bf k}_1- {\bf k}_2 - {\bf k}_3)
   \, . 
\label{5.17}
\end{eqnarray}
Here $S^{\alpha}({\bf q})$ are the $n$ components of the spin 
variables. The $d$-dimensional wave vector ${\bf k} = ({\bf p},q)$ is
decomposed into $q$, the component along the uniaxial direction, and 
${\bf p}$, the remaining (d-1) components. The bare reduced temperature is 
given by $r={(T-T_c^0) / T_c^0}$ and $g^2$ is a measure of the relative 
strength of the dipolar interaction. The dipolar term $g^2 {q^2 \over k^2}$
suppresses the fluctuations in the $z$--direction.

In Ref.~\cite{Folk77} the dynamics of uniaxial dipolar magnets was described
in terms of a simple Time Dependent Ginzburg Landau Model (TDGL) with the 
equation of motion
\begin{equation}
  {\partial \over \partial t} S^{\alpha} ({\bf k},t) =
  - \Gamma_0 {\delta {\cal H} \over \delta S^{\alpha} (-{\bf k},t)}
  + \eta ({\bf k},t)  \, .
\label{5.18}
\end{equation}
Since the order parameter is a non-conserved quantity in the presence of the
dipolar interaction, the kinetic coefficient $\Gamma_0$ is wave vector
independent. The random forces $\eta ({\bf k},t)$ represent the uncritical
degrees of freedom and are characterized by a Gaussian probability
distribution. A renormalization group analysis for the above TDGL gives
for the dynamic critical exponent~\cite{Folk77}
\begin{equation}
  z = 2 + c \eta \, , \quad {\rm with} \quad c = 0.92 \, .
\label{5.19}
\end{equation}
This may be compared with the value for the TDGL model for spin systems with
short range exchange interaction~\cite{hhm72,hhm76}, where $c=0.73$ and the
van Hove prediction $c=-1$.

Since for uniaxial dipolar magnets the fourth order coupling is marginal at
$d=3$, there are logarithmic corrections to the classical behavior. For the
statics one finds~\cite{Larkin69,Aharony73c} for the susceptibility $\chi$
and the specific heat$C$ 
\begin{equation}
  \chi \propto \tau^{-1} | \ln \tau |^{1/3} \, , \quad
  C \propto | \ln \tau |^{1/3} \, ,
\label{5.20}
\end{equation}
where $\tau = (T-T_c)/T_c$ is the reduced temperature. The effect of the
logarithmic corrections on the dynamics susceptibility can be approximated by
~\cite{Folk77}
\begin{equation}
  \chi(\omega, {\bf p}, q, \tau) =
  \left[ 
  -{i \omega \over \Gamma_0} + p^2 + g {q^2 \over p^2} +
  \mu_1(0) \mid {1 \over 2 l_0} \ln \mu_1(0) \mid^{-1/3}
  \right]^{-1} \, ,
\label{5.21}
\end{equation}
where $\mu_1(0)$ is proportional to the reduced temperature $\tau$, and $l_0$
is some constant. In the hydrodynamic region $q \ll p \ll \xi^{-1}$ the 
relaxation coefficient implied by the latter result is
\begin{equation}
  \Gamma \propto \tau \left( 
                      \mid \ln \tau \mid + {\rm const.}
                      \right)^{-1/3} \, .
\label{5.22}
\end{equation}
This result has also been found by Maleev~\cite{Maleev75} using mode coupling
arguments. A detailed theoretical study of the dynamics of uniaxial dipolar
ferromagnets, which takes into account not only the relaxational dynamics
on the basis of TDGL model, but also includes nonlinear terms resulting from 
the Larmor precession in the local magnetic field is still lacking. A mode
coupling analysis of this problem is in progress~\cite{Henneberger94}.

\subsection{Two--dimensional Systems}
\label{s5.3}

Two--dimensional systems are interesting, because matter behaves 
qualitatively different as compared to three dimensions. We recall
that the Bloch argument~\cite{Bloch30,Peierls36,Landau37}, which is proven on a
rigorous basis by the Hohenberg--Mermin--Wagner 
theorem~\cite{Hohenberg67,Mermin66}, excludes conventional long--range order 
in isotropic systems with a continuous symmetry and short--ranged 
interaction such 
as the Heisenberg ferromagnet, superfluid Helium films, and two--dimensional 
crystals~\cite{Nelson83}. Any hypothetical broken continuous magnetic or 
translational symmetry would be overwhelmed by long wavelength spin or phonon 
excitations, since the phase space for these Goldstone--like modes is enhanced
in two dimensions. Two dimensions seem to be the borderline dimension where 
thermal fluctuations are just strong enough to prevent the appearance of a 
finite order parameter. 

However, since the dipolar forces reduce the fluctuations longitudinal to the
wave vector, related to the anisotropy of the dipolar interaction with
respect to the separation of two spins one expects that a finite order
parameter could exist in two--dimensional isotropic systems. In fact we will
review in the following spin--wave theories, which show that the dipolar
interaction leads to the existence of long--range order in two-dimensional
ferromagnets and antiferromagnets and other systems which belong to the same 
universality class.

\subsubsection{Ferromagnets}
\label{s5.3.1}
The possibility of a finite order parameter in two--dimensional ferromagnets 
was shown first by Maleev~\cite{Maleev76}, who evaluated the low temperature 
properties, in particular the magnon excitation spectrum for Heisenberg 
ferromagnets in the presence of dipolar forces. Linear spin--wave 
theory~\cite{Maleev76} results in a magnon dispersion relation
\begin{equation}
  E_{\bf k} = \sqrt{ \left( Dk^2 + \Omega_0 \alpha  \right)  
  \left( Dk^2 + \Omega_0 k a \sin \varphi_{\bf k} \right) } \, ,
\label{5.23}
\end{equation}
where $D k^2 = 2S (J_0 - J_{\bf k})$ at $ka \ll 1$, $\Omega_0 = 2 \pi S (g_L
\mu_B)^2 / v_2 a \ll 2 S J_0$, $\alpha = S A_0 / \Omega_0 \sim 1$, and 
$\varphi_{\bf k}$ is the angle between the wave vector ${\bf k}$ and the 
magnetization ${\bf M}$.
At sufficiently small $k$ (and for ${\bf k}$ not parallel to the magnetization)
the energy of the spin--waves has the form
\begin{equation}
 E_{\bf k} \propto \Omega_0 \, |\sin \varphi_{\bf k}| \, \sqrt{k} \, .
\label{5.24}
\end{equation}
An analysis of the relative deviations of the magnetization from saturation (in
the framework of linear spin--wave theory) results in an order of magnitude
estimate for the transition temperature (for small dipolar couplings)
\begin{equation}
  T_c \propto {2 J_0 S^2 \over 
               \ln \left( {2 J_0 S \over \Omega_0} \right)} \, .
\label{5.25}
\end{equation}
It was also shown~\cite{Herring51,Doering61,Bander88} that an 
exchange anisotropy leads to a suppression of thermal interaction 
and in turn to the existence of long--range order
two--dimensional ferromagnets.

In two--dimensional ferromagnets, the dipolar interaction results in an
easy-plane anisotropy~\cite{Maleev76,Pokrovsky79a}. This is due to the fact
that the dipolar energy is minimized when all spins are aligned in plane. 
Thus the problem of a three-component Heisenberg ferromagnet with dipolar 
interaction is reduced to the problem of a two--component (planar) ferromagnet with 
renormalized values of the temperature and of the dipole--dipole interaction 
constant. The magnetic dipolar forces change fundamentally the nature of
the low--temperature phase. This can be seen as follows. In the absence of 
dipolar interaction any long--range order is impossible since it gets destroyed
by thermal fluctuations; the mean square of the spin fluctuations is given by 
$\langle \mid \delta S_{\bf q} \mid^2 \rangle \propto k_{_{B}} T / q^2$. A 
planar (pure) exchange ferromagnet would undergo a 
Kosterlitz--Thouless~\cite{Kosterlitz72,Kosterlitz73} transition. The 
Kosterlitz--Thouless phase is characterized by local but no long--range order
and by bound vortex pairs. There is a divergent susceptibility throughout this
phase. However, due
to the dipolar interaction the spectrum of the spin fluctuations at small 
wave numbers is changed. Including the dipolar interaction, Eq.~(\ref{5.7}), 
one obtains the following reduced Hamiltonian 
\begin{equation}
 {\cal H} = - {H \over k_BT} =
 -{J S^2 \over k_B T} \, v_2 a^2  \, \int {d^2 q \over (2 \pi)^2}
  \left[ 
  \left( q^2 - g_2 {q_y^2 \over |{\bf q}|} \right) 
  \phi_{\bf q}  \phi_{- \bf q} 
  + {\cal O} (\phi^3)
  \right]
\label{5.26}
\end{equation}
in terms of the azimuthal angle $\phi$, where we have assumed a fixed length 
${\bf S}^2 = 1$ and an ordering along the ${\hat x}$--direction. 
The relative strength of the dipolar interaction in two dimensions is
characterized by the parameter $g_2 =  A^{(1)} / J a^2$. In the
derivation of Eq.~(\ref{5.26}) the in--plane components of the dipolar 
interaction tensor have been used~\cite{Pelcovits79}
$A^{\alpha \beta}_{\bf q}  = {A^{(0)} \over 3} \delta^{\alpha \beta} - 
A^{(1)} {q^\alpha q^\beta \over q} + {\cal O} (q^2)$
( $A_0$ and $A_1$ see Eqs.~(\ref{5.8}) and (\ref{5.9})). Neglecting all but
the terms quadratic in $\phi_{\bf q}$, the mean square fluctuations
about the local ordering are given by~\cite{Pelcovits79}
\begin{equation}
  \langle \phi^2 \rangle \propto
  {k_B T \over J} \int {d^2 q \over (2 \pi)^2} \, 
  { 1 \over q^2 + g_2 |{\bf q}| \sin^2 \varphi_{\bf q} }\, ,
\label{5.27}
\end{equation}
where $\varphi_{\bf q}$ is the angle between ${\bf q}$ and the total
magnetization. The above integral no longer diverges due to the
presence of the dipolar interaction. Hence the dipolar interaction
stabilizes ferromagnetic long--range order in more than one dimension
~\cite{Pokrovsky77,Maleev76}. Besides stabilizing the long--range order,
the dipolar forces alter the vortex behavior~\cite{Pelcovits79}. 
Ref.~\cite{Pelcovits79} also contains an analysis of the static critical
behavior. More recently the critical behavior has been reanalyzed in
Ref.~\cite{DeBell89}. In passing we note that there are two reviews on 
excitations in low--dimensional systems by Pokrovsky et
al.~\cite{Pokrovsky79,Pokrovsky88}.

Actually the situation is more complicated in thin magnetic films. 
Mono-- and two--layer iron orders perpendicularly to the planes for
$T = 0 K$, only for layer numbers above $3$ the expected parallel
orientation is observed~\cite{Allensbach94}. 
This comes from the spin--orbit interaction, which effectively leads 
to a perpendicular anisotropy, which overcomes the dipolar interaction
for the very thinnest layers. Upon raising the temperature a reorientation
phase transition~\cite{Pappas90} from a perpendicular to an in--plane 
orientation of the magnetization has been observed experimentally.
It was argued by Pescia et al.~\cite{Pescia90,Pescia93} that this transition is
driven by the competition between spin--orbit and dipolar interaction.
The (classical) transition temperature has been 
estimated~\cite{Politi93,Pescia93} to be of the order 
$T_R \approx (\lambda - \Omega) {4 c J a^3 \over 7 (g_L \mu_B)^2}$, 
where $c$ is the number of nearest neighbors, 
$\Omega = 2 \pi (g_L \mu_B S)^2/a^3$ 
characterizes the strength of the dipolar interaction, and $\lambda$ is 
the single--ion anisotropy constant favoring perpendicular orientation.
Since the single--ion and the dipolar anisotropy parameters scale the
same under a renormalization group transformation~\cite{Pescia93,Politi93},
it depends on the relative values of the single--ion anisotropy and the 
dipolar parameters, whether the transition temperature for the 
reorientation phase transition is smaller or larger than the Curie 
temperature. However, since the dipolar interaction parameter increases with 
the number of layers, the value of $T_R$ is reduced for thicker films, as indeed 
observed experimentally~\cite{Pappas90}. Some of the above theoretical work 
is still rather controversial~\cite{Levanyuk93} and alternative mechanisms 
have been proposed for the experimentally observed reorientation phase 
transition~\cite{Levanyuk92}.
Due to the combination of various anisotropies, dipole--dipole interaction, 
reduced dimensionality and enhanced importance of thermal fluctuations the phase 
diagram of thin magnetic films shows a variety of new and interesting 
phases~\cite{Pappas90,Allensbach92,Allensbach94} with properties still 
awaiting a theoretical explanation.

\subsubsection{Antiferromagnets}
\label{s5.3.2}
Recently it has been shown~\cite{Pich93} that long--range order is also
possible in two--dimensional antiferromagnets due to the anisotropy of
the dipolar interaction. The existence of the long--range order is a
consequence of a subtle interplay between exchange and dipolar
interaction. The classical ground state of an isotropic pure exchange
antiferromagnet has a continuous degeneracy and hence no long--range order.
The dipolar interaction lifts the continuous degeneracy of the ground state
such that a spin alignment perpendicular to the plane is energetically 
favored (see Fig.~5.3). In other words, whereas the exchange interaction 
imposes the antiferromagnetic order, the dipolar interaction prevents thermal
fluctuations from its destruction. This can be seen most easily by
considering the equations of motion at zero wave vector. Upon
approximating the longitudinal spin components by their equilibrium expectation
value $S^z_l \approx S e^{i {\bf q}_0 \cdot {\bf x}_l}$ the
linearized form of Eqs.~(\ref{5.10}) (see Section~\ref{5.1}) reads
\begin{eqnarray}
  {d \over d t} S_0^x = 
  &&2 S \left( 
      A_{{\bf q}_0}^{zz} -  A_{{\bf q}_0}^{yy}
      \right) 
  S_{{\bf q}_0}^y \, , 
\label{5.28} \\
 {d \over d t} S_{{\bf q}_0}^y = 
  &&2 S \left[ 
      \left( J_{{\bf q}_0} - J_0 \right) +
      \left( A_{{\bf q}_0}^{zz} -  A_0^{xx} \right)
      \right] 
  S_0^x \, , 
\label{5.29}
\end{eqnarray}
and an analogous set of equations for $S_0^y$ and $S_{{\bf q}_0}^x$. The wave 
vector ${{\bf q}}_0={\pi\over a}(1,1,0)$ characterizes the antiferromagnetic, 
staggered modulation of the ground state. 
The coefficient on the right hand side of Eq.~(\ref{5.28}) 
is nonzero due to the anisotropy of the dipolar interaction in 2D. The 
resulting energy gap at zero wave vector then is 
\begin{equation} 
  E_0 = 2S \sqrt{A^{zz}_{{\bf q}_0}-A^{xx}_{{\bf q}_0}} \, 
           \sqrt{(J_{{\bf q}_0}-J_0)-
                 (A^{xx}_0-A^{zz}_{{\bf q}_0})} \, .
\label{5.30}
\end{equation}
A more detailed information on the spin--wave spectrum can be obtained
by linear spin--wave theory~\cite{Pich93}. Introducing Bose operators by 
a Holstein-Primakoff transformation~\cite{Keffer66}, given here only up to 
harmonic terms,
\begin{equation}
S_l^x  = \sqrt{S\over 2}(a_l+{a_l}^{\dag})~,~~
S_l^y = \mp i\sqrt{S\over 2}(a_l-{a_l}^{\dag})~,~~
S_l^z = \pm(S-{a_l}^{\dag}a_l) \, ,
\label{5.31}
\end{equation}
where the upper (lower) sign refers to the first (second) sublattice,
the Hamiltonian in the harmonic approximation is given by
\begin{eqnarray}
H  = & \sum_{{\bf q}}&\{A_{{\bf q}}~a_{{\bf q}}^{\dag}a_{{\bf q}} + 
       {1\over 2} B_{{\bf q}}~(a_{{\bf q}}~a_{-{\bf q}}+
                              a_{{\bf q}}^{\dag}a_{-{\bf q}}^{\dag})+
       \nonumber \\
     && ~~ C_{{\bf q}}~a_{{\bf q}}~a_{-{{\bf q}}-{{\bf q}}_0}+
           C_{{\bf q}}^{\ast}a_{{\bf q}}^{\dag}a_{-{{\bf q}}-{{\bf q}}_0}^{\dag}+
           C_{{\bf q}}~a_{{\bf q}}^{\dag}a_{{{\bf q}}+{{\bf q}}_0}+C_{{\bf q}}^{\ast}~
           a_{{{\bf q}}+{{\bf q}}_0}^{\dag}a_{{\bf q}}\}
\label{5.32}
\end{eqnarray}
with the coefficients 
\begin{mathletters}
\begin{eqnarray}
  A_{{\bf q}}  & = & S(2J_{{\bf q}_0}-J_{{\bf q}}-J_{{\bf q}+{\bf q}_0})+
  S(2A^{zz}_{{\bf q}_0}-A^{xx}_{{\bf q}}-A^{yy}_{{\bf q}+{\bf q}_0})\\ 
  B_{{\bf q}}  & = & S(J_{{\bf q}+{\bf q}_0}-J_{{\bf q}}) +
  S(A^{yy}_{{\bf q}+{\bf q}_0}-A^{xx}_{{\bf q}})\\
  C_{{\bf q}}  & = &  iSA^{xy}_{{\bf q}}.
\label{5.33}
\end{eqnarray}
\end{mathletters}
In this description the primitive cell is the crystallographic, which is half 
the magnetic. Diagonalization of the Hamiltonian, achieved by a Bogoliubov 
transformation, results in a spin--wave spectrum with two branches
\begin{equation}
  {E_{{\bf q}}^i} = \sqrt{ {1\over 2}(\Omega_1 \pm \Omega_2) }
\label{5.34}
\end{equation}
where
\[
\Omega_1 = A_{{\bf q}}^2 - B_{{\bf q}}^2 + A_{{\bf q}+{\bf q}_0}^2 -
B_{{\bf q}+{\bf q}_0}^2 + 8C_{{\bf q}}~C_{{\bf q}+{\bf q}_0}
\]
and
\[\Omega_2^2 = (A_{{\bf q}}^2-B_{{\bf q}}^2-A^2_{{\bf q}+{\bf q}_0}+B_{{\bf q}+
{\bf q}_0}^2)^2+
16[C_{{\bf q}+{\bf q}_0}(A_{{\bf q}+{\bf q}_0}-B_{{\bf q}+{\bf q}_0})-C_{{\bf q}}~
(A_{{\bf q}}-B_{{\bf q}})]
\]
\[\times [C_{{\bf q}}~(A_{{\bf q}+{\bf q}_0}+B_{{\bf q}+{\bf q}_0})-
C_{{\bf q}+{\bf q}_0}(A_{{\bf q}}+B_{{\bf q}})].\]
In Fig.~5.4 the dispersion relation is shown for three values 
for the ratio of dipolar and exchange energy $\kappa = {(g\mu_B)^2\over
4|J|a^3}$ with isotropic nearest-neighbor exchange interaction ($J<0$). The 
two branches can be resolved only for large values of $\kappa$. 

In particular the gap is proportional to the square root of the difference 
of the magnetostatic energy between the configurations of in-plane and 
out-of-plane magnetization. In a three--dimensional simple cubic lattice 
the first root in Eq.~(\ref{5.30}) vanishes because of the symmetry, 
but in two--dimensional systems there is a finite gap for perpendicular 
antiferromagnetic order. We  note that for sufficiently large exchange 
energy the gap is the geometric 
mean of dipole and exchange energy, which in turn implies that the gap is 
much larger than the dipolar energy for $\kappa \ll 1$.

The N\'eel temperature $T_N$ has been determined via 
a high--temperature expansion, by employing Callen's
method~\cite{Pich93,Pich94a,Pich94b}. If the dipolar interaction is weak in 
comparison with the exchange interaction one can give an order of magnitude 
estimate for the transition temperature in the 
framework of linear spin--wave theory based on the Holstein--Primakoff 
transformation
\begin{equation}
  T_N \propto { |J| \over \ln \left( {|J| \over E_0} \right)} \, .
\label{5.35}
\end{equation}
This, however, gives an overestimate for the transition temperature, because 
it uses a temperature--independent dispersion relation. Actually, the magnon 
frequency softens with increasing temperature. Those effects have been 
accounted for in Ref.~\cite{Pich94a} by an extension of the Tyablikov 
decoupling scheme due to Callen~\cite{Callen63}. In essence this leads to a 
replacement of the saturation magnetization of the spins $S$ by the 
temperature--dependent order parameter $\sigma$. The resulting transition 
temperature is lowered with respect to the estimate from linear spin--wave 
theory. $E_0$ has been obtained by linear spin--wave theory
and $E_0^\sigma$ and $T_N^{\rm th}$ by the method of
Callen. The results are summarized in Table~\ref{table52} and show a 
quite satisfactory agreement with experimental data. This theory has been
extended to antiferromagnets on a honeycomb~\cite{Pich94c} and several other 
lattices~\cite{Pich94d}.

\newpage

\centerline{\bf Tables:}

\begin{table}
\setdec 0.00
\caption{Behavior of the scaling functions of the line widths of an 
isotropic dipolar antiferromagnet in asymptotic regions.}
\bigskip \bigskip
\begin{tabular}{lcccc}
& IC & DC & IH & DH \\
\tableline
$\gamma_M^{L,T}$ &$1$ &$y^{3/2}$  &$x^{-1/2}$ &$r^{3/2}$\\
$\gamma_N^{L,T}$ &$1$ &$y^{-1/2}$ &$x^{3/2}$  &$r^{3/2}$\\
\end{tabular}
\label{table51}
\end{table}

\vskip 1cm

\begin{table}
\setdec 0.00
\caption{Exchange energy $|J|$, lattice constant $a$, energy gap $E_0$,
spin-flop field $H_{\rm sf}$, N\'eel temperature $T_N$ and zero temperature
order parameter $\sigma_0$.}
\bigskip \bigskip
\begin{tabular}{l|rc|ccc|cc|c}
& $|J|$ & $a $ & $E_0^{\rm exp} $ & $E_0^\sigma$ & $E_0$ 
& $T_N^{\rm exp} $ & $T_N^{\rm th}$
& $\sigma_0$ \\
&[K]&[\AA]&[\rm K]&[K]& [K] & [K]&[K]&\\ 
\tableline
$K_2MnF_4$   & 8.5$^{\rm a}$ & 4.17$^{\rm a}$ & 7.4$^{\rm b}$ & 7.1 & 7.6 &
42$^{\rm a}$ & 41 & 2.33 \\ 
$Rb_2MnF_4$  & 7.4$^{\rm c}$  & 4.20$^{\rm g}$ & 7.3$^{\rm b}$ & 6.5 & 7.0 &
38$^{\rm g}$  & 36 & 2.33 \\ 
$Rb_2MnCl_4$ & 11.2$^{\rm f}$ & 5.05$^{\rm e}$ & 7.5$^{\rm f}$ & 6.1 & 6.6 &
56$^{\rm f}$ & 48 & 2.32 \\
$(CH_3NH_3)_2MnCl_4$ & 9.0$^{\rm f}$ & 5.13$^{\rm e}$ & & 5.3 & 5.7 &  
45$^{\rm f}$ & 39 & 2.32 
\end{tabular}
{$^a$ Reference \cite{Bir73}, 
$^b$ Reference  \cite{Win73}, 
$^c$ Reference  \cite{Win731},
$^d$ Reference  \cite{Win},
$^e$ Reference  \cite{LB},
$^f$ Reference  \cite{sch},
$^g$ Reference  \cite{Bir70}.}
\label{table52}
\end{table}

\vskip 1cm

\newpage

\centerline{\bf Captions to the figures:}
\bigskip
\bigskip

\noindent {\bf Figure 5.1: } Scaling function for the line width of the 
transverse magnetization above $T_N$ versus $x=1/q\xi$ for
different values of $\varphi=N\pi/30$ (N=0,1,2,...,14).
\bigskip

\noindent {\bf Figure 5.2: } Scaling functions for the line width of the
transverse and longitudinal staggered magnetization above $T_N$
versus $x=1/q\xi$ for different values of $\varphi=N\pi/30$
(N=0,1,2,...,14).
\bigskip

\noindent {\bf Figure 5.3: } 
Classical ground state of a two--dimensional dipolar antiferromagnet with 
a dipolar interaction much smaller than the exchange interaction on a 
quadratic lattice. Taken from Ref.~\cite{Pich94b}. 
\bigskip

\noindent {\bf Figure 5.4: } The spin-wave dispersion relation 
(Eq.~(\ref{5.34})) of pure exchange antiferromagnets on a quadratic
lattice with nearest-neighbor interaction (solid line) and with 
additional dipolar interaction ($S=1/2$), for the ratios of dipolar energy 
to exchange energy $\kappa={(g\mu_B)^2\over 4|J|a^3}$ along the 
${\pi\over a}[\xi,\xi,0]$ direction: $\kappa=0.1$ (long dashed), $\kappa=0.01$ 
(dashed) and $\kappa=0.001$ (dotted). The splitting of the two
magnon branches is visible only for $\kappa = 0.1$.

\newpage
\vfill
\begin{figure}[h]
\centerline{\rotate[r]{\epsfysize=5in \epsffile{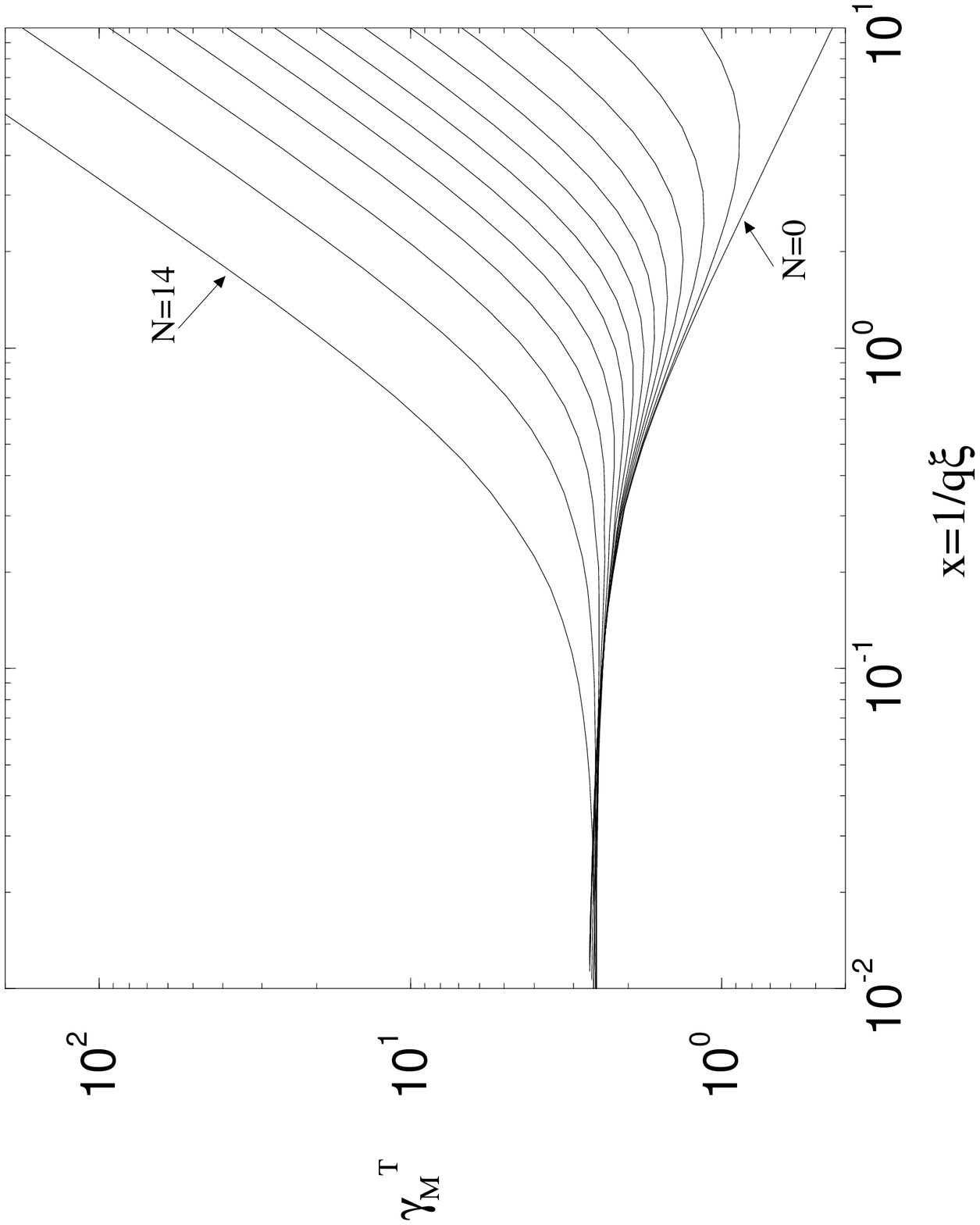}}}
\bigskip \bigskip
{\bf Figure 5.1: } Scaling function for the line width of the 
transverse magnetization above $T_N$ versus $x=1/q\xi$ for
different values of $\varphi=N\pi/30$ (N=0,1,2,...,14).
\label{fig51}
\end{figure}

\newpage
\vfill
\begin{figure}[h]
\centerline{\rotate[r]{\epsfysize=5in \epsffile{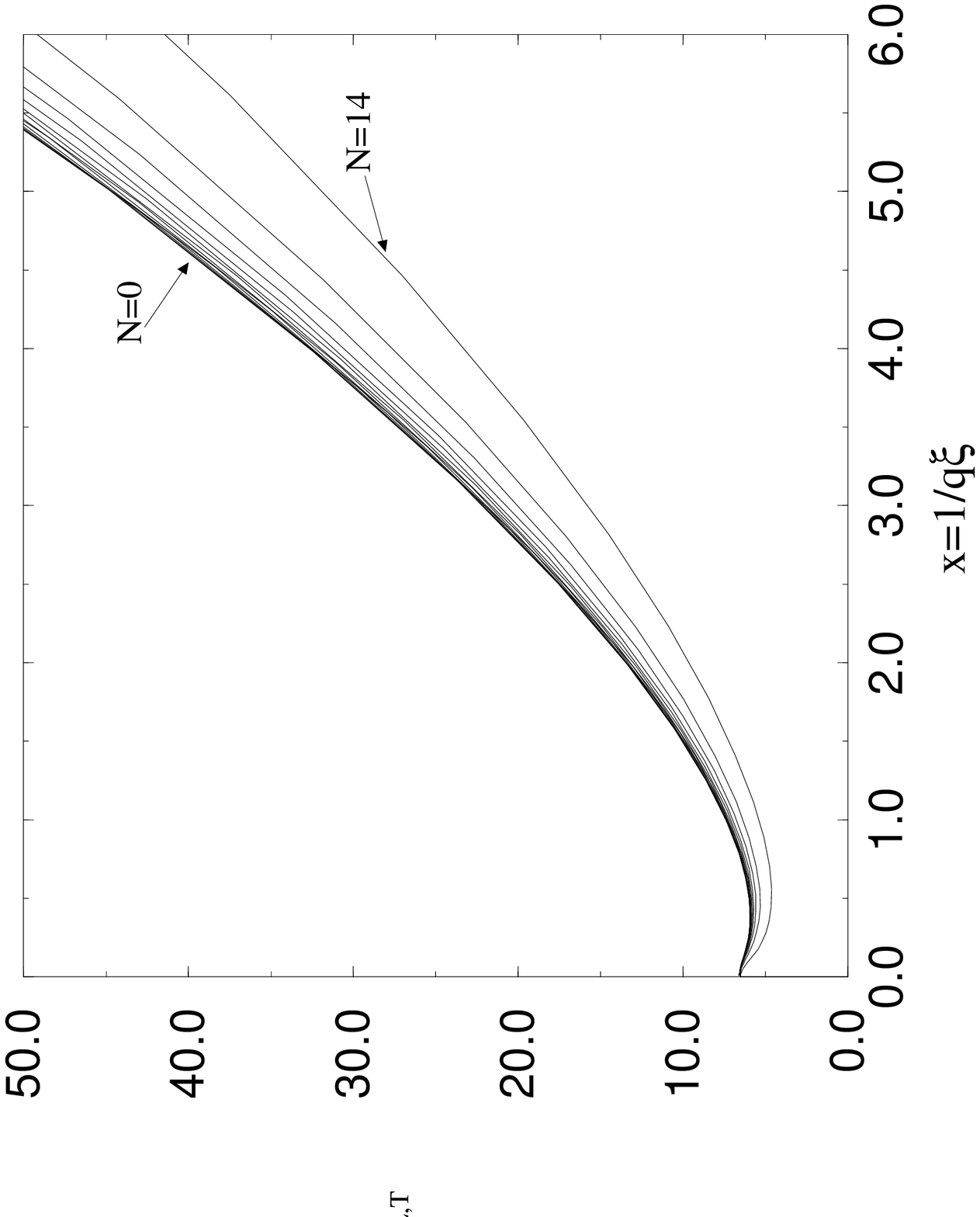}}}
\bigskip \bigskip
{\bf Figure 5.2: } Scaling functions for the line width of the
transverse and longitudinal staggered magnetization above $T_N$
versus $x=1/q\xi$ for different values of $\varphi=N\pi/30$
(N=0,1,2,...,14).
\label{fig52}
\end{figure}

\newpage
\vfill
\begin{figure}[h]
\centerline{\rotate[r]{\epsfysize=5in \epsffile{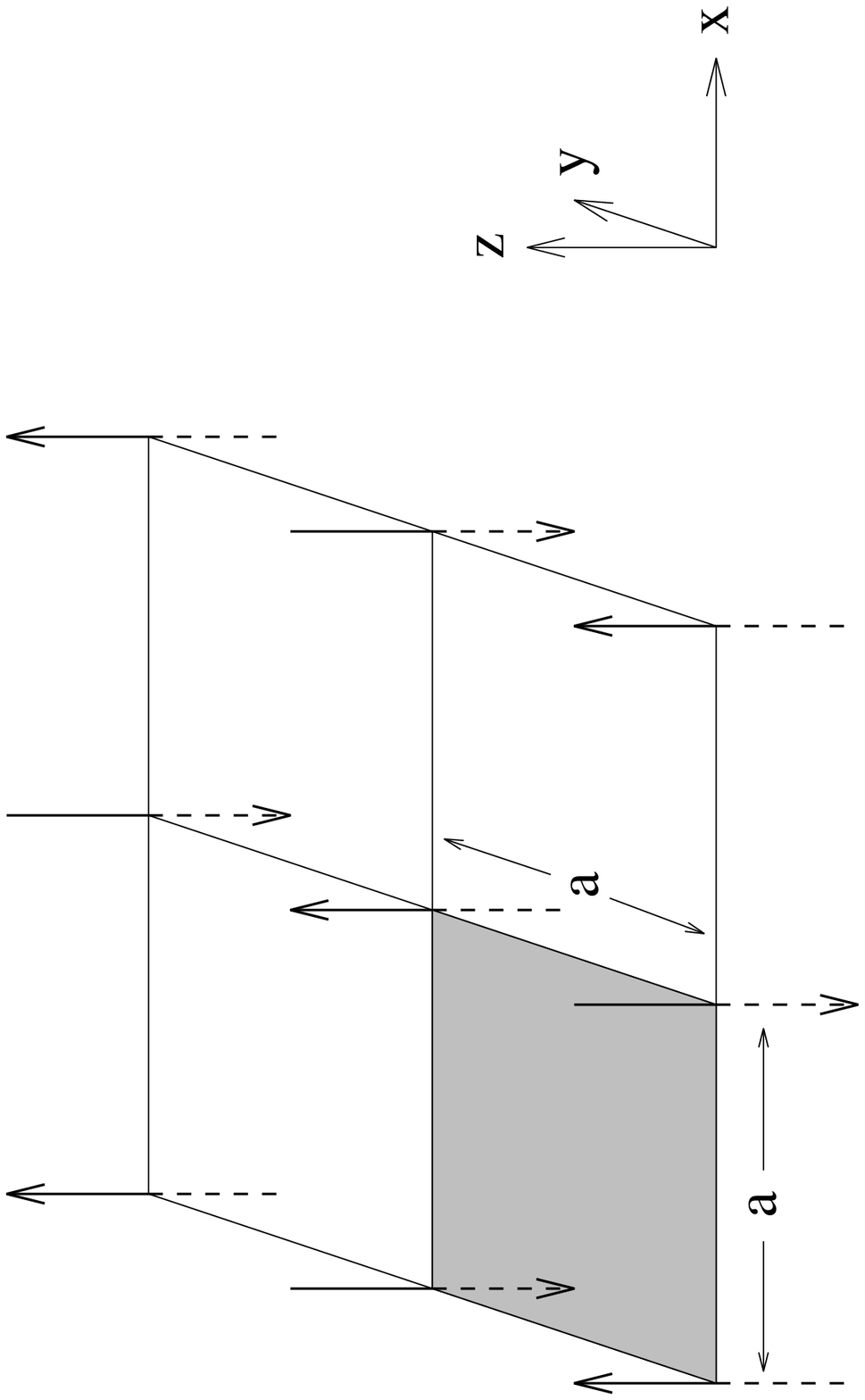}}}
\bigskip \bigskip
{\bf Figure 5.3: } 
Classical ground state of a two--dimensional dipolar antiferromagnet with 
a dipolar interaction much smaller than the exchange interaction on a 
quadratic lattice. Taken from Ref.~\cite{Pich94b}. 
\label{fig53}
\end{figure}
 
\newpage
\vfill
\begin{figure}[h]
\centerline{\rotate[r]{\epsfysize=5in \epsffile{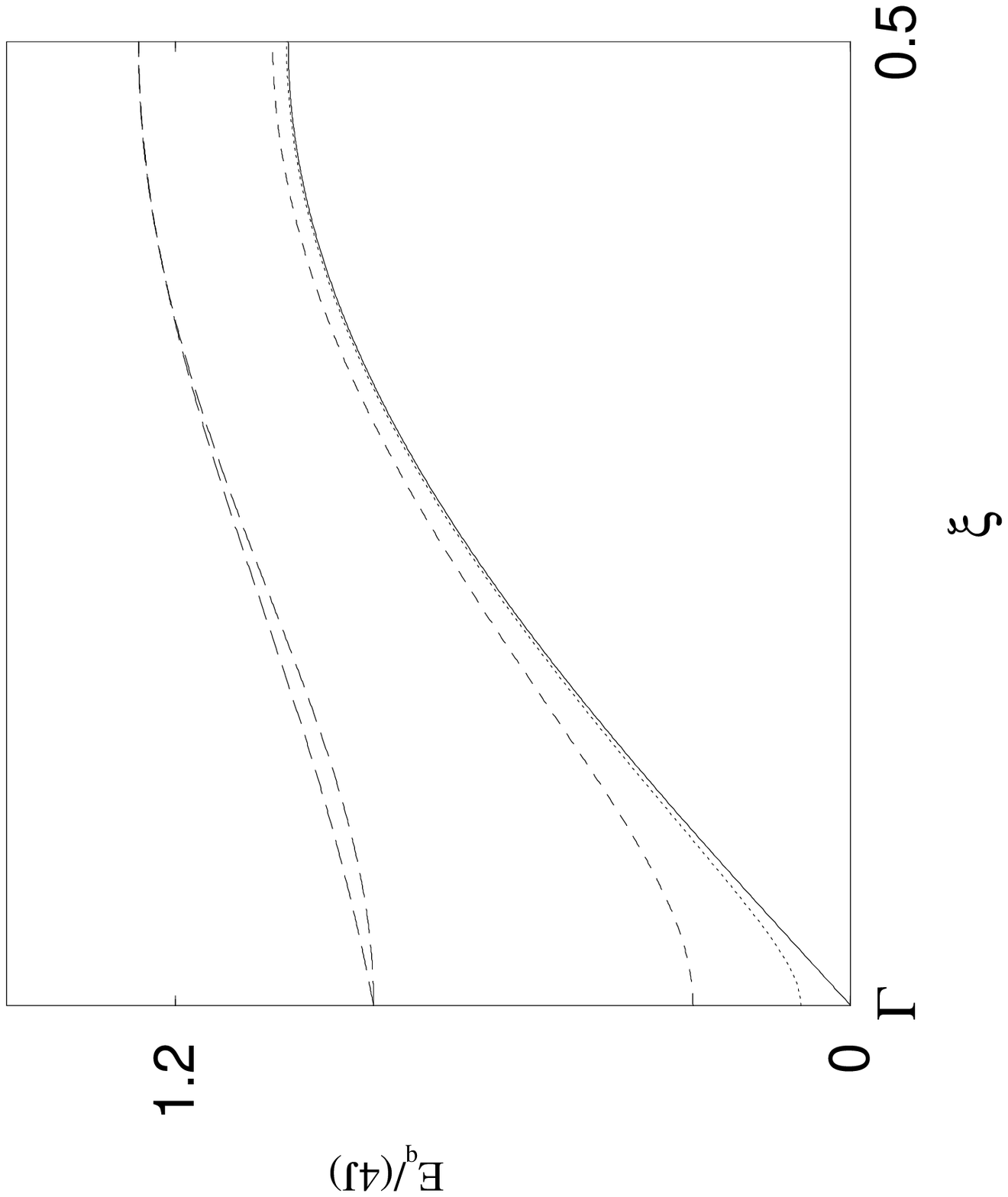}}}
\bigskip \bigskip
{\bf Figure 5.4: } The spin-wave dispersion relation (Eq.~(\ref{5.34})) of 
pure exchange antiferromagnets  on a quadratic
lattice with nearest-neighbor interaction (solid line) and with 
additional dipolar interaction ($S=1/2$), for the ratios of dipolar energy 
to exchange energy $\kappa={(g\mu_B)^2\over 4|J|a^3}$ along the 
${\pi\over a}[\xi,\xi,0]$ direction: $\kappa=0.1$ (long dashed), $\kappa=0.01$ 
(dashed) and $\kappa=0.001$ (dotted). The splitting of the two
magnon branches is visible only for $\kappa = 0.1$.
\label{fig54}
\end{figure}
 
%%%%%%%%%%%%%%%%%%%%%%%%%%%%%%%%%%%%%%%%%%%%%%%%%%%%%%%%%%%%%%%%%%%%%%%%%%%%%%
\section{STOCHASTIC THEORY}
\label{s6}

\subsection{Classical field theory and dynamic functional}
\label{s6.1}

If one tries to describe the critical dynamics of ferromagnets, one
is in general not interested in the complete, highly complicated,
microscopic time evolution. Rather, one is usually interested in
the dynamics on time scales characteristic of its slowly varying
dynamical quantities, such as those associated with hydrodynamic,
order parameter or Goldstone modes.

A method to extract from the microscopic dynamics the equations of
motion appropriate to these slowly varying quantities is the
projection operator formalism of Zwanzig~\cite{z61} and Mori~\cite{m65}.
Since the semi-phenomenological equations derived from this
formalism form a common basis of
mode coupling and dynamic renormalization group theories we briefly
review the main ideas and results of this approach. 

The idea is to eliminate the fast variables by introducing a
projection operator ${\cal P}$, which projects onto the subspace
of slowly varying modes. Descriptions of this formalism can be
found in Refs.~\cite{kk73,mf73}. A short account is given in Appendix~B. 
After having chosen an appropriate set of
slow variables - which is crucial for the validity of the dynamics - one 
can derive formally  exact non linear equations of motion for these modes 
(here these modes are denoted by $\{ S_\alpha (t) \}$)
\begin{eqnarray}
 {d \over dt} S_\alpha (t) = \, \, && v_\alpha (  \{ S(t) \} ) \nonumber \\ +
  &&\sum_\beta \int_0^t d \tau 
  P_{\rm eq}^{-1} ( \{ S(t-\tau) \} )
  { \partial M_{\alpha \beta} (\tau; \{ S(t-\tau) \} ) 
      P_{\rm eq} ( \{ S(t-\tau) \} ) \over
     \partial S_\beta^\star (t - \tau)} +
   \zeta_\alpha (t) \, ,  
\label{6.1}
\end{eqnarray}
where 
\begin{equation} 
  P_{\rm eq} ( \{ S \} ) = 
  \exp \left[- {1 \over k_B T} {\cal H} ( \{ S \} ) \right]
\label{6.2}
\end{equation}
is the equilibrium probability distribution function.  The first
term in these equations,
\begin{equation}v_\alpha (  \{ S(t) \} ) = < i {\cal L} S_\alpha; 
 \{ S(t) \} > \, ,
\label{6.3}
\end{equation}          
is called the streaming term and describes the systematic part of the
driving forces. This term can be expressed in terms of Poisson
brackets. (The conditional average $< X ; \{ a \} >$ is defined by
$< X ; \{ a \} > = < X \delta (S-a) >/ P_{\rm eq} ( \{ S(t) \} $ where the
average is taken in the microcanonical ensemble \cite{kk76}.) 
The second term contains the memory kernel
\begin{equation}M_{\alpha \beta} (t; \{ S(t) \} ) = 
  <\zeta_\alpha (t) \zeta_\beta^\star (0);
  \{ S(t) \} > \, , 
\label{6.4}
\end{equation}          
and characterizes the damping of this adiabatic motion by frictional 
effects arising from the ``random'' forces.
The ``random'' forces are given in terms of the projection operator
${\cal Q} = 1 - {\cal P}$ and the Liouville operator ${\cal L}$ 
\begin{equation}\zeta_\alpha (t) =
  \exp \left[ i t {\cal Q} {\cal L} \right] i  {\cal Q} {\cal L} S
  \, .
\label{6.5}
\end{equation}          
Whether this force can be regarded as random or not depends
crucially on the choice of the projection operator, i.e., whether
${\cal P}$ contains all the slowly varying quantities of the system
under consideration.

The semi-phenomenological equations of motion are obtained from the
above exact equations by making the following  basic, and
at least plausible assumptions. First one assumes that one can
neglect memory effects in $M_{\alpha \beta}$ and makes the
Markovian approximation 
\begin{equation}
M_{\alpha \beta} (t; \{ S(t) \} ) \approx 2 L_{\alpha \beta} 
     ( \{ S(t) \} ) \delta(t) \, .
\label{6.6}
\end{equation}          
Second one takes the kinetic coefficients to be independent of the
slow variables
\begin{equation}L_{\alpha \beta} ( \{ a \} ) \approx  L_{\alpha \beta} \, ,
\label{6.7}
\end{equation}
and third one assumes that all slowly varying variables are
contained in the choice of the projection operator so that the forces
$f_\alpha$ are really random and have a Gaussian probability
distribution
\begin{equation}w( \{ \zeta \} | t_0 \leq t \leq t_1 ) \sim
  \exp \left[ -{1 \over 4} \int_{t_0}^{t_1} dt \zeta_\alpha (t)
  L_{\alpha \beta}^{-1} \zeta_\beta (t) \right] \, .
\label{6.8}
\end{equation}
Then the exact equations reduce to the following
semi-phenomenological equations of the Langevin form
\begin{equation}{d \over dt} S_\alpha (t) = v_\alpha (  \{ S(t) \} )  -
  \sum_\beta L_{\alpha \beta}^0 
   {\delta {\cal H} ( \{ S(t) \} ) \over
    \delta S^\star_\beta (t) }
   +\zeta_\alpha (t) \, ,
\label{6.9}
\end{equation}
where the streaming term,
\begin{equation}v_\alpha (  \{ S(t) \} )  = 
  - \lambda f \sum_\beta 
  \left[
  {\delta \over \delta S_\beta} Q_{\alpha \beta} ( \{ S \} ) -
  Q_{\alpha \beta} ( \{ S \} ) {\delta {\cal H} ( \{ S \} ) \over
                                \delta S^\star_\beta (t) }
  \right] \, , 
\label{6.10}
\end{equation}
can be expressed in terms of Poisson brackets
\begin{equation}Q_{\alpha \beta} ( \{ S \} ) = \{ S_\alpha, S_\beta \}_P = 
  - Q_{\beta \alpha} ( \{ S \} ) \, .
\label{6.11}
\end{equation}
The equilibrium distribution does not
determine the driving force in the Langevin equations uniquely. 
Only the dissipative couplings which are generated by the derivative
of the effective Hamiltonian ${\cal H}$ are related to the static
properties of the system. As a direct consequence with each static
universality class are associated several dynamic universality
classes depending on the structure of the Poisson brackets.

After these general preliminaries we can now start to set up
a stochastic equation of motion for the dynamics of dipolar
ferromagnets.
If the interaction between the spins in an isotropic Heisenberg
ferromagnet is solely given by the short range exchange interaction the
dynamics are described by precession of the spins in a local
magnetic field generated by the surrounding spins. The rotation
invariance of the Heisenberg Hamiltonian implies that the total spin
is conserved. As shown explicitly in chapter~\ref{s3.1} this is no longer
the case if one takes into account the long--ranged dipole-dipole
interaction. Hence the spin dynamics of a isotropic dipolar
ferromagnet are described by the following semi-phenomenological
Langevin equations
\begin{equation}{\partial {\bf S}({\bf x},t) \over \partial t} =
   \lambda f  {\bf S}({\bf x},t) \times 
   { \delta {\cal H}([S]) \over \delta {\bf S} ({\bf x},t) }
 - {\hat L} { \delta {\cal H}([S]) \over 
                \delta {\bf S} ({\bf x},t) }
 + {\gvect \zeta} ({\bf x},t) \, ,
\label{6.12}
\end{equation}
where ${\bf S}({\bf x},t)$ is the local spin density. The local
magnetic field is determined by the functional derivative of the 
Landau-Ginzburg functional ${\cal H}[S]$ with respect to the spin
density. For the present case the Landau-Ginzburg functional is
given by 
\begin{eqnarray}
  &&{\cal H} [S] = -{1 \over 2} \inq{q}
  \left[ \left( r + q^2 \right) \delta^{\alpha \beta}
  + g {q^{\alpha} q^{\beta} \over q^2} \right] 
  S^{\alpha} ({\bf q}) S^{\beta} (- {\bf q}) \nonumber \\
  &&-{u \over 4!} \inq{q_1} \inq{q_2} \inq{q_3} 
  S^{\alpha} ({\bf q}_1) S^{\alpha} ({\bf q}_2)
  S^{\beta} ({\bf q}_3) S^{\beta} (-{\bf q}_1 - {\bf q}_2 - {\bf q}_3)   \, . 
\label{6.13}
\end{eqnarray}
$S^{\alpha}({\bf q})$ ($\alpha$=1,2,...,$n$) are the
components of the bare spin variable with $n$ equal to the space
dimensionality $d$. We further used the abbreviation $\int
\limits_q = \int {d^d q / (2 \pi)^d}$, $r$ is the reduced
temperature--deviation and $g$ denotes the relative
strength of the dipolar interaction. 

The dipolar interaction in Eq.(\ref{6.13}) breaks the symmetry of the
spin fluctuations transverse and longitudinal to the wave vector
${\bf q}$, which is reflected in the free propagator
\begin{equation}G^{\alpha \beta} =
  {q^{\alpha} q^{\beta} \over q^2} G^L+
  \left(\delta^{\alpha \beta} - {q^{\alpha} q^{\beta} \over q^2}
  \right) G^T \, ,
  \label{6.14}
\end{equation}
where $ G^L(r,g,q) = (r + g + q^2)^{-1}$ and   
$G^T(r,q) = (r + q^2)^{-1}$. 

The time scale is characterized by the quantity
$\lambda$. The mode coupling coefficient $f$ determines the 
strength of the coupling of the spin density to the local magnetic 
field. Since the order parameter is a non conserved quantity the
leading term in the Onsager operator
\begin{equation}{\hat L} =
  \lambda ( \gamma  - \nabla^2 )
\label{6.15}
\end{equation}
is wave vector independent, $\lambda \gamma$, and not
$-\lambda \nabla^2$ as for a conserved order parameter. The quantity
$\gamma$ characterizes - like $g$ in statics - the relative strength
of the dipolar interaction. The random forces ${\gvect \zeta}({\bf x},t)$
have a Gaussian probability distribution 
\begin{equation}w_\zeta ( [\zeta_i({\bf x},t)] \mid t_0 \leq t \leq t_1) 
  \sim
  \exp \left( - {1 \over 4} \sum_i \int_{t_0}^{t_1} dt \int d^dx
  \zeta_i({\bf x},t) {\hat L}^{-1} \zeta_i({\bf x},t) \right) 
  \, ,
\label{6.16}
\end{equation}
which is uniquely determined by the lowest moments
\begin{mathletters}
\begin{eqnarray}
&&\langle {\gvect \zeta}({\bf x},t) \rangle = 0 \, ,
\label{6.17.a} \\
&&\langle \zeta_i({\bf x},t) 
  \zeta_j({\bf x}^{\prime},t^{\prime}) \rangle =   
  2 {\hat L} 
  \delta ({\bf x} - {\bf x}^{\prime}) 
  \delta (t-t^{\prime})   
  \delta_{ij} \, .
\label{6.17.b}
\end{eqnarray}
\end{mathletters}
Note that we have set $\beta={1 / k_{_B} T} =1$. 

In particular for the implementation of the dynamic renormalization group 
theory in a way analogous to static critical phenomena it is convenient to
introduce a functional which generates the perturbation expansion for the 
frequency dependent correlation and response functions, which is
equivalent to the equations of motion. We shall use the functional integral 
formulation~\cite{j76,bjw76},
which converts the Langevin equations into a dynamic functional with 
one additional field~\cite{msr73}. The idea is that, 
instead of solving the Langevin equations (\ref{6.12}) 
for the stochastic
spin fields ${\bf S}({\bf k},t)$ in terms of the random forces 
${\gvect \zeta}({\bf k},t)$ and then averaging over the Gaussian weight,
one can eliminate the random forces in favor of the spin
variables by introducing a path probability density
$W(\{S\})$ via
\begin{equation}W(\{S\}) {\cal D}[S]  =
  w(\{\zeta\}) {\cal D}[\zeta ] \, .
\label{6.18}
\end{equation}
Furthermore, it is convenient to perform a Gaussian transformation
in order to "linearize" the dynamic functional. This is accomplished 
by introducing response fields ${\tilde S}$ \cite{j76,msr73} by
\begin{equation}W(\{S\}) =
  \int{\cal D}[i \tilde S]  
  \exp\left\{ {\cal J} [S, {\tilde S}] \right\} \, .
\label{6.19}
\end{equation}
\hbox{For more details on the general formalism we refer
the} reader to \hbox{Refs.~\cite{j76,d76,bjw76,msr73}}. 
The generating functional for the correlation functions is
\begin{equation}Z[h,{\tilde h}] =
  {1 \over {\cal N}} 
  \int {\cal D} [S] {\cal D} [\imath {\tilde S}] 
  \exp 
  \left[ {\cal J}_{t_0}^{t_1} [S,{\tilde S}] 
       + \sum_i \int d^dx \int_{t_0}^{t_1} dt
         \left( h_{i} S_{i} + {\tilde h}_{i} {\tilde S}_{i}  
       \right)   
  \right] \, ,
\label{6.20}
\end{equation}
where the normalization factor ${\cal N}$ is chosen such that
$Z[h=0,{\tilde h}=0] = 1$. The symbols ${\cal D} [S]$ and
${\cal D} [\imath {\tilde S}]$ denote functional measures of the path
integral. The Janssen-Dominicis functional is given by
\begin{equation}{\cal J}_{t_0}^{t_1}  [S,{\tilde S}]  =
   \sum_i \int_{t_0}^{t_1} dt \int d^dx
   \left[
     {\tilde S}_{i} {\hat L} {\tilde S}_{i}
   - {\tilde S}_{i} 
     \left({\partial S_{i} \over \partial t}-K_i[S]\right)
   - {1 \over 2} { \delta K_i[S] \over \delta S_{i} }  
   \right] \, ,
\label{6.21}
\end{equation}
where the functional of the forces is 
\begin{equation}{{\bf K}} [S] = 
   \lambda f  {\bf S}({\bf x},t) \times 
   { \delta {\cal H}([S]) \over \delta {\bf S} ({\bf x},t) }
 - {\hat L} { \delta {\cal H}([S]) \over 
                \delta {\bf S} ({\bf x},t) }  \, . 
\label{6.22}
\end{equation}

In formulating the perturbation theory one splits the functional,
Eqs.~(\ref{6.21}) and (\ref{6.22}), into a harmonic part
\begin{eqnarray}
  {\cal J}_{\rm harm} [S,{\tilde S}] =
  &&\int_k \int_{\omega} 
  \biggl[
        \lambda (\gamma + k^2) 
        {\tilde S}^{\alpha} ({\bf k},\omega)
        {\tilde S}^{\alpha} (- {\bf k},- \omega)
        \, + \nonumber \\     
      - &&{\tilde S}^{\alpha} ({\bf k},\omega)
        \left[
              i \omega \delta^{\alpha \beta}
            + \lambda (\gamma + k^2) 
              G^{\alpha \beta} (r,g,{\bf k})^{-1}
        \right]
        S^{\alpha} ({\bf k},\omega)
  \biggr] \, , 
\label{6.23}
\end{eqnarray}
which is bilinear in the stochastic spin fields $S$ and ${\tilde
S}$  and a part ${\cal J}_{\rm int} [S,{\tilde S}]$ containing the
interaction terms. Here $G^{\alpha \beta} (r,g,{\bf k})$ denotes
the static propagator, Eq.~(\ref{6.14}). Furthermore we have
introduced the short hand notations 
$\int_k = \int {d^d k / (2 \pi )^d}$,
$\int_{\omega} = \int {d \omega / 2 \pi}$.
The Fourier transform of the spin density is defined by
${\bf S} ({\bf x},t) = \int_k \int_\omega {\bf S}({\bf k}, \omega)
e^{ i({\bf k} \cdot {\bf x} - \omega t)}$.
Hence the generating functional can be written as
\begin{eqnarray}
  Z[h,{\tilde h}] =
  &&{1 \over {\cal N}} 
  \exp \left[ {\cal J}_{\rm int} \left( {\delta \over \delta h},
                             {\delta \over \delta {\tilde h}}
                      \right)
       \right] \nonumber \\
  &&\times
  \int {\cal D} [S] {\cal D} [\imath {\tilde S}] 
  \exp 
  \left[ {\cal J}_{\rm harm} [S,{\tilde S}] 
       + \sum_i \int d^dx \int_{t_0}^{t_1} dt
         \left( h_{i} S_{i} + {\tilde h}_{i} {\tilde S}_{i}  
       \right)   
  \right] \, ,
\label{6.24}
\end{eqnarray}
We start with the discussion of the harmonic part, Eq.~(\ref{6.23}). It can
be written in matrix notation as
\begin{equation}{\cal J}_{\rm harm} [S,{\tilde S}] =
  - {1 \over 2} \int_k \int_{\omega} 
    \left( {\tilde S}^{\alpha} ({\bf k},\omega),
           S^{\alpha} ({\bf k},\omega) 
    \right)
    {\bf A}^{\alpha \beta} ({\bf k},\omega)
    \left( {\tilde S}^{\beta} (- {\bf k},- \omega),
           S^{\beta} (- {\bf k},- \omega) 
    \right)^T \, .
\label{6.25}
\end{equation}
The matrix ${\bf A}^{\alpha \beta} ({\bf k},\omega)$ is given by
\begin{equation}{\bf A}^{\alpha \beta} ({\bf k},\omega) = 
  \pmatrix{ -2 L(k) 
            &&i \omega - A^{\alpha \beta} ({\bf k})\cr
            -i \omega - A^{\alpha \beta} ({\bf k})
             && 0\cr } \, ,
\label{6.26}
\end{equation}
where we have defined
\begin{equation}A^{\alpha \beta} ({\bf k}) =
  -L (k) [G^{\alpha \beta} (r,g,{\bf k})]^{-1} \, ,
\label{6.27}
\end{equation}
\begin{equation}L(k) =
  \lambda (\gamma + k^2) \, .
\label{6.28}
\end{equation}
The harmonic part of the partition function
\begin{equation}Z_0[h,{\tilde h}] =
  \int {\cal D} [S] {\cal D} [\imath {\tilde S}] 
  \exp 
  \left[ {\cal J}_{\rm harm} [S,{\tilde S}] 
       + \sum_i \int d^dx \int_{t_0}^{t_1} dt
         \left( h_{i} S_{i} + {\tilde h}_{i} {\tilde S}_{i}  
       \right)   
  \right] \, ,
\label{6.29}
\end{equation}
can be calculated explicitly with the result
\begin{equation}Z_0[h,{\tilde h}] =
  \exp 
  \left[ {1 \over 2} \int_k \int_{\omega} 
         \left( {\tilde h}^{\alpha} ({\bf k},\omega),
                         h^{\alpha} ({\bf k},\omega) 
         \right)
         ({\bf A}^{\alpha \beta} ({\bf k},\omega))^{-1}
         \left( {\tilde h}^{\beta} (- {\bf k},- \omega),
                         h^{\beta} (- {\bf k},- \omega) 
         \right)^T
  \right] \, .   
\label{6.30}
\end{equation}
Hence the free propagators are found to be 
\begin{equation}\pmatrix{ \langle {\tilde S}^{\alpha} ({\bf k},\omega) 
  {\tilde S}^{\beta} ({\bf k}^{\prime},\omega^{\prime})
  \rangle_0 
  &&\langle {\tilde S}^{\alpha} ({\bf k},\omega) 
  S^{\beta} ({\bf k}^{\prime},\omega^{\prime})
  \rangle_0 \cr
  \langle S^{\alpha} ({\bf k},\omega) 
  {\tilde S}^{\beta} ({\bf k}^{\prime},\omega^{\prime})
  \rangle_0 
  &&\langle S^{\alpha} ({\bf k},\omega) 
  S^{\beta} ({\bf k}^{\prime},\omega^{\prime})
  \rangle_0 \cr} = 
  \delta (\omega + \omega^{\prime})
        \delta ({\bf k} + {\bf k}^{\prime})
  \left({\bf A}^{\alpha \beta} (-{\bf k},-\omega)\right)^{-1} \, ,
\label{6.31}
\end{equation}
where
\begin{equation}\left({\bf A}^{\alpha \beta} ({\bf k},\omega)\right)^{-1} =
  {\bf A}_L^{-1} ({\bf k},\omega) P_L^{\alpha \beta} ({\bf k}) +
  {\bf A}_T^{-1} ({\bf k},\omega) P_T^{\alpha \beta} ({\bf k}) \, .
\label{6.32}
\end{equation}
Hence the longitudinal and transverse propagators are given by  
\begin{equation}{\bf A}_{\alpha}^{-1}({\bf k},\omega) =
  \pmatrix{ 0& {1 / [-i\omega - A_{\alpha}(k)]}\cr
            {1 /[i\omega - A_{\alpha}(k)]} 
            &{2L(k) /[\omega^2 + A_{\alpha}(k)^2]} \cr}  \, ,    
\label{6.33}
\end{equation}
where
\begin{equation}A_{L,T} (k) = - L(k) 
  \cases{ r + k^2 &for $\alpha = T$,\cr
          r + g + k^2 &for $\alpha = L$. \cr}
\label{6.34}
\end{equation}
In summary, one gets for the response propagators
\begin{equation}R^{\alpha \beta} ({\bf k},\omega) =
  \langle {\tilde S}^{\alpha} ({\bf k},\omega) 
  S^{\beta} (-{\bf k},-\omega) \rangle_0 =
  R^T (k,\omega) P_T^{\alpha \beta} ({\bf k}) +  
  R^L (k,\omega) P_L^{\alpha \beta} ({\bf k})
\, ,
\label{6.35}
\end{equation}
where the transverse and the longitudinal parts are given by
\begin{equation}R^{L,T}(k,\omega) \equiv {1 \over - i\omega - A_{L,T}(k)} \, ,
\label{6.36}
\end{equation}
and the projection operators are defined by
$P_T^{\alpha \beta} = \delta^{\alpha \beta} -q^\alpha q^\beta /q^2$ and
$P_L^{\alpha \beta} = q^\alpha q^\beta /q^2$.
For the correlation propagators one obtains
\begin{equation}C^{\alpha \beta} ({\bf k},\omega) =
  \langle S^{\alpha} ({\bf k},\omega) 
          S^{\beta}  (-{\bf k},-\omega)
  \rangle_0 =
  C^T (k,\omega) P_T^{\alpha \beta} ({\bf k}) +  
  C^L (k,\omega) P_L^{\alpha \beta} ({\bf k})
\, ,
\label{6.37}
\end{equation}
where
\begin{equation}
  C^{L,T}(k,\omega) \equiv 
  {2L(k) \over \omega^2 + A_{L,T}(k)^2} \, .
\label{6.38}
\end{equation}
The diagrammatic representations are depicted in Figs.6.1.a.

The interaction part of the dynamic functional 
\begin{equation}{\cal J}_{\rm int} [S,{\tilde S}] =
  {\cal J}_{\rm MC} [S,{\tilde S}] + {\cal J}_{\rm RE} [S,{\tilde S}] 
  \, ,
\label{6.39}
\end{equation}
consists of a mode coupling vertex  ${\cal J}_{\rm MC}
[S,{\tilde S}]$, originating from the Larmor term
${\bf S} \times {\delta {\cal H} \over \delta {\bf S}}$ 
in Eq.~(\ref{6.12}), and
relaxation vertices ${\cal J}_{\rm RE} [S,{\tilde S}]$, originating from
the nonlinear terms of the static Hamiltonian, 
${\hat L} {\delta {\cal H} \over \delta {\bf S}}$.
The mode coupling vertex is given  by
\begin{eqnarray}
  {\cal J}_{\rm MC} [S,{\tilde S}] = 
  &&\lambda f \epsilon_{\alpha \beta \gamma}
  \int_k \int_{\omega} \int_p \int_{\nu}
  \left[ ({\bf p} + {{\bf k} \over 2})^2 \delta^{\gamma \mu} 
       + g { (p^{\mu} + {k^{\mu} \over 2})
               (p^{\gamma} + {k^{\gamma} \over 2}) \over 
               ({\bf p} + {{\bf k} \over 2})^2 } 
  \right]  \nonumber \\
  &&\times
  {\tilde S}^{\alpha} ({\bf k},\omega)
  S^{\beta} ({\bf p} - {{\bf k} \over 2},\nu - {\omega \over 2})
  S^{\mu} (- {\bf p} - {{\bf k} \over 2},
             - \nu - {\omega \over 2})
  \, .
\label{6.40}
\end{eqnarray}
It consists of an isotropic part  
\begin{equation}{\cal J}_{\rm MCI} [S,{\tilde S}] = 
  \lambda f \epsilon_{\alpha \beta \gamma}
  \int_k \int_{\omega} \int_p \int_{\nu}
  ({\bf p} \cdot {\bf k} )
  {\tilde S}^{\alpha} ({\bf k},\omega)
  S^{\beta} ({\bf p}_-,\nu_-)
  S^{\gamma} ({\bf p}_+,\nu_+) \, ,
\label{6.41}
\end{equation}
and a dipolar part
\begin{equation}{\cal J}_{\rm MCD} [S,{\tilde S}] =   
  \lambda f g \epsilon_{\alpha \beta \gamma}
  \int_k \int_{\omega} \int_p \int_{\nu}
  { p_+^{\mu} p_+^{\gamma} \over  p_+^2 } 
  {\tilde S}^{\alpha} ({\bf k},\omega)
  S^{\beta} ({\bf p}_-,\nu_-)
  S^{\mu}   ({\bf p}_+,\nu_+) \, ,
\label{6.42}
\end{equation}
whose graphical representations are shown in Figs.6.1.b-c. 
We have used the short hand notations
${\bf p}_{\pm} = \mp \left( {\bf p} \pm {{\bf k} / 2} \right)$ and
$\nu_{\pm} =  \mp \left( \nu \pm {\omega / 2} \right)$.

The dipolar mode coupling vertex can be symmetrized by the
substitution ${\bf p} \rightarrow - {\bf p}$ in Eq.~(\ref{6.41}) leading to
\begin{eqnarray}
  {\cal J}_{\rm MCD} [S,{\tilde S}]    
  = {1 \over 2}\lambda f g \epsilon_{\alpha \beta \gamma}
  \int_k \int_{\omega} \int_p \int_{\nu}
  \biggl[
  &&{ p_+^{\mu} p_+^{\gamma} \over  p_+^2 } 
  {\tilde S}^{\alpha} ({\bf k},\omega)
  S^{\beta} ({\bf p}_-,\nu_-)
  S^{\mu}   ({\bf p}_+,\nu_+)
  + \nonumber \\
  -&&{ p_-^{\mu} p_-^{\beta} \over p_-^2 } 
  {\tilde S}^{\alpha} ({\bf k},\omega)
  S^{\mu}    ({\bf p}_-,\nu_-)
  S^{\gamma} ({\bf p}_+,\nu_+)
  \biggr]
  \, . 
\label{6.43}
\end{eqnarray}
The relaxation vertex is given by
\begin{eqnarray}
  {\cal J}_{\rm RE} [S,{\tilde S}] = 
  &&-{u \over 3!} F^{\alpha \beta \gamma \delta}
   \int_{k_1} \int_{k_2} \int_{k_3} 
   \int_{\omega_1} \int_{\omega_2} \int_{\omega_3} 
   \lambda \left[ \gamma + (\sum_i {\bf k}_i)^2 \right]   \nonumber \\
  &&\times
   {\tilde S}^{\alpha} (\sum_i {\bf k}_i,\sum_i \omega_i )
   S^{\beta}   ( - {\bf k}_1, - \omega_1)
   S^{\gamma}  ( - {\bf k}_2, - \omega_2)
   S^{\delta}  ( - {\bf k}_3, - \omega_3)  
   \, . 
\label{6.44}
\end{eqnarray}
It contains a diffusive part
\begin{eqnarray}
  {\cal J}_{\rm Diff} [S,{\tilde S}] = 
  &&-{u \over 3!} \lambda 
    F^{\alpha \beta \gamma \delta}
    \int_{k_1} \int_{k_2} \int_{k_3} 
    \int_{\omega_1} \int_{\omega_2} \int_{\omega_3} 
    (\sum_i {\bf k}_i)^2   \nonumber \\
  &&\times
   {\tilde S}^{\alpha} (\sum_i {\bf k}_i,\sum_i \omega_i )
   S^{\beta}   ( - {\bf k}_1, - \omega_1)
   S^{\gamma}  ( - {\bf k}_2, - \omega_2)
   S^{\delta}  ( - {\bf k}_3, - \omega_3)  
   \, , 
\label{6.45}
\end{eqnarray}
and a relaxational part
\begin{eqnarray}
  {\cal J}_{\rm Rel} [S,{\tilde S}] = 
  &&-{u \over 3!} \lambda  
   F^{\alpha \beta \gamma \delta}
   \int_{k_1} \int_{k_2} \int_{k_3} 
   \int_{\omega_1} \int_{\omega_2} \int_{\omega_3} 
   \gamma \nonumber \\
  &&\times
   {\tilde S}^{\alpha} (\sum_i {\bf k}_i,\sum_i \omega_i )
   S^{\beta}   ( - {\bf k}_1, - \omega_1)
   S^{\gamma}  ( - {\bf k}_2, - \omega_2)
   S^{\delta}  ( - {\bf k}_3, - \omega_3)  
   \, ,
\label{6.46}
\end{eqnarray}
whose diagrammatic representations are given in Figs.~6.1.d-e.
This decomposition is not important from a technical point of view,
since both terms have the same tensorial structure. But, it emphasizes
that the relaxational vertices due to the exchange interaction and
dipolar interaction are diffusive and relaxational, respectively.

With the above perturbation theory at hand, we can now calculate the 
Greens functions defined by 
\begin{eqnarray}
  &&G_{N,{\tilde N}}^{\alpha_1,..., \alpha_N, 
                             \beta_1, ..., \beta_{\tilde N} } 
  ( {\bf k}_1,\omega_1;...;
    {\bf k}_N,\omega_N;
    {\bf k}_{N+1},\omega_{N+1};...;
    {\bf k}_{N+{\tilde N}},\omega_{N+{\tilde N}} )
    = \nonumber \\
    &&=
    \langle 
    S^{\alpha_1}({\bf k}_1,\omega_1) \, ... \, 
    S^{\alpha_N}({\bf k}_N,\omega_N)
    {\tilde S}^{\beta_1}({\bf k}_{N+1},\omega_{N+1}) \, ... \,
    {\tilde S}^{\beta_{\tilde N}}
               ({\bf k}_{N+{\tilde N}},\omega_{N+{\tilde N}})
    \rangle = \nonumber \\
    &&={\delta^{N+{\tilde N}} \over 
       \delta h^{\alpha_1}({\bf k}_1,\omega_1)\cdot ... \cdot 
       \delta {\tilde h}^{\beta_{\tilde N}}
               ({\bf k}_{N+{\tilde N}},\omega_{N+{\tilde N}})}
      Z[h,{\tilde h}] \mid_{h,{\tilde h} = 0} \, . 
\label{6.47}
\end{eqnarray}
It is convenient to consider the vertex functions
\begin{equation}\Gamma_{N,{\tilde N}}^{\alpha_1,..., \alpha_N, 
                             \beta_1, ..., \beta_{\tilde N} } 
  ( {\bf k}_1,\omega_1;...;
    {\bf k}_N,\omega_N;
    {\bf k}_{N+1},\omega_{N+1};...;
    {\bf k}_{N+{\tilde N}},\omega_{N+{\tilde N}} ) \, ,
\label{6.48}
\end{equation}
which can be obtained from the cummulants by a Legendre
transformation \cite{bjw76}.

\subsection{Self consistent one loop theory}

In the preceding sections we have derived a path integral
representation of the dynamics starting from semi-phenomenological
Langevin equations. Within this method a perturbation theory for the
correlation and response functions could be formulated which is
similar to the usual field theoretic procedure for static
critical phenomena. In each finite order of perturbation theory
infrared divergences arise  reflecting the infrared
divergences in the second derivatives of the free energy close to
a critical point. In order to remove these divergences (which appear combined
with the ultraviolet divergencies at the upper
critical dimension of the model) and to have a perturbation theory
with a small parameter one can use the concepts of
renormalized field theories, i.e., upon introducing renormalization 
factors and expanding around the upper critical dimension. 

Instead we follow here a alternative route, which consists in 
the resummation of certain classes
of diagrams to infinite order in perturbation theory. Such methods
are frequently used in condensed matter physics quite successfully.
These methods are characterized by the fact that they lead to self
consistent equations for the correlation functions.

In this section we formulate such a self consistent procedure for
critical dynamics. It will turn out that the resulting equations are
equivalent to mode coupling theory.

\subsubsection{Self consistent determination of the line width
in Lorentzian approximation}

As we have already seen in the discussion of the mode coupling theory
one can obtain quite reasonable results for the line width by
assuming that the line shape is given by a Lorentzian.  In the above
formulation  in terms of a dynamical functional the line width is
given by \cite{bjw76}
\begin{equation}\Gamma_{\rm lor}^{\alpha} = 
  - 2 { [\Gamma_{11}^{\alpha} (q,0)]^2 \over
         \Gamma_2^{\alpha} (q) \Gamma_{02}^{\alpha} (q,0) }
\label{6.49}
\end{equation}
where $\Gamma_2^{\alpha} (q)$ is the inverse static susceptibility 
($\alpha = L,T$). Upon using the Fluctuation Dissipation Theorem (FDT)
\begin{equation}\Gamma_{11}^{\alpha} (q,0) = 
  \left[
  \lambda (\gamma + q^2) + \lambda f \Gamma_{X,10}^{\alpha} (q,0)
  \right]
  \Gamma_2^\alpha (q)
\label{6.50}
\end{equation}
one obtains to one loop order
$-2\Gamma_{11}^{\alpha} (q,0) = \Gamma_2^\alpha (q) \Gamma_{02}^\alpha (q,0)$
and consequently
$\Gamma_{\rm lor}^{\alpha} = \Gamma_{11}^{\alpha} (q,0) = 
-{1 \over 2} \Gamma_2^{\alpha} (q) \Gamma_{02}^{\alpha} (q,0)$.
To zeroth order we have $\Gamma_{02}^{\alpha} (q,0) = -2 \lambda (\gamma +  
q^2)$. The one loop contributions to $\Gamma_{02}^{\alpha \beta} (q,0)$ 
are shown in Fig.~6.2. The transverse and longitudinal Lorentzian
line width is found to be
\begin{eqnarray}
  \Gamma_{\rm lor}^{T} (q) \chi^T(q) = 
  \lambda (\gamma + q^2) 
  + 2 (\lambda f)^2 \int_p \int_{\omega}
  \biggl[ 
  && \upsilon_{TT}^{T}(g,{\bf p},{\bf q}) 
  C^T({\bf p}_-,\omega) C^T({\bf p}_+,\omega) + \nonumber \\
  +&&\upsilon_{TL}^{T}(g,{\bf p},{\bf q}) 
  C^T({\bf p}_-,\omega) C^L({\bf p}_+,\omega) + \cr
  +&& \upsilon_{LL}^{T}(g,{\bf p},{\bf q}) 
  C^L({\bf p}_-,\omega) C^L({\bf p}_+,\omega)
  \biggr] \, ,
\label{6.51}
\end{eqnarray}
and
\begin{eqnarray}
  \Gamma_{\rm lor}^{L} (q) \chi^L(q) = 
  \lambda (\gamma + q^2) 
  + 2 (\lambda f)^2 \int_p \int_\omega
  \biggl[ 
  && \upsilon_{TT}^{L}(g,{\bf p},{\bf q}) 
  C^T({\bf p}_-,\omega) C^T({\bf p}_+,\omega) + \nonumber \\
  +&& \upsilon_{TL}^{L}(g,{\bf p},{\bf q}) 
  C^T({\bf p}_-,\omega) C^L({\bf p}_+,\omega) 
  \biggr] \, ,
\label{6.52}
\end{eqnarray}
where we have introduced the notation
${\bf p}_{\pm} = {\bf p} \pm {{\bf q} \over 2}$. The vertex functions
$\upsilon_{\alpha \beta}^{\sigma}(g,{\bf p},{\bf q})$ are given by
\begin{mathletters}
\begin{equation}
  \upsilon_{TT}^{T}(g,{\bf p},{\bf q}) =
  {1 \over d-1}
  \left[
  d^2 - 4d + 6 - {({\bf p}_+ \cdot {\bf p}_-)^2 \over p_+^2 p_-^2} -
  2 {({\bf p}_+ {\bf q})^2 \over p_+^2 q^2}
  \right] 
  ({\bf p} \cdot {\bf q})^2
\label{6.53a}
\end{equation}
\begin{equation}\upsilon_{TL}^{T}(g,{\bf p},{\bf q}) =
  {4 \over d-1}
  \left[
  d - 3 + {({\bf p}_+  \cdot {\bf p}_-)^2 \over p_+^2 p_-^2} +
  {({\bf p}_+ \cdot  {\bf q})^2 \over p_+^2 q^2}
  \right] 
  ({\bf p}  \cdot   {\bf q} + {g \over 2})^2
\label{6.53b}
\end{equation}
\begin{equation}\upsilon_{LL}^{T}(g,{\bf p},{\bf q}) =
  {1 \over d-1}
  \left[
  1 - {({\bf p}_+  \cdot {\bf p}_-)^2 \over p_+^2 p_-^2} 
  \right] 
  ({\bf p} \cdot  {\bf q})^2
\label{6.53c}
\end{equation}
\begin{equation}\upsilon_{TT}^{L}(g,{\bf p},{\bf q}) =
  \left[
  d - 3 + 2 {({\bf p}_+ \cdot  {\bf q})^2 \over p_+^2 q^2}
  \right] 
  ({\bf p}  \cdot {\bf q})^2
\label{6.53d}
\end{equation}
\begin{equation}\upsilon_{TL}^{L}(g,{\bf p},{\bf q}) =
  4 \left[
  1 -  {({\bf p}_+ \cdot  {\bf q})^2 \over p_+^2 q^2}
  \right] 
  ({\bf p}  \cdot {\bf q} + {g \over 2})^2
\label{6.53e}
\end{equation}
\end{mathletters}
Upon substituting in the integrals (\ref{6.51}) and (\ref{6.52})
${\bf p}_+ = {\bf p} + {{\bf q} \over 2} \rightarrow {\bf p}$ the
above vertex functions are identical to the vertex functions in
section III, Eqs.~(\ref{3.20})-(\ref{3.24}). The frequency integration 
can readily be done leading to 
\begin{equation}\int_{\omega} C^{\alpha} ({\bf p}_-,\omega) 
                C^{\beta}  ({\bf p}_+,\omega)  =
  {\chi^{\alpha} ({\bf p}_-) \chi^{\beta} ({\bf p}_+) 
   \over
   \lambda \left[ (\gamma + p_-^2)/\chi^{\alpha} ({\bf p}_-)      
                + (\gamma + p_+^2)/\chi^{\beta} ({\bf p}_+)     
           \right]} \, ,
\label{6.54}
\end{equation}
i.e., the product of the static susceptibilities divided by the sum
of the line width to zeroth order. This structure corresponds to the
first step in an iteration procedure and suggests to define a 
self-consistent approximation by replacing the bare line width by 
the full line width. 
\begin{equation}{\chi^{\alpha} ({\bf p}_-) \chi^{\beta} ({\bf p}_+) 
   \over
   \lambda \left[ (\gamma + p_-^2)/\chi^{\alpha} ({\bf p}_-)      
                + (\gamma + p_+^2)/\chi^{\beta} ({\bf p}_+)     
           \right]}
  \rightarrow
  {\chi^{\alpha} ({\bf p}_-) \chi^{\beta} ({\bf p}_+) 
   \over
   \left[ \Gamma_{\rm lor}^{\alpha} ({\bf p}_-)               
                + \Gamma_{\rm lor}^{\beta} ({\bf p}_+)     
           \right]}
\label{6.55}
\end{equation}
This self consistent approach corresponds to a partial resummation
of the perturbation series as shown in Fig.~6.3. The resulting
equations are identically with the mode coupling equations in
Lorentzian approximation found by the conventional derivation in
section III. The factorization approximation (two mode
approximation) in the conventional derivation corresponds to the structure 
of the one loop diagrams here.

\subsubsection{Self consistent equation for the Kubo
relaxation function}

In the preceding section we derived for the sake of simplicity the
mode coupling equations in the so called Lorentzian approximation for
the line shape. Here we go beyond this approximation and 
derive a self consistent theory for the full wave vector and
frequency dependent relaxation functions. The dynamic
susceptibility can be written as \cite{bjw76} ($\alpha = L,T$)
\begin{equation}\chi^{\alpha}(q,\omega) = 
  {\lambda (\gamma + q^2) 
   + \lambda f \Gamma_{X,10}^{\alpha}(q,\omega)
   \over
   \Gamma_{11}^{\alpha} (-q,-\omega)} \, .       
\label{6.56}
\end{equation}
The  X-insertion is defined by~\cite{bjw76} 
\begin{equation}X^{\alpha} (\vec q, \omega) =
  \epsilon_{\alpha \beta \gamma} \int_k \int_{\nu}
  {\tilde S}^{\beta} (\vec k - \vec q/2, \nu - \omega/2)
  S^{\gamma} (-\vec k - \vec q/2, -\nu - \omega/2) \, .
\label{6.57}
\end{equation}
Upon using the (explicit) structure of the dynamic propagators
$C^{\alpha}(q,t) \Theta(t) = \chi^{\alpha}(q) R^{\alpha}(q,t)$
one can show that the relation
\begin{equation}\Gamma_{11}^{\alpha} (-q,-\omega) =
  -i\omega + 
  \left[
  \lambda (\gamma + q^2)
  + \lambda f \Gamma_{X,10}^{\alpha} (q,\omega)
  \right]
  \Gamma_2^\alpha (q)
\label{6.58}            
\end{equation}
holds to one loop order. This implies for the Kubo relaxation
function $\Phi^{\alpha}(q,\omega)$, which is related to the dynamic
susceptibility by 
\begin{equation}\Phi^{\alpha}(q,\omega) =
  { 1 \over i \omega}
  \left[ \chi^{\alpha}(q,\omega) - \chi(q) \right] \, ,
\label{6.59}            
\end{equation}
a structure analogous to Eq.~(\ref{3.25})
\begin{equation}\Phi^{\alpha}(q,\omega) =
  { \chi^{\alpha}(q)  \over 
  \Gamma_{11}^{\alpha} (-q,-\omega) } =
  { \chi^{\alpha}(q) \over - i \omega 
  + \left[ \lambda (\gamma + q^2)
           + \lambda f \Gamma_{X,10}^{\alpha} (q,\omega)
    \right] \Gamma_2^{\alpha}(q)} \, .
\label{6.60}
\end{equation}
To one loop order one obtains for $\Gamma_{X,10}^{\alpha} (q,t)$ 
\begin{mathletters}
\begin{eqnarray}
  \Gamma_{X,10}^{T} (q,t) = 
  2 \lambda f \int_p \Theta(t)
  \biggl[ 
  &&\upsilon_{TT}^{T}(g,{\bf p},{\bf q}) 
  C^T({\bf p}_-,t) C^T({\bf p}_+,t) + \nonumber \\
  +&&\upsilon_{TL}^{T}(g,{\bf p},{\bf q}) 
  C^T({\bf p}_-,t) C^L({\bf p}_+,t) + \nonumber \\
  +&&\upsilon_{LL}^{T}(g,{\bf p},{\bf q}) 
  C^L({\bf p}_-,t) C^L({\bf p}_+,t)
  \biggr] \, ,
\label{6.61a}
\end{eqnarray}
\begin{eqnarray}
  \Gamma_{X,10}^{L} (q,t) = 
  2 \lambda f \int_p \Theta(t)
  \biggl[ 
  && \upsilon_{TT}^{L}(g,{\bf p},{\bf q}) 
  C^T({\bf p}_-,t) C^T({\bf p}_+,t) + \nonumber \\
  +&& \upsilon_{TL}^{L}(g,{\bf p},{\bf q}) 
  C^T({\bf p}_-,t) C^L({\bf p}_+,t) 
  \biggr] \, , 
\label{6.61b}
\end{eqnarray}
\end{mathletters}
where the vertex functions are identical to Eqs.~(\ref{6.53a}-\ref{6.53e}). 
We have $C^{\alpha}(q,t) \Theta(t) = \chi^{\alpha}(q) R^{\alpha}(q,t)$.
Eqs.~(\ref{6.60}-\ref{6.61b}) can be written in a self-consistent form by 
replacing the propagator $C^{\alpha} (q,t)$ by the full Kubo relaxation 
function $\Phi^{\alpha} (q,t)$ on the right hand side of 
Eqs.~(\ref{6.61a},\ref{6.61b}). This corresponds, as
in the preceding section, to a partial resummation of the
perturbation series. The resulting equations are identical with the
mode coupling equations Eqs.~(\ref{3.19})-(\ref{3.25}) in section III . 
Hence we have shown that mode coupling theory is equivalent to a self 
consistent one loop theory. The factorization approximation in the 
conventional derivation of mode coupling theories is here a direct 
consequence of the structure of the one loop theory. Furthermore the present
procedure has the advantage to be extendible to a self consistent
formulation of higher loop order. In order to justify the validity
of the mode coupling approach one has to ask in what sense higher
order terms are small compared with the self consistent one loop
theory. This question will be addressed in Appendix~C, where we
restrict ourselves to the isotropic case (in order to simplify the
discussion).

\newpage

\centerline{\bf Captions to the figures:}
\bigskip
\bigskip

\noindent {\bf Figure 6.1: } 
Basic elements of the dynamical perturbation theory. a) Correlation and 
response propagators, b) isotropic mode coupling  vertex, c) dipolar 
mode coupling  vertex, d) isotropic relaxational vertex, and e) dipolar 
relaxational vertex.
\bigskip

\noindent {\bf Figure 6.2: } 
One-loop diagrams for the vertex function $\Gamma_{02}^{\alpha \beta}
({\bf q}, \omega)$.
\bigskip

\noindent {\bf Figure 6.3: } 
Partial summation of the perturbation theory for $\Gamma_{02}^{\alpha \beta}
({\bf q}, \omega)$, where vertex corrections are
neglected. This resummation leads to the mode coupling equations.
\bigskip

\newpage

\vfill
\begin{figure}[h]
  \centerline{\rotate[r]{\epsfysize=5in \epsffile{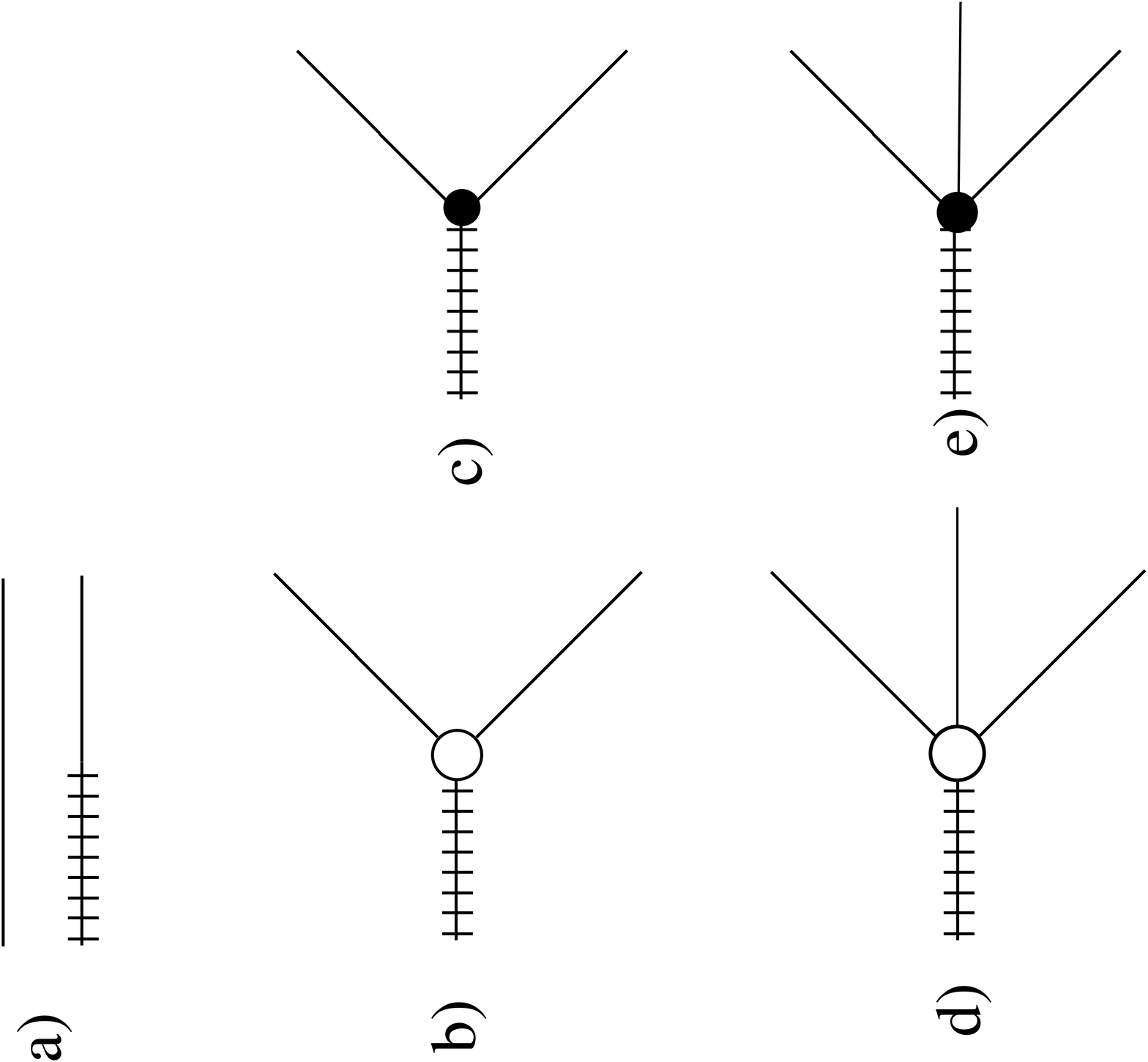}}}
\bigskip \bigskip
 {\bf Figure 6.1: } 
Basic elements of the dynamical perturbation theory. a) Correlation and 
response propagators, b) isotropic mode coupling  vertex, c) dipolar 
mode coupling  vertex, d) isotropic relaxational vertex, and e) dipolar 
relaxational vertex.
\label{fig61}
\end{figure}

\newpage
\vfill
\begin{figure}[h]
  \centerline{\rotate[r]{\epsfysize=5in \epsffile{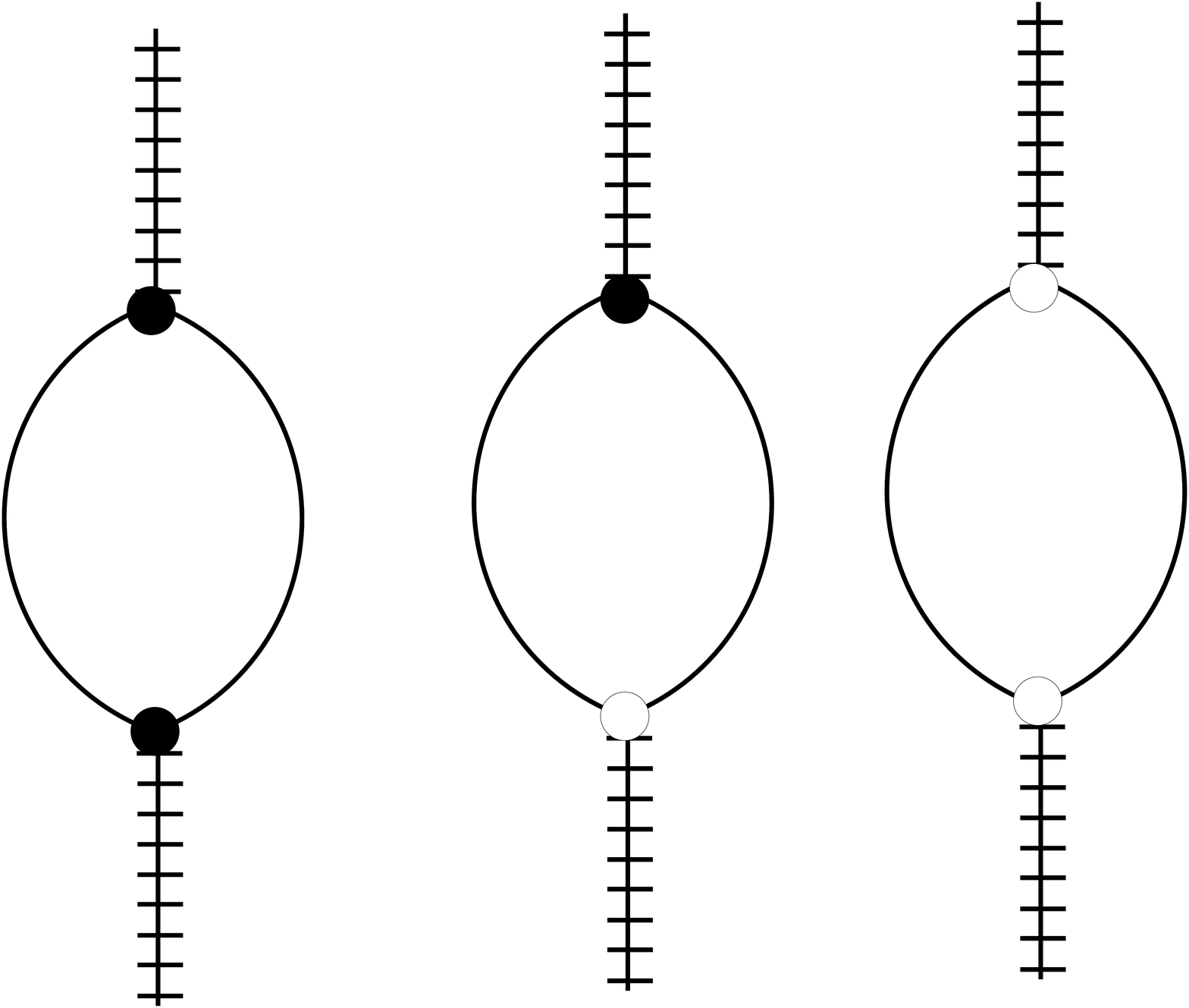}}}
\bigskip \bigskip
 {\bf Figure 6.2: } 
One-loop diagrams for the vertex function $\Gamma_{02}^{\alpha \beta}
({\bf q}, \omega)$.
\label{fig62}
\end{figure}

\newpage
\vfill
\begin{figure}[h]
  \centerline{\rotate[r]{\epsfysize=6in \epsffile{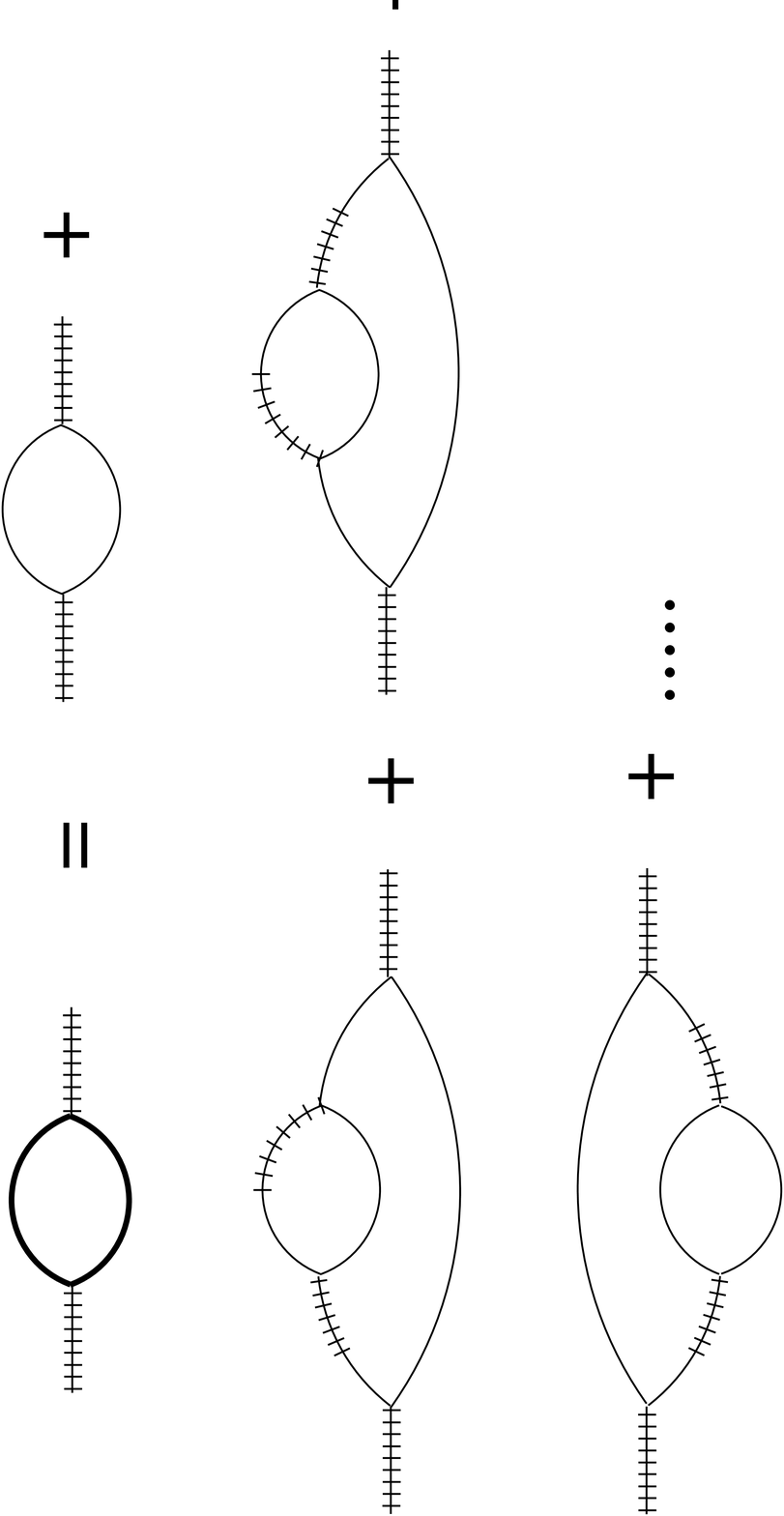}}}
\bigskip \bigskip
 {\bf Figure 6.3: } 
Partial summation of the perturbation theory for $\Gamma_{02}^{\alpha \beta}
({\bf q}, \omega)$, where vertex corrections are
neglected. This resummation leads to the mode coupling equations.
\label{fig63}
\end{figure}

%%%%%%%%%%%%%%%%%%%%%%%%%%%%%%%%%%%%%%%%%%%%%%%%%%%%%%%%%%%%%%%%%%%%%%%%%%%%%%%%
\section{CONCLUSIONS AND OUTLOOK}

The past decade has witnessed substantial progress in understanding the
critical dynamics of real ferromagnets, both on the experimental and the
theoretical side. In this review we have concentrated on the interplay
between  exchange and dipole--dipole interaction. We have presented theoretical 
and experimental evidence that many
aspects of the critical dynamics of ferromagnets such as $Fe$, $Ni$, $EuO$,
$EuS$, and many other magnetic materials are now fairly well understood on 
the basis of a mode coupling theory which takes into account exchange as 
well as the dipole--dipole interaction. In this final section we conclude 
our overview by summarizing some of the main theoretical and experimental
achievements and by briefly discussing a number of open problems in the 
field of critical dynamics of magnets.

One of the most important theoretical advances towards our understanding of
the critical dynamics of real ferromagnets was the realization that
additional interactions such as the dipole--dipole interaction can lead to a
qualitatively new behavior of the frequency and wave vector dependence of 
the spin--spin correlation functions~\cite{fe86,fs87}. A description
of a mode coupling theory, which on top of the exchange interaction takes 
into account the dipole-dipole interaction, has been given in section III.
The main results of this mode coupling analysis can be summarized as
follows. The dipolar interaction leads to an anisotropy of the spin
fluctuations with respect to the direction of the wave vector and 
introduces a second length scale $q_{_{D}}^{-1}$ besides the correlation 
length $\xi$. The presence of a second length scale leads to  generalized 
dynamic scaling laws, containing  two scaling variables, $x=1/q\xi$ and 
$y=q_{_{D}}/q$ for the length scale and one scaling variable 
$\tau_{\alpha} = \Lambda q^z \Omega^{\alpha} (x,y) t$ for the time scale. 
Due to the dipolar anisotropy the characteristic time scales $1/[\Lambda
q^z \Omega^{\alpha} (x,y)]$ are different for the longitudinal
and transverse modes ($\alpha = L,\,T$). This is mainly due to the non 
critical longitudinal 
static susceptibility implying that the longitudinal characteristic 
frequency $\Lambda q^z \Omega^L (x,y)$ shows no critical slowing down 
asymptotically. Furthermore, the dipolar interaction leads to a quite
interesting crossover in the line width and line shape of the spin-spin
correlation functions.

{\it Line width crossover.} 
Precisely at the critical temperature the dynamic critical exponent for the
transverse line width undergoes a crossover from $z = {5 / 2}$ to $z = 2$,
which is displaced with respect to the static crossover to wave vectors
smaller by almost one order of magnitude. This explains why up to now this 
crossover escaped detection by neutron scattering experiments right at
$T_c$. For the longitudinal line width the crossover is predicted to
be from $z = {5 / 2}$ to an uncritical behavior $z = 0$ and, in contrast 
to the transverse width, it occurs in the immediate vicinity of the static 
crossover, characterized by $q_{_{D}}$.  If the line shapes are approximated 
by Lorentzians the concomitant Lorentzian line widths obey the 
scaling law $\Gamma^{\alpha} (q,\xi,g) = \Lambda q^z 
\gamma^{\alpha} (x,y)$. For vanishing dipolar coupling $g$, 
the scaling functions coincide with the Resibois-Piette scaling function.
If the strength of the dipolar interaction $g$ is finite, the curves
approach the Resibois-Piette scaling function for small values of the 
scaling variable $x$ and deviate therefrom with increasing $x$. Since the
dipolar interaction leads to a non conserved order parameter, the line
width in the hydrodynamic limit is given by $\Gamma \propto q^0$ instead of
$\Gamma \propto q^2$ as for an isotropic exchange ferromagnet.

{\it Line shape crossover.} 
Close to $T_c$ the line shapes of the longitudinal and transverse 
relaxation function coincide in the isotropic Heisenberg limit, i.e., for 
large values of the wave vector $q$ ($q \gg q_{_{D}}$). In this limit the 
dipolar interaction becomes negligible and the shape is of the 
Hubbard-Wegner type as discussed in section II. Upon increasing the value 
of the scaling variable $r=\sqrt{ (1/q\xi)^2 + (q_{_{D}}/q)^2 }$, the line 
shapes of the transverse and longitudinal relaxation function become
drastically different. Whereas the transverse relaxation function shows a 
nearly exponential temporal decay, over-damped oscillations show up for 
the longitudinal relaxation function. The line shape crossover sets in in the 
vicinity of the dipolar wave vector $q_{_{D}}$ in contrast to the line width
crossover, which starts at a wave vector almost one order of magnitude 
smaller. The situation is quite different for temperatures well separated from
the critical temperature. Then, the difference in the shape crossover of the 
longitudinal and transverse relaxation function diminishes with decreasing 
$q_{_{D}} \xi$. For $q_{_{D}} \xi \ll 1$ the shape crossover as a function
of $r$ corresponds to the crossover from the critical (Hubbard-Wegner) 
shape to the hydrodynamic shape as discussed in section II.

On the experimental side, an important development began with the observation 
that the data for the line widths above the transition 
temperature~\cite{m82,m84,bsbz88} in magnetic materials such as $EuO$and $Fe$ 
could not be described by the Resibois-Piette scaling function resulting from 
a mode coupling theory~\cite{rp70} as well as a renormalization group 
theory~\cite{mm75,i87}, which take into account the short range exchange 
interaction only. Even more, the data could not be collapsed onto any other
single scaling function. Those experimental results were explained quantitatively on
the basis of the mode coupling theory~\cite{fe86,fs87,fs88a} described in 
section III. Clear indications of the importance of the dipolar interaction 
have also been observed in hyperfine interaction experiments on $Fe$ and $Ni$,
where one found a crossover in the dynamical critical exponent from 
$z={5/2}$ to $z=2$~\cite{rh72,gh73,hcs82,hrk89}, and with electron spin 
resonance and magnetic relaxation experiments 
\cite{kkw76,ksbk78,kp78} \cite{ks78,dg80,k88}, where a non vanishing 
Onsager coefficient at zero wave vector was found. For all these data 
now there exists a quantitative theoretical description (see section IV).

With advances in the neutron scattering technique subsequent experiments
investigated the line shape of the spin-spin correlation function.
This progress made it possible to test the predictions for the line shape 
from MC--theory~\cite{w68,h71a} and RG--theory~\cite{d76,bf85}, which take 
into account solely the short--range exchange interaction. Originally 
those theories for the critical dynamics were expected to fit the experimental 
data in an increasing quantitative way as one was able to measure the 
correlation functions at smaller and smaller wave vectors, since the 
influence of additional irrelevant interactions should diminish as one 
moves closer to criticality. It came as quite a surprise when Mezei found 
a nearly exponential decay of the Kubo relaxation function by spin  echo 
experiments on $EuO$ at $q=0.024 \AA^{-1}$ and $T=T_c$~\cite{m86},
which was in drastic disagreement with the bell--like shape
predicted~\cite{w68,h71a,d76,bf85}. The anomalous exponential decay found 
by Mezei~\cite{m86}, was shown again to be a dipolar 
effect~\cite{fst88,fst89}. Recently, significant experiments have been 
performed using polarized neutrons~\cite{bgkm91,Goerlitz92a}, which
allowed for the first time to measure the longitudinal and transverse 
spin-spin correlation functions separately. The results confirmed the 
theoretical predictions concerning the line width~\cite{fs87} as well as 
the line shape of the longitudinal relaxation function~\cite{fst88,fst89}.
The experimental data, de-convoluted by a maximum entropy method, agree quite 
well with the theoretical predictions, especially the double peak structure 
corresponding to the over-damped oscillations in time have been observed. 
This can be regarded as a success of the mode coupling theory.

For the critical dynamics below $T_c$ experimental investigations are scare. 
In contrast to the very satisfactory situation above $T_c$,
where quantitative agreement between experiment and theory has been achieved,
the situation in the ferromagnetic phase is by far less clear. Theoretical
investigations so far have concentrated on the dynamics of isotropic exchange
ferromagnets, neglecting dipolar effects~\cite{fs88a,schinz94a}. In this context 
it has been shown within the framework of mode coupling theory that the amplitude of 
the scaling function for the spin--wave frequency of the isotropic Heisenberg
Hamiltonian is universal~\cite{schinz94a}. A comparison of the
theory~\cite{fs88a}, taking into account the so determined universal 
amplitude~\cite{schinz94a}, with recent measurements of the longitudinal
line width on $Ni$~\cite{bmt91} below and not too close to the critical
point gives quantitative agreement. 

But also in the ferromagnetic phase profound effects of the
dipolar interaction on the spin dynamics have been revealed in recent experiments.
Using polarized neutron scattering the spin excitations in $EuS$ and
$Pd_2MnSn$ have been investigated below $T_c$ by B\"oni et
al.~\cite{Boeni93b,Boeni94}.
These investigations indicate that there are deviations from the theoretical
correlation functions of isotropic exchange ferromagnets.
Especially, an anisotropy of the spin fluctuations with respect to the
polarization relative to the wave vector ${\bf q}$ has been observed. Still
further experimental and theoretical progress is needed in order to
reach a thorough understanding of the spin dynamics in the ferromagnetic
phase. The most convincing experimental indication for the importance of the
dipolar interaction below $T_c$ has been obtained recently in a measurement of
the homogeneous magnetization dynamics~\cite{Dombrowski94}. It is found that 
the scaling function for the kinetic coefficient below $T_c$ coincides
with the one observed earlier above $T_c$. Up to now, there is no theoretical
explanation for these interesting experimental findings, but work 
in the framework of mode coupling theory is in progress~\cite{schinz94b}. 
 
Besides three--dimensional ferromagnetic materials with isotropic exchange 
interaction there are other magnetic systems, where the dipole--dipole 
interaction influences the critical behavior, such as antiferromagnets or 
ferromagnets with a uniaxial or planar exchange anisotropy. The critical 
dynamics of isotropic dipolar antiferromagnets is affected by the dipolar 
interaction featuring a crossover from diffusive to relaxational dynamics of
the magnetization and a crossover in the dynamic exponent~\cite{a73}. The 
effect on the staggered magnetization is much less pronounced than for the 
corresponding order 
parameter modes in the ferromagnetic case. For ferromagnets with a uniaxial
exchange anisotropy the dipolar interaction leads first of all to a reduction
of the upper critical dimension from $d_c = 4$ to $d_c = 3$. As a consequence
the critical statics is described by mean field theory with logarithmic
corrections~\cite{Larkin69}. A time dependent Ginzburg Landau 
model~\cite{Folk77} for the dynamics results in logarithmic corrections to 
the conventional van Hove theory of critical dynamics.

We conclude with some comments on two--dimensional (2D) magnetic systems. 
These are interesting because the Hohenberg--Mermin--Wagner 
theorem~\cite{Hohenberg67,Mermin66} excludes conventional long--range order 
in isotropic systems with a continuous symmetry and short--ranged interaction.
Two dimensions seem to be the borderline dimension where thermal fluctuations 
are just strong enough to prevent the appearance of a finite order parameter. 
However, the anisotropy of the dipolar interaction with respect to the
separation of two spins leads to a suppression of the longitudinal
fluctuations and thus a finite order parameter may  exist in two dimensions 
(see section V). The dipolar interaction is responsible for the existence of long 
range order in both ferromagnetic~\cite{Maleev75} and antiferromagnetic thin 
films~\cite{Pich93,Pich94a}. There is quite an interesting phase diagram with
various spin flop and intermediate phases 
for two--dimensional dipolar antiferromagnets in a
finite external magnetic field~\cite{Pich94b}.

Two--dimensional ferromagnets are interesting for practical reasons as magnetic 
storage devices as well as from a fundamental point of view.
With the recent advances in thin film technology it becomes possible to study 
the critical behavior of two-dimensional magnetic systems~\cite{Allensbach94}.
Due to the combined effect of anisotropy, dipole-dipole interaction, and
increased importance of thermal fluctuations in two dimensions, the phase 
diagram of ferromagnetic thin films shows a variety of new 
phases~\cite{Pappas90,Allensbach92,Allensbach94}. The theoretical
predictions in this field are still rather 
controversial~\cite{Pescia90,Pescia93,Levanyuk93,Kashuba93}. A still open and
very interesting question is the critical dynamics of all those phases. 

\acknowledgements{It is a pleasure to acknowledge helpful discussions with P.
B\"oni, D. G\"orlitz, J. K\"otzler, F. Mezei, C. Pich, H. Schinz, U.C. T\"auber, and 
A. Yaouanc. The work of E.F. has been supported by the Deutsche 
Forschungsgemeinschaft (DFG) under Contracts No. Fr. 850/2-1,2. This work has 
also been supported by the German Federal Ministry for Research and 
Technology (BMFT) under the contract number 03-SC3TUM.}

\newpage
%%%%%%%%%%%%%%%%%%%%%%%%%%%%%%%%%%%%%%%%%%%%%%%%%%%%%%%%%%%%%%%%%%%%%%%%%%%%%%%%
\appendix

\section{Fluctuation-dissipation relations}
\label{app.a}

In this appendix we collect several fluctuation dissipation theorems, which
are of importance for the dynamics of systems described by nonlinear 
Langevin equations.

Let ${\cal T}$ be the time reversal operation: $t \rightarrow - t$.
Then detailed balance (time reversal symmetry) implies~\cite{j79}  
that
\begin{equation}
  {\cal T} \exp \left[ {\cal J}_{t_1}^{t_2} - {\cal H}_{t_1} \right] =
  \exp \left[ {\cal J}_{-t_2}^{-t_1} - {\cal H}_{-t_2} \right] \, ,
\label{A.1}
\end{equation}
where ${\cal H}_t = {\cal H} (S(t))$ is the stationary probability 
distribution function and  ${\cal J}_{t_1}^{t_2} [S,{\tilde S}]$ the dynamic 
functional. For simplicity we have assumed a one-component field $S$.
Without loss of generality one can assume that the field $S(x,t)$ is
even or odd under time reversal, i.e.
\begin{equation}
  {\cal T} S = \epsilon S, \quad \epsilon = \pm 1 \, .
\label{A.2}
\end{equation}
In the case of spins $S$ is odd under time reversal. The stationary 
distribution $P_{\rm st} [S] = e^{- {\cal H}[S]}$ is characterized by the 
``free energy'' ${\cal H}[S]$. The time reversal symmetry implies that
\begin{equation}
  {\cal T} {\tilde S} (t) =
  - \epsilon \left( {\tilde S} (-t) - 
   {\delta {\cal H}[S(-t)] \over \delta S(-t)} \right)\, .
\label{A.3}
\end{equation}
Now one uses the causality property of the response functions
\begin{equation}
  <S(t_1) S(t_2) ... S(t_k) {\tilde S} ({\tilde t}_1) 
  {\tilde S} ({\tilde t}_2) ... {\tilde S} ({\tilde t}_k) > = 0,
  \quad {\rm if} \, {\rm one} \, \, {\tilde t}_j > \, {\rm all} \, \, t_i \, .
\label{A.4}
\end{equation}
Then for example 
\begin{equation}
  <S(t) {\tilde S} (0)> = 0 \, \, {\rm for} \, \, t < 0 \, .
\label{A.5}
\end{equation}
With the time reversal operation and Eq.~(\ref{A.3}) it then follows from
Eq.~(\ref{A.5}) for $t<0$
\begin{equation}
  <S(-t) \left( {\tilde S} (0) - 
  {\delta {\cal H}[S(0)] \over \delta S (0)} \right)> = 0 \, .
\label{A.6}
\end{equation}
Upon redefining $t = - t$ one obtains for $t>0$
\begin{equation}
  <S(t) {\tilde S} (0)> =  
  \Theta(t) <S(t) {\delta {\cal H}[S(0)] \over \delta S (0)}>
\label{A.7}
\end{equation}
The same arguments can be repeated for 
$<S(t_1) S(t_2) ... S(t_k){\tilde S} ({\tilde t}_1) >$ with 
${\tilde t}_1 > \, {\rm all} \, t_j$.
The result is
\begin{equation}
  <S(t_1) S(t_2) ... S(t_k){\tilde S} ({\tilde t}_1) > = 
  \Theta({\tilde t}_1,\{ t_j \} ) 
  <S(t_1) S(t_2) ... S(t_k) 
  {\delta {\cal H}[S({\tilde t}_1)] \over \delta S ({\tilde t}_1)}> \, .
\label{A.8}
\end{equation}
where $\Theta({\tilde t}_1,\{ t_j \} ) $ is an obvious generalization of the 
$\Theta$-function. Note that these generalized FDT's hold for the cummulants 
and not for the vertex functions. In particular we get 
\begin{equation}
  G_{11} (k,t) = \Theta (t) { \nu k^2 \over D} G_{02} (k,t) \, .
\label{A.9}
\end{equation}

In a completely analogous way one can derive the following
identities~\cite{bjw76,j79}
\begin{mathletters}
\begin{equation}
\chi_{\alpha \beta} (x-x^\prime,t-t^\prime) = 
 - \Theta(t-t^\prime) {d \over dt} < S^\alpha (x,t)
                                     S^\beta (x^\prime , t^\prime) >
\label{A.10.a}
\end{equation}
\begin{equation}
C_{\alpha \beta} (k,\omega ) = {2 k_B T \over \omega} 
  Im[ \chi_{\alpha \beta} (k, \omega ) ]
\label{A.10.b}
\end{equation}
\begin{equation}
-\Theta(t-t^\prime ) {d \over dt}
  < S^\alpha (x,t) S^\beta (x^\prime , t^\prime) > =
  L_{\gamma \beta} 
 < S^\alpha (x,t) {\tilde S}^\gamma (x^\prime , t^\prime) > -
  \lambda f 
 < S^\alpha (x,t) X^\beta (x^\prime , t^\prime) > 
\label{A.10.c}
\end{equation}
\end{mathletters}
with $X^\beta = \epsilon_{\beta \mu \nu} {\tilde S}^\mu S^\nu$.
Therefrom one can deduce the following identities for the vertex functions
\begin{mathletters}
\begin{equation}
\Gamma_{11}^{\alpha \beta} (q, \omega =0) =
  \left[ 
    \lambda q^a \delta^{\alpha \gamma} + 
    \lambda f \Gamma_{X,10}^{\alpha \gamma} (q, \omega =0 )
  \right]
  \Gamma_2^{\gamma \beta} (q) \, ,
\label{A.11.a}
\end{equation}
\begin{eqnarray}
-i \omega \Gamma_{20}^{\alpha \beta}  (q, \omega) = 
  &&\lambda q^a \left[ 
                  \Gamma_{11}^{\alpha \beta}  (q, \omega ) - 
                  \Gamma_{11}^{\alpha \beta}  (-q,- \omega ) 
                \right] 
  \nonumber \\ +
  &&\lambda f \left[
                 \Gamma_{11}^{\alpha \gamma}  (q, \omega ) 
                 \Gamma_{X, 10}^{\gamma \beta}  (q, \omega ) -
                 \Gamma_{11}^{\alpha \gamma}  (-q,- \omega ) 
                 \Gamma_{X, 10}^{\gamma \beta}  (-q,- \omega )
               \right] \, ,
\label{A.11.b}
\end{eqnarray}
\end{mathletters}
where  $\Gamma_2^{\alpha \beta} (q) = < S^\alpha(-q) S^\beta(q) >^{-1}$ is 
the static two-point vertex  function, and 
$\Gamma_{X,{\tilde N} N}^{\alpha \beta}$ denotes a vertex function with one
$X$-insertion.
If the Hamiltonian is quadratic in the fields $\{ S \}$, i.e., 
${\cal H} = {1 \over 2} \int d^d x c_{\alpha \beta} S^\alpha S^\beta$
then one has a further identity
\begin{equation}
<S^\alpha (t) {\tilde S}^\beta (t^\prime ) > =
  \Theta (t - t^\prime ) c_{\alpha \beta} 
  <S^\alpha (t) S^\beta (t^\prime ) > \, .
\label{A.12}
\end{equation}

\section{Derivation of non--linear Langevin equations} 
\label{app.b}

In this Appendix we give a short derivation of non--linear Langevin equations
based on the projector formalism of Mori~\cite{m65} and Zwanzig~\cite{z61}. 
In our presentation we follow closely the papers by Kawasaki~\cite{kk73} and 
Mori et al.~\cite{mf73,mfs74}. 

As a first step in deriving non--linear Langevin equations for a particular
physical system, one has to choose the relevant slow variables. This choice is
in general dictated by the conservation laws and spontaneously broken
symmetries in the system under consideration. Let us assume that we have found
an appropriate set of slowly varying variables $\{ S \} = 
(S_1, S_2,\cdots ,S_n)$. Then, the next step consists in separating an
arbitrary dynamical variable $X$ into a part that is associated with the slow
variables $\{ S \}$ and the rest. This is accomplished by introducing an
projection operator by
\begin{equation}
  {\cal P} X(t) = \sum_j \left( X, \Psi_j \right)
  \Phi_j ( \{ S \} ) \, ,
\label{B.1} 
\end{equation}
where one defines the inner product as the value of the Kubo relaxation
function at $t=0$
\begin{equation}
  (A,B) = \Phi_{AB} (t=0) = i \lim_{\epsilon \rightarrow 0}
  \int_t^\infty d \tau e^{-\epsilon \tau} 
  < [A,B^\dagger] > |_{t=0} \, ,
\label{B.2}
\end{equation}
which is identical with the static susceptibility.
The functions $\Phi_j ( \{ S \} )$ and $\Psi_j ( \{ S \} )$ are two suitable 
sets of functions which are orthonormal with respect to the inner product 
defined in Eq.~(\ref{B.2})
\begin{equation}
  \left(\Phi_i, \Psi_j \right) = \delta_{ij} \, . 
\label{B.3}
\end{equation}
The microscopic dynamics of an arbitrary dynamic variable $X$ following from
the Heisenberg equations of motion can formally be presented by the 
Liouville equation
\begin{equation}
  {d \over dt} X(t) = i {\cal L} X(t) \, .
\label{B.4} 
\end{equation}
It was shown in Ref.~\cite{kk73} that with the aid of the operator identity,
\begin{eqnarray}
  {d \over dt} e^{ it {\cal L} }= 
  &&e^{ it {\cal L} }  i {\cal L}_0 +
    \int_0^t ds  e^{ i(t-s) {\cal L} }          i {\cal L}_0 
                 e^{ it ({\cal L}-{\cal L}_0) } i ({\cal L}-{\cal L}_0) 
  \nonumber \\ +
  &&e^{ it ({\cal L}-{\cal L}_0) } i ({\cal L}-{\cal L}_0)  \, ,    
\label{B.5} 
\end{eqnarray}
one can derive a formally exact non--linear Langevin equation
\begin{eqnarray}
  {d \over dt} X (t) = 
  &&\sum_\beta 
    \left( i {\cal L} X, \Psi_\beta \right)
    \Phi_\beta ( \{ S(t) \} ) \nonumber \\ -
  &&\sum_\beta \int_0^t d \tau 
    \left( f_X (\tau), \tilde f_\beta \right) 
    \Phi_\beta ( \{ S(t-\tau \} ) + f_X (t) \, ,
\label{B.6}
\end{eqnarray}
with the ``random'' forces
\begin{mathletters}
\begin{eqnarray}
  f_X (t) = &&e^{i t {\cal Q} {\cal L} } i {\cal Q} {\cal L} X   \, ,
  \label{B.7a} \\
  \tilde f_\alpha = &&\tilde {\cal Q} i \tilde {\cal L}  
                    \Psi_\alpha ( \{ S(t) \} ) \, ,
  \label{B.7b} 
\end{eqnarray}
\end{mathletters}
where ${\cal Q} = 1 - {\cal P}$.
Note that the adjoint operators of ${\cal L}$ and ${\cal P}$ are denoted by 
$\tilde {\cal L}$ and $\tilde {\cal P}$. 

The first term in Eq.~(\ref{B.6}) is the so called ``adiabatic'' term, and it
describes the time variation of the dynamic variable $X$ which follows
adiabatically the changes in the slow variables $\{ S(t) \}$. The second term 
represents the damping of this adiabatic motion. The last term is the
stochastic force acting upon the dynamical variable $X$. Note, however, that
this interpretation depends on the appropriate choice of the slow variables 
and the sets of functions $\Phi_\alpha ( \{ S(t) \} )$ and 
$\Psi_\alpha ( \{ S(t) \} )$. Upon choosing (we assume that the statics is diagonal
in the variables $S^\alpha$) 
\begin{mathletters}
\begin{eqnarray}
   \Phi_\alpha ( \{ S(t) \} ) = &&\chi_\alpha^{-1/2} S_\alpha \, ,
   \label{B.8a} \\
   \Psi_\alpha ( \{ S(t) \} ) = &&\chi_\alpha^{-1/2} S_\alpha \, ,
   \label{B.8b}
\end{eqnarray}
\end{mathletters}
with $(S_\alpha,S_\beta)=\delta_{\alpha \beta} \chi_\alpha$ one obtains
Mori's generalized Langevin equations~\cite{m65}
\begin{eqnarray}
  {d \over dt} S_\alpha (t) = 
  \sum_\beta i \omega_{\alpha \beta} S_\beta (t) -
  \sum_\beta \int_0^t d \tau 
  \left( f_\alpha (\tau), f_\beta \right) 
  S_\beta (t-\tau) + f_\alpha (t) \, ,
\label{B.9}
\end{eqnarray}
where
\begin{mathletters}
\begin{eqnarray}
 i \omega_{\alpha \beta} = && (i {\cal L} S_\alpha,S_\beta) \chi_\alpha^{-1} \, ,
 \label{B.10a} \\
 f_\alpha (t) = && e^{it {\cal Q} {\cal L}} {\cal Q} i {\cal L} S_\alpha \, .
 \label{B.10b}
\end{eqnarray}
\end{mathletters}
The latter equations are the basis for the mode coupling theory discussed in
sections II and III. However, with the above choice of a linear projection 
operator the ``random'' forces are orthogonal with respect to the slow
variables $S_\alpha$ only. It has been shown in Refs.~\cite{kk70,kk76,z72} that the
``random'' forces may not be really random since they contain products of
the slow variables $S_\alpha$. Another choice for the projection operator is
to include all the suitable symmetrized polynomials of $\{ S \}$ among the
sets of functions $\{ \Psi \}$ and $\{ \Phi \}$. Now we restrict ourselves to the case 
of classical mechanics, where one has $\{ \Psi \} = \{ \Phi \}$. This choice
corresponds to the projection operator introduced by Zwanzig~\cite{z60,z61}.
The completeness relation for the Zwanzig projection operator reads
\begin{equation}
  \sum_\alpha \Phi_\alpha ( \{ S \} ) 
              \Phi_\alpha^\star ( \{ S^\prime \} ) =
  {\delta ( S - S^\prime ) \over P_{eq} ( \{ S \} ) }
\label{B.11} 
\end{equation}
with the equilibrium probability distribution function
\begin{equation}
  P_{eq} ( \{ S \} ) = \exp \left[- {1 \over k_B T} {\cal H} ( \{ S \} ) \right] \, .
\label{B.12} 
\end{equation}
The resulting generalized Langevin equation takes the form
\begin{eqnarray}
  {d \over dt} S_\alpha (t) = 
  &&v_\alpha (  \{ S(t) \} ) \nonumber \\ +
  &&\sum_\beta \int_o^t d \tau 
  P_{eq}^{-1} ( \{ S(t-\tau) \} )
  { \partial M_{\alpha \beta} (\tau; \{ S(t-\tau) \} ) 
      P_{\rm eq} ( \{ S(t-\tau) \} ) \over
     \partial S_\beta^\star (t - \tau)} +
   \zeta_\alpha (t) \, ,
\label{B.13}
\end{eqnarray}
with the so called mode coupling term
\begin{mathletters}
\begin{equation}
  v_\alpha (  \{ S(t) \} ) = < i {\cal L} S_\alpha; \{ a \} >  \, ,
\label{B.14a} 
\end{equation}
and the memory kernel
\begin{equation}
  M_{\alpha \beta} (t; \{ a \} ) = <f_\alpha (t) f_\beta^\star (0); \{ a \} >  \, ,
\label{B.14b} 
\end{equation}
\end{mathletters}
where we have introduced the notation 
$<X; \{ a \} > = <X \delta(S-a)> / P_{eq} ( \{ a \} )$ for
the conditional average.

The semi-phenomenological equations of motion are obtained from the
above exact equations by making three basic, plausible assumptions. (i)
First one makes a Markovian approximation for the kinetic coefficients. This
is justified by the fact that one has included all suitable symmetrized
polynomials of the slow variables in the projection operator.
\begin{mathletters}
\begin{equation}
  M_{\alpha \beta} (t; \{ a \} ) \approx 
  2 L_{\alpha \beta} ( \{ a \} ) \delta(t) \, .
\label{B.15a} 
\end{equation}
(ii) Second one assumes that all kinetic coefficients are independent of the 
slow variables
\begin{equation}
  L_{\alpha \beta} ( \{ a \} ) \approx  L_{\alpha \beta} \, .
\label{B.15b} 
\end{equation}
(iii) Third one takes the random forces $\zeta$ as Gaussian white noise
\begin{equation}
  w( \{ \zeta \} | t_0 \leq t \leq t_1 ) \sim
  \exp \left[ -{1 \over 4} \int_{t_0}^{t_1} dt \zeta_\alpha (t)
  L_{\alpha \beta}^{-1} \zeta_\beta (t) \right]
\label{B.15c} 
\end{equation}
\end{mathletters}
Then the generalized Langevin equations reduce to
\begin{equation}
  {d \over dt} S_\alpha (t) = v_\alpha (  \{ S(t) \} ) \\ -
  \sum_\beta L_{\alpha \beta}^0 
  {\delta {\cal H} ( \{ S(t) \} ) \over
  \delta S^\star_\beta (t) }
  +\zeta_\alpha (t) \, ,
\label{B.16}
\end{equation}
where
\begin{mathletters}
\begin{equation}
 {\cal H}( \{ S \} )  = - k_B T \ln \left( P_{\rm eq} ( \{ S \} ) \right) \, ,
\label{B.17a} 
\end{equation}
\begin{equation}
  v_\alpha (  \{ S(t) \} )  = 
  - \lambda f \sum_\beta 
  \left[
  {\delta \over \delta S_\beta} Q_{\alpha \beta} ( \{ S \} ) -
  Q_{\alpha \beta} ( \{ S \} ) {\delta H ( \{ S \} ) \over
                                \delta S^\star_\beta (t) }
  \right] \, .
\label{B.17b} 
\end{equation}
\end{mathletters}
The Poisson brackets are defined by
\begin{equation}
  Q_{\alpha \beta} ( \{ S \} ) = \{ S_\alpha, S_\beta \}_P = 
  - Q_{\beta \alpha} ( \{ S \} ) \, .
\label{B.18} 
\end{equation}
We close this appendix by presenting the generalized Langevin equation for 
the example of an isotropic ferromagnet
\begin{equation}
  {\cal H} = {1 \over 2} \int d^d x [ r S^2 ({\bf x},t) + 
  (\bfnabla S({\bf x},t))^2 ] + {u \over 4!} \int d^d x 
  (S^2 ({\bf x},t))^2 \, .
\label{B.19} 
\end{equation}
The Poisson brackets are given by
\begin{equation}
  Q_{\alpha \beta} ({\bf k}, {\bf k}^\prime) = \epsilon_{\alpha \beta
  \gamma} S^\gamma ({\bf k} + {\bf k}^\prime) \, .
\label{B.20} 
\end{equation}
Since the order parameter is conserved the kinetic coefficient is
\begin{equation}
  L({\bf k}) = \lambda k^2 \, ,
\label{B.21} 
\end{equation}
(in general $L({\bf k}) = \lambda k^a$). Hence one obtains the following
generalized Langevin equation for isotropic ferromagnets
\begin{equation}{d \over dt} {\bf S} ({\bf x},t) =
  \lambda f {\bf S} \times 
  {\delta {\cal H} \over \delta {\bf S} ({\bf x},t)} -
  \lambda (i \bfnabla)^a  {\delta {\cal H} \over \delta {\bf S} ({\bf x},t)}
  + {\gvect \zeta} ({\bf x},t) \, .
\label{B.22} 
\end{equation}
Note that the conventional (van Hove) theory of critical dynamics is obtained 
by making the following additional assumptions:
(i) The Onsager coefficient $L$ remains finite at the
critical point. (ii) The mode coupling term is ignored: $f=0$. (iii) A 
quadratic approximation is made for the free energy
functional ${\cal H} = {1 \over 2} \int_k \chi^{-1} ({\bf k}) 
{\bf S}({\bf k},t) {\bf S}(-{\bf k},t)$, where $\chi ({\bf k})$ is the
static susceptibility.

One should also note that in mode coupling theory solely the latter
approximation is made (and further additional approximations like
two-mode approximation and neglecting the vertex corrections).

\section{Validity of mode coupling theory, higher orders in perturbation 
theory}
\label{app.c}

In this section we analyze how higher orders in perturbation theory can be 
incorporated in the self consistent approach. For the sake of
simplicity and clarity we restrict ourselves to the case of
isotropic ferromagnets with exchange interaction only. The extension
to dipolar ferromagnets is straightforward.

First we consider the corrections from two-loop contributions to the vertex 
functions  $\Gamma_{02}$ and $\Gamma_{11}$ shown in Figs.~A.1 and A.2. 
Due to the tensorial structure of the relaxation and mode 
coupling vertices they can not be contracted ($F^{\alpha \beta
\gamma \delta} \epsilon_{\alpha^\prime \gamma \delta} = 0$).

The 2-loop diagrams of $\Gamma_{11}$ and $\Gamma_{02}$ can be
divided into two categories.
The first category consists of ``true'' 2-loop diagrams connecting two
relaxation vertices. The second type consists of diagrams whose
structure is - up to vertex corrections - identical to the one-loop 
diagrams. Upon defining a renormalized mode coupling vertex, as
shown in Fig.~A.3, the 2-loop contributions to the vertex functions 
$\Gamma_{11}$ and $\Gamma_{02}$ are given by the diagrams in 
Figs.~A.4 and A.5.

The modifications of the self consistent equations resulting from
these types of two-loop diagrams are now discussed separately.

(I) The vertex corrections in Fig.~A.3 correspond to the vertex
function $\Gamma_{12}(q,p,\nu=0,\omega)$. Since all diagrams
have a bare mode coupling vertex with an external response line (see
e.g. \cite{dh74}), the one-loop contribution to 
$\Gamma_{12}(q,p,\nu=0,\omega)$ is
proportional to ${\bf p} \cdot {\bf q}$, i.e., it has the same wave
vector dependence as the original bare mode coupling vertex.
The remaining frequency integral is a function  $f(\omega)$ of
the external frequency $\omega$. Because only the long time behavior
is of importance for the critical dynamics the limit $\omega
\rightarrow 0$ merely leads to a renormalization of the amplitude,$\lambda f$,
of the mode coupling vertex and not to a change in its wave vector
dependence. (But, see the discussion of higher order vertex
corrections later)

(II) The two-loop diagram connecting two relaxation vertices 
$\Gamma_{02}^{RR}$ is proportional to 
\begin{equation}
  \Gamma_{02}^{RR} \sim 
 \lambda^2 q^4 \int_p \int_k \int_{\omega}
  C({\bf p},\omega) C({\bf k},\omega) C({\bf q}-{\bf p}-{\bf k},\omega) \, .
\label{C.1}
\end{equation}
Upon introducing the dimensionless wave vector variables
${\hat {\bf p}} = {\bf p} / q$ and ${\hat {\bf k}} = {\bf k} / q$, 
carrying out the frequency integrals, and replacing  the
correlation propagators according to Eqs. (\ref{6.55}) one finds
\begin{equation}
  \Gamma_{02}^{RR} \sim
  q^{4-z} \, .
\label{C.2}
\end{equation}
In comparing this result with the wave vector dependence, $q^{3-z}$, 
of the one-loop diagram, connecting two mode coupling vertices, one
recognizes that the contribution of $\Gamma_{02}^{RR}$ can be
neglected in the long wave length limit. Hence this two-loop
diagram does not change the self-consistent equations as well.

The above discussion shows that in a self consistent theory 
the two-loop diagrams are small compared to the one-loop diagrams in
the long time and long wave length limit.

Now we discuss higher order vertex corrections, where an internal line of a 
one-loop vertex correction is decorated with a static insertion.
It can be shown that the renormalization of the mode coupling vertex is of 
purely static origin~\cite{Frey_PhD}. This is due to an exact relation
between the renormalization factor of the mode coupling vertex and
the time scale and field renormalization. To lowest order the mode
coupling vertex is given by
\begin{equation}
  \Gamma_{12}^{(0)} ({\bf p}, {\bf q}) = \upsilon_3({\bf p}, {\bf q}) =
  \lambda_0 f_0 \epsilon_{\alpha \beta \gamma} {\bf p} \cdot {\bf q} 
  \, .
\label{C.3}
\end{equation}
This expression can also be written in terms of the bare static
susceptibilities as
\begin{equation}
  \upsilon_3({\bf p}, {\bf q}) =
  {\lambda_0 f_0 \over 2} \epsilon_{\alpha \beta \gamma} 
  \left[ \chi_B^{-1} ({\bf p} + {\bf q} /2) -
         \chi_B^{-1} ({\bf p} - {\bf q} /2) \right] 
\label{C.4}
\end{equation}
showing that the wave vector dependence of the mode coupling 
vertex is determined by the static susceptibilities. In order to take into
account the static IR divergences correctly, one way to proceed is to replace
the bare static susceptibilities by the fully renormalized static
susceptibilities
\begin{equation}
   \lambda_0 f_0 {\bf p} \cdot  {\bf q} \rightarrow
   {\lambda f \over 2}
  \left[ \chi^{-1} ({\bf p} + {\bf q} /2) -
         \chi^{-1} ({\bf p} - {\bf q} /2) \right] 
\label{C.5}
\end{equation}
in the mode coupling vertex. This replacement is suggested by the
exact relation between the renormalization factors. It is also equivalent to
replacing the nonlinear Landau-Ginzburg functional by
\begin{equation}
  {\cal H}_{eff} = \int_k \chi^{-1} ({\bf k}) {\bf S} ({\bf k})
  \cdot {\bf S} (-{\bf k})
\label{C.6}
\end{equation}
where $\chi^{-1} ({\bf k})$ is the fully renormalized static
susceptibility. This approximation is often used in mode coupling
theories without any justification. Here it is a consequence of the exact 
relation between the renormalization factors.

Besides these static vertex renormalizations (due to the non linearity
in the Landau-Ginzburg functional) there could also be dynamic
vertex renormalizations. But this is not the case here as we will
show next. This can be seen from the diagrammatic representation of
the dynamic vertex renormalization (see Fig.~A.3.
Any diagram contributing to the vertex corrections starts with a
bare mode coupling vertex. Therefore to leading
order in the wave number one can replace the remaining wave
vectors (in the integrals of the corresponding diagrams) by
their values at ${\bf p} = {\bf q} = 0$. This leads then only to
changes of the amplitude. 
Hence we have
\begin{equation}
  \upsilon_3 ({\bf p}, {\bf q}) =
  {\lambda { f} \over 2}
  \left[ \chi^{-1} ({\bf p} + {\bf q} /2) -
         \chi^{-1} ({\bf p} - {\bf q} /2) \right]
  + O(\omega,p) \, .
\label{C.7}
\end{equation}
This gives a precise specification of the approximations involved by
neglecting the vertex corrections.

As we have seen above, the effects of the four point coupling can be
taken into account by a static renormalization of the mode coupling
vertex, Eq.~(\ref{C.7}).
Then one is left with a dynamic theory with a harmonic (effective)
Landau-Ginzburg functional and a mode coupling vertex. In this case 
the fluctuation-dissipation relation (\ref{A.12}) applies (see Appendix A), 
which in the case
of a purely quadratic Landau-Ginzburg functional reduces to
\begin{equation}
  G_{11} ({\bf k},t) = \Theta(t) \chi^{-1} ({\bf k}) G_{02}({\bf k},t)
\label{C.8}
\end{equation}
This implies for the Fourier transform
$G_{11} (\omega) + G_{11} (- \omega) = \chi^{-1} G_{02} (\omega)$
and consequently
\begin{equation}2 {\it Re} [ G_{11} (\omega)] = \chi^{-1} G_{02} (\omega)
\label{C.9}
\end{equation}
By using the generalized FDT's in Appendix A we find
\begin{equation}\Phi({\bf q}, \omega) = 
  { 1 \over i \omega}
  [ \chi({\bf q}, \omega) - \chi({\bf q}) ] =
  {\chi({\bf q}) \over - i \omega + [ \lambda q^2 + \lambda f 
  \Gamma_{X,10} ({\bf q}, \omega)  ] / \chi({\bf q})} \, .
\label{C.10}
\end{equation}
This is exactly the same structure as the non linear Langevin
equations (with the linear projection operator) (see section III)
with the memory kernel
\begin{equation}M({\bf q}, \omega) = {\lambda q^2 + \lambda f 
  \Gamma_{X,10} ({\bf q}, \omega) \over  \chi({\bf q}) }
\label{C.11}
\end{equation}
Neglecting the mode coupling contribution ($f=0$) leads immediately
to the conventional theory (van Hove theory) of critical dynamics.
The mode coupling contribution to the memory function is given by
\begin{equation}\Gamma_{X,10} ({\bf q},t) = 
  \lambda f \int_k \left[ {1 \over \chi({\bf q} + {\bf k} /2)} -
                          {1 \over \chi({\bf q} - {\bf k} /2)} \right]
   {1 \over \chi({\bf q} + {\bf k} /2)}
   \Phi({\bf q} + {\bf k} /2,t) \Phi({\bf q} - {\bf k} /2,t) \, .
\label{C.12}
\end{equation}
In the strong coupling limit, where the van Hove term $\lambda q^2$
(in general there will be a crossover from van Hove to strong
coupling) can be neglected, a scaling analysis of the above self consistent
equations shows that the dynamic exponent is given by
\begin{equation}
  z = {1 \over 2} ( d + 2 -\eta)
\label{C.13}
\end{equation}
i.e., with this generalized mode coupling theory we find the correct
dynamic exponent. This resolves a long standing problem with the
conventional derivation of the mode coupling theory, which gives a
dynamic exponent with the wrong sign of $\eta$. The correct
expression, Eq.~(\ref{C.13}), is a consequence of the fact that we have
correctly taken into account the vertex renormalizations of the mode
coupling vertex by static decorations of internal lines.

The above analysis is not a rigorous derivation of a self consistent
theory. It shows, however, that the neglected terms are small in
the long time and long wave length limit. Furthermore, taking into
account static corrections of the mode coupling vertex - suggested by
an exact relation between the renormalization factors of the mode
coupling vertex and the static field renormalization - we could show
that the self consistent theory (generalized mode coupling theory)
is capable of giving the exact result for the dynamic critical
exponent with the correct sign of $\eta$.

\newpage

\centerline{\bf Captions to the figures:}
\bigskip
\bigskip

\noindent {\bf Figure A.1: }
One--loop and two--loop diagrams  contributing to the vertex function 
$\Gamma_{11}$. \par
\bigskip

\noindent {\bf Figure A.2: }
One--loop and two--loop diagrams  contributing to the vertex function 
$\Gamma_{02}$ (the self energy).\par
\bigskip

\noindent {\bf Figure A.3: }
Diagrams contributing to the one--loop vertex correction.\par
\bigskip

\noindent {\bf Figure A.4: }
One--loop and two--loop diagrams  contributing to the vertex function 
$\Gamma_{11}$, drawn in terms of the renormalized vertex.\par
\bigskip

\noindent {\bf Figure A.5: }
One--loop and two--loop diagrams  contributing to the vertex function 
$\Gamma_{02}$, drawn in terms of the renormalized vertex.\par
\bigskip

\newpage
\vfill
\begin{figure}[h]
\centerline{\rotate[r]{\epsfysize=5in \epsffile{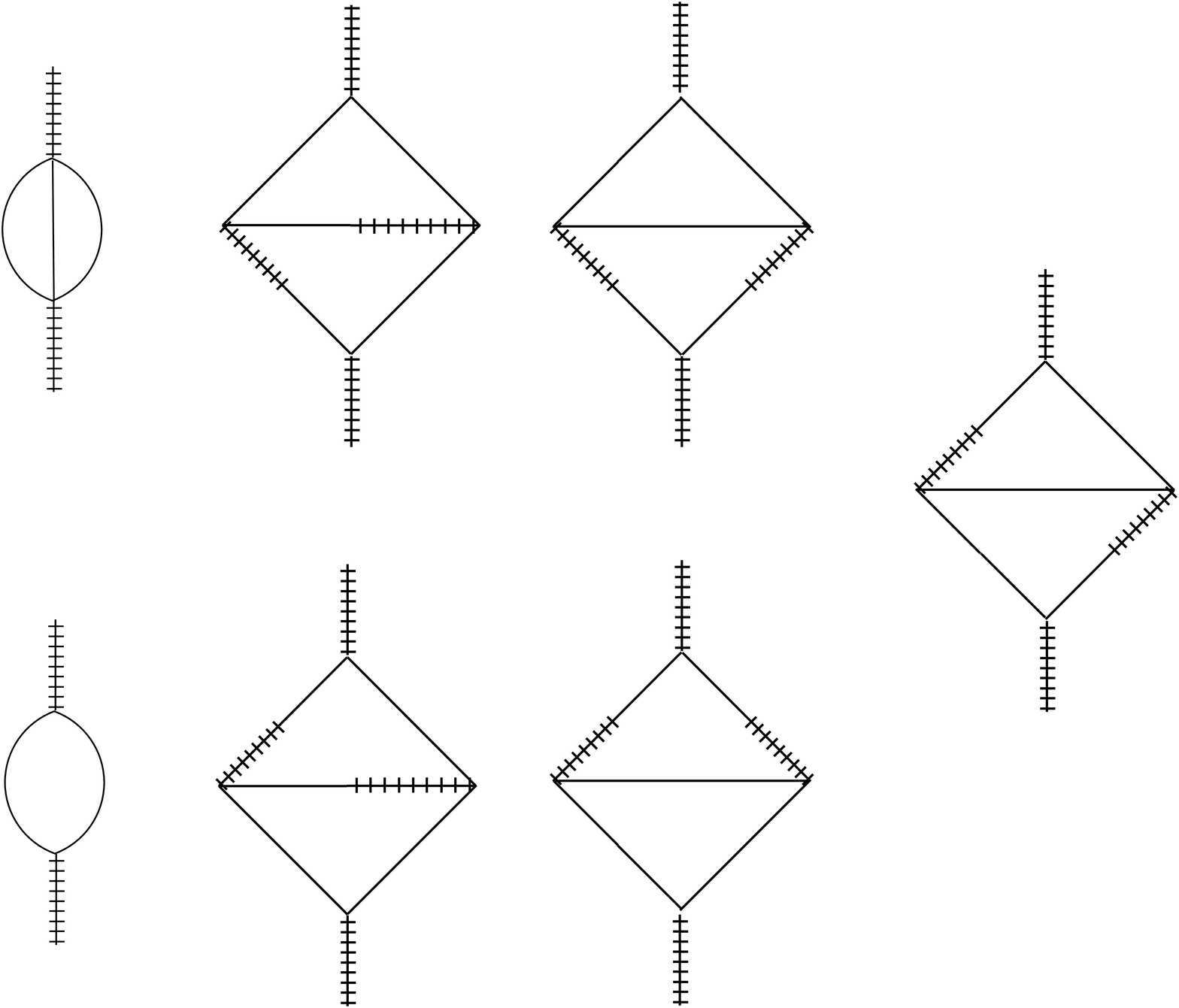}}}
\bigskip \bigskip
\noindent {\bf Figure A.1: }
One--loop and two--loop diagrams  contributing to the vertex function 
$\Gamma_{11}$.
\label{figa1}
\end{figure}
\newpage
\vfill
\begin{figure}[h]
\centerline{\rotate[r]{\epsfysize=5in \epsffile{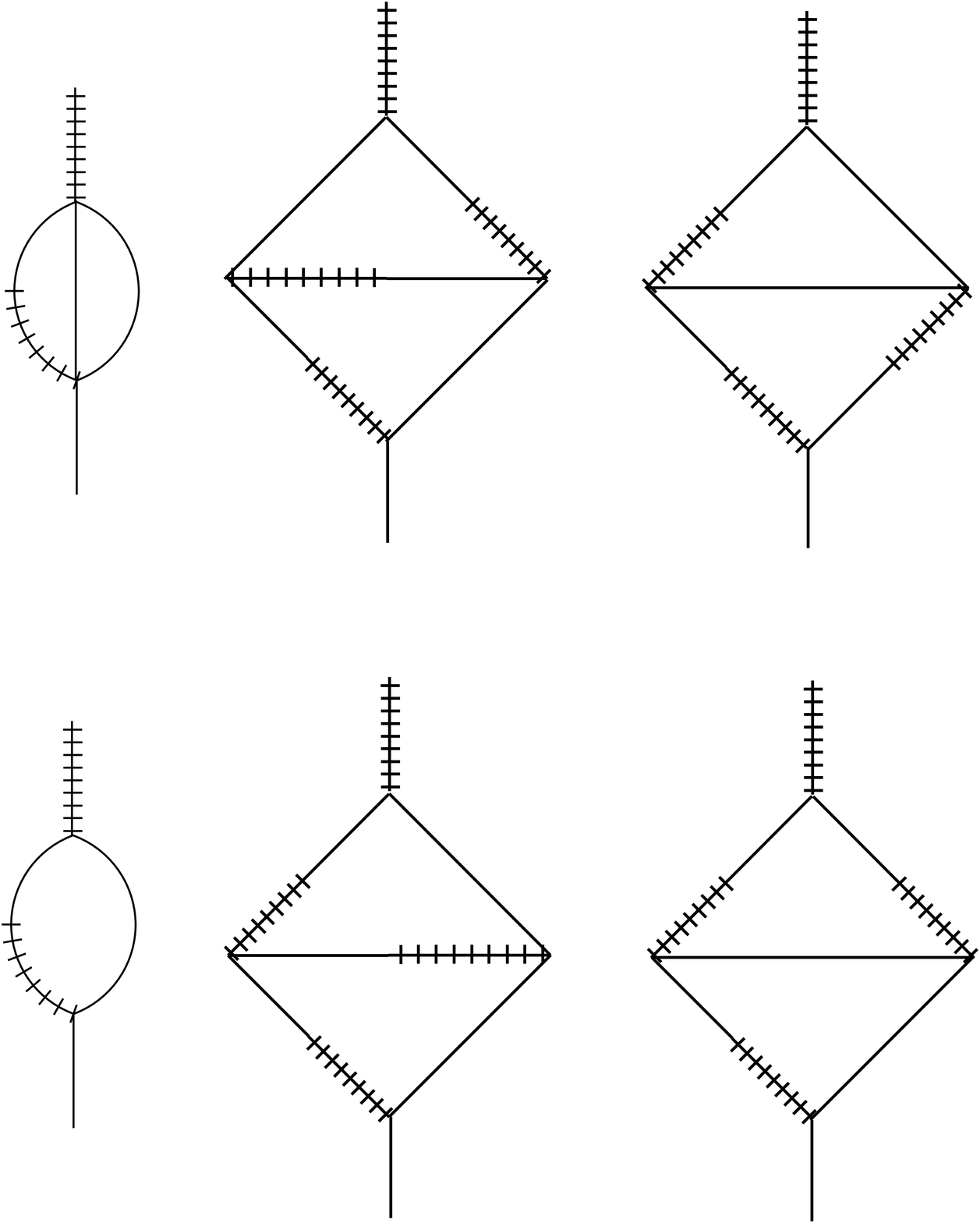}}}
\bigskip \bigskip
\noindent {\bf Figure A.2: }
One--loop and two--loop diagrams  contributing to the vertex function 
$\Gamma_{02}$ (the self energy).
\label{figa2}
\end{figure}
\newpage
\vfill
\begin{figure}[h]
\centerline{\rotate[r]{\epsfysize=5in \epsffile{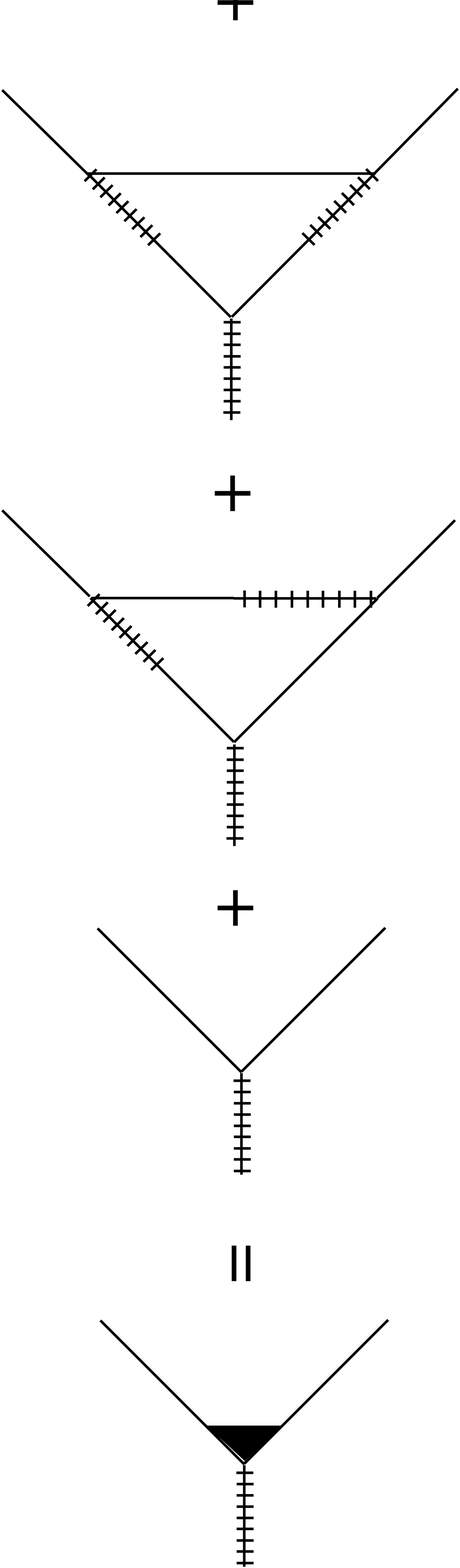}}}
\bigskip \bigskip
\noindent {\bf Figure A.3: }
Diagrams contributing to the one--loop vertex correction.
\label{figa3}
\end{figure}
\newpage
\vfill
\begin{figure}[h]
\centerline{\rotate[r]{\epsfysize=5in \epsffile{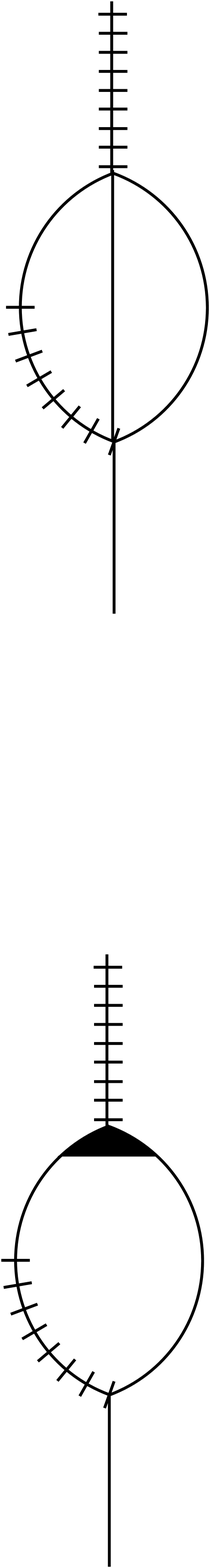}}}
\bigskip \bigskip
\noindent {\bf Figure A.4: }
One--loop and two--loop diagrams  contributing to the vertex function 
$\Gamma_{11}$, drawn in terms of the renormalized vertex.
\label{figa4}
\end{figure}
\newpage
\vfill
\begin{figure}[h]
\centerline{\rotate[r]{\epsfysize=5in \epsffile{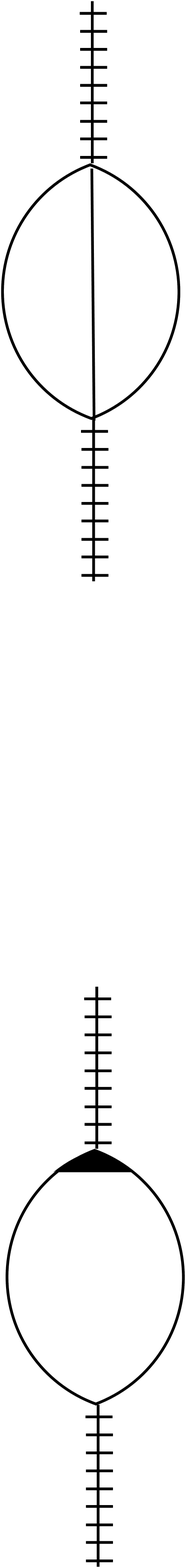}}}
\bigskip \bigskip
\noindent {\bf Figure A.5: }
One--loop and two--loop diagrams  contributing to the vertex function 
$\Gamma_{02}$, drawn in terms of the renormalized vertex.
\label{figa5}
\end{figure}

\end{document}